\documentclass{article}
\usepackage{epsf}
\usepackage{amsmath,amssymb,amsfonts}
\usepackage{longtable}
\usepackage{epubtk}
\usepackage{graphicx}

\newcommand{\mat}{{\rm{mat}}}
\newcommand{\baryon}{{\rm{b}}}

\newcommand{\etal}{\emph{et al.}}
\newcommand{\dd}{{\rm d}}
\newcommand{\ppn}{{\rm PPN}}
\newcommand{\bx}{{\bf x}}
\newcommand{\br}{{\bf r}}
\newcommand{\bk}{{\bf k}}
\newcommand{\bn}{{\bf n}}
\newcommand{\aem}{\alpha_{_{\rm EM}}}
\newcommand{\aw}{\alpha_{_{\rm W}}}
\newcommand{\ag}{\alpha_{_{\rm G}}}
\newcommand{\as}{\alpha_{_{\rm S}}}
\newcommand{\gfermi}{G_{_{\rm F}}}
\def\ga{\mathrel{\raise.3ex\hbox{$>$\kern-.75em\lower1ex\hbox{$\sim$}}}}
\def\la{\mathrel{\raise.3ex\hbox{$<$\kern-.75em\lower1ex\hbox{$\sim$}}}}
\newcommand{\dslash}{D\!\!\!\!/}

\begin{document}

\title{Varying constants, Gravitation and Cosmology}

\author{%
\epubtkAuthorData{Jean-Philippe Uzan\\}{%
Institut d'Astrophysique de Paris, \\
UMR-7095 du CNRS, Universit\'e Pierre et Marie Curie,\\
              98 bis bd Arago, 75014 Paris (France)\\
              and\\
Department of Mathematics and Applied Mathematics,\\
Cape Town University,\\
Rondebosch 7701 (South Africa)\\
	      and\\
National Institute for Theoretical Physics (NITheP),\\
Stellenbosch 7600 (South Africa).\\
              }{%
uzan@iap.fr\\}{%
http//www2.iap.fr/users/uzan/}%
}
\date{\today}
\maketitle

\begin{abstract}
Fundamental constants are a cornerstone of our physical laws. Any constant
varying in space and/or time would reflect the existence of an almost massless
field that couples to matter. This will induce a violation of the universality of free
fall. It is thus of utmost importance for our understanding of gravity and of the
domain of validity of general relativity to test for their constancy. We thus detail the relations between
the constants, the tests of the local position invariance and of the
universality of free fall. 
We then review the main experimental and observational constraints
that have been obtained from atomic clocks, the Oklo phenomenon,
Solar system observations,
meteorites dating, quasar absorption spectra, stellar physics, pulsar timing, the
cosmic microwave background and big bang nucleosynthesis. 
At each step we describe the basics of each system, its dependence with respect to the
constants, the known systematic effects and the most recent constraints that have been
obtained.
We then describe
the main theoretical frameworks in which the low-energy constants may actually be varying
and we focus on the unification mechanisms and the relations between
the variation of different constants.
To finish, we discuss the more speculative possibility of understanding their numerical
values and the apparent fine-tuning that they confront us with.
\end{abstract}

\epubtkKeywords{general theory of gravitation, fundamental constants, theoretical cosmology}
\newpage
\tableofcontents
\newpage

\section{Introduction}\label{section:intro}

Fundamental constants appear everywhere in the mathematical laws
we use to describe the phenomena of Nature. They
seem to contain some truth about the
properties of the physical world while their real nature seem to
evade us.

The question of the constancy of the constants of physics was probably
first addressed by Dirac~\cite{dirac37,dirac38} who expressed, in his
``Large Numbers hypothesis'', the opinion that very large (or small)
dimensionless universal constants cannot be pure mathematical numbers
and must not occur in the basic laws of physics. He suggested, on the
basis of this numerological principle, that these large numbers should
rather be considered as variable parameters characterizing the state
of the universe.  Dirac formed five dimensionless ratios 
among which $\delta\equiv H_0\hbar/m_{\rm p}c^2\sim
2h\times10^{-42}$ and $\epsilon\equiv G\rho_0/H_0^2\sim 5h^{-2}\times
10^{-4}$ and asked the question of which of these ratio is constant as
the universe evolves. Usually, $\delta$ and $\epsilon$ vary as
the inverse of the cosmic time. Dirac then noticed that $\ag/\mu\aem$,
representing the relative magnitude of electrostatic and gravitational
forces between a proton and an electron, was of the same order as
$H_0e^2/m_{\rm e}c^2=\delta\aem\mu$ representing the age of the
universe in atomic units so that his five numbers can be
``harmonized'' if one assumes that $\ag$ and $\delta$ vary with time
and scale as the inverse of the cosmic time.

This argument by Dirac is indeed not a physical theory but it
opened many doors in the investigation on physical constants, both
on questioning whether they are actually constant and on trying
to understand the numerical values we measure.

First, the implementation of Dirac's phenomenological idea into a
field-theory framework was proposed by
Jordan~\cite{jordan37} who realized that the constants have to
become dynamical fields and proposed
a theory where both the gravitational and
fine-structure constants can vary (Ref.~\cite{unzicker}
provides some summary of some earlier attempts to
quantify the cosmological implications of Dirac argument).  Fierz~\cite{fierz56} then realized that
in such a case, atomic spectra will be spacetime-dependent
so that these theories can be observationally tested. Restricting
to the sub-case in which only $G$ can vary led to definition of
the class of scalar-tensor theories which were further
explored by Brans and Dicke~\cite{brans61}. This kind of theory was further
generalized to obtain various functional dependencies for $G$ in the
formalization of scalar-tensor theories of gravitation (see
e.g. Ref.~\cite{damour92}).

Second, Dicke~\cite{dicke61} pointed out that in fact the density of the universe is
determined by its age, this age being related to the time needed to
form galaxies, stars, heavy nuclei... This led him to formulate that
the presence of an observer in the universe places constraints on the
physical laws that can be observed. In fact, what is meant by observer
is the existence of (highly?) organized systems and this
principle can be seen as a rephrasing of the question ``why is the
universe the way it is?'' (see Ref.~\cite{hogan00}). Carter~\cite{carter74,carter83},
who actually coined the term ``anthropic principle'' for it, showed that the
numerological coincidences found by Dirac can be derived from physical
models of stars and the competition between the weakness of gravity
with respect to nuclear fusion. Carr and Rees~\cite{carr79} then showed how
one can scale up from atomic to cosmological scales only by using
combinations of $\aem$, $\ag$ and $m_{\rm e}/m_{\rm p}$.

To summarize, Dirac's insight was to question whether some numerical
coincidences between very large numbers, that cannot be themselves
explained by the theory in which they appear, was a mere coincidence or whether it can reveal
the existence of some new physical laws. This gives three main roads
of investigations
\begin{itemize}
 \item how do we construct theories in which what were
thought to be constants are in fact dynamical fields,
 \item how can we constrain, experimentally or
 observationally, the spacetime dependencies of the
 constants that appear in our physical laws
 \item how can we explain the values of the fundamental
 constants and the fine-tuning that seems to exist
 between their numerical values.
\end{itemize}
While ``Varying constants'' may seem, at first glance, to be an oxymoron, it 
has to be considered merely as jargon to be
understood as ``revealing new degrees of freedom,
and their coupling to the known fields of our theory''. The tests on the constancy
of the fundamental constants are indeed very important tests of fundamental
physics and of the laws of Nature we are currently using. Detecting any such
variation will indicate the need for new physical degrees of freedom in our theories,
that is new physics.

The necessity of theoretical physics on deriving bounds on their variation
is, at least, threefold:
\begin{enumerate}
 \item it is necessary to understand and to model the physical
 systems used to set the constraints. In particular one needs to
 determine the effective parameters that can be
 observationally constrained to a set of fundamental constants;
 \item it is necessary to relate and compare different constraints
 that are obtained at different spacetime positions. This often
 requires a spacetime dynamics and thus to specify a model
 as well as a cosmology;
 \item it is necessary to relate the variations of different
 fundamental constants.
\end{enumerate}
We shall thus start in \S~\ref{section1} by recalling the link between the constants
of physics and the theories in which they appear, as well as with metrology.
From a theoretical point of view, the constancy of the fundamental constants
is deeply linked with the equivalence principle and general relativity. In
\S~\ref{section1} we will recall this relation and in particular the link with the
universality of free fall. We will then summarize the
various constraints that exist on such variations, mainly for the fine
structure constant and for the gravitational constant in \S~\ref{section3}
and~\ref{section4} respectively. We will then turn to the
theoretical implications in \S~\ref{section6} in describing some
of the arguments backing up the fact that constants are expected to vary,
the main frameworks used in the literature and the various ways
proposed to explain why they have the values we observe today. We shall
finish by a discussion on their spatial variations in \S~\ref{subsec23} and by
discussing the possibility to understand their numerical values in \S~\ref{section5}.

Various reviews have been written on this topic. We will refer to
the review~\cite{jpu-revue} as FVC and we mention the following
later reviews~\cite{barrow05,flamrev,bronni,damour-issi,gbisern,karshen-rev,karshen-metro,olive-rio,jpu-rio,jpu-issi}
and we refer to Ref.~\cite{codatalist} for the numerical values of the constants
adopted in this review.

\section{Constants and fundamental physics}\label{section1}

\subsection{About constants}\label{subsec11}

Our physical theories introduce various structures to describe the phenomena of Nature.
They involve various fields, symmetries and constants. These structures are postulated in order
to construct a mathematicaly consistent description of the known physical phenomena
in the most unified and simple way. 

We define the fundamental constants of a physical theory as
{\it any parameter that cannot be explained by this theory}. Indeed,
we are often dealing with other constants that in principle can be expressed
in terms of these fundamental constants. 
The existence of these two sets of constants is important and arises from
two different considerations. From a theoretical point of view we would like
to extract the minimal set of fundamental constants, but often these constants
are not measurable. From a more practical point of view, we need to
measure constants, or combinations of constants which allow to 
reach the highest accuracy.

These fundamental constants are thus contingent quantities that can only be measured.
Such parameters have to be assumed constant in this theoretical framework for two reasons:
\begin{itemize}
 \item from a theoretical point of view: the considered framework does not provide any
 way to compute these parameters, i.e. it does not have any equation of evolution for them since otherwise it
 would be considered as a dynamical field,
 \item from an experimental point of view: these parameters can only be measured. If the theories
 in which they appear have been validated experimentally, it means that, at the precisions
 of these experiments, these parameters have indeed been checked to be constant,
 as required by the necessity of the reproductibility of experimental results.
\end{itemize}
This means that testing for the constancy of these parameters is a test of the theories
in which they appear and allow to extend our knowledge of their domain of validity.
This also explains the definition chosen by Weinberg~\cite{weinberg83} who stated
that they cannot be calculated in terms of other constants
``... not just because the calculation is too complicated (as for the viscosity of
water) but because we do not know of anything more fundamental''.

This has a series of implications. First, the list of fundamental constants 
to consider depends on our theories of physics and thus on time. Indeed,
when introducing new, more unified or more fundamental, theories the number
of constants may change so that this list reflects both our knowledge of physics and, more
important, our ignorance.
Second, it also implies that some of these fundamental constants can become dynamical quantities
in a more general theoretical framework so that the tests of the constancy
of the fundamental constants are tests of fundamental physics which can reveal
that what was thought to be a fundamental constant is actually a field
whose dynamics cannot be neglected. If such fondamental constants are
actually dynamical fields it also means that the equations we are using are only approximations
of other and more fundamental equations, in an adiabatic limit, and that an
equation for the evolution of this new field has to be obtained.

The reflections on the nature of the constants and their role in physics are numerous.
We refer to the books~\cite{barrow-book,fritzsch-book,ul-book,uzanleclercqbook} 
as well as Refs.~\cite{duff01,hfcte,okun96,volovik09,weinberg83,wilczek07}
for various discussions on this issue that we cannot develop at length here. This
paragraph summarizes some of the properties of the fundamental constants that
have attracted some attention.

\subsubsection{Characterizing the fundamental constants}

Physical constants seem to play a central role in our physical
theories since, in particular, they determined the magnitudes
of the physical processes. Let us sketch briefly some of their properties.\\

{\it How many fundamental constants should be considered?} 
The set of constants which are conventionally considered as fundamental~\cite{flower01}
consists of the electron charge $e$, the electron
mass $m_{\rm e}$, the proton mass $m_{\rm p}$, the reduced Planck constant
$\hbar$, the velocity of light in vacuum $c$, the Avogadro constant
$N_{_{\rm A}}$, the Boltzmann constant $k_{_{\rm B}}$, the Newton
constant $G$, the permeability and permittivity of space,
$\varepsilon_0$ and $\mu_0$. The latter has a fixed value in the SI
system of unit ($\mu_0=4\pi\times10^{-7}\,{\rm H}\,{\rm m}^{-1}$) which
is implicit in the definition of the Ampere; $\varepsilon_0$ is then
fixed by the relation $\varepsilon_0\mu_0=c^{-2}$.

It is however clear that this cannot corresponds to the list of the fundamental
constants, as defined earlier as the free parameters of the theoretical
framework at hand. To define such a list we must specify this framework.
Today, gravitation is described by general relativity, and the
three other interactions and the matter fields are described by the standard model
of particle physics.  It follows that one has to consider 22 unknown constants (i.e. 19
unknown dimensionless parameters):
the Newton constant $G$, 6 Yukawa couplings for the quarks ($h_{\rm u},h_{\rm d},h_{\rm c},h_{\rm s},h_{\rm t},h_{\rm b}$) 
and 3 for the leptons ($h_{\rm e},h_\mu,h_\tau$),
2 parameters of the Higgs field potential ($\mu,\lambda$), 4 parameters for
the Cabibbo-Kobayashi-Maskawa matrix (3 angles $\theta_{ij}$ and a phase $\delta_{_{\rm CKM}}$), 3 coupling constants 
for the gauge groups $SU(3)_c\times SU(2)_L\times U(1)_Y$ ($g_1,g_2,g_3$ or equivalently
$g_2,g_3$ and the Weinberg angle $\theta_{\rm W}$), and
a phase for the QCD vacuum ($\theta_{{\rm QCD}}$), to which one must add the speed of light $c$  and the Planck constant
$h$. See Table~\ref{tab-list} for a summary and their numerical values.

\begin{table}[t]
{\small
\begin{center}
\begin{tabular}{p{6cm}ll}
 \hline\hline
 Constant & Symbol & Value \\
 \hline
 Speed of light & $c$ & $299 792 458\, {\rm m\cdot s}^{-1}$ \\
 Planck constant (reduced) & $\hbar$ & $1.054 571 628(53)\times10^{-34}\, {\rm J\cdot s}$ \\
 Newton constant & $G$ & $6.674 28(67)\times10^{-11}\, {\rm m^2\cdot kg^{-1}\cdot s^{-2}}$\\
 \hline
 Weak coupling constant (at $m_Z$) & $g_2(m_Z)$ & $ 0.6520\pm0.0001$\\
 Strong coupling constant (at $m_Z$) & $g_3(m_Z)$ & $1.221\pm0.022$\\ 
 Weinberg angle & $ \sin^2\theta_{_{\rm W}}(91.2{\rm GEV})_{_{\bar{\rm MS}}}$ & $0.23120\pm0.00015$\\
 \hline
 Electron Yukawa coupling & $h_{\rm e}$ & $2.94\times10^{-6}$\\
 Muon Yukawa coupling & $h_\mu$ & $0.000607$\\
 Tauon Yukawa coupling & $h_\tau$ & $0.0102156$\\
 Up Yukawa coupling & $h_{\rm u}$ & $0.000016\pm0.000007$\\
 Down Yukawa coupling & $h_{\rm d}$ & $0.00003\pm0.00002$\\
 Charm Yukawa coupling & $h_{\rm c}$ & $0.0072\pm0.0006$\\
 Strange Yukawa coupling & $h_{\rm s}$ & $0.0006\pm0.0002$\\
 Top Yukawa coupling & $h_{\rm t}$ & $1.002\pm0.029$\\
 Bottom Yukawa coupling & $h_{\rm b}$ & $0.026\pm0.003$\\
 \hline
 Quark CKM matrix angle & $\sin\theta_{12}$ & $0.2243\pm0.0016$\\
                                          & $\sin\theta_{23}$ & $0.0413\pm0.0015$\\
                                           & $\sin\theta_{13}$ & $0.0037\pm0.0005$\\
 Quark CKM matrix phase & $\delta_{_{\rm CKM}}$ & $1.05\pm0.24$\\
 \hline
 Higss potential quadratic coefficient & $\mu^2$ & ? \\
 Higss potential quartic coefficient & $\lambda$ &  ?\\
 QCD vacuum phase & $\theta_{{\rm QCD}}$ & $<10^{-9}$\\
\hline 
\end{tabular}
\caption{\it List of the fundamental constants of our standard model. The numerical
values are given in the Planck system of units (see below) defined by the requirement
that the numerical value of $G$, $c$ and $\hbar$ is 1 in this system of units.}
\label{tab-list}
\end{center}}
\end{table}

Again, this list of fundamental constants relies on what we accept as
a fundamental theory. Today we have many hints that the standard
model of particle physics has to be extended, in particular
to include the existence of massive neutrinos. Such an extension
will introduce at least seven new constants (3 Yukawa couplings and 4 Maki-Nakagawa-Sakata (MNS)
parameters, similar to the CKM parameters). On the other hand, the number of constants
can decrease if some unifications between various interaction exist (see \S~\ref{subsecGUT} for more
details)
since the various coupling constants may be related to a unique
coupling constant $\alpha_U$ and an energy scale of unification
$M_U$ through
$$
 \alpha_i^{-1}(E) = \alpha_U^{-1} + \frac{b_i}{2\pi}\ln \frac{M_U}{E},
$$
where the $b_i$ are numbers which depend on the explicit model
of unification. Note that this would also imply that the variations,
if any, of various constants shall be correlated.\\

{\it Relation to other usual constants.} 
These parameters of the standard model are related to various constants that
will appear in this review (see Table~\ref{tab-list2}). First, the quartic and quadratic coefficients of the Higgs
field potential are related to the Higgs mass and vev, $m_H=\sqrt{-\mu^2/2}$ and
$v=\sqrt{-\mu^2/\lambda}$. The latter is related to the Fermi constant $\gfermi=(v^2\sqrt{2})^{-1}$
which imposes that $v=(246.7\pm0.2)$~GeV while the Higgs mass is badly constrained.
The masses of the quarks and leptons are related to their Yukawa coupling and the
Higgs vev by $m=hv/\sqrt{2}$. The values of the gauge couplings depend on energy
via the renormalisation group so that they are given at a chosen energy scale, here
the mass of the $Z$-boson, $m_Z$.  $g_1$ and $g_2$ are related by the Weinberg angle
as $g_1=g_2\tan \theta_{_{\rm W}}$. The electromagnetic coupling constant is not
$g_1$ since $SU(2)_L\times U(1)_Y$ is broken to $U(1)_{\rm elec}$ so that it is given by
\begin{equation}
 g_{_{\rm EM}}(m_Z)=e=g_2(m_Z) \sin\theta_{_{\rm W}}.
\end{equation} 
Defining the fine-structure constant as $\aem= g_{_{\rm EM}}^2/\hbar c$, the
(usual) zero energy electromagnetic fine structure constant is $\aem=1/137.03599911(46)$ is related 
to $\aem(m_Z)=1/(127.918\pm0.018)$ by the renormalisation group equations.
In particular, it implies that $\aem\sim\alpha(m_Z)+
\frac{2}{9\pi}\ln\left(\frac{m_Z^{20}}{m_{\rm u}^4m_{\rm c}^4m_{\rm d}m_{\rm s}m_{\rm b}m_{\rm e}^3m_\mu^3m_\tau^3}\right)$.
We define the QCD energy scale, $\Lambda_{_{\rm QCD}}$, as the
energy at which the strong coupling constant diverges. Note that it implies that $\Lambda_{_{\rm QCD}}$ also
depends on the Higgs and fermion masses through threshold effects.

More familiar constants, such as the masses of the proton and the neutron are, as we shall discuss
in more details below (see \S~\ref{subsubmass}), more difficult to relate to the fundamental parameters because they
depend not only on the masses of the quarks but also on the electromagnetic and strong binding
energies.\\

\begin{table}[t]
\begin{center}
{\small
\begin{tabular}{p{6cm}ll}
 \hline\hline
 Constant & Symbol & Value \\
 \hline
  Electromagnetic coupling constant & $g_{_{\rm EM}}=e=g_2 \sin\theta_{_{\rm W}}$ & $0.313429\pm0.000022$\\
 Higss mass & $m_H$ & $>100$~GeV \\
 Higss vev & $v$ & $(246.7\pm0.2)$~GeV\\
 Fermi constant & $\gfermi=1/\sqrt{2}v^2$ & $1.166 37(1)\times10^{-5}\, {\rm GeV}^{-2}$\\
 Mass of the $W^\pm$ & $m_W$ & $80.398\pm0.025$~GeV\\
 Mass of the $Z$ & $m_Z$ & $91.1876\pm0.0021$~GeV\\
 Fine structure constant & $\aem$ & $1/137.035 999 679(94)$\\
 Fine structure constant at $m_Z$ & $\aem(m_Z)$ & $1/(127.918\pm0.018)$\\
 Weak structure constant at $m_Z$ & $\aw(m_Z)$ & $0.03383\pm0.00001$\\
 Strong structure constant at $m_Z$ & $\as(m_Z)$ & $0.1184\pm0.0007$\\
 Gravitational structure constant &$ \ag= Gm_{\rm p}^2/\hbar c$ & $\sim5.905\times10^{-39}$ \\
 Electron mass & $m_{\rm e}= h_{\rm e}v/\sqrt{2}$ & $510.998910\pm 0.000013$~keV\\
 Mu mass & $m_\mu= h_\mu v/\sqrt{2}$ & $105.658367\pm0.000004$~MeV\\
 Tau mass & $m_\tau= h_\tau v/\sqrt{2}$ & $1776.84 \pm0.17$~MeV\\
 Up quark mass & $m_{\rm u}= h_{\rm u}v/\sqrt{2}$ & $(1.5-3.3)$~MeV\\
 Down quark mass & $m_{\rm d}= h_{\rm d}v/\sqrt{2}$ & $(3.5-6.0)$~MeV\\
 Strange quark mass & $m_{\rm s}= h_{\rm s}v/\sqrt{2}$ & $105^{+25}_{-35}$~MeV \\
 Charm quark mass & $m_{\rm c}= h_{\rm c}v/\sqrt{2}$ & $1.27^{+0.07}_{-0.11}$~GeV\\
  Bottom quark mass & $m_{\rm b}= h_{\rm b}v/\sqrt{2}$ & $4.20^{+0.17}_{-0.07}$~GeV\\
 Top quark mass & $m_{\rm t}= h_{\rm t}v/\sqrt{2}$ & $171.3\pm2.3$~Gev\\
 QCD energy scale & $\Lambda_{{\rm QCD}}$ & $(190-240)$~MeV\\
 Mass of the proton & $m_{\rm p}$ & $938.272013 \pm 0.000023$~MeV \\
 Mass of the neutron & $m_{\rm n}$ & $939.565346 \pm 0.000023$~MeV \\
 proton-neutron mass difference & $Q_{\rm np}$ & $1.2933321\pm0.0000004$~MeV \\
 proton-to-electron mass ratio & $\mu=m_{\rm p}/m_{\rm e}$ & $1836.15$\\
 electron-to-proton mass ratio & $\bar\mu=m_{\rm e}/m_{\rm p}$ & $1/1836.15$\\
 $d-u$ quark mean mass & $m_{\rm q}=(m_{\rm u}+m_{\rm d})/2$ & $(2.5-5.0)$~MeV\\
 $d-u$ quark mass difference & $\delta m_{\rm q} = m_{\rm d}-m_{\rm u}$&$(0.2-4.5)$~MeV \\
 proton gyromagnetic factor &$g_{\rm p}$ & 5.586 \\
 neutron gyromagnetic factor &$g_{\rm n}$ & -3.826 \\
 Rydberg constant & $R_\infty$ & $10 973 731.568 527(73)\,{\rm m}^{-1}$\\
\hline 
\end{tabular}
\caption{\it List of some related constants that appear in our discussions.}
\label{tab-list2}
}
\end{center}
\end{table}

{\it Are some constants more fundamental?}
As pointed-out by Levy-Leblond~\cite{levy79}, all constants of
physics do not play the same role, and some have a much deeper role
than others. Following Ref.~\cite{levy79}, we can define three
classes of fundamental constants, {\em class A} being the class of the
constants characteristic of a particular system, {\em class B} being
the class of constants characteristic of a class of physical
phenomena, and {\em class C} being the class of universal
constants. Indeed, the status of a constant can change with time. For
instance, the velocity of light was initially a class A constant
(describing a property of light) which then became a class B constant
when it was realized that it was related to electromagnetic phenomena
and, to finish, it ended as a type C constant (it enters special relativity and
is related to the notion of causality, whatever the physical phenomena).
It has even become a much more fundamental constant
since it now enters in the definition of the metre~\cite{petley83} (see
Ref.~\cite{ul-book} for a more detailed discussion). This has to
be constrasted with the proposition of Ref.~\cite{wilczek07} 
to distinguish the standard model free parameters as  the gauge
and gravitational couplings (which are associated to internal and
spacetime curvatures) and the other parameters entering
the accomodation of inertia in the Higgs sector.\\

{\it Relation with physical laws}.
Levy-Leblond~\cite{levy79} thus proposed to rank the constants in terms of their universality
and he proposed that only three constants be considered to be
of class C, namely $G$, $\hbar$ and $c$. He pointed out two
important roles of these constants in the laws of physics. First, they act as
{\em concept synthetizer} during the process of our understanding
of the laws of nature: contradictions between existing theories have often been resolved by
introducing new concepts that are more general or more
synthetic than older ones. Constants build bridges between quantities
that were thought to be incommensurable and thus allow new
concepts to emerge. For example $c$ underpins the synthesis
of space and time while the Planck constant allowed to related the
concept of energy and frequency and the gravitational constant creates
a link between matter and space-time. Second, it follows that
this constants are related the domains of validity of
these theories. For instance, as soon as velocity approaches $c$,
relativistic effects become important, relativistic effects cannot be negligible.
On the other hand, for speed much below $c$, Galilean kinematics
is sufficient. Planck constant also acts as a referent, since if the
action of a system greatly exceeds the value of that constant, classical
mechanics will be appropriate to describe this system. While the
place of $c$ (related to the notion of causality) and $\hbar$ 
(related to the quantum) in this list are well argumented, the place of $G$ remains debated
since it is thought that it will have to be replaced by some mass scale.\\

{\it Evolution}. There are many ways the list of constants can change
with our understanding of physics. First, new constants
may appear when new systems or new physical laws are discovered; this
is for instance the case of the charge of the electron or more recently
the gauge couplings of the nuclear interactions. A constant can also
move from one class to a more universal class. An example is that of the electric 
charge, initially of class A (characteristic of the electron) which then
became class B when it was understood that it characterizes the
strength of the electromagnetic interaction. A constant can also
disappear from the list, because it is either replaced by
more fundamental constants (e.g. the Earth acceleration due
to gravity and the proportionality constant entering Kepler law
both disappeared because they were ``explained'' in terms
of the Newton constant and the mass of the Earth or the Sun)
or because it can happen that a better understanding of physics
teaches us that two hitherto distinct quantities have to be considered
as a single phenomenon (e.g. the understanding by Joule that heat
and work were two forms of energy led to the fact that
the Joule constant, expressing the proportionality between work
and heat, lost any physical meaning and became a simple
conversion factor between units used in the measurement of heat
(calories) and work (Joule)). Nowadays the calorie has fallen in disuse.
Indeed demonstrating that a constant is varying will have direct
implications on our list of constants.\\

In conclusion, the evolution of the number, status of the constants
can teach us a lot about the evolution of the ideas and theories
in physics since it reflects the birth of new concepts, their 
evolution and unification with other ones.

\subsubsection{Constants and metrology}

Since we cannot compute them in the theoretical framework in which
they appear, it is a crucial property of the fundamental constants
(but in fact of all the constants) that their value can be measured.
The relation between constants and metrology is a huge
subject to which we just draw the attention on some selected aspects.
For more discussions, see Refs.~\cite{qed-karshen1,karshen-metro}.

The introduction of constants in physical laws is also closely related
to the existence of systems of units. For instance, Newton's law
states that the gravitational force between two masses is
proportional to each mass and inversely proportional to the square of their
separation. To transform the proportionality to an equality one
requires the use of a quantity with dimension of ${\rm
m}^3\cdot{\rm kg}^{-1}\cdot {\rm s}^{-2}$ independent of the separation
between the two bodies, of their mass, of their composition
(equivalence principle) and on the position (local position
invariance). With an other system of units the numerical value of
this constant could have simply been anything. Indeed, the numerical value of
any constant crucially depends on the definition of the system of units.\\

{\it Measuring constants}. The determination of the laboratory value of constants relies
mainly on the measurements of lengths, frequencies, times,... (see
Ref.~\cite{petley85} for a treatise on the measurement of constants and
Ref.~\cite{flower01} for a recent review). Hence, any
question on the variation of constants is linked to the definition
of the system of units and to the theory of measurement. 
The behavior of atomic matter is determined by the value of many
constants. As a consequence, if e.g. the fine-structure
constant is spacetime dependent, the comparison between several
devices such as clocks and rulers will also be spacetime
dependent. This dependence will also differ from one clock to another
so that metrology becomes both device and spacetime dependent,
a property that will actually be used to construct tests of the constancy of
the constants.

Indeed a measurement is always a comparison between two
physical systems of the same dimensions. This is thus
a relative measurement which will give as result a pure number.
This trivial statement is oversimplifying
since in order to compare two similar quantities measured
separately, one needs to perform a number of
comparisons. In order to reduce this number of comparisons
(and in particular to avoid creating every time a chain of comparisons), 
a certain set of them has been included in the definitions
of units. Each units can then be seen
as an abstract physical system, which has to be realised effectively
in the laboratory, and to which another physical system is compared.
A measurement in terms of these units is usually called an absolute 
measurement.
Most fundamental constants are related to microscopic physics
and their numerical values can be obtained either from a
pure microscopic comparison (as is e.g. the case for $m_{\rm e}/m_{\rm p}$)
or from a comparison between microscopic and macroscopic values
(for instance to deduce the value of the mass of the electron in kilogram).
This shows that the choice of the units has an impact on the
accuracy of the measurement since the pure microscopic comparisons
are in general more accurate than those involving macroscopic physics.

It is also important to stress that in order to deduce the value of constants
from an experiment, one usually needs to use theories and models. An
example~\cite{karshen-metro} is provided by the Rydberg constant. It can
easily be expressed in terms of some fundamental constants as
$R_\infty = {\aem^2m_{\rm e}c}/{2 h}$.
It can be measured from e.g. the triplet $1s-2s$ transition in hydrogen, the
frequency of which is related to the Rydberg constant and other constants
by assuming QED so that the accuracy of $R_\infty$  is much
lower than that of the measurement of the transition. This could be solved 
by defining $R_\infty$ as $4\nu_{\rm H}(1s-2s)/3c$ but then the relation
with more fundamental constants would be more complicated and actually
not exactly known. This illustrates the relation between a practical and a fundamental
approach and the limitation arising from the fact that we often cannot
both exactly calculate and directly measure some quantity. Note also that
some theoretical properties are plugged in the determination of the constants.

As a conclusion, let us recall that (i) in general, the values of the
constants are not determined by a direct measurement but by a chain
involving both theoretical and experimental steps, (ii) they depend on
our theoretical understanding, (iii) the determination of a
self-consistent set of values of the fundamental constants results
from an adjustment to achieve the best match between theory and a
defined set of experiments (which is important because we
actually know that the theories are only good approximation and have
a domain of validity) (iv) that the
system of units plays a crucial role in the measurement chain, since
for instance in atomic units, the mass of the electron could have been
obtained directly from a mass ratio measurement (even more precise!)
and (v) fortunately the test of the variability of the constants does
not require {\it a priori} to have a high-precision value of the
considered constants.\\

{\it System of units}. One thus need to define a coherent system of units.
This has a long, complex and interesting history that was driven by 
simplicity and universality but also
by increasing stability and accuracy~\cite{barrow-book,uzanleclercqbook}. 

Originally, the sizes of the human body were mostly used to measure the length
of objects (e.g. the foot and the thumb gave feet and inches) and some of these
units can seem surprising to us nowaday (e.g. the {\it span} was the
measure of hand with fingers fully splayed, from the tip of the thumb to the tip of the
little finger!). Similarly weights were related to what could be carried in the hand: the pound, 
the ounce, the dram\ldots Needless to say that this system had a few disadvantages since
each country, region has its own system (for instance in France there was more
than 800 different units in use in 1789). The need to define a system
of units based on natural standard led to several propositions to define
a standard of length (e.g. the {\it mille} by Gabriel Mouton in 1670 defined
as the length of one angular minute of a great circle on the Earth or
the length of the pendulum that oscillates once a second by Jean Picard
and Christiaan Huygens). The  real change happened during the French Revolution
during which the idea of a universal and non anthropocentric system of units
arose. In particular, the Assembl\'ee adopted the principle of a uniform
system of weights and measures on the 8th of May 1790 and, on March 1791
a decree (these texts are reprinted in Ref.~\cite{ul-book})
was voted, stating that a quarter of the terrestrial meridian
would be the basis of the definition of the {\it metre} (from the Greek
metron, as proposed by Borda): a metre would henceforth be one
ten millionth part of a quarter of the terrestrial meridian. Similarly the {\it gram}
was defined as the mass of one cubic centimetre of distilled water (at
a precise temperature and pressure) and the second was defined from
the property that a mean Solar day must last 24 hours.

To make a long story short, this led to the creation of the metric system and then
of the signature of {\it La convention du m\`etre} in 1875. Since then, the
definition of the units have evolved significantly. First, the definition of the metre 
was related to more immutable systems than our planet which, as pointed out
by Maxwell in 1870, was  an arbitrary and inconstant reference. He then 
suggested that atoms may be such a universal reference. In 1960, the BIPM
established a new definition of the metre as the length equal
to 1650763 wavelengths, in a vacuum, of the transition line between
the levels $2p_{10}$ and $5d_{5}$ of krypton-86. Similarly the rotation of
the Earth was not so stable and it was proposed in 1927 by Andr\'e Danjon
to use the tropical year as a reference, as adopted in 1952. In 1967, the
second was also related to an atomic transition, defined as the
duration of 9162631770 periods of the transition between the two hyperfine
levels of the ground state of caesium-133. To finish, it was decided in 1983,
that the metre shall be defined by fixing the value of the speed of light
to $c=299792458 {\rm m}\cdot{\rm s}^{-1}$ and we refer to Ref.~\cite{bipm}
for an up to date description of the SI system. Today, the possibility to
redefine the kilogram in terms of a fixed value of the Planck
constant is under investigation~\cite{karshen-kilo}.

This summary illustrates that the system of units is a human product and
all SI definitions are historically based on non-relativistic classical physics.
The changes in the definition were driven by the will to use more
stable and more fundamental quantities so that they closely follow
the progress of physics. This system has been created for legal use and indeed
the choice of units is not restricted to SI.\\

{\it SI systems and the number of basic units}.
The International System of Units defines seven basic units: the
metre (m), second (s) and kilogram (kg), the Ampere (A), Kelvin (k), mole (mol)  and candela (cd),
from which one defines secondary units. While needed
for pragmatic reasons, this system of units is unnecessarily
complicated from the point of view of theoretical physics.
In particular, the Kelvin, mole and candela are derived from
the four other units since temperature is actually a measure
of energy, the candela is expressed in terms of energy flux
so that both can be expressed in mechanical units of
length [L], mass [M] and time [T]. The mole is merely a unit denoting numbers
of particule and has no dimension.

The status of the Ampere is interesting in itself. The discovery
of the electric charge [Q] led to the introduction of a new units,
the Coulomb. The Coulomb law describes the force between
two charges as being proportional to the product of the
two charges and to the inverse of the distance squared.
The dimension of the force being known as [MLT$^{-2}$], this
requires the introduction of a new constant $\varepsilon_0$
(which is only a conversion factor), with
dimensions [Q$^2$M$^{-1}$L$^{-3}$T$^2$] in the Coulomb law,
and that needs to be measured. Another route could have been 
followed since the Coulomb law tells us that no new constant
is actually needed if one uses [M$^{1/2}$L$^{3/2}$T$^{-1}$] as
the dimension of the charge. In this system of units,
known as Gaussian units, the numerical
value of $\varepsilon_0$ is 1. Hence the Coulomb can be
expressed in terms of the mechanical units  [L],  [M] and [T],
and so will the Ampere. This reduces the number of conversion
factors, that need to be experimentally determined, but this choice of units
assumes the validity of the Coulomb law so that keeping
a separate unit for the charge may be a more robust attitude.\\

{\it Natural units.} The previous discussion tends to show that all
units can be expressed in terms of the three mechanical units. It follows,
as realized by Johnstone-Stoney in 1874, that these three basic units can be
defined in terms of 3 independent constants.  He proposed~\cite{barrowston,stoney}
to use three constants: the Newton constant, the velocity of light and the
basic units of electricity, i.e. the electron charge, in order to define,
from dimensional analysis a ``natural series of physical units'' defined
as
\begin{eqnarray}
 t_{\rm S}&=&\sqrt{\frac{Ge^2}{4\pi\varepsilon_0 c^6}}\sim 4.59\times10^{-45}\,{\rm s},
 \nonumber\\
  \ell_{\rm S}&=&\sqrt{\frac{G e^2}{4\pi\varepsilon_0 c^4}}\sim 1.37\times10^{-36}\,{\rm m},
 \nonumber\\
  m_{\rm S}&=&\sqrt{\frac{e^2}{4\pi\varepsilon_0 G}}\sim 1.85\times10^{-9}\,{\rm kg},
  \nonumber
\end{eqnarray}
where the $\varepsilon_0$ factor has been included because we are using the SI definition
of the electric charge. In such a system of units, by construction, the numerical value
of $G$, $e$ and $c$ is 1, i.e. $c=1\times  \ell_{\rm S}\cdot t_{\rm S}^{-1}$ etc.

A similar approach to the definition of the units was independently proposed
by Planck~\cite{planck1} on the basis of the two constants $a$ and
$b$ entering the Wien law and $G$, which he reformulated later~\cite{planck2} in terms
of $c$, $G$ and $\hbar$ as
\begin{eqnarray}
 t_{\rm P}&=&\sqrt{\frac{G \hbar}{c^5}}\sim 5.39056\times10^{-44}\,{\rm s},
 \nonumber\\
  \ell_{\rm P}&=&\sqrt{\frac{G\hbar}{c^3}}\sim 1.61605\times10^{-35}\,{\rm m},
 \nonumber\\
  m_{\rm P}&=&\sqrt{\frac{\hbar c}{G}}\sim 2.17671\times10^{-8}\,{\rm kg}.
  \nonumber
\end{eqnarray}
The two systems are clearly related by the fine-structure constant since
$e^2/4\pi\varepsilon_0=\aem hc$.

Indeed, we can construct many such systems since the choice of the 3 constants
is arbitrary. For instance, we can construct a system based on ($e, m_{\rm e},
h)$, that we can call the {\em Bohr units} which will be suited to the study of the atom.
The choice may be dictated by the system which is studied (it is indeed far
fetched to introduce $G$ in the construction of the units when studying
atomic physics) so that the system is well adjusted in the sense that the numerical
values of the computations are expected to be of order unity in these units.

Such constructions are very useful for theoretical computations but not adapted
to measurement so that one needs to switch back to SI units. More important,
this shows that, from a theoretical point of view, one can
define the system of units from the laws of nature, which are
supposed to be universal and immutable.\\

{\it Do we actually need 3 natural units?} is an issue
debated at length. For instance, Duff, Okun and
Veneziano~\cite{duff01} respectively argue
for none, three and two (see also Ref.~\cite{wignall00}). Arguing for no
fundamental constant leads to consider them simply as conversion
parameters. Some of them are, like the Boltzmann constant, but some
others play a deeper role in the sense that when a physical quantity
becomes of the same order of this constant new phenomena appear, this
is the case e.g. of $\hbar$ and $c$ which are associated respectively
to quantum and relativistic effects. Okun~\cite{okun91} considered that only
three fundamental constants are necessary, as
indicated by the International System of Units.   In the framework of quantum field
theory + general relativity, it seems that this set of three constants
has to be considered and it allows to classify the physical theories (with the famous
{\em cube of physical theories}).
However, Veneziano~\cite{veneziano86} argued that in the
framework of string theory one requires only two dimensionful
fundamental constants, $c$ and the string length $\lambda_s$. The use
of $\hbar$ seems unnecessary since it combines with the string tension
to give $\lambda_s$. In the case of the Goto-Nambu action
$S/\hbar=(T/\hbar)\int\dd(Area)\equiv \lambda_s^{-2}\int\dd(Area)$ and
the Planck constant is just given by $\lambda_s^{-2}$. In this view,
$\hbar$ has not disappeared but has been promoted to the role of a UV
cut-off that removes both the infinities of quantum field theory and
singularities of general relativity. This situation is analogous to
pure quantum gravity~\cite{novikov82} where $\hbar$ and
$G$ never appear separately but only in the combination $\ell_{_{\rm
Pl}}=\sqrt{G\hbar/c^{3}}$ so that only $c$ and $\ell_{_{\rm Pl}}$ are
needed. Volovik~\cite{volovik02} made an analogy with quantum
liquids to clarify this. There, an observer knows both the effective and microscopic
physics so that he can judge whether the fundamental constants of the
effective theory remain fundamental constants of the microscopic
theory. The status of a constant depends on the considered theory
(effective or microscopic) and, more interestingly, on the observer
measuring them, i.e. on whether this observer belongs to the world of
low-energy quasi-particles or to the microscopic world.\\

{\it Fundamenal parameters}. Once a set of three independent constants
has been chosen as natural units, then all other constants are dimensionless
quantities. The values of these combinations of constants does not depend
on the way they are measured,~\cite{sansdim1,sansdim3,sansdim2}, on the definition of the units etc... It follows
that any variation of constants that will let these numbers unaffected is actually just
a redefinition of units.

These dimensionless numbers represent e.g. the mass ratio, relative magnitude of strength etc... 
Changing their values will indeed have an impact on the intensity 
of various physical phenomena, so that they encode some properties
of our world. They have specific values (e.g. $\aem\sim1/137$, $m_{\rm p}/m_{\rm e}
\sim1836$, etc.) that we may hope to understand. Are all these numbers
completely contingent, or are some (why not all?) of them related by relations arising from
some yet unknown and more fundamental theories. In such theories, some
of these parameters may actually be dynamical quantities and
thus vary in space and time. These are our potential varying constants.

\subsection{The constancy of constants as a test of general relativity}\label{subsec12}

The previous paragaphs have yet emphasize why testing for the consistency of the
constants is a test of fundamental physics since it can reveal the need for new
physical degrees of freedom in our theory. We now want to stress the relation
of this test with other tests of general relativity and with cosmology.

\subsubsection{General relativity}

The tests of the constancy of fundamental constants take all their
importance in the realm of the tests of the equivalence principle~\cite{will-book}. 
Einstein general relativity is based on two independent hypotheses, which
can conveniently be described by decomposing the action of the 
theory as $S = S_{\rm grav}+S_{\rm matter}$.\\

The equivalence principle has strong implication for the
functional form of $S_{\rm grav}$. This principles include
three hypothesis: 
\begin{itemize}
 \item the universality of free fall, 
 \item the local position invariance,
 \item the local Lorentz invariance. 
\end{itemize} 
In its weak form (that is for all interactions but gravity), it is satisfied by any metric theory of gravity 
and general relativity is conjectured to satisfy it in its strong form (that is for all interactions including gravity). We
refer to Ref.~\cite{will-book} for a detailed description of these
principles. The weak equivalence principle
can be mathematically implemented by assuming that all matter
fields are minimally coupled to a single metric tensor
$g_{\mu\nu}$. This metric defines the length and times measured by
laboratory clocks and rods so that it can be called the {\it
physical metric}. This implies that the action for any matter
field, $\psi$ say, can be written as 
$$
 S_{\rm matter}(\psi,g_{\mu\nu}).
$$
This so-called {\it metric coupling} ensures in particular the validity of the universality of
free-fall. Since locally, in the neighborhood of the worldline, there always exists
a change of coordinates so that the metric takes a Minkowskian form at
lowest order, the gravitational field can be locally
``effaced''  (up to tidal effects). If we identify this neighborhood to a small lab, this means that
any physical properties that can be measured in this lab must be independent
of where and when the experiments are carried out. This is indeed the
assumption of {\it local position invariance} which implies that the constants
must take the same value independent of the spacetime point where they
are measured. Testing the constancy of fundamental constants is thus
a direct test of this principle and thus of the metric coupling. Interestingly,
the tests we are discussing in this review allow to extend them much further than
the Solar scales and even in the early universe, an important
information to check the validity of relativity in cosmology.

As an example, the action of a point-particle reads
\begin{equation}\label{Spp}
  S_{\rm matter} = -\int mc\sqrt{-g_{\mu\nu}({\bf x}) v^\mu v^\nu}\dd t,
\end{equation}
with $v^\mu\equiv\dd x^\mu/\dd t$. The equation of motion that derives from
this action is the usual geodesic equation
\begin{equation}\label{geo1}
 a^\mu \equiv u^\nu\nabla_\nu u^\mu = 0,
\end{equation}
where $u^\mu= \dd x^\mu/c\dd\tau$, $\tau$ being the proper time;
$\nabla_\mu$ is the covariant derivative associated with the metric $g_{\mu\nu}$
and $a^\nu$ is the 4-acceleration.
Any metric theory of gravity will enjoy such a matter Lagrangian and the worldline of any test
particle shall be a  geodesic of the spacetime with metric $g_{\mu\nu}$,
as long as there is no other long range force acting on it (see Ref.~\cite{gefmotion}
for a detailed review of motion in alternative theories of gravity).

Note that in
the Newtonian limit $g_{00}=-1+2\Phi_N/c^2$ where $\Phi_N$ is the Newtonian
potential. It follows that, in the slow velocity limit, the geodesic equation reduces to
\begin{equation}\label{eq.inertie}
 \dot{\bf v} =  {\bf a} = -\nabla\Phi_N \equiv  {\bf g}_N, 
\end{equation}
hence defining the Newtonian acceleration ${\bf g}_N$.
Remind that the proper time of a clock is related to the coordinate time
by $\dd\tau = \sqrt{-g_{00}}\dd t$. Thus, if one exchanges electromagnetic
signals between two identical clock in a stationary situation, the apparent
difference between the two clocks rates will be
$$
 \frac{\nu_1}{\nu_2} = 1 + \frac{\Phi_N(2)-\Phi_N(1)}{c^2},
$$
at lowest order. This is the so called universality of gravitational redshift.

The assumption of a metric coupling is actually well tested in the Solar
system:
\begin{itemize}
\item First, it implies that all non-gravitational constants are
spacetime independent, which have been tested to a very high
accuracy in many physical systems and for various fundamental
constants; this the subject of this review. 
\item Second, the isotropy has been tested
from the constraint on the possible quadrupolar shift of nuclear
energy levels~\cite{isotes3,isotest2,isotest}
proving that different matter fields couple to a unique metric tensor
at the $10^{-27}$ level. 
\item Third, the universality of free fall can be tested by comparing the
accelerations of two test bodies in an external gravitational field. The
parameter $\eta_{12}$ defined as
\begin{equation}
 \eta_{12}\equiv 2 \frac{|{\bf a}_1-{\bf a}_2|}{|{\bf a}_1+{\bf a}_2|},
\end{equation}
can be constrained experimentally, e.g. in the
laboratory by comparing the acceleration of a Beryllium and a Copper
mass in the Earth gravitational field~\cite{uff2} to get
\begin{equation}
 \eta_{\rm Be,Cu} = (-1.9\pm2.5)\times10^{-12}.
\end{equation}
Similarly the comparison of Earth-core-like and Moon-mantle-like
bodies gave~\cite{uff1} 
\begin{equation}
 \eta_{\rm Earth,Moon} = (0.1\pm2.7\pm1.7)\times10^{-13}.
\end{equation}
 The Lunar Laser ranging experiment~\cite{uff3},
which compares the relative acceleration of the Earth and Moon in
the gravitational field of the Sun, also set the constraints
\begin{equation}
 \eta_{\rm Earth,Moon} = (-1.0\pm1.4)\times10^{-13}.
\end{equation}
Note that since the core represents only 1/3 of the mass of the Earth, and
since the Earth mantle has the same composition of the Moon (and
thus shall fall in the same way), one looses a factor 3 so that this
constraint is actually similar as the one obtained in the lab.
Further constraints are summarized in Table~\ref{tab-eta12}.
The latter constraint also contains some contribution from the gravitational
binding energy and thus includes the strong equivalence principle.
When the laboratory result of Ref.~\cite{uff1} is combined with the
LLR results of Refs.~\cite{G-llr0} and~\cite{muller91}, one gets a
constraints on the strong equivalence principle parameter, respectively
$$
\eta_{_{\rm SEP}} = (3\pm6)\times10^{-13}\quad{\rm and}\quad
\eta_{_{\rm SEP}} = (-4\pm5)\times10^{-13}.
$$

Large improvements are expected thanks to existence of two
dedicated space mission projects: Microscope~\cite{touboul01} and
STEP~\cite{mester01}.

\begin{table}[t]
\begin{center}
{\small
\begin{tabular}{p{4 cm}ccc}
 \hline\hline
 Constraint  & Body 1 & Body 2  & Ref. \\
 \hline
  $(-1.9\pm2.5)\times10^{-12}$ & Be & Cu & \cite{uff2}\\
  $(0.1\pm2.7\pm1.7)\times10^{-13}$ & Earth-like rock & Moon-like rock & \cite{uff1} \\
  $(-1.0\pm1.4)\times10^{-13}$ & Earth & Moon &\cite{uff3} \\
  $(0.3 \pm1.8)\times 10^{-13}$ & Te & Bi & \cite{tebi}\\
  $(-0.2 \pm 2.8) \times 10^{-12} $ & Be & Al & \cite{suetal}\\
   $(-1.9 \pm 2.5) \times 10^{-12} $ & Be & Cu & \cite{suetal}\\
   $(5.1 \pm 6.7) \times 10^{-12} $ & Si/Al & Cu & \cite{suetal}\\
  \hline\hline
\end{tabular}
\caption{\it Summary of the constraints on the violation of the universality
of free fall}
\label{tab-eta12}
}
\end{center}
\end{table}
\item Fourth, the Einstein
effect (or gravitational redshift) has been
measured at the $2\times10^{-4}$ level~\cite{clock1}.
\end{itemize}
We can conclude that the hypothesis of metric coupling is extremely well-tested
in the Solar system.\\

The second building block of general relativity 
is related to the dynamics of the gravitational sector,
assumed to be dictated by the
Einstein-Hilbert action
\begin{equation}\label{Sgrav}
 S_{\rm grav} = \frac{c^3}{16\pi G}\int\sqrt{-g_*}R_*\dd^4x.
\end{equation}
This defines the dynamics of a massless spin-2 field $g^*_{\mu\nu}$,
called the Einstein metric. General relativity then assumes that both 
metrics coincide, $g_{\mu\nu}=g^*_{\mu\nu}$ (which is related
to the strong equivalence principle), but it is possible
to design theories in which this indeed not the case (see the example
of scalar-tensor theories below; \S~\ref{subsecST}) so that general relativity is
one out of a large family of metric theories.

The variation of the total action with respect to the metric yields the
Einstein equations
\begin{equation}
 R_{\mu\nu}-\frac{1}{2}R g_{\mu\nu} = \frac{8\pi G}{c^4}T_{\mu\nu},
\end{equation}
where $T^{\mu\nu}\equiv (2/\sqrt{-g})\delta S_{\rm matter}/\delta g_{\mu\nu}$
is the matter stress-energy tensor. The coefficient $8\pi G/c^4$ is
determined by the weak-field limit of the theory that should reproduce
the Newtonian predictions.

The dynamics of general relativity can be tested in the
Solar system by using the parameterized post-Newtonian formalism (PPN).
Its  is a general
formalism that introduces 10 phenomenological parameters to
describe any possible deviation from general relativity at the
first post-Newtonian order~\cite{will-book,will-llr} (see also Ref.~\cite{blanchet-lrr}
for a review on higher orders). The formalism assumes
that gravity is described by a metric and that it does not involve
any characteristic scale. In its simplest form, it reduces to the
two Eddington parameters entering the metric of the Schwartzschild
metric in isotropic coordinates
$$
 g_{00} = - 1 + \frac{2Gm}{rc^2} +
 2\beta^\ppn\left(\frac{2Gm}{rc^2}\right)^2,
 \qquad
 g_{ij} = \left(1+2\gamma^\ppn\frac{2Gm}{rc^2}\right)\delta_{ij}.
$$
Indeed, general relativity predicts $\beta^\ppn=\gamma^\ppn=1$.

These two phenomenological parameters are constrained (1) by the
shift of the Mercury perihelion~\cite{mercure} which
implies that $|2\gamma^\ppn-\beta^\ppn-1|<3\times10^{-3}$, (2) the
Lunar laser ranging experiments~\cite{uff3} which
implies that $|4\beta^\ppn-\gamma^\ppn-3|=(4.4\pm4.5)\times10^{-4}$ and
(3) by the deflection of electromagnetic signals which are all
controlled by $\gamma^\ppn$. For instance the very long baseline
interferometry~\cite{vlbi} implies that
$|\gamma^\ppn-1|=4\times10^{-4}$ while the measurement of the time
delay variation to the Cassini spacecraft~\cite{cassini}
sets $\gamma^\ppn-1=(2.1\pm2.3)\times10^{-5}$.

The PPN formalism does not allow to test finite range effects that
could be caused e.g. by a massive degree of freedom. In that case
one expects a Yukawa-type deviation from the Newton potential,
$$
 V=\frac{Gm}{r}\left(1+\alpha\hbox{e}^{-r/\lambda}\right),
$$
that can be probed by ``fifth force'' experimental searches.
$\lambda$ characterizes the range of the Yukawa deviation of
strength $\alpha$. The constraints on
$(\lambda,\alpha)$ are summarized in Ref.~\cite{5force} which
typically shows that $\alpha<10^{-2}$ on scales ranging from the
millimetre to the Solar system size.\\

General relativity is also tested  with pulsars~\cite{gefpul2,gefpul}
and in the strong field regime~\cite{psaltis}. For more details we refer to
Refs.~\cite{lilley,tury,will-book,will-llr}.
Needless to say that any extension of general relativity has to
pass these constraints. However, deviations from general
relativity can be larger in the past, as we shall see, which makes
cosmology an interesting physical system to extend these
constraints.

\subsubsection{Varying constants and the universality of free fall}

As the previous description shows, the constancy of the fundamental constants
and the universality are two pillars of the equivalence principle. Dicke~\cite{dicke64} realized that
they are actually not independent and that if the coupling constants are
spatially dependent then this will induce a violation of the universality of free fall.

The connection lies in the fact that the mass of any composite body, starting
e.g. from nuclei, includes the mass of the elementary particles that constitute
it (this means that it will depend on the Yukawa couplings and on the Higgs
sector parameters) but also a contribution, $E_{\rm binding}/c^2$, arising from the binding
energies of the different interactions (i.e. strong, weak and electromagnetic)
but also gravitational for massive bodies. Thus the mass of any body is a complicated
function of all the constants, $m[\alpha_i]$.

It follows that the action for a point particle is no more given by Eq.~(\ref{Spp})
but by
\begin{equation}\label{Sppvar}
  S_{\rm matter} = -\int m_A[\alpha_j]c\sqrt{-g_{\mu\nu}({\bf x}) v^\mu v^\nu}\dd t,
\end{equation}
where $\alpha_j$ is a list of constant including $\aem$ but also many others
and where the index $A$ in $m_A$ recalls that the
dependency in these constant is a priori different for body of
different chemical composition. The variation of this action
gives the equation of motion
\begin{equation}
 u^\nu\nabla_\nu u^\mu = - \left(\sum_i\frac{\partial \ln m_A}{\partial \alpha_i} 
 \frac{\partial \alpha_i}{\partial x^\beta} \right)
 \left( g^{\beta\mu}+u^\beta u^\mu\right).
\end{equation}
It follows that a test body will not enjoy a geodesic motion. In
the Newtonian limit $g_{00}=-1+2\Phi_N/c^2$, and at first order
in $v/c$, the equation of motion of a test particle reduces to
\begin{equation}\label{eq.inertie2}
 {\bf a} =  {\bf g}_N  +  \delta {\bf a}_A, \qquad
  \delta {\bf a}_A = -c^2\sum_if_{A,i}\left(\nabla\alpha_i+\dot\alpha_i\frac{{\bf v}_A}{c^2}\right)
\end{equation}
so that in the slow velocity (and slow variation) limit it reduces
to
$$
\delta {\bf a}_A = -c^2\sum_i f_{A,i}\nabla\alpha_i.
$$
where we have introduce the sensitivity of the mass $A$ with respect to
the variation of the constant $\alpha_i$ by
\begin{equation}\label{deffai}
 f_{A,i} \equiv \frac{\partial\ln m_A}{\partial\alpha_i}.
\end{equation}
This simple argument shows that if the constants depend on time then
there must exist an anomalous acceleration that will depend
on the chemical composition of the body $A$.

This anomalous acceleration is generated by the change in the
(electromagnetic, gravitational,...) binding energies~\cite{dicke64,haugan76,nordtvedt90}
but also in the Yukawa couplings and in the Higgs sector parameters so that
the $\alpha_i$-dependencies are a priori composition-dependent. 
As a consequence, any variation of the
fundamental constants will entail a violation of the universality of
free fall: the total mass of the body being space dependent, an
anomalous force appears if energy is to be conserved. The variation of
the constants, deviation from general relativity and violation of the
weak equivalence principle are in general expected together.

On the other hand, the composition dependence of $\delta{\bf a}_A$ and thus
of $\eta_{AB}$ can be used to optimize the choice of materials
for the experiments testing the equivalence principle~\cite{damourmat}
but also to distinguish between several models if data from
the universality of free fall and atomic clocks are combined~\cite{dentuff}.\\

From a theoretical point of view, the computation of $\eta_{AB}$
will requires the determination of the coefficients $f_{Ai}$.
This can be achieved in two steps by first relating the new degrees
of freedom of the theory to the variation of the fundamental constants 
and then relating them to the variation of the masses.
As we shall see in \S~\ref{section6}, the first issue is very
model dependent while the second is especially difficult, particularly
when one wants to understand the effect of the quark mass, since
it is related to the intricate structure of QCD and its role in low
energy nuclear reactions.

As an example, the mass of a nuclei
of charge $Z$ and atomic number $A$ can be expressed as
$$
m(A,Z)=Zm_{\rm p}+(A-Z)m_{\rm n}+Zm_{\rm e}+E_{_{\rm S}}+E_{_{\rm EM}},
$$
where $E_{_{\rm S}}$ and $E_{_{\rm EM}}$ are respectively the
strong and electromagnetic contributions to the binding energy.
The Bethe-Weiz\"acker formula allows to estimate the latter as
\begin{equation}\label{bethe}
E_{_{\rm EM}}=98.25\frac{Z(Z-1)}{A^{1/3}}\aem\,{\rm MeV}.
\end{equation}
If we decompose the proton and neutron masses as~\cite{gasser82}
$m_{({\rm p,n})}=u_3 +b_{({\rm u,d})}m_{\rm
u}+b_{({\rm d,u})}m_{\rm d}+B_{({\rm p,n})}\aem$ where $u_3$ is the
pure QCD approximation of the nucleon mass ($b_{\rm u}$, $b_{\rm d}$
and $B_{({\rm n,p})}/u_3$ being pure numbers), it reduces to
\begin{eqnarray}\label{mass}
m(A,Z)&=&\left(Au_3+E_{_{\rm S}}\right)+
(Zb_{\rm u}+Nb_{\rm d})m_{\rm u}+(Zb_{\rm d}+Nb_{\rm u})m_{\rm d}\nonumber\\
&+&\left(ZB_{\rm p}+NB_{\rm n}+98.25\frac{Z(Z-1)}{A^{1/3}}\,{\rm
MeV}\right)\aem,\label{mAZ}
\end{eqnarray}
with $N=A-Z$, the neutron number. For an
atom, one would have to add the contribution of the electrons,
$Zm_{\rm e}$. This form depends on strong, weak and
electromagnetic quantities. The numerical coefficients $B_{({\rm
n,p})}$ are given explicitly by~\cite{gasser82}
\begin{equation}\label{gl}
B_{\rm p}\aem=0.63\,{\rm MeV}\quad
B_{\rm n}\aem=-0.13\,{\rm MeV}.
\end{equation}
Such estimations were used in the first analysis of the relation between variation of the
constant and the universality of free fall~\cite{damour94a,dvaliZ} but the dependency on the quark mass
is still not well understood and we refer to Refs.~\cite{damourdono,bbn-dimi2,donomass,oklo-14} 
for some attempts to refine this description.

For macroscopic bodies, the mass has also a negative
contribution
\begin{equation}\label{llr8}
  \Delta m(G)=-\frac{G}{2c^2}\int\frac{\rho(\vec r)\rho(\vec r')}{|\vec r-\vec r'|}
\dd^3\vec r\dd^3\vec r'
\end{equation}
from the gravitational binding energy. As a conclusion, from
(\ref{mass}) and (\ref{llr8}), we expect the mass to depend on all the
coupling constant, $m(\aem,\aw,\as,\ag,...)$.\\

We shall discuss this issue in more details in \S~\ref{subsubmass}.

\subsubsection{Relations with cosmology}\label{subseccosmo}

Most constraints on the time variation of the fundamental constants will not be local
and related to physical systems at various epochs of the evolution of the universe.
It follows that the comparison of different constraints requires a full cosmological
model.\\

Our current cosmological model is known as the $\Lambda$CDM (see 
Ref.~\cite{peteruzanbook} for a detailed description).
It is important to recall that its construction relies on 4 main
hypotheses: (H1) a theory of gravity; (H2) a description of the
matter components contained in the Universe and their non-gravitational
interactions; (H3) symmetry hypothesis; and (H4) a hypothesis on the
global structure, i.e. the topology, of the Universe.
These hypotheses are not on the same footing since H1 and H2 refer
to the physical theories. These hypotheses are however not
sufficient to solve the field equations and we must make an
assumption on the symmetries (H3) of the solutions describing our
Universe on large scales while H4 is an assumption on some global
properties of these cosmological solutions, with same local
geometry. But the last two hypothesis are unavoidable
because the knowledge of the fundamental theories is not
sufficient to construct a cosmological model~\cite{jpu-model}.

The $\Lambda$CDM model assumes that gravity is described by general relativity (H1), that
the Universe contains the fields of the standard model of particle
physics plus some dark matter and a cosmological constant, the
latter two having no physical explanation at the moment. It also
deeply involves the Copernican principle as a symmetry hypothesis
(H3), without which the Einstein equations usually cannot been
solved, and assumes most often that the spatial sections are simply
connected (H4). H2 and H3 imply that the description of the standard
matter reduces to a mixture of a pressureless and a radiation
perfect fluids. This model is compatible with all astronomical
data which roughly indicates that $ \Omega_{\Lambda0} \simeq
0.73$, $\Omega_{\rm mat0} \simeq 0.27,$ and $\Omega_{K0} \simeq
0$. Cosmology thus roughly imposes that $|\Lambda_0|\leq H_0^2$,
that is $\ell_\Lambda \leq H_0^{-1} \sim 10^{26}\,\mathrm{m} \sim
10^{41}\,\mathrm{GeV}^{-1}$. 

Hence, the analysis of the cosmological dynamics of the universe
and of its large scale structures requires the introduction of a new constant,
the {\it cosmological constant}, associated with
a recent acceleration of the cosmic expansion,  
that can be introduced by modifying the Einstein-Hilbert
action to
$$
 S_{\rm grav} = \frac{c^3}{16\pi G}\int\sqrt{-g}(R-2\Lambda)\dd^4x.
$$
Note however that it is disproportionately
large compared  to the natural scale fixed by the Planck length
\begin{equation}
 \rho_{\Lambda_0}\sim10^{-120}M_{\rm
 Pl}^4\sim10^{-47}\,\mathrm{GeV}^4.
\end{equation}
Classically, this value is no problem but it was pointed out that at
the quantum level, the vacuum energy should scale as $M^4$,
where $M$ is some energy scale of high-energy physics. 
In such a case, there is a discrepancy of 60-120
order of magnitude between the cosmological conclusions
and the theoretical expectation. This is the {\it
cosmological constant problem}~\cite{weinberg89}.

\begin{table}[t]
\begin{center}
{\small
\begin{tabular}{p{6cm}ll}
 \hline\hline
 Parametre & Symbol & Value \\
 \hline
  Reduced Hubble constant & $h$ & $0.73(3)$ \\
  baryon-to-photon ratio & $\eta = n_\baryon/n_\gamma$ & $6.12(19)\times10^{-10}$ \\
  Photon density  & $\Omega_\gamma h^2$ & $2.471\times10^{-5}$\\
  Dark matter density  & $\Omega_{\rm CDM}h^2$ & $0.105(8)$ \\
  Cosmological constant & $\Omega_\Lambda$& $0.73(3)$\\
  Spatial curvature & $\Omega_K$& $0.011(12)$\\
  Scalar modes amplitude & $Q$ & $(2.0\pm0.2)\times10^{-5}$  \\
  Scalar spectral index & $n_S$ & $0.958(16)$ \\
 \hline 
  Neutrino density  & $\Omega_\nu h^2$ & $(0.0005-0.023)$ \\
  Dark energy equation of state & $w$ & $-0.97(7)$ \\
  Scalar running spectral index & $\alpha_S$ & $-0.05\pm0.03$\\
  Tensor-to-scalar ratio & T/S&  $<0.36$\\
  Tensor spectral index & $n_T$& $<0.001$ \\
  Tensor running spectral index & $\alpha_T$&  ?\\
  \hline
    Baryon density  & $\Omega_\baryon h^2$ & $0.0223(7)$\\
  \hline\hline
\end{tabular}
\caption{\it Main cosmological parameters in the standard $\Lambda$-CDM model. There
are 7 main parameters (because $\sum\Omega_i=0$) to which one can add 6 more to include
dark energy, neutrinos and gravity waves. Note that often the spatial curvature is set to $\Omega_K=0$.}
\label{tab-cosmo}
}
\end{center}
\end{table}

Two solutions are considered. Either one accepts such a constant
and such a fine-tuning and tries to explain it on anthropic ground. 
Or, in the same spirit as Dirac, one interprets it as an indication
that our set of cosmological hypotheses have to be extended,
by either abandoning the Copernican principle~\cite{uce} or
by modifying the local physical laws (either gravity or the 
matter sector). The way to introduce such new physical
degrees of freedom were classified in Ref.~\cite{jpu-revu3}.
In that latter approach, the tests of the constancy of the fundamental
constants are central since they can reveal the coupling of this
new degree of freedom to the standard matter fields. Note
however that the cosmological data still favor a pure cosmological
constant.

Among all the proposals quintessence involves a scalar 
field rolling down a runaway potential
hence acting as a fluid with an effective equation
of state in the range $- 1 \leq w \leq 1$ if the field is
minimally coupled. It was proposed that the quintessence field is also 
the dilaton~\cite{gasperini02,riazuelo2001,uzan99}.
The same scalar field then drives the time variation of the cosmological constant
and of the gravitational constant and it has the property to also have
tracking solutions~\cite{uzan99}.
One of the underlying motivation to replace the cosmological
constant by a scalar field comes from superstring models in which
any dimensionful parameter is expressed in terms of the string
mass scale and the vacuum expectation value of a scalar field.
However, the requirement of slow roll (mandatory to have a
negative pressure) and the fact that the quintessence field
dominates today imply, if the minimum of the potential is zero,
that it is very light, roughly of order $m\sim 10^{- 33}\,{\rm
eV}$~\cite{carroll98}. 

Such a light field can lead to observable violations
of the universality of free fall if it is non-universally coupled
to the matter fields.
Carroll~\cite{carroll98} considered the effect of the coupling of this very
light quintessence field to ordinary matter via a coupling to
the electromagnetic field as
$\phi F^{\mu\nu}\widetilde F_{\mu\nu}$. Chiba and Kohri~\cite{chiba2001} also
argued that an ultra-light quintessence field induces a time variation
of the coupling constant if it is coupled to ordinary matter and
studied a coupling of the form $\phi F^{\mu\nu}F_{\mu\nu}$, as
e.g. expected from Kaluza-Klein theories (see below). This
was generalized to quintessence models with  a couplings of the form 
$Z(\phi) F^{\mu\nu}F_{\mu\nu}$~\cite{anchor03,copQ,doran,lee,lee2,marra,parkinson,xyz3}
and then to models of runaway dilaton~\cite{damourrunaway,damourrunaway2} inspired by string theory
(see \S~\ref{subsub1}).
The evolution of the scalar field drives both the acceleration of the universe
at late time and the variation of the constants.
As pointed in Refs.~\cite{chiba2001,dvaliZ,wetterich02} such models
are extremely constrained from the bound on the universality of free-fall (see \S~\ref{subsec22}).

We thus have two ways of investigation
\begin{itemize}
 \item The field driving the time variation of the fundamental constants does
not explain the acceleration of the universe (either it does not dominate the
matter content today or its equation of state is not negative enough). In such
a case, the variation of the constants is disconnected from
the dark energy problem. Cosmology allows to determine the
dynamics of this field during the whole history of the universe and
thus to compare local constraints and cosmological constraints.
An example is given by scalar-tensor theories (see \S~\ref{subsecST}) for
which one can compare e.g. primordial nucleosynthesis to local
constraints~\cite{bbn-Gpichon}. In such a situation, one
should however take into account the effect of the variation
of the constants on the astrophysical observations since it
can affect local physical processes and bias e.g. the luminosity
of supernovae and indirectly modify the distance luminosity-redshift
relation derived from these observations~\cite{barrow01,cmb-G1}.
\item The field driving the time variation of the fundamental constants is
also responsible for the acceleration of the universe. It follows that
the dynamics of the universe, the level of variation
of the constants and the other deviations from general relativity are
connected~\cite{msu} so that the study of the variation of the
constants can improve the reconstruction of the equation state of the
dark energy~\cite{avelino06,doran,nunes09,parkinson}. 
\end{itemize}

In conclusion, cosmology seems to require a new constant. It also provides
a link between the microphysics and cosmology, as forseen by Dirac. The
tests of fundamental constants can discriminate between various explanations
of the acceleration of the universe. When a model is specified, cosmology also
allows to set stringer constraints since it relates observables that cannot
be compared otherwise.

\section{Experimental and observational constraints}\label{section3}

This section focuses on the experimental and observational
constraints on the non-gravitational constants, that is assuming
$\ag$ remains constant.

\begin{figure}[hptb]
  \def\epsfsize#1#2{0.5#1}
  \centerline{\includegraphics[scale=0.55]{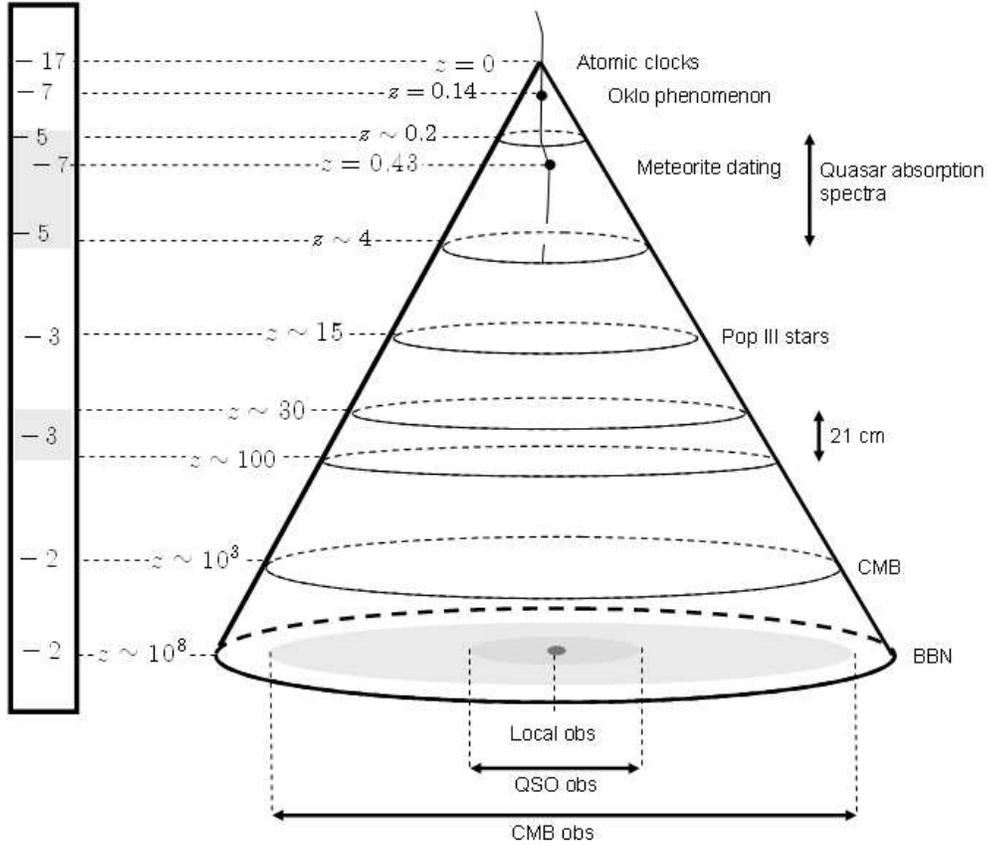}}
  \vskip1cm
  \centerline{\includegraphics[scale=0.7]{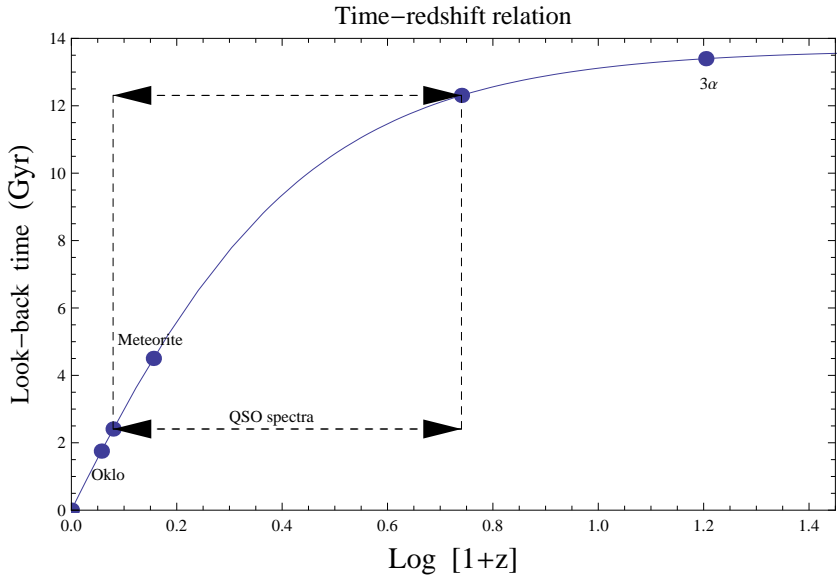}}
  \caption{\it (Top): Summary of the systems that have been used to probe the constancy of
  the fundamental constants and their position in a space-time diagram in which the cone
  represents our past light cone. The shaded areas represents the comoving space probed
  by different tests.
  (Bottom): The look-back time-redshift relation for the standard $\Lambda$CDM model.}
  \label{fig-systems}
\end{figure}

The various physical systems that have been considered can be
classified in many ways. We can classify them according to their
look-back time and more precisely their space-time position
relative to our actual position. This is a summarized on
Fig.~\ref{fig-systems}. Indeed higher redshift systems offer the
possibility to set constraints on an larger time scale, but this
is at the expense of usually involving other parameters such as the
cosmological parameters. This is in particular the case of the cosmic
microwave background or of primordial nucleosynthesis. The systems
can also be classified in terms of the physics they involve. For
instance, atomics clocks, quasar absorption spectra and the cosmic
microwave background require only to use quantum electrodynamics
to draw the primary constraints while the Oklo phenomenon,
meteorites dating and nucleosynthesis require nuclear physics.

\begin{table}[t]
\begin{center}
{\small
\begin{tabular}{p{2.8 cm}lll}
 \hline\hline
 System  & Observable & Primary constraints  & Other hypothesis \\
 \hline
 Atomic clock      & $\delta\ln\nu$           &  $g_i,\aem,\mu$ & - \\
 Oklo phenomenon   & isotopic ratio           &  $E_r$          & geophysical model \\
 Meteorite dating  & isotopic ratio           &  $\lambda$      &  -\\
 Quasar spectra    & atomic spectra           &  $g_{\rm p},\mu,\aem$ & cloud physical properties\\
 Stellar physics   & element abundances       &  $B_D$          & stellar model  \\
 21 cm             & $T_b/T_{\rm CMB}$                    &  $g_{\rm p},\mu,\aem$ & cosmological model\\
 CMB               & $\Delta T/T$                      &  $\mu,\aem$     & cosmological model \\
 BBN               & light element abundances &  $Q_{\rm np},\tau_{\rm n},m_{\rm e},m_{\rm N},\aem,B_D$
          & cosmological model     \\
 \hline\hline
\end{tabular}
\caption{\it Summary of the systems considered to set constraints
on the variation of the fundamental constants. We summarize the
observable quantities, the primary constants used to interpret the
data and the other hypothesis required for this interpretation.}
\label{tab-sum}
}
\end{center}
\end{table}

For any system, setting constraints goes through several steps.
First we have some observable quantities from which we can draw
constraints on primary constants, which may not be fundamental
constants (e.g. the BBN parameters, the lifetime of
$\beta$-decayers,...). This primary parameters must then be related to
some fundamental constants such as masses and couplings. In a last
step, the number of constants can be reduced by relating them in
some unification schemes. Indeed each step requires a specific
modelisation and hypothesis and has its own limitations. This is
summarized on Table~\ref{tab-sum}.

\subsection{Atomic clocks}\label{subsec31}

\subsubsection{Atomic spectra and constants}\label{subsec31-e.h}

The laboratory constraints on the time variation of fundamental
constants are obtained by comparing the long-term behavior of
several oscillators and rely on frequency measurements. The atomic
transitions have various dependencies in the fundamental
constants. For instance, for the hydrogen atom, the gross, fine
and hyperfine-structures are roughly given by
$$
 2p-1s:\,\, \nu\propto cR_\infty,\qquad
 2p_{3/2}-2p_{1/2}:\,\, \nu\propto cR_\infty\aem^2,\qquad
 1s:\,\,\propto cR_\infty\aem^2 g_{\rm p}\bar\mu,
$$
respectively, where the Rydberg constant set the dimension. $g_{\rm p}$ is the proton gyromagnetic
factor and $\bar\mu=m_{\rm e}/m_{\rm p}$. In the non-relativistic approximation, the transitions of
all atoms have similar dependencies but two effects have to be taken
into account. First, the hyperfine-structures involve a
gyromagnetic factor $g_i$ (related to the nuclear magnetic moment
by $\mu_i=g_i\mu_{\rm N}$, with $\mu_{\rm N}= e\hbar/2m_{\rm
p}c$) which are different for each nuclei. Second, relativistic
corrections (including the Casimir contribution) which also
depend on each atom (but also on the type of the transition) can
be included through a multiplicative function $F_{\rm rel}(\aem)$.
It has a strong dependence on the atomic number $Z$, which can be
illustrated on the case of alkali atoms, for which
$$
 F_{\rm rel}(\aem) =\left[1-(Z\aem)^2\right]^{-1/2}\left[1-\frac43(Z\aem)^2\right]^{-1}
 \simeq 1 +\frac{11}{6}(Z\aem)^2.
$$
The developments of highly accurate atomic clocks using different
transitions in different atoms offer the possibility to test a
variation of various combinations of the fundamental
constants.\\

It follows that at the lowest level of description, we can
interpret all atomic clocks results in terms of the g-factors of
each atoms, $g_i$, the electron to proton mass ration $\mu$ and
the fine-structure constant $\aem$. We shall thus parameterize the
hyperfine and fine-structures frequencies as follows.

The hyperfine frequency in a given electronic state of an
alkali-like atom, such as $^{133}$Cs, $^{87}$Rb, $^{199}$Hg$^+$, is
\begin{equation}\label{e.hf}
 \nu_{\rm hfs} \simeq R_\infty c \times A_{\rm hfs} \times g_i
 \times \aem^2\times\bar\mu \times F_{\rm hfs}(\alpha)
\end{equation}
where $g_i=\mu_i/\mu_{\rm N}$ is the nuclear $g$ factor. $A_{\rm hfs}$ is a numerical
factor depending on each particular atom and we have set $F_{\rm
rel}=F_{\rm hfs}(\alpha)$. Similarly, the frequency of an
electronic transition is well-approximated by
\begin{equation}\label{e.elec}
 \nu_{\rm elec} \simeq R_\infty c \times A_{\rm elec} \times F_{\rm
 elec}(Z,\alpha),
\end{equation}
where, as above, $A_{\rm elec}$ is a numerical factor depending on
each particular atom and $F_{\rm elec}$ is the function accounting
for relativistic effects, spin-orbit couplings and many-body
effects. Even though an electronic transition should also include
a contribution from the hyperfine interaction, it is generally
only a small fraction of the transition energy and thus should not
carry any significant sensitivity to a variation of the
fundamental constants.

The importance of the relativistic corrections was probably first
emphasized in Ref.~\cite{clock-prestage} and their computation
through relativistic $N$-body calculations was carried out for
many transitions in
Refs.~\cite{kappa-dzuba2,kappa-dzuba3,kappa-dzuba,kappa-flamb}.
They can be characterized by introducing the sensitivity of the
relativistic factors to a variation of $\aem$,
\begin{equation}\label{clock-sensitivity}
 \kappa_\alpha \equiv \frac{\partial \ln F}{\partial \ln\aem}.
\end{equation}
Table~\ref{tab0} summarizes the values of some of them, as computed
in Refs.~\cite{kappa-dzuba,kappa-tedesco}. Indeed a reliable
knowledge of these coefficients at the 1\% to 10\% level is
required to deduce limits to a possible variation of the
constants. The interpretation of the spectra in this context
relies, from a theoretical point of view, only on
quantum electrodynamics (QED), a theory which is well tested
experimentally~\cite{qed-karshen1} so that we can safely obtain
constraints on $(\aem,\mu,g_i)$, still keeping in mind that the
computation of the sensitivity factors required numerical $N$-body
simulations.\\

\begin{table}[t]
\begin{center}
{\small
\begin{tabular}{p{4.0 cm}cc}
 \hline\hline
 Atom  & Transition & sensitivity $\kappa_\alpha$ \\
 \hline
 $^{1}$H         & $1s-2s$                    &  0.00 \\
 $^{87}$Rb       &       hf                   & 0.34   \\
 $^{133}$Cs      & ${}^2S_{1/2}(F=2)-(F=3)$   & 0.83   \\
 $^{171}$Yb$^+$  & ${}^2S_{1/2}-{}^2D_{3/2}$  & 0.9    \\
 $^{199}$Hg$^+$  & ${}^2S_{1/2}-{}^2D_{5/2}$  & $-3.2$   \\
 $^{87}$Sr       & ${}^1S_0-{}^3P_0$          & 0.06       \\
 $^{27}$Al$^+$   & ${}^1S_0-{}^3P_0$          & 0.008    \\
 \hline\hline
\end{tabular}
\caption{\it Sensitivity of various transitions on a variation of
the fine-structure constant.} \label{tab0}
}
\end{center}
\end{table}

From an experimental point of view, various combinations of clocks
have been performed. It is important to analyze as much species as
possible in order to rule-out species-dependent systematic
effects. Most experiments are based on a frequency comparison to
caesium clocks. The hyperfine splitting frequency between the
$F=3$ and $F=4$ levels of its ${}^{2}S_{1/2}$ ground state at
9.192~GHz has been used for the definition of the second since
1967. One limiting effect, that contributes mostly to the
systematic uncertainty, is the frequency shift due to cold
collisions between the atoms. On this particular point, clocks
based on the hyperfine frequency of the ground state of the
rubidium at 6.835~GHz, are more favorable.

\subsubsection{Experimental constraints}

\begin{table}[t]
\begin{center}
{\small
\begin{tabular}{p{2.0 cm}crrc}
\hline\hline
 Clock 1 & Clock 2             & Constraint  (yr$^{-1}$) & Constants dependence &  Reference \\
               & $\frac{\dd}{\dd t} \ln\left(\frac{\nu_{\rm clock_1}}{\nu_{\rm clock_2}}
                  \right)\quad$ &  &  \\
 \hline\hline
 $^{87}$Rb & $^{133}$Cs & $(0.2\pm7.0)\times10^{-16}$ &
                          $\frac{g_{\rm Cs}}{g_{\rm Rb}}\aem^{0.49}$ & \cite{clock-marion03} \\
 $^{87}$Rb & $^{133}$Cs & $(-0.5\pm5.3)\times10^{-16}$
                          & & \cite{clock-bize05}\\
 $^{1}$H & $^{133}$Cs & $(-32\pm63)\times10^{-16}$
                         & $g_{\rm Cs}\bar\mu\aem^{2.83}$ & \cite{clock-fisher04}\\
 $^{199}$Hg$^+$ & $^{133}$Cs & $(0.2\pm7)\times10^{-15}$
                         & $g_{\rm Cs}\bar\mu\aem^{6.05}$& \cite{clock-bize03}\\
 $^{199}$Hg$^+$ & $^{133}$Cs & $(3.7\pm3.9)\times10^{-16}$
                         &  & \cite{clock-fortier07}\\
 $^{171}$Yb$^+$ & $^{133}$Cs & $(-1.2\pm4.4)\times10^{-15}$ & $g_{\rm Cs}\bar\mu\aem^{1.93}$
 & \cite{clock-peik04}\\
 $^{171}$Yb$^+$ & $^{133}$Cs & $(-0.78\pm1.40)\times10^{-15}$ & & \cite{clock-peik06}\\
 $^{87}$Sr  & $^{133}$Cs &     $(-1.0\pm1.8)\times10^{-15}$       &  $g_{\rm Cs}\bar\mu\aem^{2.77}$& \cite{clock-blatt}\\
 $^{87}$Dy  & $^{87}$Dy &   $(-2.7\pm2.6)\times10^{-15}$         &$\aem$ & \cite{clock-cingoz}\\
 $^{27}$Al$^+$ & $^{199}$Hg$^+$ &   $(-5.3\pm7.9)\times10^{-17}$   & $\aem^{-3.208}$
  & \cite{clock-rosen08}\\
\hline\hline
\end{tabular}
\caption{\it Summary of the constraints obtained from the
comparisons of atomic clocks. For each constraint on the relative
drift of the frequency of the two clocks, we provide the
dependence in the various constants, using the numbers of
Table~\ref{tab0}.} \label{tab1}
}
\end{center}
\end{table}

\begin{figure}[hptb]
  \def\epsfsize#1#2{0.5#1}
  \centerline{\includegraphics[scale=0.7]{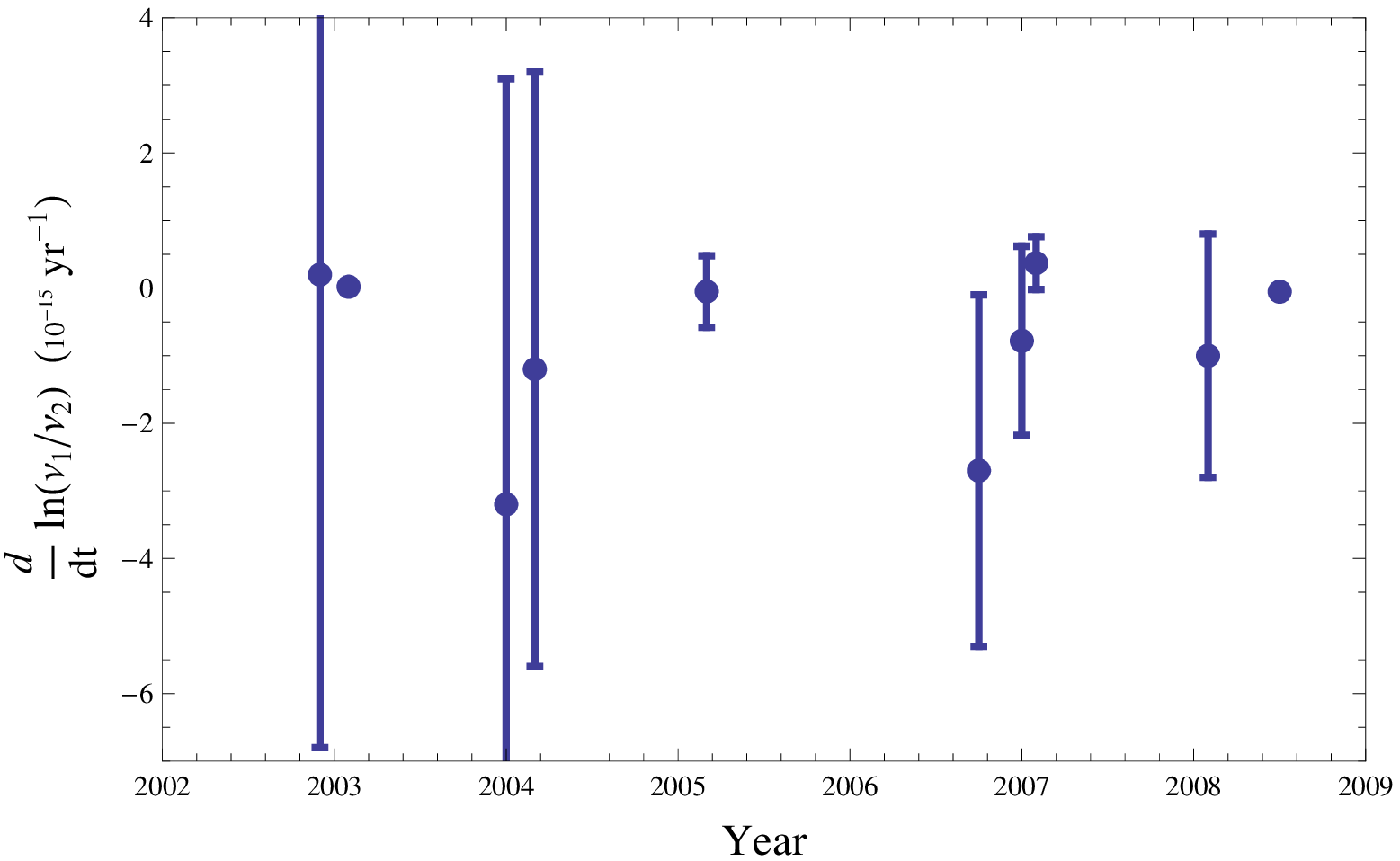}}
  \caption{\it Evolution of the comparison of different atomic clocks summarized in Table~\ref{tab1}.}
  \label{fig-clock}
\end{figure}

We present the latest results that have been obtained and refer to
\S~III.B.2 of FCV~\cite{jpu-revue} for earlier studies. They all
rely on the developments of new atomic clocks, with the primarily
goal to define better frequency standards.
\begin{itemize}
 \item {\em Rubidium}: The comparison of the
 hyperfine frequencies of the rubidium and caesium in their electronic
 ground state between 1998 and 2003, with an accuracy of order
 $10^{-15}$, leads to the constraint~\cite{clock-marion03}
  \begin{equation}
  \frac{\dd}{\dd t}\ln\left(\frac{\nu_{\rm Rb}}{\nu_{\rm Cs}}\right)
    = (0.2\pm7.0)\times10^{-16}\,\rm{yr}^{-1}.
  \end{equation}
  With one more year of experiment, the constraint dropped
  to~\cite{clock-bize05}
  \begin{equation}
  \frac{\dd}{\dd t}\ln\left(\frac{\nu_{\rm Rb}}{\nu_{\rm Cs}}\right)
    = (-0.5\pm5.3)\times10^{-16}\,\rm{yr}^{-1}.
  \end{equation}
  From Eq.~(\ref{e.hf}), and using the values of the sensitivities
  $\kappa_\alpha$, we deduce that this comparison constrains
  $$
   \frac{\nu_{\rm Cs}}{\nu_{\rm Rb}}\propto\frac{g_{\rm Cs}}{g_{\rm Rb}}\aem^{0.49}.
  $$

 \item{\em Atomic hydrogen}: The $1s-2s$
 transition in atomic hydrogen was compared tp the ground state hyperfine
 splitting of caesium~\cite{clock-fisher04} in 1999 and 2003,
 setting an upper limit on the variation of $\nu_{\rm H}$ of
 $(-29\pm57)$~Hz within 44~months. This can be translated in a
 relative drift
 \begin{equation}\label{clock-H}
  \frac{\dd}{\dd t}\ln\left(\frac{\nu_{\rm H}}{\nu_{\rm Cs}}\right)
    = (-32\pm63)\times10^{-16}\,\rm{yr}^{-1}.
  \end{equation}
  Since the relativistic correction for the atomic hydrogen
  transition nearly vanishes, we have $\nu_{\rm H}\sim R_\infty$
  so that
  $$
  \frac{\nu_{\rm Cs}}{\nu_{\rm H}}\propto g_{\rm Cs}\bar\mu\aem^{2.83}.
  $$

  \item {\em Mercury}: The $^{199}$Hg$^+$ ${}^2S_{1/2}-
  {}^2D_{5/2}$ optical transition has a high sensitivity to $\aem$
  (see Table~\ref{tab0}) so that it is well suited to test its variation.
  The frequency of the $^{199}$Hg$^+$ electric quadrupole transition at 282~nm
  was thus compared to the ground state hyperfine transition of caesium
  during a two year period, which lead to~\cite{clock-bize03}
  \begin{equation}
   \frac{\dd}{\dd t}\ln\left(\frac{\nu_{\rm Hg}}{\nu_{\rm
   Cs}}\right)=
    (0.2\pm7)\times10^{-15}\,\rm{yr}^{-1}.
  \end{equation}
  This was improved by a comparison over a 6 year period~\cite{clock-fortier07}
  to get
  \begin{equation}
   \frac{\dd}{\dd t}\ln\left(\frac{\nu_{\rm Hg}}{\nu_{\rm
   Cs}}\right)=
    (3.7\pm3.9)\times10^{-16}\,\rm{yr}^{-1}.
  \end{equation}
  While $\nu_{\rm Cs}$ is still given by Eq.~(\ref{e.hf}), $\nu_{\rm
  Hg}$ is given by Eq.~(\ref{e.elec}). Using the sensitivities of
  Table~\ref{tab0}, we conclude that this comparison test the
  stability of
  $$
 \frac{\nu_{\rm Cs}}{\nu_{\rm Hg}}\propto g_{\rm Cs}\bar\mu\aem^{6.05}.
  $$

  \item {\em Ytterbium}: The ${}^2S_{1/2}-
  {}^2D_{3/2}$ electric quadrupole transition at 688~THz of $^{171}$Yb$^+$
  was compared to  the ground state hyperfine transition of
  caesium. The constraint of Ref.~\cite{clock-peik04} was updated,
  after comparison over a six year period, which lead to~\cite{clock-peik06}
  \begin{equation}
   \frac{\dd}{\dd t}\ln\left(\frac{\nu_{\rm Yb}}{\nu_{\rm
   Cs}}\right)=
    (-0.78\pm1.40)\times10^{-15}\,\rm{yr}^{-1}.
  \end{equation}
 Proceeding as previously, this tests the stability of
  $$
 \frac{\nu_{\rm Cs}}{\nu_{\rm Yb}}\propto g_{\rm Cs}\bar\mu\aem^{1.93}.
  $$

  \item {\em Strontium}:  The comparison of the ${}^1S_0-{}^3P_0$
  transition in neutral ${}^{87}$Sr with a caesium clock was
  performed in three independent laboratories. The combination of
  these three experiments~\cite{clock-blatt} leads to the
  constraint
  \begin{equation}
   \frac{\dd}{\dd t}\ln\left(\frac{\nu_{\rm Sr}}{\nu_{\rm Cs}}\right)=
    (-1.0\pm1.8)\times10^{-15}\,\rm{yr}^{-1}.
  \end{equation}
 Proceeding as previously, this tests the stability of
  $$
 \frac{\nu_{\rm Cs}}{\nu_{\rm Sr}}\propto g_{\rm Cs}\bar\mu\aem^{2.77}.
  $$

  \item {\em Atomic dyprosium}: It was suggested in Refs.~\cite{kappa-dzuba,kappa-dzuba3}
  (see also Ref.~\cite{Dy-df} for a computation of the transition amplitudes of the low
  states of dyprosium)
  that the electric dipole (E1) transition between two nearly degenerate opposite-parity
  states in atomic dyprosium should be highly sensitive to the variation of $\aem$.
  It was then demonstrated~\cite{nguyen04} that a constraint of the order of
  $10^{-18}$/yr can be reached. The frequencies of nearly of two isotopes of dyprosium
  were monitored over a 8 months period~\cite{clock-cingoz} showing
  that the frequency variation of the 3.1-MHz transition in $^{163}$Dy and the 235-MHz transition in
  $^{162}$Dy are 9.0$\pm$6.7 Hz/yr and -0.6$\pm$6.5 Hz/yr, respectively. These provides
  the constraint
   \begin{equation}\label{clock-dypro}
  \frac{\dot\aem}{\aem}
  =(-2.7\pm2.6)\times 10^{-15}\,\rm{yr}^{-1},
  \end{equation}
  at 1$\sigma$ level, without any assumptions on the constancy of other fundamental constants.  

  \item {\em Aluminium and mercury single-ion optical clocks}:
  The comparison of the ${}^1S_{0}- {}^3P_{0}$ transition in
  ${}^{27}$Al$^+$ and ${}^2S_{1/2}-{}^2D_{5/2}$ in ${}^{199}$Hg$^+$
  over a year allowed to set the constraint~\cite{clock-rosen08}
  \begin{equation}
   \frac{\dd}{\dd t}\ln\left(\frac{\nu_{\rm Al}}{\nu_{\rm
   Hg}}\right)=
    (-5.3\pm7.9)\times10^{-17}\,\rm{yr}^{-1}.
  \end{equation}
  Proceeding as previously, this tests the stability of
  $$
 \frac{\nu_{\rm Hg}}{\nu_{\rm Al}}\propto \aem^{-3.208},
  $$
  which directly set the constraint
  \begin{equation}\label{clock-bound1}
  \frac{\dot\aem}{\aem}
  =(-1.6\pm2.3)\times10^{-17}\,\rm{yr}^{-1},
  \end{equation}
  since it depends only on $\aem$.
\end{itemize}
While the constraint~(\ref{clock-bound1}) was obtained directly
from the clock comparison, the other studies need to be combined
to disentangle the contributions of the various constants. As an
example, we first use the bound~(\ref{clock-bound1}) on $\aem$, we
can then extract the two following bounds
\begin{equation}\label{clock-gtest}
 \frac{\dd}{\dd t}\ln\left(\frac{g_{\rm Cs}}{g_{\rm Rb}}\right)=
    (0.48\pm6.68)\times10^{-16}\,\rm{yr}^{-1}, \qquad
 \frac{\dd}{\dd t}\ln\left(g_{\rm Cs}\bar\mu\right)=
    (4.67\pm5.29)\times10^{-16}\,\rm{yr}^{-1},
\end{equation}
on a time scale of a year. We cannot lift the degeneracies further
with this clock comparison since that would require a constraint
on the time variation of $\mu$. All these constraints are summarized
in Table~\ref{tab1} and Fig.~\ref{fig-clock}.\\

A solution is to consider diatomic molecules since, as first
pointed out by Thomson~\cite{mu-theorie}, molecular lines can
provide a test of the variation of $\mu$. The energy difference
between two adjacent rotational levels in a diatomic molecule is
inversely proportional to $M r^{-2}$, $r$ being the bond length
and $M$ the reduced mass, and the vibrational transition of the
same molecule has, in first approximation, a $\sqrt{M}$
dependence. For molecular hydrogen $M=m_{\rm p}/2$ so that the
comparison of an observed vibro-rotational spectrum with a
laboratory spectrum gives an information on the variation
of $m_{\rm p}$ and $m_{\rm n}$. Comparing pure rotational
transitions with electronic transitions gives a measurement of
$\mu$. It follows that the frequency of vibro-rotation transitions
is, in the Born-Oppenheimer approximation, of the form
\begin{equation}\label{mu1}
\nu\simeq E_I\left(c_{_{\rm elec}} +c_{_{\rm vib}}\sqrt{\bar\mu}
+c_{_{\rm rot}}\bar\mu\right)
\end{equation}
where $c_{_{\rm elec}}$, $c_{_{\rm vib}}$ and $c_{_{\rm rot}}$ are
some numerical coefficients.

The comparison of the vibro-rotational transition in the molecule
SF6 was compared to a caesium clock over a two-year period,
leading to the constraint~\cite{clock-mu}
\begin{equation}
   \frac{\dd}{\dd t}\ln\left(\frac{\nu_{\rm SF6}}{\nu_{\rm
   Cs}}\right)=
    (1.9\pm0.12\pm2.7)\times10^{-14}\,\rm{yr}^{-1},
\end{equation}
where the second error takes into account uncontrolled
systematics. Now, using again Table~\ref{tab0}, we deduce that
$$
 \frac{\nu_{\rm SF6}}{\nu_{\rm Cs}}\propto \bar\mu^{1/2}\aem^{-2.83}(g_{\rm Cs}\bar\mu)^{-1}.
$$
It can be combined with the constraint~(\ref{clock-H}) which
enjoys the same dependence to caesium to infer that
\begin{equation}\label{clock-bound2}
   \frac{\dot\mu}{\mu}=(-3.8\pm5.6)\times10^{-14}\,\rm{yr}^{-1}.
\end{equation}
Combined with Eq.~(\ref{clock-gtest}), we can thus obtain
independent constrains on the time variation of $g_{\rm Cs}$,
$g_{\rm Rb}$ and $\mu$.

\subsubsection{Physical interpretation}\label{sec-clock-phy}

The theoretical description must be pushed further if ones wants
to extract constraints on constant more fundamental than the
nuclear magnetic moments. This requires to use quantum
chromodynamics. In particular, it was argued than within this
theoretical framework, one can relate the nucleon $g$-factors in
terms of the quark mass and the QCD scale~\cite{kappa-flamb}.
Under the assumption of a unification of the three
non-gravitational interaction (see \S~\ref{subsec22}), the
dependence of the magnetic moments on the quark masses was
investigated in Ref.~\cite{kappa-tedesco}. The magnetic moments,
or equivalently the $g$-factors, are first related to the ones of
the proton and a neutron to derive a relation of the form
$$
 g \propto g_{\rm p}^{a_{\rm p}}g_{\rm n}^{a_{\rm n}}.
$$
Refs.~\cite{kappa-flamb,kappa-tedesco} argued that these
$g$-factors mainly depend on the light quark mass $m_{\rm
q}=\frac12(m_{\rm u}+m_{\rm d})$ and $m_{\rm s}$, respectively for
the up, down and strange quarks, that is in terms of $X_{\rm
q}=m_{\rm q}/\Lambda_{\rm QCD}$ and $X_{\rm s}=m_{\rm
s}/\Lambda_{\rm QCD}$. Using a chiral perturbation theory, it was
deduced that
$$
 g_{\rm p}\propto X_{\rm q}^{-0.087}X_{\rm s}^{-0.013},\qquad
 g_{\rm n}\propto X_{\rm q}^{-0.118}X_{\rm s}^{0.0013},
$$
so that for a hyperfine transition
$$
 \nu_{\rm hfs} \propto \aem^{2+\kappa_\alpha}X_{\rm q}^{\kappa_{\rm q}}X_{\rm s}^{\kappa_{\rm s}}
 \bar\mu.
$$
Both coefficients can be computed, leading to the possibility to
draw constraints on the independent time variation of $X_{\rm q}$,
$X_{\rm s}$ and $X_{\rm e}$.

To simplify, we may assume that $X_{\rm q}\propto X_{\rm s}$,
which is motivated by the Higgs mechanism of mass generation, so
that the dependence in the quark masses reduces to
$\kappa=\frac12(\kappa_{\rm q}+\kappa_{\rm s})$. For instance, we
have
$$
 \kappa_{\rm Cs} =0.009,\qquad
 \kappa_{\rm Rb} =-0.016,\qquad
 \kappa_{\rm H} = -0.10.
$$
For hyperfine transition, one further needs to take into account
the dependence in $\mu$ that can be described~\cite{clock-muq} by
$$
 m_{\rm p}\sim 3\Lambda_{\rm QCD} X_{\rm q}^{0.037} X_{\rm
 s}^{0.011},
$$
so that the hyperfine frequencies behaves as
$$
 \nu_{\rm hfs}\propto \aem^{2+\kappa_\alpha} X_{\rm q}^{\kappa
 -0.048} X_{\rm e},
$$
in the approximation $X_{\rm q}\propto X_{\rm s}$ and where
$X_{\rm e}\equiv m_{\rm e}/\Lambda_{\rm QCD}$. This allows to get
independent constraints on the independent time variation of
$X_{\rm e}$, $X_{\rm q}$ and $\aem$. Indeed, these constraints are
model-dependent and, as an example, Table~III of
Ref.~\cite{kappa-tedesco} compares the values of the sensitivity
$\kappa$ when different nuclear effects are considered. For
instance, it can vary from 0.127, 0.044 to 0.009 for the caesium
according to whether one includes only valence nucleon,
non-valence non-nucleon or effect of the quark mass on the
spin-spin interaction. It is thus a very promising framework which
still needs to be developed and the accuracy of which must be
quantified in details.

%

\subsubsection{Future evolutions}

Further progresses in a near future are expected mainly through
three types of developments:
\begin{itemize}
 \item{\em New systems}: Many new systems with enhanced
 sensitivity~\cite{dzubapolo,dzubapolo2,mu-N2,th4,yb2}
 to some fundamental constants have recently been
 proposed. Other atomic systems are considered, such as e.g.
 the hyperfine transitions in the electronic ground state of
 cold, trapped, hydrogen-like highly charged
 ions~\cite{hci00,clock-further6,clock-further2}, or ultra-cold atom and molecule
 systems near the Feshbach resonances~\cite{clock-further4}, where the
 scattering length is extremely sensitive to $\mu$.

  Concerning diatomic molecules, it was shown that this
  sensitivity can be enhanced in transitions between narrow close
  levels of different nature~\cite{clock-further1}. In such transitions, the fine
  structure mainly depends on the fine-structure constant,
  $\nu_{\rm fs}\sim (Z\aem)^2R_\infty c$, while the vibrational
  levels depend mainly on the electron-to-proton mass ratio and
  the reduced mass of the molecule, $\nu_{\rm v}\sim M_r^{-1/2}
  \bar\mu^{1/2}R_\infty c$. There could be a cancellation between the
  two frequencies when $\nu=\nu_{\rm hf}-n\nu_{\rm v}\sim0$ with
  $n$ a positive integer. It follows that $\delta\nu/\nu$ will be
  proportional to $K=\nu_{\rm hf}/\nu$ so that the sensitivity to
  $\aem$ and $\mu$ can be enhanced for these particular
  transitions. A similar effect between with hyperfine-structures,
  for which the sensitivity to $\aem$ can reach 600 for instance
  for ${}^{139}$La${}^{32}$S or silicon monobrid~\cite{silicon}
  that allows to constrain $\aem\bar\mu^{-1/4}$.

  Nuclear transitions, such as an optical clock based on a very narrow ultraviolet nuclear
  transition between the ground and first excited states in the
  ${}^{229}$Th, are also under consideration. Using a Walecka model for the nuclear potential,
  it was concluded~\cite{clock-further6} that the sensitivity of
  the transition to the fine-structure constant and quark mass was
  typically
  $$
  \frac{\delta\omega}{\omega}\sim 10^5\left(4\frac{\delta\aem}{\aem}
  + \frac{\delta X_{\rm q}}{X_{\rm q}} - 10 \frac{\delta X_{\rm s}}{X_{\rm
  s}}\right),
  $$
 which roughly provides a 5 order of magnitude amplification, which can
 lead to a constraint at the level of $10^{-24}\,{\rm yr}^{-1}$
 on the time variation of $X_{\rm q}$. Such
 a method is promising and would offer different sensitivities to
 systematic effects compared to atomic clocks. However, this
 sensibility is not clearly established since different
 nuclear calculations do not agree~\cite{berengutTh,hayes}.
 
 \item{\em Atomic clocks in space (ACES)}: An improvement of at least an
 order of magnitude on current constraints can be achieved in
 space with the PHARAO/ACES project~\cite{reynaud09,clock-ACES} of the
 European Spatial Agency. PHARAO (Projet d'Horloge Atomique par Refroidissement
 d'Atomes en Orbite) combines laser cooling techniques and a
 microgravity environment in a satellite orbit. It aims at
 achieving time and frequency transfer with stability better than
 $10^{-16}$.

  The SAGAS (Search for anomalous gravitation using atomic sensor)
  project aims at flying highly sensitive optical atomic clocks
  and cold atom accelerometres on a Solar system trajectory on a
  time scale of 10 years. It could test the constancy of the
  fine-structure constant along the sattelite wordline which, in
  particular, can set a constraint on its spatial variation of the
  order of $10^{-9}$~\cite{reynaud09,wolf09}.

 \item{\em Theoretical developments}: We remind one more time that
 the interpretation of the experiments requires a good
 theoretical understanding of the systems but also that the
 constraints we draw on the fundamental constants such as the
 quark masses are conditional to our theoretical modelling, hence
 on hypothesis on a unification scheme as well as nuclear physics.
 The accuracy and the robustness of these steps need to be determined,
 e.g. by taking the dependence in the nuclear radius~\cite{dinh}.
\end{itemize}

\subsection{The Oklo phenomenom}

\subsubsection{A natural nuclear reactor}

Oklo is the name of a town in the Gabon republic (West Africa)
where an open-pit uranium mine is situated. About
$1.8\times10^{9}$~yr ago (corresponding to a redshift of $\sim
0.14$ with the cosmological concordance model), in one of the rich
vein of uranium ore, a natural nuclear reactor went critical,
consumed a portion of its fuel and then shut a few million years
later (see e.g. Ref.~\cite{uzanleclercqbook} for more details).
This phenomenon was discovered by the French Commissariat \`a
l'\'Energie Atomique in 1972 while monitoring for uranium
ores~\cite{oklo-1}. Sixteen natural uranium reactors have been
identified. Well studied reactors include the zone RZ2 (about 60
bore-holes, 1800~kg of ${}^{235}{\rm U}$ fissioned during
$8.5\times10^5$~yr) and zone RZ10 (about 13 bore-holes, 650~kg of
${}^{235}{\rm U}$ fissioned during $1.6\times10^5$~yr).

The existence of such a natural reactor was predicted by P.
Kuroda~\cite{oklo-3} who showed that under favorable conditions, a
spontaneous chain reaction could take place in rich uranium
deposits. Indeed, two billion years ago, uranium was naturally
enriched (due to the difference of decay rate between
${}^{235}{\rm U}$ and ${}^{238}{\rm U}$) and ${}^{235}{\rm U}$
represented about 3.68\% of the total uranium (compared with
0.72\% today and to the 3-5\% enrichment used in most commercial
reactors). Besides, in Oklo the conditions were favorable: (1) the
concentration of neutron absorbers, which prevent the neutrons
from being available for the chain fission, was low; (2) water
played the role of moderator (the zones RZ2 and RZ10 operated at a
depth of several thousand metres, so that the water pressure and
temperature was close to the pressurized water reactors of 20~Mpa
and 300~C) and slowed down fast neutrons so that they can interact
with other ${}^{235}{\rm U}$ and (3) the reactor was large enough
so that the neutrons did not escape faster than they were
produced. It is estimated that the Oklo reactor powered 10 to
50~kW. This explanation is backed up by the substantial depletion
of ${}^{235}$U as well as a correlated peculiar distribution of
some rare-earth isotopes. These rare-earth isotopes are abundantly
produced during the fission of uranium and, in particular, the
strong neutron absorbers like ${}^{149}_{62}{\rm Sm}$,
${}^{151}_{63}{\rm Eu}$, ${}^{155}_{64}{\rm Gd}$ and
${}^{155}_{64}{\rm Gd}$ are found in very small quantities in the
reactor.

From the isotopic abundances of the yields, one can extract
informations about the nuclear reactions at the time the reactor
was operational and reconstruct the reaction rates at that time.
One of the key quantity measured is the ratio ${}^{149}_{62}{\rm
Sm}/{}^{147}_{62}{\rm Sm}$ of two light isotopes of samarium which
are not fission products. This ratio of order of 0.9 in normal
samarium, is about 0.02 in Oklo ores. This low value is
interpreted~\cite{oklo-2} by the depletion of ${}^{149}_{62}{\rm
Sm}$ by thermal neutrons produced by the fission process and to
which it was exposed while the reactor was active. The capture
cross-section of thermal neutron by ${}^{149}_{62}{\rm Sm}$
\begin{equation}\label{oklo1}
n+{}^{149}_{62}{\rm Sm}\longrightarrow {}^{150}_{62}{\rm
Sm}+\gamma
\end{equation}
is dominated by a capture resonance of a neutron of energy of
about 0.1 eV ($E_r=97.3$~meV today). The existence of this
resonance is a consequence of an almost cancellation between the
electromagnetic repulsive force and the strong interaction.

Shlyakhter~\cite{oklo-2} pointed out that this phenomenon can be
used to set a constraint on the time variation of fundamental
constants. His argument can be summarized as follows.
\begin{itemize}
\item First, the cross-section $\sigma_{(n,\gamma)}$ strongly
 depends on the energy of a resonance at $E_{r}=97.3$~meV.
\item Geochemical data allow to determine the isotopic
 composition of various element, such as uranium, neodynium,
 gadolinium and samarium. Gadolinium and neodium allow to
 determine the fluence (integrated flux over time) of the neutron
 while both gadolinium and samarium are strong neutron absorbers.
\item From these data, one deduces the value of the averaged value
 of the cross-section on the neutron flux, $\hat\sigma_{149}$. This
 value depends on hypothesis on the geometry of the reactor
 zone.
\item The range of allowed value of $\hat\sigma_{149}$ was
 translated into a constraint on $E_r$. This step involves an
 assumption on the form and temperature of the neutron spectrum.
\item $E_r$ was related to some fundamental constant,
 which involve a model of the nucleus.
\end{itemize}
In conclusion, we have different steps, which all involve
assumptions:
\begin{itemize}
 \item Isotopic compositions and geophysical parameters are
 measured in a given set of bore-hold in each zone. A choice has
 to be made on the sample to use, in order e.g. to ensure that
 they are not contaminated.
 \item With hypothesis on the geometry of the reactor, on the
 spectrum and temperature of the neutron flux, one can deduce the
 effective value of the cross-sections of neutron absorbers (such
 as samarium and gadolinium). This requires to solve a network of
 nuclear reaction describing the fission.
 \item One can then infer the value of the resonance energy $E_r$,
 which again depends on the assumptions on the neutron spectrum.
 \item $E_r$ needs to be related to fundamental constant, which
 involves a model of the nucleus and high energy physics
 hypothesis.
\end{itemize}
We shall now detail the assumptions used in the various analysis
that have been performed since the pioneering work of
Ref.~\cite{oklo-2}.

\subsubsection{Constraining the shift of the resonance energy}

\paragraph{Cross sections}
The cross-section of the neutron capture (\ref{oklo1}) strongly
depends on the energy of a resonance at $E_{r}=97.3$~meV and is
well described by the Breit-Wigner formula
\begin{equation}\label{oklo2}
 \sigma_{(n,\gamma)}(E)=\frac{g_0\pi}{2}\frac{\hbar^2}{m_{\rm n}E}
  \frac{\Gamma_{\rm n}\Gamma_\gamma}{(E-E_r)^2+\Gamma^2/4}
\end{equation}
where $g_0\equiv(2J+1)(2s+1)^{-1}(2I+1)^{-1}$ is a statistical
factor which depends on the spin of the incident neutron $s=1/2$,
of the target nucleus $I$, and of the compound nucleus $J$. For the
reaction~(\ref{oklo1}), we have $g_0=9/16$. The total width
$\Gamma\equiv\Gamma_{\rm n}+\Gamma_\gamma$ is the sum of the
neutron partial width $\Gamma_{\rm n}=0.533$~meV (at
$E_r=97.3$~meV and it scales as $\sqrt{E}$ in the center of mass)
and of the radiative partial width $\Gamma_\gamma=60.5$~meV.
${}^{155}_{64}{\rm Gd}$ has a resonance at $E_r=26.8$~meV with
$\Gamma_{\rm n}=0.104$~meV, $\Gamma_\gamma=108$~meV and $g=5/8$
while ${}^{157}_{64}{\rm Gd}$ has a resonance at $E_r=31.4$~meV
with $\Gamma_{\rm n}=0.470$~meV, $\Gamma_\gamma=106$~meV and
$g=5/8$.

As explained in the previous section, this cross-section cannot be
measured from the Oklo data, which allow only to measure its value
averaged on the neutron flux $n(v,T)$, $T$ being the temperature
of the moderator. It is conventionally defined as
\begin{equation}\label{hatsig}
 \hat\sigma =\frac{1}{nv_0}\int\sigma_{(n,\gamma)}n(v,T)
 v\dd v,
\end{equation}
where the velocity $v_0=2200\,{\rm m\cdot s}^{-1}$ corresponds to
an energy $E_0=25.3$~meV and $v=\sqrt{2E/m_{\rm n}}$, instead of
$$
  \bar\sigma = \frac{\int\sigma_{(n,\gamma)}n(v,T)v\dd v}{\int n(v,T)
 v\dd v}.
$$
When the cross-section behaves as $\sigma=\sigma_0v_0/v$, which is
the case for nuclei known as ``$1/v$-absorbers'',
$\hat\sigma=\sigma_0$ and does not depend on the temperature,
whatever the distribution $n(v)$. In a similar way, the effective
neutron flux defined
\begin{equation}
 \hat\phi=v_0\int n(v,T)\dd v
\end{equation}
which differs from the true flux
$$
 \phi= \int n(v,T)v\dd v.
$$
However, since $\bar\sigma\phi = \hat\sigma\hat\phi$, the reaction
rates are note affected by these definitions.

\paragraph{Extracting the effective cross-section from the data}
To ``measure'' the value of $\hat\sigma$ from the Oklo data, we
need to solve the nuclear reaction network that controls the
isotopic composition during the fission.

The samples of the Oklo reactors were exposed~\cite{oklo-1} to an
integrated effective fluence $\int\hat\phi\dd t$ of about
$10^{21}$~neutron$\cdot{\rm cm}^{-2}=1~{\rm kb}^{-1}$. It implies
that any process with a cross-section smaller than 1~kb can safely
be neglected in the computation of the abundances. This includes
neutron capture by ${}^{144}_{62}{\rm Sm}$ and ${}^{148}_{62}{\rm
Sm}$, as well as by ${}^{155}_{64}{\rm Gd}$ and ${}^{157}_{64}{\rm
Gd}$. On the other hand, the fission of ${}^{235}_{92}{\rm U}$,
the capture of neutron by ${}^{143}_{60}{\rm Nd}$ and by
${}^{149}_{62}{\rm Sm}$ with respective cross-sections
$\sigma_{5}\simeq0.6$~kb, $\sigma_{143}\sim0.3$~kb and
$\sigma_{149}\geq70$~kb are the dominant processes. It follows
that the equations of evolution for the number densities
$N_{147}$, $N_{148}$, $N_{149}$ and $N_{235}$ of
${}^{147}_{62}{\rm Sm}$, ${}^{148}_{62}{\rm Sm}$,
${}^{149}_{62}{\rm Sm}$ and ${}^{235}_{92}{\rm U}$ takes the form
\begin{eqnarray}
 \frac{\dd N_{147}}{\hat\phi\dd t}&=&-\hat\sigma_{147} N_{147}+
       \hat\sigma_{f235}y_{147} N_{235}\label{e1}\\
\frac{\dd N_{148}}{\hat\phi\dd t}&=&\hat\sigma_{147} N_{147}\label{e2}\\
\frac{\dd N_{149}}{\hat\phi\dd t}&=&-\hat\sigma_{149} N_{149}+
       \hat\sigma_{f235}y_{149} N_{235}\label{e3}\\
\frac{\dd N_{235}}{\hat\phi\dd t}&=&-\sigma_5 N_{235},\label{e4}
\end{eqnarray}
where $y_i$ denotes the yield of the corresponding element in the
fission of ${}^{235}_{92}{\rm U}$ and $\hat\sigma_5$ is the
fission cross-section. This system can be integrated under the
assumption that the cross-sections and the neutron flux are
constant and the result compared with the natural abundances of
the samarium to extract the value of $\hat\sigma_{149}$ at the
time of the reaction. Here, the system has been closed by
introducing a modified absorption cross-section~\cite{oklo-4}
$\sigma_5^*$ to take into account both the fission, capture but
also the formation from the $\alpha$-decay of ${}^{239}_{94}{\rm
Pu}$. One can instead extend the system by considering
${}^{239}_{94}{\rm Pu}$, and ${}^{235}_{92}{\rm U}$ (see
Ref.~\cite{oklo-9}). While most studies focus on the samarium,
Ref.~\cite{oklo-5} also includes the gadolinium even though it is
not clear whether it can reliably be measured~\cite{oklo-4}. They
give similar results.

By comparing the solution of this system with the measured
isotopic composition, one can deduce the effective cross-section.
At this step, the different
analysis~\cite{oklo-2,oklo-2bis,oklo-4,oklo-5,oklo-7,oklo-8,oklo-9}
differ from the choice of the data. The measured values of
$\hat\sigma_{149}$ can be found in these articles. They are given
for a given zone (RZ2, RZ10 mainly) with a number that correspond
to the number of the bore-hole and the depth (e.g. in Table~2 of
Ref.~\cite{oklo-4}, SC39-1383 means that we are dealing with the
bore-hole number 39 at a depth of 13.83~m). Recently, another
approach~\cite{oklo-8,oklo-9} was proposed in order to take into
account of the geometry and details of the reactor. It relies on a
full-scale Monte-Carlo simulation and a computer model of the
reactor zone RZ2~\cite{oklo-8} and both RZ2 and
RZ10~~\cite{oklo-9} and allows to take into account the spatial
distribution of the neutron flux.

\begin{table}[t]
\begin{center}
{\small
\begin{tabular}{p{1.5 cm} ccccc}
\hline\hline
 Ore & neutron spectrum & Temperature ($^{\rm o}$C) & $\hat\sigma_{149}$ (kb)  & $\Delta E_r$ (meV) & Ref. \\
 \hline
     ?        &  Maxwell      &  20      & $55\pm8$ & $0\pm20$
                                 & \cite{oklo-2}\\
   RZ2 (15)  &  Maxwell      &  180-700    & $75\pm18$ &
                     $-1.5\pm10.5$             & \cite{oklo-4}\\
    RZ10         &  Maxwell      &   200-400   & $91\pm6$ &
    $4\pm16$                               & \cite{oklo-5}\\
       RZ10         &      &   & & $-97\pm8$
                                 & \cite{oklo-5}\\
    -  &  Maxwell  + epithermal    &  327    & $91\pm6$ &  $-45^{+7}_{-15}$& \cite{oklo-7}\\
    RZ2     &  Maxwell + epithermal     &      & $73.2\pm9.4$ & $-5.5\pm67.5$ & \cite{oklo-8}\\
    RZ2    &  Maxwell + epithermal     &  200-300    & $71.5\pm10.0$
    &-
    & \cite{oklo-9}\\
    RZ10            &  Maxwell + epithermal     &  200-300    & $85.0\pm6.8$
    &-
    & \cite{oklo-9}\\
  RZ2+RZ10    &      &     & & $7.2\pm18.8 $ & \cite{oklo-9}\\
    RZ2+RZ10            &     &    & & $90.75\pm11.15 $ & \cite{oklo-9}\\

\hline\hline
\end{tabular}
\caption{\it Summary of the  analysis of the Oklo data. The
principal assumptions to infer the value of the resonance energy
$E_r$ are the form of the neutron spectrum and its temperature.}
\label{tab-oklo}
}
\end{center}
\end{table}

\paragraph{Determination of $E_r$}

To convert the constraint on the effective cross-section, one
needs to specify the neutron spectrum. In the earlier
studies~\cite{oklo-2,oklo-2bis}, a Maxwell distribution,
$$
 n_{\rm th}(v,T) = \left(\frac{m_{\rm n}}{2\pi T}
 \right)^{3/2}\hbox{e}^{-\frac{m v^2}{2 k_{\rm B}T}},
$$
was assumed for the neutron with a temperature of $20^{\rm o}$~C,
which is probably too small. Then $v_0$ is the mean velocity at a
temperature $T_0=m_{\rm n}v_0^2/2k_{\rm B}=20.4^{\rm o}$~C.
Refs.~\cite{oklo-4,oklo-5} also assume a Maxwell distribution but
let the moderator temperature vary so that they deduce an
effective cross-section $\hat\sigma(R_r,T)$. They respectively
restricted the temperature range to $180^{\rm o}$~C$<T<700^{\rm
o}$~C and $200^{\rm o}$~C$<T<400^{\rm o}$~C, based on geochemical
analysis. The advantage of the Maxwell distribution assumption is
that it avoids to rely on a particular model of the Oklo reactor
since the spectrum is determined solely by the temperature.

It was then noted~\cite{oklo-7,oklo-8} that above an energy of
several eV, the neutrons spectrum shifted to a $1/E$ tail because
of the absorption of neutrons in uranium resonances. The
distribution was thus adjusted to include an epithermal
distribution
$$
 n(v) = (1 - f) n_{\rm th}(v,T) + f n_{\rm epi}(v),
$$
with $n_{\rm epi}=v_c^2/v^2$ for $v>v_c$ and vanishing otherwise.
$v_c$ is a cut-off velocity that also needs to be specified. The
effective cross-section can then be parameterized~\cite{oklo-9} as
\begin{equation}
 \hat\sigma = g(T)\sigma_0 + r_0 I,
\end{equation}
where $g(T)$ is a measure of the departure of $\sigma$ from the
$1/v$ behavior, $I$ is related to the resonance integral of the
cross-section and $r_0$ is the Oklo reactor spectral index. It
characterizes the contribution of the epithermal neutrons to the
cross-section. Among the unknown parameters, the most uncertain
is probably the amount of water present at the time of the
reaction. Ref.~\cite{oklo-9} chooses to adjust it so that $r_0$
matches the experimental values.\\

These hypothesis on the neutron spectrum and on the temperature,
as well as the constraint on the shift of the resonance energy,
are summarised in Table~\ref{tab-oklo}. Many
analysis~\cite{oklo-5,oklo-8,oklo-9} find two branches for $\Delta
E_r=E_r - E_{r0}$, with one (the left branch) indicating a
variation of $E_r$. Note that these two branches disappear when
the temperature is higher since $\hat\sigma(E_r,T)$ is more peaked
when $T$ decreases but remain in any analysis at low
temperature. This shows the importance of a good determination of
the temperature. Note that the analysis of Ref.~\cite{oklo-8}
indicates that the curves $\hat\sigma(T,E_r)$ lie appreciably
lower than for a Maxwell distribution and that Ref.~\cite{oklo-5}
argues that the left branch is hardly compatible with the
gadolinium data.

\subsubsection{From the resonance energy to fundamental constants}

The energy of the resonance depends a priori on many constants
since the existence of such resonance is mainly the consequence of
an almost cancellation between the electromagnetic repulsive force
and the strong interaction. But, since no full analytical
understanding of the energy levels of heavy nuclei is available,
the role of each constant is difficult to disentangle.

In his first analysis, Shlyakhter~\cite{oklo-2} stated that for
the neutron, the nucleus appears as a potential well with a depth
$V_0\simeq 50\,{\rm MeV}$. He attributed the change of the
resonance energy to a modification of the strong interaction
coupling constant and concluded that $\Delta g_{_{\rm S}}/g_{_{\rm
S}}\sim \Delta E_r/V_0$. Then, arguing that the Coulomb force
increases the average inter-nuclear distance by about 2.5\% for
$A\sim150$, he concluded that $\Delta\aem/\aem\sim20\Delta
g_{_{\rm S}}/g_{_{\rm S}}$, leading to
$|\dot\aem/\aem|<10^{-17}\,{\rm yr}^{-1}$, which can be translated
to
\begin{equation}
|\Delta\aem/\aem|<1.8\times10^{-8}.
\end{equation}

The following analysis focused on the fine-structure constant and
ignored the strong interaction. Damour and Dyson~\cite{oklo-4}
related the variation of $E_r$ to the fine-structure constant by
taking into account that the radiative capture of the neutron by
${}^{149}_{62}{\rm Sm}$ corresponds to the existence of an excited
quantum state of ${}^{150}_{62}{\rm Sm}$ (so that
$E_r=E_{150}^*-E_{149}-m_{\rm n}$) and by assuming that the
nuclear energy is independent of $\aem$. It follows that the
variation of $\aem$ can be related to the difference of the
Coulomb binding energy of these two states. The computation of
this latter quantity is difficult and requires to be related to
the mean-square radii of the protons in the isotopes of samarium.
In particular this analysis~\cite{oklo-4} showed that the
Bethe-Weiz\"acker formula overestimates by about a factor the 2
the $\aem$-sensitivity to the resonance energy.  It follows from
this analysis that
\begin{equation}\label{okdamdy}
   \aem\frac{\Delta E_r}{\Delta\aem}\simeq-1.1\,{\rm MeV},
\end{equation}
which, once combined with the constraint on $\Delta E_r$, implies
\begin{equation}
  -0.9\times10^{-7}<\Delta\aem/\aem<1.2\times10^{-7}
\end{equation}
at $2\sigma$ level, corresponding to the range
$-6.7\times10^{-17}\,{\rm yr}^{-1}< \dot\aem/\aem
<5.0\times10^{-17} \,{\rm yr}^{-1}$ if $\dot\aem$ is assumed
constant. This tight constraint arises from the large
amplification between the resonance energy ($\sim0.1$~eV) and the
sensitivity ($\sim1$~MeV). The re-analysis of these data and also including
the data of Ref.~\cite{oklo-5} with gadolinium, found the
favored result $\dot\aem/\aem=(-0.2\pm0.8)\times10^{-17}\,{\rm
yr}^{-1}$ which corresponds to
\begin{equation}
\Delta\aem/\aem=(-0.36\pm1.44)\times10^{-8}
\end{equation}
and the other branch (indicating a variation; see
Table~\ref{tab-oklo}) leads to $\dot\aem/\aem =(4.9\pm0.4)
\times10^{-17}\,{\rm yr}^{-1}$. This non-zero result cannot be
eliminated.

The more recent analysis, based on a modification of the neutron
spectrum lead respectively to~\cite{oklo-8}
\begin{equation}
 \Delta\aem/\aem=(3.85\pm5.65)\times10^{-8}
\end{equation}
and~\cite{oklo-9}
\begin{equation}
 \Delta\aem/\aem=(-0.65\pm1.75)\times10^{-8},
\end{equation}
at a 95\% confidence level, both using the formalism of
Ref.~\cite{oklo-4}.

Olive {\em et al.}~\cite{oklo-11}, inspired by grand unification
model, reconsider the analysis of Ref.~\cite{oklo-4} by letting
all gauge and Yukawa couplings vary.  Working within the Fermi gas
model, the over-riding scale dependence of the terms which
determine the binding energy of the heavy nuclei was derived.
Parameterizing the mass of the hadrons as $m_i\propto\Lambda_{\rm
QCD}(1+\kappa_im_{\rm q}/\Lambda_{\rm QCD}+\ldots)$, they deduce
that the nuclear Hamiltonian was proportional to $m_{\rm
q}/\Lambda_{\rm QCD}$ at lowest order, which allows to estimate
that the energy of the resonance is related to the quark mass by
\begin{equation}\label{e33}
 \frac{\Delta E_r}{E_r}\sim (2.5-10)\times 10^{17}
 \Delta\ln\left(\frac{m_{\rm q}}{\Lambda_{\rm QCD}}\right) .
\end{equation}
Using the constraint~(\ref{okdamdy}), they first deduced that
$$
 \left|\Delta\ln\left(\frac{m_{\rm q}}{\Lambda_{\rm
 QCD}}\right)\right|<(1-4)\times10^{-8}.
$$
Then, assuming that $\aem\propto m_{\rm q}^{50}$ on the basis of
grand unification (see \S~\ref{subsec22} for details), they
concluded that
\begin{equation}
 \left|\Delta\aem/\aem \right|<(2-8)\times10^{-10}.
\end{equation}

Similarly, Refs.~\cite{oklo-12,oklo-13,oklo-15} related the
variation of the resonance energy to the quark mass.  Their first
estimate~\cite{oklo-12} assumes that it is related to the pion
mass, $m_\pi$, and that the main variation arises from the
variation of the radius $R\sim 5{\rm fm}+1/m_\pi$ of the nuclear
potential well of depth $V_0$, so that
$$
 \delta E_r\sim -2V_0\frac{\delta R}{R}
 \sim 3\times10^8\frac{\delta m_\pi}{m_\pi},
$$
assuming that $R\simeq 1.2A^{1/3}r_0$, $r_0$ being the inter-nucleon
distance.

Then, in Ref.~\cite{oklo-13}, the nuclear potential was described
by a Walecka model which keeps only the $\sigma$ (scalar) and
$\omega$ (vector) exchanges in the effective nuclear force. Their
masses was related to the mass $m_{\rm s}$ of the strange quark to
get $m_\sigma\propto m_{\rm s}^{0.54}$ and $m_\omega\propto m_{\rm
s}^{0.15}$. It follows that the variation of the potential well
can be related to the variation of $m_\sigma$ and $m_\omega$ and thus
on $m_{\rm q}$ by $V\propto m_{\rm q}^{-3.5}$. The
constraint~(\ref{okdamdy}) then implies that
$$
 \left|\Delta\ln\left(\frac{m_{\rm s}}{\Lambda_{\rm
 QCD}}\right)\right|< 1.2\times10^{-10}.
$$
By extrapolating from light nuclei where the $N$-body calculations
can be performed more accurately, it was concluded~\cite{oklo-14}
that the resonance energy scales as $\Delta
E_r\simeq10(\Delta\ln X_{\rm q} -0.1\Delta\ln\aem)$, so that the
the constraints from Ref.~\cite{oklo-8} would imply that
$\Delta\ln (X_{\rm q}/\aem^{0.1})<7\times10^{-9}$.\\

In conclusion, this last results illustrate that a detailed
theoretical analysis and quantitative estimates of the nuclear
physics (and QCD) aspects of the resonance shift still remain to
be carried out. In particular, the interface between the
perturbative QCD description and the description in term of hadron
is not fully understand: we do not know the exact dependence of
hadronic masses and coupling constant on $\Lambda_{\rm QCD}$ and
quark masses. The second problem concerns modelling nuclear forces
in terms of the hadronic parameters.

At present, the Oklo data, while being stringent and consistent
with no variation, have to be considered carefully. While a better
understanding of nuclear physics is necessary to understand the
full constant-dependence, the data themselves require more
insight, particularly to understand the existence of the
left-branch.

\subsection{Meteorite dating}

Long-lived $\alpha$- or $\beta$-decay isotopes may be
sensitive probes of the variation of fundamental constants on
geological times ranging typically to the age of the Solar system,
$t\sim(4-5)$~Gyr, corresponding to a mean redshift of
$z\sim~0.43$. Interestingly, it can be compared with the shallow
universe quasar constraints. This method was initially pointed out
by Wilkinson~\cite{meteo0} and then revived by
Dyson~\cite{revueDyson}. The main idea is to extract the
$\aem$-dependence of the decay rate and to use geological samples
to bound its time variation.

The sensitivity of the decay rate of a nucleus to a change of the
fine-structure constant is defined, in a similar way as for atomic
clocks [Eq.~(\ref{clock-sensitivity})], as
\begin{equation}\label{L-sensitivity}
 s_\alpha \equiv \frac{\partial \ln \lambda}{\partial \ln\aem}.
\end{equation}
$\lambda$ is a function of the decay energy $Q$. When $Q$ is
small, mainly due to an accidental cancellation between different
contributions to the nuclear binding energy, the sensitivity
$s_\alpha$ maybe strongly enhanced. A small variation of the
fundamental constants can either stabilize or destabilize certain
isotopes so that one can extract bounds on the time variation of
their lifetime by comparing laboratory data to geophysical ans Solar system
probes.

Assume some meteorites containing an isotope $X$ that decays into
$Y$ are formed at a time $t_*$. It follows that
\begin{equation}\label{adecT}
 N_X(t) = N_{X*}\hbox{e}^{-\lambda(t-t_*)},\qquad
 N_Y(t) = N_{X*}\left[1-\hbox{e}^{-\lambda(t-t_*)}\right] + N_{Y*}
\end{equation}
if one assumes the decay rate constant. If it is varying then
these relations have to be replaced by
$$
 N_X(t) = N_{X*}\hbox{e}^{\int_{t_*}^{t}\lambda(t')\dd t'}
$$
so that the value of $N_X$ today can be interpreted with
Eq.~(\ref{adecT}) but with an effective decay rate
\begin{equation}\label{Leff}
 \bar\lambda = \frac{1}{t_0-t_*}\int_{t_*}^{t_0}\lambda(t')\dd t'.
\end{equation}
From a sample of meteorites, we can measure $\{ N_X(t_0),N_Y(t_0)
\}$ for each meteorite. These two quantities are related by
$$
N_Y(t_0) = \left[\hbox{e}^{\bar\lambda(t_0-t_*)} - 1 \right]
N_X(t_0) + N_{Y*},
$$
so that the data should lie on a line (since $N_{X*}$ is a priori
different for each meteorite), called an ``isochron'', the slope
of which determines $\bar\lambda(t_0-t_*)$. It follows that
meteorites data only provides an {\it average} measure of the decay
rate, which complicates the interpretation of the constraints (see
Refs.~\cite{meteo-fuji2,meteo-fuji1} for explicit examples). To
derive a bound on the variation of the constant we also need a
good estimation of $t_0-t_*$, which can be obtained from the same
analysis for an isotope with a small sensitivity $s_\alpha$, as
well as an accurate laboratory measurement of the decay rate.

\subsubsection{Long lived $\alpha$-decays}

The $\alpha$-decay rate, $\lambda$, of a nucleus ${}^A_Z{\rm X}$
of charge $Z$ and atomic number $A$,
\begin{equation}
{}_{Z+2}^{A+4}{\rm X}\longrightarrow {}_Z^A{\rm X}+ {}_2^4{\rm
He},
\end{equation}
is governed by the penetration of the Coulomb barrier that can be
described by the Gamow theory. It is well approximated by
\begin{equation}
\lambda\simeq\Lambda(\aem,v)\exp\left(-4\pi Z\aem
\frac{c}{v}\right),
\end{equation}
where $v/c=\sqrt{Q/2m_{\rm p}c^2}$ is the escape velocity of the
$\alpha$ particle. $\Lambda$ is a function that depends slowly on
$\aem$ and $Q$. It follows that the sensitivity to the
fine-structure constant is
\begin{equation}
 s_\alpha \simeq -4\pi Z \frac{\aem}{\sqrt{Q/2m_{\rm p}}}
   \left(1- \frac{1}{2}\frac{\dd\ln Q}{\dd\ln\aem}\right).
\end{equation}
The decay energy is related to the nuclear binding energies
$B(A,Z)$ of the different nuclei by
$$
 Q = B(A,Z) + B_\alpha -B(A+4,Z+2)
$$
with $B_\alpha=B(4,2)$. Physically, an increase of $\aem$ induces
an increase in the height of the Coulomb barrier at the nuclear
surface while the depth of the nuclear potential well below the
top remains the same. It follows that $\alpha$-particle escapes
with a greater energy but at the same energy below the top of the
barrier. Since the barrier becomes thiner at a given energy below
its top, the penetrability increases.  This computation indeed
neglects the effect of a variation of $\aem$ on the nucleus that
can be estimated to be dilated by about 1\% if $\aem$ increases by
1\%.

As a first insight, when focusing on the fine-structure constant,
one can estimate $s_\alpha$ by varying only the Coulomb term of
the binding energy. Its order of magnitude can be estimated from
the Bethe-Weiz\"acker formula
\begin{equation}\label{bethe}
  E_{_{\rm EM}}=98.25\frac{Z(Z-1)}{A^{1/3}}\aem\,{\rm MeV}.
\end{equation}
Table~\ref{tab-alphadec} summarizes the most sensitive isotopes,
with the sensitivities derived from a semi-empirical analysis for
a spherical nucleus~\cite{oklo-11}. They are in good agreement
with the ones derived from Eq.~(\ref{bethe}) (e.g., for
${}^{238}$U, one would obtain $s_\alpha=540$ instead of
$s_\alpha=548$).

\begin{table}[t]
\begin{center}
{\small
\begin{tabular}{p{3.0 cm} ccccc}
\hline\hline
 Element   &  $Z$ & $A$ & Lifetime (yr) & $Q$ (MeV)  & $s_\alpha$ \\
 \hline
 Sm   &  62 & 147 & $1.06\times10^{11}$ & 2.310 &   774 \\
 Gd   &  64 & 152 & $1.08\times10^{14}$ & 2.204 &   890 \\
 Dy   &  66 & 154 & $3\times10^{6}$     & 2.947 &    575 \\
 Pt   &  78 & 190 & $6.5\times10^{11}$  & 3.249 &   659 \\
 Th   &  90 & 232 & $1.41\times10^{10}$ & 4.082 &    571 \\
 U    &  92 & 235 & $7.04\times10^{8}$  & 4.678 &    466 \\
 U    &  92 & 238 & $4.47\times10^{9}$  & 4.270 &  548\\
\hline\hline
\end{tabular}
\caption{\it Summary of the main nuclei and their physical
properties that have been used in $\alpha$-decay studies.}
\label{tab-alphadec}
}
\end{center}
\end{table}

The sensitivities of all the nuclei of Table~\ref{tab-alphadec}
are similar, so that the best constraint on the time variation of
the fine-structure constant will be given by the nuclei with the
smaller $\Delta\lambda/\lambda$.

Wilkinson~\cite{meteo0} considered the most favorable case, that is the decay of
$^{238}_{92}{\rm U}$ for which $s_\alpha=548$ (see
Table~\ref{tab-alphadec}). By comparing the geological dating of
the Earth by different methods, he concluded that the decay
constant $\lambda$ of $^{238}{\rm U}$, $^{235}{\rm U}$ and
${}^{232}{\rm Th}$ have not changed by more than a factor 3 or 4
during the last $3-4\times10^{9}$~years from which it follows
\begin{equation}
  \left|{\Delta\aem}/{\aem}\right|<8\times10^{-3}.
\end{equation}
This constraint was revised by Dyson~\cite{revueDyson} who claimed
that the decay rate has not changed by more than 20\%, during the
past $2\times10^9$ years, which implies
\begin{equation}
  \left|{\Delta\aem}/{\aem}\right|<4\times10^{-4}.
\end{equation}
Uranium has a short lifetime so that it cannot be used to set
constraints on a longer time scales. It is also used to calibrate
the age of the meteorites. It was thus suggested~\cite{oklo-11} to
consider ${}^{147}$Sm. Assuming that $\Delta\lambda_{\rm
147}/\lambda_{\rm 147}$ is smaller than the fractional uncertainty
of $7.5\times10^{-3}$ of its half-life
\begin{equation}
  \left|{\Delta\aem}/{\aem}\right|\la\times10^{-5}.
\end{equation}

As for the Oklo phenomena, the effect of other constants has not
been investigated in depth. It is clear that at lowest order both
$Q$ and $m_{\rm p}$ scales as $\Lambda_{\rm QCD}$ so that one
needs to go beyond such a simple description to determine the
dependence in the quark masses. Taking into account the
contribution of the quark masses, in the same way as for
Eq.~(\ref{e33}), it was argued that $\lambda\propto X_{\rm
q}^{300-2000}$, which leads to $|\Delta\ln X_{\rm q}|\la10^{-5}$.
In a grand unify framework, that could lead to a constraint of the
order of $|\Delta\ln\aem|\la2\times10^{-7}$.

\subsubsection{Long lived $\beta$-decays}

Dicke~\cite{meteoDicke} stressed that the comparison of the
rubidium-strontium and potassium-argon dating methods to uranium
and thorium rates constrains the variation of $\aem$.

As long as long-lived $\beta$-decay isotopes are concerned for
which the decay energy $Q$ is small, we can use a non-relativistic
approximation for the decay rate
\begin{equation}
\lambda=\Lambda_\pm Q^{p_\pm}
\end{equation}
respectively for $\beta^-$-decay and electron capture.
$\Lambda_\pm$ are functions that depend smoothly on $\aem$ and
which can thus be considered constant, $p_+=\ell+3$ and
$p_-=2\ell+2$ are the degrees of forbiddenness of the transition.
For high-$Z$ nuclei with small decay energy $Q$, the exponent $p$
becomes $p=2+\sqrt{1-\aem^2Z^2}$ and is independent of $\ell$. It
follows that the sensitivity to a variation of the fine-structure
constant is
\begin{equation}
 s_\alpha=p\frac{\dd\ln Q}{\dd\ln\aem}.
\end{equation}
The second factor can be estimated exactly as for $\alpha$-decay.
We note that $\Lambda_\pm$ depends on the Fermi constant and on
the mass of the electron as $\Lambda_\pm\propto\gfermi^2m_{\rm
e}^5 Q^p$. This dependence is the same for any $\beta$-decay so
that it will disappear in the comparison of two dating methods
relying on two different $\beta$-decay isotopes, in which case
only the dependence on the other constants appear again through
the nuclear binding energy. Note however that comparing a
$\alpha$- to a
$\beta$- decay may lead to interesting constraints.\\

We refer to \S~III.A.4 of FVC~\cite{jpu-revue} for earlier
constraints derived from rubidium-strontium, potassium-argon and
we focus on the rhenium-osmium case,
\begin{equation}
{}^{187}_{75}{\rm Re}\longrightarrow{}^{187}_{76}{\rm
Os}+\bar\nu_e+e^-
\end{equation}
first considered by Peebles and Dicke~\cite{meteoPD}. They noted
that the very small value of its decay energy $Q=2.6$~keV makes it
a very sensitive probe of the variation of $\aem$. In that
case $p\simeq 2.8$ so that $s_\alpha\simeq-18000$; a change of
$10^{-2}$\% of $\aem$ will induce a change in the decay energy of
order of the keV, that is of the order of the decay energy itself.
Peebles and Dicke~\cite{meteoPD} did not have reliable laboratory
determination of the decay rate to put any constraint.
Dyson~\cite{meteoDyson} compared the isotopic analysis of
molybdenite ores ($\lambda_{187}=(1.6\pm0.2)\times10^{-11}\,{\rm
yr}^{-1}$), the isotopic analysis of 14 iron meteorites
($\lambda_{187}=(1.4\pm0.3)\times10^{-11}\,{\rm yr}^{-1}$) and
laboratory measurements of the decay rate ($\lambda_{187}=(1.1
\pm0.1)\times10^{-11}\,{\rm yr}^{-1}$). Assuming that the
variation of the decay energy comes entirely from the variation of
$\aem$, he concluded that $\left|{\Delta\aem}/{\aem}\right|
<9\times10^{-4}$ during the past $3\times10^9$ years. Note that
the discrepancy between meteorite and lab data could have been
interpreted as a time-variation of $\aem$, but the
laboratory measurement were complicated by many technical issues so
that Dyson only considered a conservative upper limit.

The modelisation and the computation of $s_\alpha$ were improved
in Ref.~\cite{oklo-11}, following the same lines as for
$\alpha$-decay.

$$
\frac{\Delta\lambda_{187}}{\lambda_{187}} = p \frac{\Delta Q}{Q}
 \simeq p\left(\frac{20\,{\rm MeV}}{Q} \right)\frac{\Delta\aem}{\aem}
 \sim -2.2\times10^4\frac{\Delta\aem}{\aem}
$$
if one considers only the variation of the Coulomb energy in $Q$.
A similar analysis~\cite{dent1} leads to $\Delta\ln \lambda_{187}
\simeq10^4\Delta\ln[\aem^{-2.2}X_{\rm q}^{-1.9}(X_{\rm d}-X_{\rm
u})^{0.23}X_{\rm e}^{-0.058}]$.\\

The dramatic improvement in the meteoric analysis of the Re/Os
ratio~\cite{meteodata} led to a recent reanalysis of the
constraints on the fundamental constants. The slope of the
isochron was determined with a precision of 0.5\%. However, the
Re/Os ratio is inferred from iron meteorites the age of which is
not determined directly. Models of formation of the Solar system
tend to show that iron meteorites and angrite meteorites form
within the same 5~million years. The age of the latter can be
estimated from the ${}^{207}$Pb-${}^{208}$Pb method which gives
4.558 Gyr~\cite{meteodata2} so that $\lambda_{187}= (1.666\pm
0.009)\times 10^{-11}\,{\rm yr}^{-1}$ . We could thus
adopt~\cite{oklo-11}
$$
 \left|\frac{\Delta \lambda_{187}}{\lambda_{187}}\right|<5\times10^{-3}.
$$
However, the meteoritic ages are determined mainly by ${}^{238}$U
dating so that effectively we have a constraint on the variation
of $ \lambda_{187}/ \lambda_{238}$. Fortunately, since the
sensitivity of ${}^{238}$U is much smaller than the one of the
rhenium, it is safe to neglect its effect. Using the recent
laboratory measurement~\cite{meteodata4} ($\lambda_{187}=
(-1.639\pm 0.025)\times 10^{-11}\,{\rm yr}^{-1}$), the variation
of the decay rate is not given by the dispersion of the meteoritic
measurement, but by comparing to its value today, so that
\begin{equation}
 \left|\frac{\Delta \lambda_{187}}{\lambda_{187}}\right|=
 -0.016\pm0.016.
\end{equation}
The analysis of Re.~\cite{meteo-olive}, following the assumption
of Ref.~\cite{oklo-11}, deduced that
\begin{equation}
 \Delta\aem/\aem = (-8\pm16)\times10^{-7},
\end{equation}
at a 95\% confidence level.

As pointed out in Ref.~\cite{meteo-fuji2,meteo-fuji1}, this
constraints really represents a bound on the average decay rate
$\bar\lambda$ since the formation of the meteorites. This implies
in particular that the redshift at which one should consider this
constraint depends on the specific functional dependence
$\lambda(t)$. It was shown that well-designed time dependence for
$\lambda$ can obviate this limit, due to the time average.

\subsubsection{Conclusions}

Meteorites data allow to set constraints on the variation of the
fundamental constants which are comparable to the ones set by the
Oklo phenomenon. Similar constraints can also bet set from
spontaneous fission (see \S~III.A.3 of FVC~\cite{jpu-revue}) but
this process is less well understood and less sensitive than the
$\alpha$- and $\beta$- decay processes and.

From an experimental point of view, the main difficulty concerns
the dating of the meteorites and the interpretation of the
effective decay rate.

As long as we only consider $\aem$, the sensitivities can be
computed mainly by considering the contribution of the Coulomb
energy to the decay energy, that reduces to its contribution to
the nuclear energy. However, as for the Oklo phenomenon, the
dependencies in the other constants, $X_{\rm q}$, $\gfermi$,
$\mu$\ldots, require a nuclear model and remain very
model-dependent.

\subsection{Quasar absorbtion spectra}\label{subsec33}

\subsubsection{Generalities}

Quasar (QSO) absorption lines provide a powerful probe of the
variation of fundamental constants. Absorption lines in
intervening clouds along the line of sight of the QSO give access
to the spectra of the atoms present in the cloud, that it is to
paleo-spectra. The method was first used by
Savedoff~\cite{q-savedo} who constrained the time variation of the
fine-structure constraint from the doublet separations seen in
galaxy emission spectra. For general introduction to these
observations, we refer to Refs.~\cite{dlmu,srintro,radiointro}.

Indeed, one cannot use a single transition compared to its
laboratory value since the expansion of the universe induces a
global redshifting of all spectra. In order to tackle down a
variation of the fundamental constants, one should resort on
various transitions and look for chromatic effects that can indeed
not be reproduce by the expansion of the universe which acts
chromatically on all wavelengths.

\begin{figure}[hptb]
  \def\epsfsize#1#2{0.5#1}
  \centerline{\includegraphics[scale=0.5]{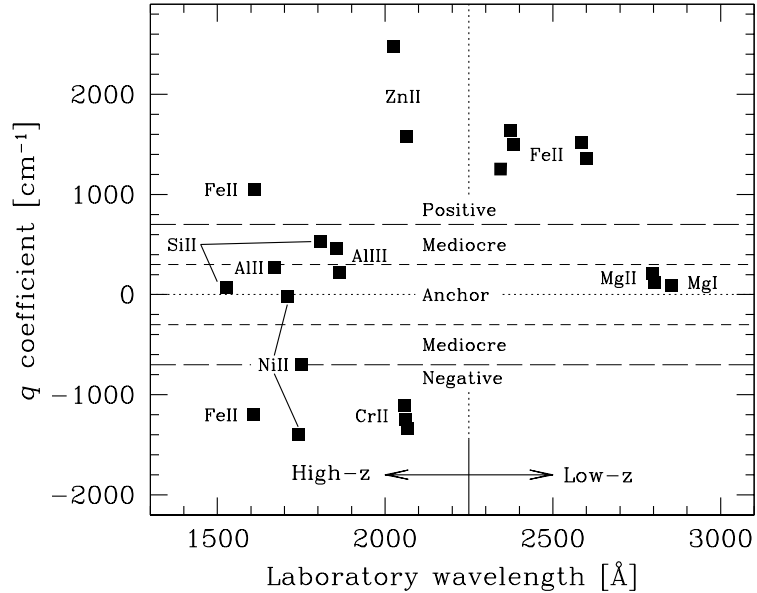}}
  \caption{\it Summary of the values of some coefficients entering
  the
  parameterization~(\ref{qpara}) and necessary to interpret the QSO absorption spectra
  data. From Ref.~\cite{q-murphyMMlast}}
  \label{fig-qdata}
\end{figure}

To achieve such a test, one needs to understand the dependencies
of different types of transitions, in a similar way as for atomic
clock experiments. Refs.~\cite{kappa-dzuba,quasar-q1} suggested to
use the convenient formulation
\begin{equation}\label{qpara}
 \omega = \omega_0 + q\left[\left(\frac{\aem}{\aem^{(0)}}\right)^2-1\right]
  + q_2\left[\left(\frac{\aem}{\aem^{(0)}}\right)^4-1\right],
\end{equation}
in order to take into account the dependence of the spectra on the
fine-structure constant. $\omega$ is the energy in the rest-frame
of the cloud, that is at a redshift $z$, $\omega_0$ is the energy
measured today in the laboratory. $q$ and $q_2$ are two
coefficients that determine the frequency dependence on a
variation of $\aem$ and that arise from the relativistic
corrections for the transition under consideration. The
coefficient $q$ is typically an order of magnitude larger than
$q_2$ so that the possibility to constrain a variation of the
fine-structure constant is mainly determined by $q_1$. This
coefficients were computed for a large set of transitions, first
using a relativistic Hartree-Fock method and then  using many-body
perturbation theory. We refer to
Refs.~\cite{kappa-dzuba,q-calc1,q-calc2} for an extensive discussion
of the computational methods and a list of the $q$-coefficients
for various transitions relevant for both quasar spectra and
atomic clock experiments. Fig.~\ref{fig-qdata} summarizes some of
these results. The uncertainty in $q$ are typically smaller than
$30\,{\rm cm}^{-1}$ for Mg, Si, Al and Zn, but much larger
for Cr, Fe and Ni due to their more
complicated electronic configurations. The accuracy for $\omega_0$
from dedicated laboratory measurements now reach $0.004\,{\rm
cm}^{-1}$. It is important to stress that the form~(\ref{qpara})
ensures that errors in the $q$-coefficients cannot lead
to a non zero detection of $\Delta\aem$.

The shift between two lines is easier to
measure when the difference between the $q$-coefficients of the
two lines is large, which occurs e.g. for two levels with large
$q$ of opposite sign. Many methods were developed to take this
into account. The {\it alkali doublet method} (AD) focuses on the
fine-structure doublet of alkali atoms. It was then generalized to
the {\it many-multiplet method} (MM) which uses correlations between
various transitions in different atoms. As can be seen on
Fig.~\ref{fig-qdata}, some transitions are almost insensitive to a
variation of $\aem$. This is the case of Mg{\sc ii}, which can be used
as an {\it anchor}, i.e. a reference point. To obtain strong
constraints one can either compare transitions of light atoms with
those of heavy atoms (because the $\aem$ dependence of the ground
state scales as $Z^2$) or compare $s-p$ and $d-p$ transitions in
heavy elements (in that case, the relativistic correction will be
of opposite signs). This latter effect increases the sensitivity
and strengthens the method against systematic errors. The results
of this method relies however on two assumptions: (i) ionization
and chemical homogeneity and (ii) isotopic abundance of Mg{\sc ii}
close to the terrestrial value. Even though these are reasonable
assumptions, one cannot completely rule out systematic biases that
they could induce. The AD method completely avoids the assumption
of homogeneity because, by construction, the two lines of the
doublet must have the same profile. Indeed the AD method avoids
the implicit assumption of the MM method that chemical and
ionization inhomogeneities are negligible. Another way to avoid
the influence of small spectral shift due to ionization
inhomogeneities within the absorber and due to possible non-zero
offset between different exposures was to rely on different
transitions of a single ion in individual exposure. This method has been
called the {\it Single ion differential alpha measurement method}
(SIDAM).

Most studies are based on
{\it optical techniques} due to the profusion of strong UV transitions 
that are redshifted into the optical band (this includes AD, MM, SIDAM and it
implies that they can be applied only above a given redshift, e.g. Si{\sc iv} at $z>1.3$,
Fe{\sc ii}$\lambda 1608$ at $z>1$)
or on {\it radio techniques} since radio transitions arise from many
different physical effects (hyperfine splitting and in particular H{\sc i} 21~cm hyperfine
transition, molecular rotation, Lambda-doubling, etc).
In the latter case, the line frequencies and their comparisons yield 
constraints on different sets of fundamental constants
including $\aem$, $g_{\rm p}$ and $\mu$. These techniques
are thus complementary since systematic effects are different
in optical and radio regimes. Also the radio techniques offer some
advantages: (1) to reach high spectral resolution ($<1$~km/s),
alleviating in particular problems with line blending and the use
of e.g. masers allow to reach a frequency calibration better
than roughly 10~m/s; (2) in general, the sensitivity of the line position to a variation
of a constant is higher; (3) the isotopic lines are observed separately, while in optical there
is a blend with possible differential saturations (see e.g. Ref.~\cite{calma} for a discussion).

Let us first emphasize that the shifts in the absorption lines
to be detected are extremely small. For instance a change 
of $\aem$ of order $10^{-5}$ corresponds a shift of at most
$20\, {\rm m\AA}$ for a redshift of $z\sim2$, which would
corresponds to a shift of order $\sim 0.5\,{\rm km/s}$, or
to about a third of a pixel at a spectral resolution of $R\sim40000$,
as achievd with Keck/HIRES or VLT/UVES. As we shall discuss
later, there are several sources of uncertainty that hamper the 
measurement. In particular, the absorption lines have
complex profiles (because they result from the
propagation of photons through a highly inhomogeneous
medium) that are fitted using a combination of
Voigt profiles. Each of these components depends on
several parameters including the redshift, the column density and the
width of the line (Doppler parameter) to which one now needs
to add the constants that are assumed to be varying. These parameters
are constrained assuming that the profiles are the same for all transitions,
which is indeed a non-trivial assumption for transitions from
different species (this was one of the driving motivation to use
transition from a single species and of the SIDAM method). More
important, the fit is usually not unique. This is not a problem when the lines are
not saturated but it can increase the error on $\aem$ by a factor 2
in the case of strongly saturated lines~\cite{q-chandAD}.

\subsubsection{Alkali doublet method (AD)}

The first method used to set constraint on the time variation of
the fine-structure constant relies on fine-structure doublets
splitting for which
$$
 \Delta\nu \propto\frac{\aem^2 Z^4 R_\infty}{2 n^3}.
$$
It follows that the relative separation is proportional $\aem$,
$\Delta\nu/\bar\nu\propto\aem^2$ so that the variation of the fine
structure constant at a redshift $z$ can be obtained as
$$
  \left(\frac{\Delta\aem}{\aem}\right)(z)= \frac{c_r}{2}
 \left[\left(\frac{\Delta\lambda}{\bar\lambda}\right)_z/\left(
 \frac{\Delta\lambda}{\bar\lambda}\right)_0 -1\right],
$$
where $c_r\sim1$ is a number taking into account the relativistic
corrections. This expression is indeed a simple approach of the
alkali doublet since one should, as for atomic clocks, take into
account the relativistic corrections more precisely. Using the
formulation~(\ref{qpara}), one can deduce that
$$
 c_r = \frac{\delta q +\delta q_2}{\delta q + 2\delta q_2},
$$
where the $\delta q$ are the differences between the
$q$-coefficients for the doublet transitions.

Several authors have applied the AD method to doublets of several
species such as e.g. C{\sc iv}, N{\sc v}, O{\sc vi}, Mg{\sc ii}, Al{\sc iii}, Si{\sc ii}, Si{\sc iv}.
We refer to \S~III.3 of FVC~\cite{jpu-revue} for a summary of
their results (see also Ref.~\cite{Levseul}) and focus on the three most recent analysis, based
on the Si{\sc iv} doublet. In this
particular case, $q=766$ (resp. 362)~cm$^{-1}$ and $q_2=48$ (resp.
$-8$)~cm$^{-1}$ for Si{\sc iv} $\lambda1393$ (resp. $\lambda1402$) so
that $c_r=0.8914$. The method is based on a $\chi^2$ minimization
of multiple component Voigt profile fits to the absorption
features in the QSO spectra. In general such a profile depends on
three parameters, the column density $N$, the Doppler width ($b$)
and the redshift. It is now extended to include $\Delta\aem/\aem$.
The fit is carried out by simultaneously varying these parameters
for each component.
\begin{itemize}
 \item Murphy \etal~\cite{q-murphyAD} analyzed  21 Keck/HIRES
 Si{\sc iv} absorption systems toward 8 quasars to obtain the weighted
 mean of the sample,
 \begin{equation}\label{AD-m}
 \Delta\aem/\aem = (-0.5\pm1.3)\times10^{-5},\qquad
 2.33< z < 3.08,
 \end{equation}
 with a mean redshift of $z=2.6$. The S/N ratio of these data is in the
 range 15-40 per pixel and the spectral resolution is
 $R\sim34000$.
 \item Chand \etal~\cite{q-chandAD} analyzed  15
 Si{\sc iv} absorption systems selected from a ESO-UVES sample
 containing 31 systems (eliminating contaminated, saturated or
 very broad systems; in particular a lower limit on the
 column density was fixed so that both lines of the doublets
 are detected at more than $5\sigma$) to get the weighted  mean,
 \begin{equation}\label{a-AD1}
 \Delta\aem/\aem = (-0.15\pm0.43)\times10^{-5},\qquad
 1.59< z < 2.92.
 \end{equation}
 The improvement of the constraint arises mainly from a better S/N
 ratio, of order 60-80 per pixel, and resolution $R\sim45000$. Note that combining this result with the previous
 one~(\ref{AD-m} in a weighted mean would lead
 to $\Delta\aem/\aem = (-0.04\pm0.56)\times10^{-5}$ in the range
 $1.59< z < 3.02$
 \item The analysis~\cite{q-martinezAD} of seven C{\sc iv} systems and two Si{\sc iv} systems in
 the direction of a single quasar, obtained by the VLT-VES (during
 the science verification) has led to
 \begin{equation}
 \Delta\aem/\aem = (-3.09\pm8.46)\times10^{-5},\qquad
 1.19< z < 1.84.
 \end{equation}
This is less constraining than the two previous analysis, mainly because
the $q$-coefficients are smaller for C{\sc iv} (see Ref.~\cite{ppc4} for
the calibration of the laboratory spectra)
\end{itemize}
One limitation may arise from the isotopic composition of silicium.
Silicium has three naturally occurring isotopes with terrestrial
abundances ${}^{28}{\rm Si}:{}^{29}{\rm Si}:{}^{30}{\rm Si}
=92.23:4.68:3.09$ so that each absorption line is a composite of
absorption lines from the three isotopes. It was shown that this
effect of isotopic shifts~\cite{q-murphyAD} is however negligible in the
case of Si{\sc iv}.

\subsubsection{Many multiplet method (MM)}\label{MMQSO}

A generalization of the AD method, known as the many-mulptiplet
was proposed in Ref.~\cite{q-MMmethod}. It relies on the
combination of transitions from different species. In particular,
as can be seen on Fig.~\ref{fig-qdata}, some transitions are fairly
unsensitive to a change of the fine-structure constant (e.g. Mg{\sc ii} or Mg{\sc i},
hence providing good anchors)
while others such as Fe{\sc ii} are more sensitive. The first
implementation~\cite{q-webprl99} of the method was based on a
measurement of the shift of the Fe{\sc ii} (the rest wavelengths of which
are very sensitive to $\aem$) spectrum with respect
to the one of Mg{\sc ii}. This comparison increases the sensitivity
compared with methods using only alkali doublets. Two series of
analysis were performed during the past ten years and lead to
contradictory conclusions. The accuracy of the measurements
depends on how well the absorption line profiles are modelled.

 \paragraph{\bf Keck/HIRES data.} The MM-method was first applied in Ref.~\cite{q-webprl99}
 who analyzed one transition of the Mg{\sc ii} doublet and five Fe{\sc ii}
 transitions from three multiplets. Using 30 absorption systems toward 17 quasars, they
 obtained
 \begin{eqnarray}
 &&\Delta\aem/\aem=(-0.17\pm0.39)\times 10^{-5},\qquad 0.6<z<1\nonumber\\
 &&\Delta\aem/\aem=(-1.88\pm0.53)\times 10^{-5},\qquad
 1<z<1.6.\nonumber
\end{eqnarray}
 This was the first claim that a constant may have varied during
 the evolution of the universe. It was later confirmed in a
 re-analysis~\cite{q_murphysyt2,q-webprl01} of the initial sample and by
 including new optical QSO data to reach 28 absorption systems with redshift
 $z=0.5-1.8$ plus 18 damped Lyman-$\alpha$ absorption systems
 towards 13 QSO plus 21 Si{\sc iv} absorption systems toward 13 QSO .
 The analysis used mainly the multiplets of Ni{\sc ii}, Cr{\sc ii} and Zn{\sc ii}
 and Mg{\sc i}, Mg{\sc i}, Al{\sc ii}, Al{\sc iii} and Fe{\sc ii} was also included. The
 most recent analysis~\cite{q-murphy03a} relies on 128 absorbtion
 spectra, later updated~\cite{q-murphyMMlast} to include 143
 absorption systems. The more robust estimates is the weighted
 mean
 \begin{equation}\label{a-MMw}
 \Delta\aem/\aem = (-0.57\pm0.11)\times10^{-5},\qquad 0.2<z<4.2.
 \end{equation}
 The resolution for most spectra was $R\sim45000$ and the S/N per pixel
 ranges from 4 to 240, with most spectral regions with S/N$\sim30$. The wavelength
 scale was calibrated by mean of a Thorium-argon emission lamp.
 This calibration is crucial and its quality is discussed in
 Ref.~\cite{q-thar,q_murphysyt1} for the Keck/HIRES (see also Ref.~\cite{q-griest}) 
 as well as Ref.~\cite{whit} for the VLT/UVES measurements.

 The low-$z$ ($z<1.8$) and high-$z$ rely on different ions and transitions
 with very different $\aem$-dependencies. At low-$z$, the Mg
 transitions are used as anchors against which the large positive shifts
 in the Fe{\sc ii} can be measured. At high-$z$, different transitions
 are fitted (Fe{\sc ii}, S{\sc ii}, Cr{\sc ii}, Ni{\sc ii}, Zn{\sc ii}, Al{\sc ii}, Al{\sc iii}).
 The two sub-samples respond differently to simple
 systematic errors due to their different arrangement of
 $q$-coefficients in wavelength space. The analysis for each
 sample give the weighted mean
 \begin{eqnarray}
 \Delta\aem/\aem = (-0.54\pm0.12)\times10^{-5},\qquad
 0.2<z<1.8\nonumber\\
 \Delta\aem/\aem = (-0.74\pm0.17)\times10^{-5},\qquad
 1.8<z<4.2,
 \end{eqnarray}
 with respectively 77 and 66 systems. 
 
\paragraph{Hunting systematics.}

While performing this kind of observations a number of problems
and systematic effects have to be taken into account and
controlled. (1) Errors in the determination of laboratory
wavelengths to which the observations are compared. (2) While
comparing wavelengths from different atoms one has to take into
account that they may be located in different regions of the cloud
with different velocities and hence with different Doppler
shifts. (3) One has to ensure that there is no transition not
blended by transitions of another system.
(4) The differential isotopic saturation has to be controlled.
Usually quasar absorption systems are expected to have lower
heavy element abundances. The spatial inhomogeneity of these
abundances may also play a role. (5) Hyperfine splitting can
induce a saturation similar to isotopic abundances. (6) The
variation of the velocity of the Earth during the integration of a
quasar spectrum can also induce differential Doppler shift. (7)
Atmospheric dispersion across the spectral direction of the
spectrograph slit can stretch the spectrum. It was shown that, on
average,  thisv can, for low redshift observations, mimic a negative $\Delta\aem/\aem$,
while this is no more the case for high redshift observations (hence empahasizing
the complementarity of these observations). (8) The presence of a
magnetic field will shift the energy levels by Zeeman effect. (9)
Temperature variations during the observation will change the air
refractive index in the spectrograph. In particular, flexures in the instrument
are dealt with by recording a calibration lamp spectrum before and
after the science expossure and the signal-to-noise and stability of the
lamp is crucial (10) Instrumental effects
such as variations of the intrinsic instrument profile have to be
controlled.

All these effects have been discussed in details in
Refs.~\cite{q_murphysyt1,q_murphysyt2} to argue that none of them
can explain the current detection. This was recently
complemented by a study on the calibration since adistortion of the
wavelength scale could lead to a non-zero value of $\Delta\aem$.
The quality of the calibration is discussed in Ref.~\cite{q-thar} and shown to
have a negligible effect on the measurements (a similar
result has been obtained for the VLT/UVES data~\cite{whit}).

As we pointed out earlier, one assumption of the method concerns
the isotopic abundances of Mg{\sc ii} that can affect the low-$z$
sample since any changes in the isotopic composition will alter
the value of effective rest-wavelengths. This isotopic
composition is assumed to be close to terrestrial ${}^{24}{\rm
Mg}:{}^{25}{\rm Mg}:{}^{26}{\rm Mg} =79:10:11$. No direct
measurement of $r_{\rm Mg}=({}^{26}{\rm Mg}+{}^{25}{\rm
Mg})/{}^{24}{\rm Mg}$ in QSO absorber is currently feasible due to
the small separation of the isotopic absorption lines. It was
however shown~\cite{q-gaylamber}, on the basis of molecular
absorption lines of MgH that $r_{\rm Mg}$ generally decreases with
metallicity. It was also argued that ${}^{13}{\rm C}$ is a tracer
of ${}^{25}{\rm Mg}$ and was shown to be low in the case
of HE~0515-4414~\cite{levCMg}.
However, contrary to this trend, it was
found~\cite{q-yong} that $r_{\rm Mg}$ can reach high values for
some giant stars in the globular cluster NGC~6752 with metallicity
[De/H]$\sim-1.6$. This led Ashenfelter \etal~\cite{q-aften} to
propose a chemical evolution model with strongly enhanced
population of intermediate ($2-8M_\odot$) stars which in their
asymptotic giant branch phase are the dominant factories for heavy
Mg at low metallicities typical of QSO absorption systems, as
a possible explanation of the low-$z$ Keck/HIRES observations
without any variation of $\aem$. It would require that $r_{\rm
Mg}$ reaches $0.62$, compared to $0.27$ (but then
the UVES/VLT constraints would be converted to a detection). However, such modified
nucleosynthetic history will lead to an overproduction of elements
such as P, Si, Al, P above current constraints~\cite{q-fenner}.

In conclusion, no compelling evidence for a systematic effect has
been raised at the moment.

\paragraph{\bf VLT/UVES data.} The previous results, and their
importance for fundamental physics, led another team to check this
detection using observations from UVES spectrograph operating
on the VLT. In order to avoid as much systematics as possible, and
based on numerical simulations, they apply a series of selection
criteria~\cite{q-chandMM} on the systems used to constrain the
time variation of the fine-structure constant: (1) consider only
lines with similar ionization potentials (Mg{\sc ii}, Fe{\sc ii}, Si{\sc ii} and
Al{\sc ii}) as they are most likely to originate from similar regions
in the cloud; (2) avoid absorption lines contaminated by
atmospheric lines; (3) consider only systems with hight enough
column density to ensure that all the mutiplets are detected at
more than $5\sigma$; (4) demand than at least one of the anchor
lines is not saturated to have a robust measurement of the
redshift; (5) reject strongly saturated systems  with large
velocity spread; (6) keep only systems for which the majority of
the components are separated from the neighboring by more than the
Doppler shift parameter.

The advantage of this choice is to reject most complex or
degenerate systems, which could result in uncontrolled systematics
effects. The drawback is indeed that the analysis will be based
on less systems.

Ref.~\cite{q-chandMM,q-chandMM2} analyzed the observations of 23
absorption systems, fulfilling the above criteria, in direction of
18 QSO with a S/N ranging between 50 and 80 per pixel and a
resolution $R>44000$. They concluded that
$$
 \Delta\aem/\aem = (-0.06\pm0.06)\times10^{-5},\qquad 0.4<z<2.3,
$$
hence giving a $3\sigma$ constraint on a variation of $\aem$.

This analysis was challenged by Murphy, Webb and
Flambaum~\cite{q-contr1,q-contr2,q-contr3}. Using (quoting them)
the same reduced data, using the same fits to the absorption
profiles, they claim to find different individual measurements of
$\Delta\aem/\aem$ and a weighted mean,
$$
 \Delta\aem/\aem = (-0.44\pm0.16)\times10^{-5},\qquad 0.4<z<2.3,
$$
which differs from the above cited value. The main points that were
raised are (1) the fact that some of the uncertainties on
$\Delta\aem/\aem$ are smaller than a minimum uncertainty that they
estimated and (2) the quality of the statistical
analysis (in particular on the basis of the $\chi^2$ curves). 
These arguments were responded in Ref.~\cite{q-contr3b}
The revision~\cite{q-contr3b} of the VLT/UVES constraint
rejects two more than 4$\sigma$ deviant systems that were
claimed to dominate the reanalysis~\cite{q-contr2,q-contr3} and
concludes that
\begin{equation}\label{a-MMpp}
 \Delta\aem/\aem = (0.01\pm0.15)\times10^{-5},\qquad 0.4<z<2.3,
\end{equation}
emphasizing that the errors are probably larger.

On the basis of the articles~\cite{q-contr1,q-contr2,q-contr3} and the
answer~\cite{q-contr3b}, it is indeed difficult (without having
played with the data) to engage one of the parties. This exchange
has enlightened some differences in the statistical analysis.

To finish, let us mention that Ref.~\cite{q-sidam1} reanalyzed
some systems of Refs.~\cite{q-chandMM,q-chandMM2} by means of the
SIDAM method (see below) and disagree with some of them, claiming
for a problem of calibration. They also claim that the errors
quoted in Ref.~\cite{q-murphyMMlast} are underestimated by a
factor 1.5.

\paragraph{\bf Regressional MM (RMM).}

The MM method was adapted to use a linear regression
method~\cite{q-quast}. The idea is to measure the redshift $z_i$
deduced from the transition $i$ and plot $z_i$ as a function of
the sensitivity coefficient. If $\Delta\aem\not=0$ then there
should exist a linear relation with a slope proportional to
$\Delta\aem/\aem$. On a single absorption system (VLT/UVES), on
the basis of Fe{\sc ii} transition, they concluded that
\begin{equation}\label{qrmm}
 \Delta\aem/\aem = (-0.4\pm1.9\pm2.7_{\rm syst})\times10^{-6},\qquad z=1.15,
\end{equation}
compared to $\Delta\aem/\aem = (0.1\pm1.7)\times10^{-6}$ that is
obtained with the standard MM technique on the same data. This is
also consistent with the constraint~(\ref{qch}) obtained on the
same system with the HARPS spectrograph.

\paragraph{Open controversy.}

At the moment, we have to face a situation in which two teams have
performed two independent analysis based on data sets obtained by
two instruments on two telescopes. Their conclusions do not agree,
since only one of them is claiming for a detection of a variation
of the fine-structure constant. This discrepancy between VLT/UVES
and Keck/Hires results is yet to be resolved. In particular, they
use data from a different telescopes observing a different
(Southern/Northern) hemisphere.

Note however a recent analysis~\cite{q-griest} of the wavelength accuracy 
of the Keck/HIRES spectrograph. An absolute uncertainty of 
$\Delta z\sim10^{-5}$, corresponding to $\Delta\lambda\sim0.02~{\rm \AA}$
with daily drift of $\Delta z\sim5\times10^{-6}$ and
multiday drift of $\Delta z\sim2\times10^{-5}$. While the cause of
this drift remains unknown, it is argued~\cite{q-griest} that this
level of systematic uncertainty makes it difficult to use the
Keck/HIRES to constrain the time variation of $\aem$
(at least for a single system or a small sample since
the distortion pattern pertains to the echelle orders as they are
recorded on the CDD, that is it is similar from exposure to exposure,
the effect on $\Delta\aem/\aem$ for
an ensemble of absorbers at different redshifts 
would be random since the transitions fall in different
places with respect to the pattern of the disortion). This
needs to be confirmed and investigated in more details. We refer
to Ref.~\cite{murphycal} for a discussion on the Keck wavelength
calibration error and Ref.~\cite{whit} for the VLT/UVES
as well as Ref.~\cite{centurion} for a discussion on the ThAr calibration.

On the one hand, it is sane that one team has reanalyzed the data
of the other and challenge its analysis. This would indeed lead to
an improvement of the robustness of these results. Indeed a
similar reverse analysis would also be sane. On the other hand
both teams have achieved an amazing work in order to understand
and quantify all sources of systematics. Both developments, as
well as the new techniques which are appearing, should hopefully
set this observational issue. Today, it is unfortunately premature
to choose one data set compared to the other.

A recent data~\cite{webspace} set of 60 quasar spectra (yielding 153 absorption systems) for the VLT was
used and split at $z=1.8$ to get
$$
 \left(\Delta\aem/\aem\right)_{{\rm VLT};\, z<1.8} = (-0.06\pm0.16)\times10^{-5},
$$
in agreement with the former study~\cite{q-contr3b}, while at higher
redshift
$$
 \left(\Delta\aem/\aem\right)_{{\rm VLT}\, z>1.8} = (+0.61\pm0.20)\times10^{-5}.
$$
This higher component exhibits a positive variation of $\aem$, that is
of opposite sign with respect to the previous Keck/HIRES detection~\cite{q-murphyMMlast}  
$$
 \left(\Delta\aem/\aem\right)_{{\rm Keck};\, z<1.8} = (-0.54\pm0.12)\times10^{-5},
 \qquad
  \left(\Delta\aem/\aem\right)_{{\rm Keck};\, z>1.8} = (-0.74\pm0.17)\times10^{-5}.
$$
It was pointed out that the Keck/HIRES and VLT/UVES observations can
be made consistent in the case the fine structure constant is spatially varying~\cite{webspace}.
Indeed, one can note that they do not correspond to
the same hemisphere and invoque a spatial variation. Ref.~\cite{webspace}
concludes that the distribution of $\aem$ is well represented by
a spatial dipole, significant at 4.1$\sigma$, in the direction right ascension $17.3\pm0.6$~hours
and declination $-61\pm9$~deg (see also Ref.~\cite{berenspa,berenspa2}). 
This emphasizes the difficulty to compare different data sets
and shows that the constraints can easily be combined as long as they are compatible with
no variation but one must care about a possible spatial variation otherwise.

\subsubsection{Single ion differential measurement (SIDAM)}

This method~\cite{q-sidam0} is an adaptation of the MM method in
order to avoid the influence of small spectral shifts due to
ionization inhomogeneities within the absorbers as well as to
non-zero offsets between different exposures. It was mainly used
with Fe{\sc ii} which provides transitions with positive and negative
$q$-coefficients (see Fig.~\ref{fig-qdata}). Since it relies on a
single ion, it is less sensitive to isotopic abundances, and in
particular not sensitive to the one of Mg.

The first analysis relies on the QSO HE~0515-4414 that was used in
Ref.~\cite{q-quast} to get the constraint~(\ref{qrmm}). An independent analysis~\cite{q-sidam1} 
of the same system gave a weighted mean
\begin{equation}
 \Delta\aem/\aem =(-0.12\pm1.79)\times10^{-6},\qquad z=1.15,
\end{equation}
at $1\sigma$. The same system was studied independently, using the
HARPS spectrograph mounted on the 3.6m telescope at La Silla
observatory~\cite{q-chandSIDAM}. The HARPS spectrograph has a
higher resolution that UVES; $R\sim112000$. Observations based on
Fe{\sc ii} with a S/N of about 30-40 per pixel set the constraint
\begin{equation}\label{qch}
 \Delta\aem/\aem =(0.5\pm2.4)\times10^{-6},\qquad z=1.15.
\end{equation}
The second constraint~\cite{Q1101,q-sidam1} is obtained from an
absorption system toward Q~1101-264,
\begin{equation}
 \Delta\aem/\aem =(5.66\pm2.67)\times10^{-6},\qquad z=1.84,
\end{equation}
These constraints do not seem to be compatible with the results of
the Keck/HIRES based on the MM method. A potential systematic 
uncertainty which can affect these constraints is the relative
shift of the wavelength calibration in the blue and the red arms
of UVES where the distant Fe lines are recorded simultaneously
(see e.g. Ref.~\cite{molaro} for discussion of systematics of this
analysis).

\subsubsection{H{\sc i}-21 cm vs UV: $x=\aem^2 g_{\rm p}/\mu$}

The comparison of UV heavy element transitions with the hyperfine
H{\sc i} transition allows to extract~\cite{UV-1}
$$
 x\equiv\aem^2 g_{\rm p}/\mu,
$$
since the hyperfine transition is proportional to $\aem^2 g_{\rm
p}\mu^{-1} R_\infty$ while optical transitions are simply
proportional to $R_\infty$. It follows that constraints on the time variation
of $x$ can be obtained from high resolution 21cm spectra compared
to UV lines, e.g. of Si{\sc ii}, Fe{\sc ii} and/or Mg{\sc ii}, as first performed
in Ref.~\cite{x-wolfe} in $z\sim0.524$ absorber.

Using 9 absorption systems, there was no evidence for any
variation of $x$~\cite{UV-2},
\begin{equation}
 \Delta x/x=(-0.63\pm0.99)\times10^{-5},\qquad
 0.23<z<2.35,
\end{equation}
This constraints was criticized in Ref.~\cite{x-kanekar} on the basis
that the systems have multiple components and that it is not necessary that
the strongest absorption arises in the same component in both type of lines.
However the error analysis of Ref.~\cite{UV-2} tries to estimate the
effect of the assumption that the strongest absorption arises in the same component.

Following Ref.~\cite{dent1}, we note that the systems lie in two
widely-separated ranges and that the two samples have completely
different scatter. It can thus be split in two samples of
respectively 5 and 4 systems to get
\begin{eqnarray}
 &&\Delta x/x=(1.02\pm1.68)\times10^{-5},\qquad 0.23<z<0.53,\\
 &&\Delta x/x=(0.58\pm1.94)\times10^{-5},\qquad 1.7<z<2.35.
\end{eqnarray}

In such an approach two main difficulties arise: (1) the radio and optical source must coincide
(in the optical QSO can be considered pointlike and it must be checked that
this is also the case for the radio source), (2) the clouds reponsible for the 21cm
and UV absorptions must be localized in the same place. The systems must thus
be selected with care and today the number of such systems is small
and are activily looked for~\cite{dlasearch}. 

The recent detection of 21cm and molecular hydrogen absorption lines in the
same damped Lyman-$\alpha$ system at $z_{\rm abs}=3.174$ towards 
SDSS J1337+3152 constrains~\cite{x21cm} the variation $x$ to
\begin{equation}
 \Delta x/x=-(1.7\pm1.7)\times10^{-6},\qquad z=3.174.
\end{equation}
This system is unique since it allows for 21cm, H$_2$ and UV observation so
that in principle one can measure $\aem$, $x$ and $\mu$ independently.
However, as the H$_2$ column density was low, only Werner band absorption lines are seen so that
the range of sensitivity coefficients is too narrow to provide a stringent constraint,
$\Delta\mu/\mu<4\times10^{-4}$. It was also shown that the H$_2$ and 21cm are shifted because of the inhomogeneity
of the gas, hence emphasizing this limitation.
Ref.~\cite{dlasearch} also mentioned that 4 systems
at $z=1.3$ sets $\Delta x/x = (0.0\pm1.5)\times10^{-6}$ and that another
system at $z=3.1$ gives  $\Delta x/x = (0.2\pm0.5)\times10^{-6}$. Note also that
the comparison~\cite{OH-3b} with C{\sc i} at $z \sim 1.4 - 1.6$ towards Q0458-020 and Q2337-011, 
yields $\Delta x/x = (6.8 \pm 1.0) \times 10^{-6}$ over the band o redshift $0 < <z> \le 1.46$. 
It was argued that, using the existing constraints on $\Delta \mu/\mu$, this measurement is inconsistent 
with claims of a smaller value of $\aem$ from the many-multiplet method,
unless fractional changes in $g_p$ are larger than those in $\aem$ and $\mu$.

\subsubsection{H{\sc i} vs molecular transitions: $y\equiv g_{\rm p}\aem^2$ }

The H{\sc i} 21~cm hyperfine transition frequency is proportional to
$g_{\rm p}\mu^{-1}\aem^2R_\infty$ (see \S~\ref{subsec31-e.h}).
On the other hand, the rotational transition frequencies of diatomic are
inversely proportional to their reduced mass $M$. As on the example
of Eq.~(\ref{mu1}) where we compared an electronic transition to a
vibro-rotational transition, the comparison of the hyperfine and
rotational frequencies is proportional to
$$
 \frac{\nu_{\rm hf}}{\nu_{\rm rot}} \propto g_{\rm p}\aem^2\frac{M}{m_{\rm
 p}} \simeq g_{\rm p}\aem^2 \equiv y,
$$
where the variation of $M/m_{\rm p}$ is usually suppressed by a
large factor of the order of the ratio between the proton mass and
nucleon binding energy in nuclei, so that we can safely neglect
it.

The constraint on the variation of $y$ is directly determined by
comparing the redshift as determined from H{\sc i} and molecular
absorption lines,
$$
\frac{\Delta y}{y} = \frac{z_{\rm mol}-z_{\rm H}}{1+z_{\rm mol}}.
$$

This method was first applied~\cite{y-var} to the CO molecular
absorption lines~\cite{y-combes} towards PKS~1413+135 to get
$$
 \Delta y/y=(-4\pm6)\times10^{-5}\qquad z=0.247.
$$
The most recent constraint~\cite{y-murph} relies on the comparison
of the published redshifts of two absorption systems determined
both from H{\sc i} and molecular absorption. The first is a system at
$z=0.6847$ in the direction of TXS~0218+357 for which th spectra of
CO(1-2), ${}^{13}$CO(1-2), C${}^{18}$O(1-2), CO(2-3),
HCO$^+$(1-2) and HCN(1-2) are available. They concluded that
\begin{equation}
 \Delta y/y=(-0.16\pm0.54)\times10^{-5}\qquad z=0.6847.
\end{equation}
The second system is an absorption system in direction of
PKS~1413+135 for which the molecular lines of CO($1-2$),
HCO$^+$(1-2) and HCO$^+$(2-3) have been detected. The analysis
led to
\begin{equation}
 \Delta y/y=(-0.2\pm0.44)\times10^{-5},\qquad z=0.247.
\end{equation}
Ref.~\cite{y-carilli} obtains the constraints $|\Delta y/y|<3.4\times10^{-5}$
at $z\sim 0.25$ and $z\sim0.685$.

The radio domain has the advantage of heterodyne techniques, with
a spectral resolution of $10^6$ or more, and dealing with cold gas
and narrow lines. The main systematics is the kinematical bias,
i.e. that the different lines do not come exactly from the same
material along the line of sight, with the same velocity. To
improve this method one needs to find more sources, which may be
possible with the radiotelescope
ALMA~\footnote{http://www.eso.org/sci/facilities/alma/}.

\subsubsection{OH - 18~cm: $F=g_{\rm p}(\aem^2\mu)^{1.57}$}

Using transitions originating from a single species, like with
SIDAM, allows to reduce the systematic effects. The 18~cm lines of
the OH radical offers such a possibility~\cite{OH-1,OH-2}.

The ground state, ${}^2\Pi_{3/2} J=3/2$, of OH is split into
two levels by $\Lambda$-doubling and each of these doubled level
is further split into two hyperfine-structure states. Thus it has
2 ``main'' lines ($\Delta F=0$) and two ``satellite'' lines
($\Delta F=1$). Since this four lines arise from two different
physical processes ($\Lambda$-doubling and hyperfine splitting),
they enjoy the same Rydberg dependence but different $g_{\rm p}$
and $\aem$ dependencies. By comparing the four transitions to the
H{\sc i} hyperfine line, one can have access to
\begin{equation}
 F\equiv g_{\rm p}(\aem^2\mu)^{1.57}
\end{equation} 
and it was also proposed to combine them with HCO$^+$ transitions
to lift the degeneracy.

Using the four 18~cm OH lines from the gravitational lens at
$z\sim0.765$ toward PMN~J0134-0931 and comparing the H{\sc i} 21cm
and OH absorption redshifts of the different components allowed to set
the constraint~\cite{OH-3}
\begin{equation}
 \Delta F/F=(-0.44\pm0.36\pm1.0_{\rm syst})\times10^{-5},\qquad
 z=0.765,
\end{equation}
where the second error is due to velocity offsets between OH and
H{\sc i} assuming a velocity dispersion of 3~km/s. A similar
analysis~\cite{darling} in a system in the direction of
PKS~1413+135 gave
\begin{equation}
 \Delta F/F=(0.51\pm1.26)\times10^{-5},\qquad
 z=0.2467.
\end{equation}

\subsubsection{Far infrared fine-structure lines: $F'=\aem^2\mu$}

Another combination~\cite{FIR-2} of constants can be obtained
from the comparison of far infrared fine-structure spectra with
rotational transitions, which respectively behaves as
$R_\infty\aem^2$ and $R_\infty\bar\mu=R_\infty/\mu$ so that they give access to
$$
F'=\aem^2\mu .
$$
A good candidate for the rotational lines is CO since it is the
second most abundant molecule in the Universe after H$_2$.

Using the C{\sc ii} fine-structure and CO rotational emission lines
from the quasars J1148+5251 and BR~1202-0725, it was concluded
that
\begin{eqnarray}
 &&\Delta F'/F'=(0.1\pm1.0)\times10^{-4},\qquad z=6.42,\\
 &&\Delta F'/F'=(1.4\pm1.5)\times10^{-5},\qquad z=4.69,
\end{eqnarray}
which represents the best constraints at high redshift. As usual,
when comparing the frequencies of two different species, one must
account for random Doppler shifts caused by non-identical spatial
distributions of the two species. Several other candidates for microwave and FIR
lines with good sensitivities are discussed in Ref.~\cite{kozrio}.

\subsubsection{``Conjugate" satellite OH lines: $G=g_{\rm p}(\aem\mu)^{1.85}$}

The satellite OH~18cm lines are conjugate so that the two lines have the same shape, but with one line in emission
and the other in absorption. This arises due to an inversion of the level of populations within the
ground state of the OH molecule. This behaviour has recently been discovered at cosmological
distances  and it was shown~\cite{OH-1} that a comparison between the sum and difference
of satellite line redshifts probes $G=g_{\rm p}(\aem\mu)^{1.85}$.

From the analysis of the two conjugate satellite OH systems at $z\sim0.247$ towards
PKS~1413+135 and at $z\sim0.765$ towards PMN~J0134-0931, it was
concluded~\cite{OH-1} that
\begin{equation}
 |\Delta G/G| < 1.1\times10^{-5}.
\end{equation}
It was also applied to a nearby system, Centaurus A, to give $ |\Delta G/G| < 1.6\times10^{-5}$
at $z\sim0.0018$. A more recent analysis~\cite{OH-2b} claims for a tentative evidence (with 2.6$\sigma$
significance, or at 99.1\% confidence) for a smaller value of $G$
\begin{equation}
 \Delta G/G = (-1.18\pm0.46)\times10^{-5}
\end{equation}
for the system at $z\sim0.247$ towards PKS~1413+135.

One strength of this method is that it guarantees that the satellite lines
arise from the same gas, preventing from velocity offset between the
lines. Also, the shape of the two lines must agree if they arise from the same gas.

\subsubsection{Molecular spectra and the electron-to-proton mass ratio}

As was pointed out in \S~\ref{subsec31}, molecular lines can
provide a test of the variation\footnote{Again, $\mu$ is used
either from $m_{\rm e}/m_{\rm p}$ or $m_{\rm p}/m_{\rm e}$. I have chosen
to use $\mu=m_{\rm p}/m_{\rm e}$ and $\bar\mu=m_{\rm e}/m_{\rm p}$.}~\cite{mu-theorie} 
of $\mu$ since rotational and
vibrational transitions are respectively inversely proportional to
their reduce mass and its square-root [see Eq.~(\ref{mu1})].\\

{\it Constraints with H$_2$}\\

H$_2$ is the most abundant molecule in the universe and there
were many attempts to use its absorption spectra to put constraints
on the time variation of $\mu$ despite the fact that H$_2$ is very difficult to detect~\cite{noterdame}.

As proposed in Ref.~\cite{mu-vl},
the sensitivity of a vibro-rotational wavelength to a variation of
$\mu$ can be parameterized as
$$
 \lambda_i =  \lambda_i^0(1+z_{\rm abs})\left(1+ K_i\frac{\Delta\mu}{\mu}\right),
$$
where $ \lambda_i^0$ is the laboratory wavelength (in the vacuum) and
$ \lambda_i$ is the wavelength of the transition $i$ in the rest-frame
of the cloud, that is at a redshift $z_{\rm abs}$ so that the
observed wavelength is $ \lambda_i/(1+z_{\rm abs})$. $K_i$ is a
sensitivity coefficient analogous to the $q$-coefficient
introduced in Eq.~(\ref{qpara}), but with different normalisation
since in the parameterization we would have $q_i=\omega_i^0K_i/2$,
$$
 K_i \equiv \frac{\dd\ln \lambda_i}{\dd\ln\mu}
$$
corresponding to the Lyman and Werner bands of molecular hydrogen.
From this expression, one can deduce that the observed redshift
measured from the transition $i$ is simply
$$
 z_i = z_{\rm abs} + bK_i,\qquad b\equiv -(1+z_{\rm
 abs})\frac{\Delta\mu}{\mu},
$$
which implies in particular that $z_{\rm abs}$ is not the mean of
the $z_i$ if $\Delta\mu\not=0$ . Indeed $z_i$ is measured
with some uncertainty of the astronomical measurements $ \lambda_i$
and by errors of the laboratory measurements $ \lambda_i^0$.
But if $\Delta\mu\not=0$ there must exist a correlation between $z_i$
and $K_i$ so that a linear regression of $z_i$
(measurement) as a function of $K_i$ (computed) allows to extract
$(z_{\rm abs},b)$ and their statistical significance.

We refer to \S~V.C of FVC~\cite{{jpu-revue}} for earlier studies
and we focus on the latest results. The recent constraints are
mainly based on the molecular hydrogen of two damped
Lyman-$\alpha$ absorption systems at $z=2.3377$ and $3.0249$ in
the direction of two quasars (Q~1232+082 and Q~0347-382) for which
a first analysis of VLT/UVES data showed~\cite{mu-ivan1} a slight
indication of a variation,
$$
 \Delta\mu/\mu=(5.7\pm3.8)\times10^{-5}
$$
at 1.5$\sigma$ for the combined analysis. The lines were selected
so that they are isolated, unsaturated and unblended. It follows
that the analysis relies on 12 lines (over 50 detected) for the
first quasar and 18 (over 80) for second but the two selected
spectra had no transition in common. The authors performed their
analysis with two laboratory catalogs and got different results.
They point out that the errors on the laboratory wavelengths are
comparable to those of the astronomical measurements.

It was further improved with an analysis on two absorption systems at
$z=2.5947$ and $z=3.0249$ in the directions 
of Q~0405-443 and Q~0347-383 observed with the VLT/UVES
spectrograph. The data have a resolution $R=53000$ and a S/N ratio
ranging between 30 and 70. The same selection criteria where
applied, letting respectively 39 (out of 40) and 37 (out of 42)
lines for each spectrum and only 7 transitions in common. The
combined analysis of the two systems led~\cite{mu-ivan2}
$$
 \Delta\mu/\mu=(1.65\pm0.74)\times10^{-5}\qquad\hbox{or}\qquad
 \Delta\mu/\mu=(3.05\pm0.75)\times10^{-5},
$$
according to the laboratory measurements that were used. The same
data were reanalyzed with new and highly accurate measurements of
the Lyman bands of H$_2$, which implied a reevaluation of the
sensitivity coefficient $K_i$.  It leads to the two
constraints~\cite{mu-rein}
\begin{eqnarray}
 &&\Delta\mu/\mu = (2.78\pm0.88)\times10^{-5},\qquad z= 2.59,\\
 &&\Delta\mu/\mu = (2.06\pm0.79)\times10^{-5},\qquad z = 3.02,\label{muxx0}
\end{eqnarray}
leading to a 3.5$\sigma$ detection for the weighted mean
$\Delta\mu/\mu = (2.4\pm0.66)\times10^{-5}$. The authors
of Ref.~\cite{mu-rein} do not claim for a detection and are cautious
enough to state that systematics dominate the measurements.
The data of the
$z=3.02$ absorption system were reanalyzed in Ref.~\cite{mu-wendt}
which claim that they lead to the bound
$|\Delta\mu/\mu|<4.9\times10^{-5}$ at a 2$\sigma$ level, instead
of Eq.~(\ref{muxx0}). 
Adding a new set of 6 spectra, it was concluded
that $\Delta\mu/\mu = (15\pm14)\times10^{-6}$ for the weighted fit~\cite{wendt}.

These two systems were reanalyzed~\cite{mu-king}, adding a new
system in direction of Q~0528-250,
\begin{eqnarray}
 &&\Delta\mu/\mu = (1.01\pm0.62)\times10^{-5},\qquad z= 2.59,\\
 &&\Delta\mu/\mu = (0.82\pm0.74)\times10^{-5},\qquad z= 2.8,\\
 &&\Delta\mu/\mu = (0.26\pm0.30)\times10^{-5},\qquad z = 3.02,
\end{eqnarray}
respectively with 52, 68 and 64 lines. This gives a weighted mean
of $(2.6\pm3.0)\times10^{-6}$ at $z\sim2.81$. To compare with the previous data,
the analysis of the two quasars in common was performed by using
the same lines (this implies adding 3 and removing 16 for
Q~0405-443 and adding 4 and removing 35 for Q~0347-383) to get
respectively $(-1.02\pm0.89)\times10^{-5}$ ($z=2.59$) and
$(-1.2\pm1.4)\times10^{-5}$ ($z = 3.02$). Both analysis disagree
and this latter analysis indicates a systematic shift of
$\Delta\mu/\mu$ toward 0. A second reanalysis of the same data was performed
in Refs.~\cite{mu-thom,mu-thom2} using a different analysis method to get
\begin{equation}
 \Delta\mu/\mu = (-7\pm8)\times10^{-6}.
\end{equation}
Recently discovered molecular transitions at $z=2.059$ toward the
quasar J2123-0050 observed by the Keck telescope allow to obtain
86 H$_2$ transitions and 7 HD transitions to conclude~\cite{HD-1}
\begin{equation}
 \Delta\mu/\mu= (5.6\pm5.5_{\rm stat}\pm2.7_{\rm syst})\times 10^{-6},\qquad
 z=2.059.
\end{equation}

This method is subject to important systematic errors among which
(1) the sensitivity to the laboratory wavelengths (since the use of two different
catalogs yield different results~\cite{mu-rein}), (2) the molecular lines are located in the
Lyman-$\alpha$ forest where they can be strongly blended with intervening H{\sc i}
Lyman-$\alpha$ absorption lines which requires a carfull fitting of the lines~\cite{mu-king}
since it is hard to find lines that are not contaminated.
From an observational point of view, very few damped Lyman-$\alpha$ systems
have a measurable amount of H$_2$ so that only
a dozen systems is actually known even though more systems
will be obtained soon~\cite{dlasearch}. To finish, the sensitivity coefficients
are usually low, typically of the order of $10^{-2}$. Some advantages of using H$_2$ arise
from the fact there are several hundred available H$_2$ lines so that many
lines from the same ground state can be used to eliminate different kinematics between
regions of different excitation temperatures. The overlap between Lyman and
Werner bands also allow to reduce the errors of calibration.

To conclude, the combination of all the existing observations indicate that $\mu$
is constant at the $10^{-5}$ level during the past 11 Gigayrs while an improvement
of a factor 10 can be expected in the five coming years.\\

{\it Other constraints}\\

It was recently proposed~\cite{mu-N1,mu-N2} that the inversion
spectrum of ammonia allows for a better sensitivity to $\mu$. The
inversion vibro-rotational mode is described by a double well with
the first two levels below the barrier. The tunnelling implies
that these two levels are split in inversion doublets. It was
concluded that the inversion transitions scale as $\nu_{\rm
inv}\sim \bar\mu^{4.46}$, compared with a rotational transition which
scales as $\nu_{\rm rot}\sim \bar\mu$. This implies that the redshifts
determined by the two types of transitions are modified according to
$\delta z_{\rm inv}=4.46(1+z_{\rm abs})\Delta\mu/\mu$ and $\delta
z_{\rm rot}\sim(1+z_{\rm abs})\Delta\mu/\mu$ so that
$$
 \Delta\mu/\mu = 0.289\frac{z_{\rm inv} - z_{\rm rot}}{1+z_{\rm
 abs}}.
$$
Only one quasar absorption system, at $z=0.68466$ in the direction
of B~0218+357, displaying NH$_3$ is currently known and allows for
this test. A first analysis~\cite{mu-N1} estimated from the
published redshift uncertainties that a precision of
$\sim2\times10^{-6}$ on $\Delta\mu/\mu$ can be achieved. A
detailed measurement~\cite{mu-N3} of the ammonia inversion
transitions by comparison to HCN and HCO$^+$ rotational
transitions concluded that
\begin{equation}\label{muweb}
 |\Delta\mu/\mu| < 1.8\times10^{-6}, \qquad z=0.685,
\end{equation}
at a 2$\sigma$ level. Recently the analysis of the comparison
of NH$_3$ to HC$_3$N spectra was performed toward the gravitational lens system 
PKS~1830-211 ($z\simeq0.89$), which is a much more suitable system, with
10 detected NH$_3$ inversion lines and a forest of rotational transitions. It
reached the conclusion that
\begin{equation}\label{muhenkel}
 |\Delta\mu/\mu| < 1.4\times10^{-6}, \qquad z=0.89,
\end{equation}
at a 3$\sigma$ level~\cite{mu-henkel}. From a comparison of the
ammonia inversion lines with the NH$_3$ rotational transitions, it
was concluded~\cite{menten}
\begin{equation}\label{mumentel}
 |\Delta\mu/\mu| < 3.8\times10^{-6}, \qquad z=0.89,
\end{equation}
at 95\% C.L. One strength of this analysis is to focus on lines arising from only one molecular
species but it was mentionned that the frequencies of the inversion lines
are about 25 times lower than the rotational ones, which might cause differences in the
absorbed background radio continuum.

This method was also applied~\cite{mu-lev} in the Milky Way, in
order to constrain the spatial variation of $\mu$ in the galaxy (see \S~\ref{subsec23mw}).
Using ammonia emission lines from interstellar molecular clouds
(Perseus molecular core, the Pipe nebula and the infrared dark
clouds) it was concluded that $\Delta\mu=(4-14)\times10^{-8}$. This
indicates a positive velocity offset between the ammonia inversion
transition and rotational transitions of other molecules. Two
systems being located toward the galactic center while one is in
the direction of the anti-center, this may indicate a spatial
variation of $\mu$ on galactic scales.\\

{\it New possibilities}\\

The detection of several deuterated molecular hydrogen HD transitions makes it possible to test the variation
of $\mu$ in the same way as with H$_2$ but in a completely
independent way, even though today it has been detected only in
2 places in the universe. The sensitivity coefficients have been published
in Ref.~\cite{mu-ivanov} and HD was first detected by Ref.~\cite{noterdame}.

HD was recently detected~\cite{HDdetect} together with CO and H$_2$ in a
DLA cloud at a redshift of 2.418 toward SDSS1439+11 with 5 lines
of HD in 3 components together with several H$_2$ lines in 7 components.
It allowed to set the 3$\sigma$ limit of $|\Delta\mu/\mu|<9\times10^{-5}$~\cite{dlmu}.

Even though the small number of lines does not allow to reach the level of
accuracy of H$_2$ it is a very promising system in particular to obtain
independent measurements.

\subsubsection{Emission spectra}

Similar analysis to constrain the time variation of the
fundamental constants were also performed with emission spectra.
Very few such estimates have been performed, since it is less
sensitive and harder to extend to sources with high redshift. In
particular, emission lines are usually broad as compared to absorption
lines and the larger individual errors need to be beaten by large
statistics.

The O{\sc iii} doublet analysis~\cite{em-o3} from a sample of 165
quasars from SDSS gave the constraint
\begin{equation}
 \Delta\aem/\aem = (12\pm7)\times10^{-5},\qquad 0.16<z<0.8.
\end{equation}
The method was then extended straightforwardly along the lines of
the MM method and applied~\cite{em-grup} to the fine-structure
transitions in Ne{\sc iii}, Ne{\sc v}, O{\sc iii}, O{\sc i} and S{\sc ii} multiplets from a
sample of 14 Seyfert 1.5 galaxies to derive the constraint
\begin{equation}
 \Delta\aem/\aem = (150\pm70)\times10^{-5},\qquad 0.035<z<0.281.
\end{equation}

\subsubsection{Conclusion and prospects}

\begin{table}[h]
\begin{center}
{\small
\begin{tabular}{p{3.0 cm} ccccc}
\hline\hline
 Constant   &  Method & System & Constraint $(\times10^{-5})$ &Redshift  & Ref. \\
 \hline
 $\aem$     &   AD    &   21   & $(-0.5\pm1.3)$   & 2.33 - 3.08   &   \cite{q-murphyAD}\\
            &   AD    &   15   & $(-0.15\pm0.43)$   &1.59 - 2.92 &    \cite{q-chandAD}\\
            &   AD    &   9   & $(-3.09\pm8.46)$   &1.19 - 1.84 &    \cite{q-martinezAD}\\
            &   MM    &  143  & $(-0.57\pm0.11)$ &0.2 - 4.2 &   \cite{q-murphyMMlast}\\
            &   MM    &   21  & $(0.01\pm0.15)$ & 0.4 - 2.3 &  \cite{q-chandMM}\\
             & SIDAM &    1  & $(-0.012\pm0.179)$ &1.15 &   \cite{q-sidam1}\\
            & SIDAM &    1  & $(0.566\pm0.267)$ &1.84 &   \cite{q-sidam1}\\
 \hline
 $y$   &  H{\sc i} - mol   &  1  & $ (-0.16\pm0.54)$ & 0.6847   & \cite{y-murph}  \\
       &   H{\sc i} - mol   &  1  & $(-0.2\pm0.44) $ & 0.247    & \cite{y-murph} \\
       &   CO, CHO$^+$   &    & $(-4\pm6) $ & 0.247   & \cite{y-combes}\\
  \hline
  $F$   &   OH - H{\sc i}  &  1  & $(-0.44\pm0.36\pm1.0_{\rm syst})$ &  0.765  & \cite{OH-3}\\
     &   OH - H{\sc i}  &  1  & $(0.51\pm1.26)$ &  0.2467  & \cite{darling}\\
  \hline
  $x$   &  H{\sc i} - UV   &  9  & $(-0.63\pm0.99)$ &  0.23 -2.35  & ~\cite{UV-2} \\
       &  H{\sc i} - UV   &  2  & $-(0.17\pm0.17)$ &  3.174  & ~\cite{x21cm} \\ 
  \hline
  $F'$   &  C{\sc ii} - CO   &  1  & $(1\pm10)$  &   4.69 &  \cite{FIR-1} \\
         &  C{\sc ii} - CO   &  1  & $(14\pm15)$ &   6.42 &  \cite{FIR-1} \\
   \hline
  $G$   &  OH   &  1  & $<1.1$  &   0.247, 0.765 &  \cite{OH-1} \\
         &  OH   &  1  & $<1.16$ &   0.0018 &  \cite{OH-1} \\
                 &  OH   &  1  & $(-1.18\pm0.46)$ &   0.247 &  \cite{OH-2b} \\
   \hline
  $\mu$   & H$_2$    & 1   & $(2.78\pm0.88)$ &  2.59  &  \cite{mu-rein} \\
     & H$_2$    & 1   & $(2.06\pm0.79)$ &  3.02  &  \cite{mu-rein} \\
       & H$_2$    & 1   & $(1.01\pm0.62)$ &  2.59  &  \cite{mu-king} \\
        & H$_2$    & 1   & $(0.82\pm0.74)$ &  2.8  &  \cite{mu-king} \\
     & H$_2$    & 1   & $(0.26\pm0.30)$ &  3.02  &  \cite{mu-king} \\
     & H$_2$    & 1   & $(0.7\pm0.8)$ &  3.02, 2.59  &  \cite{mu-thom} \\
      & NH$_3$    & 1   & $ <0.18$ &  0.685  &  \cite{mu-N3} \\
      & NH$_3$    & 1   & $ <0.38$ &  0.685  &  \cite{menten} \\
        & HC$_3$N    & 1   & $ <0.14$ &  0.89  &  \cite{mu-henkel} \\
        & HD   & 1   & $<9$ &  2.418  &  \cite{dlmu} \\
           & HD   & 1   & $(0.56\pm0.55_{\rm stat}\pm0.27_{\rm syst})$ &  2.059  &  \cite{HD-1} \\
\hline\hline
\end{tabular}
\caption{\it Summary of the latest constraints on the variation of
fundamental constants obtained from the analysis of quasar
absorption spectra. We recall that $y\equiv g_{\rm p}\aem^2$,
$F\equiv g_{\rm p}(\aem^2\mu)^{1.57}$, $x\equiv\aem^2 g_{\rm
p}/\mu$, $F'\equiv\aem^2\mu$ and $\mu\equiv m_{\rm p}/m_{\rm e}$, $G=g_{\rm p}(\alpha\mu)^{1.85}$.}
\label{tab-quasar}}
\end{center}
\end{table}

\begin{figure}[hptb]
  \def\epsfsize#1#2{0.5#1}
  \centerline{\includegraphics[scale=0.5]{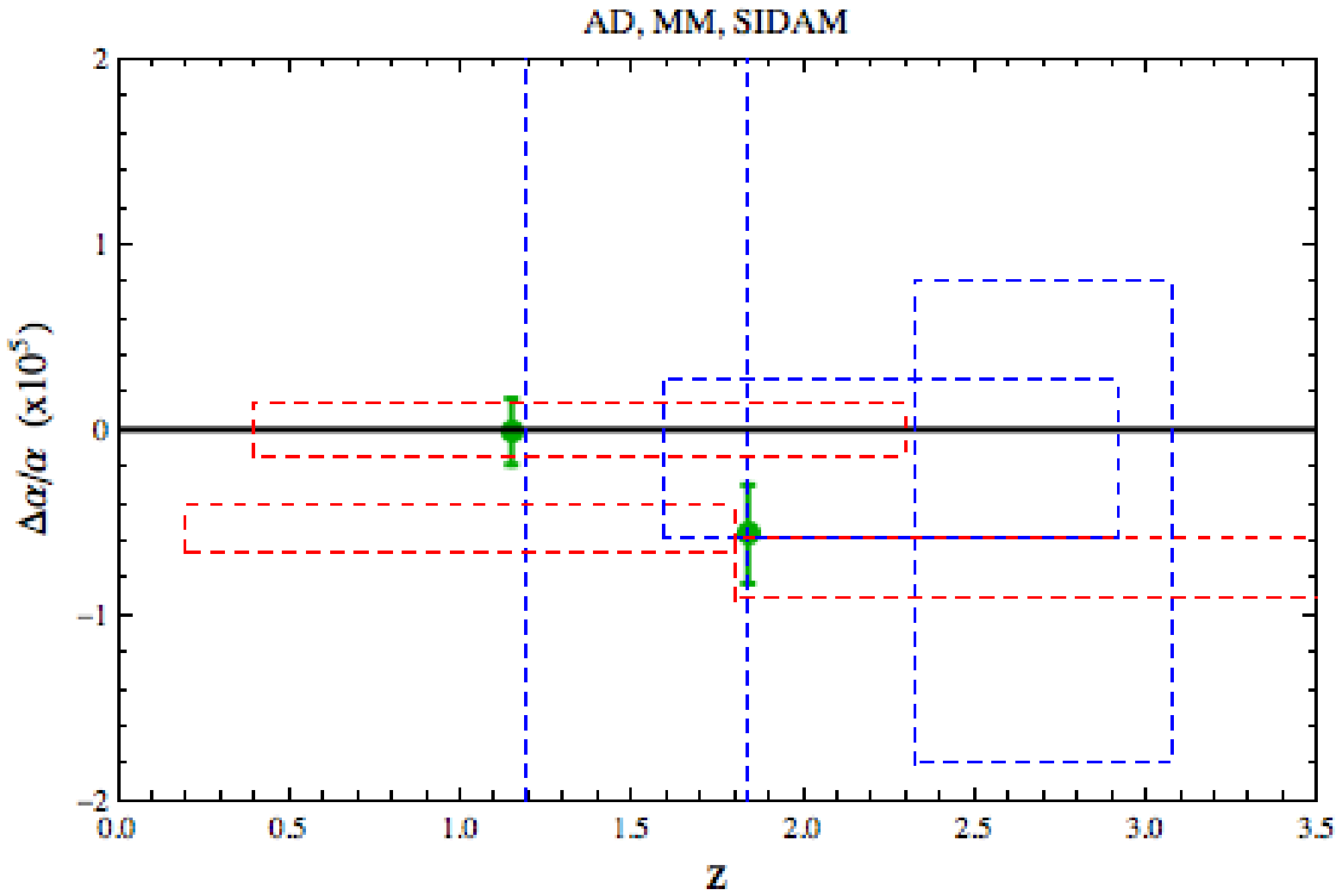},\includegraphics[scale=0.5]{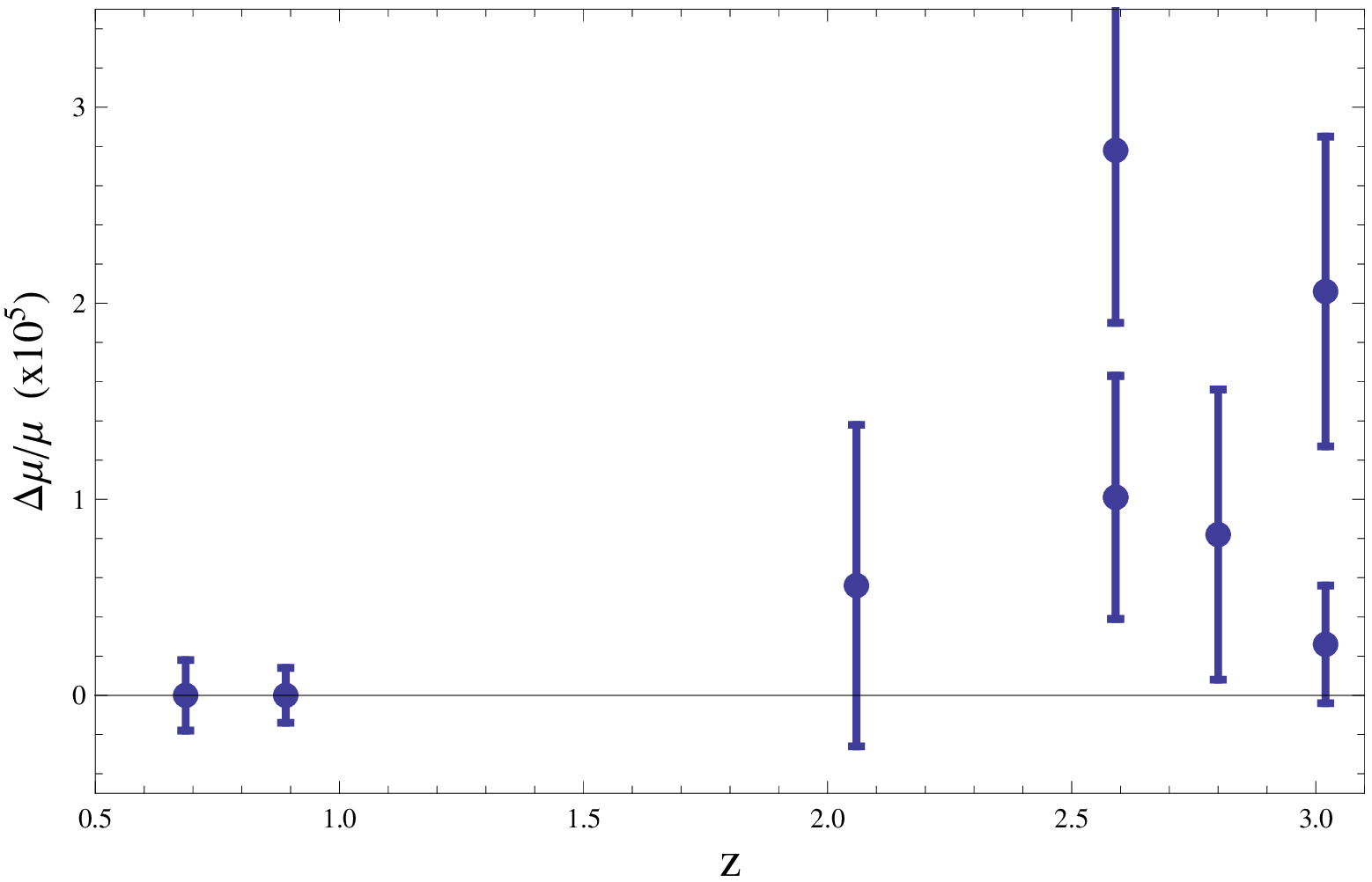}}
  \caption{\it Summary of the direct constraints on $\aem$ obtained from the AD (blue), 
  MM (red) and AD (green) methods (left) and on $\mu$ (right) that are
  summarized in Table~\ref{tab-quasar}.}
  \label{fig-qso}
\end{figure}

This paragraph illustrates the diversity of methods and the
progresses that have been achieved to set robust constraints on
the variation of fundamental constants. Many systems are now used,
giving access to different combinations of the constants. It
exploits a large part of the electromagnetic spectrum from far
infrared to ultra violet and radio bands and optical and radio 
techniques have played complementary roles. The most recent and
accurate constraints are summarized in Table~\ref{tab-quasar} and
Fig.~\ref{fig-qso}.

At the moment, only one analysis claims to have detected a variation of the fine
structure constant (Keck/HIRES) while the VLT/UVES points toward 
no variation of the fine structure constant. It has led to the proposition
that $\aem$ may be space dependent and exhibit a dipole, the
origin of which is not explained.
Needless to say that such a controversy and hypotheses are sane since it will help
improve the analysis of this data but it is premature to conclude
on the issue of this debate and the jury is still out. Most of the
systematics have been investigated in details and now seem under
control.

Let us what we can learn on the physics from these measurement. As
an example, consider the constraints obtained on $\mu$, $y$ and $F$
in the redshift band 0.6-0.8 (see Table~\ref{tab-quasar}). They
can be used to extract independent constraints on $g_{\rm p}$,
$\aem$ and $\mu$
$$
 \Delta\mu/\mu=(0\pm0.18)\times10^{-5}, \quad
 \Delta\aem/\aem=(-0.27\pm2.09)\times10^{-5}, \quad
 \Delta g_{\rm p}/g_{\rm p}=(0.38\pm4.73)\times10^{-5}.
$$
This shows that one can test the compatibility of the
constraints obtained from different kind of systems. Independently
of these constraints, we have seen in
\S~\ref{subsec22} that in grand unification theory the variation
of the constants are correlated. The former constraints show
that if $\Delta\ln\mu=R\Delta\ln\aem$ then the
constraint~(\ref{muweb}) imposes that
$|R\Delta\ln\aem|<1.8\times10^{-6}$. In general $R$ is expected to
be of the order of $30-50$. Even if its value its time-dependent,
that would mean that $\Delta\ln\aem\sim(1-5)\times10^{-7}$ which is highly
incompatible with the constraint~(\ref{a-MMw}) obtained by the
same team on $\aem$, but also on the constraints~(\ref{AD-m})
and~(\ref{a-AD1}) obtained from the AD method and on which both
teams agree. This illustrates how important the whole set of data
is since one will probably be able to constrain the order of
magnitude of $R$ in a near future, which would be a very important
piece of information for the theoretical investigations.\\

We mention in the course of this paragraph many possibilities to
improve these constraints.

Since the AD method is free of the two main assumptions of the MM
method, it seems important to increase the precision of this
method as well as any method relying only on one species. This can
be achieved by increasing the S/N ratio and spectral resolution of
the data used or by increasing the sample size and including new 
transitions (e.g. cobalt~\cite{co1,co2}).

The search for a better resolution is being investigated in many
direction. With the current resolution of $R\sim40000$, the
observed line positions can be determined with an accuracy of
$\sigma_\lambda\sim 1\,{\rm m}$\AA. This implies that the accuracy
on $\Delta\aem/\aem$ is of the order of $10^{-5}$ for lines with
typical $q$-coefficients. As we have seen this limit can be
improved to $10^{-6}$ when more transitions or systems are used
together. Any improvement is related to the possibility to measure
line positions more accurately. This can be done by increasing $R$
up to the point at which the narrowest lines in the absorption
systems are resolved. The Bohlin formula~\cite{bohlin} gives the
estimates
$$
 \sigma_\lambda\sim \Delta\lambda_{\rm pix}
     \left(\frac{\Delta\lambda_{\rm pix}}{W_{\rm obs}}\right)
     \frac{1}{\sqrt{N_e}}\left(\frac{M^{3/2}}{\sqrt{12}}\right),
$$
where $\Delta\lambda_{\rm pix}$ is the pixel size, $W_{\rm obs}$
is the observed equivalent width, $N_e$ is the mean number of
photoelectron at the continuum level and $M$ is the number of
pixel covering the line profile. The metal lines have intrinsic
width of a few km/s. One can thus expect improvements from higher
spectral resolution. Progresses consercning the
calibration are also expected, using e.g. laser comb~\cite{steinmetz}. 
Let us just mention, the EXPRESSO (Echelle
Spectrograph for PREcision Super Stable Observation)
project~\cite{Expresso} on 4 VLT units
or the CODEX (COsmic Dynamics EXplorer) on E-ELT
projects~\cite{CODEX,CODEX2,ubm}. They shall provide a resolving power of
$R=150000$ to be compared to the
HARPS\footnote{http://obswww.unige.ch/Instruments/HARPS/} (High
Accuracy Radial velocity planet Searcher) spectrograph
($R\sim112000$) has been used but it is operating on a 3.6m
telescope.

The limitation may then lie in the statistics and the calibration
and it would be useful to use more than two QSO with overlapping
spectra to cross-calibrate the line positions. This means that one
needs to discover more absorption systems suited for these
analysis. Many progresses are expected. For instance, the FIR
lines are expected to be observed by a new generation of
telescopes such as
HERSCHEL\footnote{http://sci.esa.int/science-e/www/area/index.cfm?fareaid=16}.
While the size of the radio sample is still small, surveys are
being carried out so that the number of known redshift OH, HI and
HCO+ absorption systems will increase. For instance the
future Square Kilometre Array (SKA) will be able to detect relative changes
of the order of $10^{-7}$ in $\aem$.

In conclusion, it is clear that these constraints and the
understanding of the absorption systems will increase in the
coming years.

\subsection{Stellar constraints}\label{secstellar}

\begin{figure}[hptb]
  \def\epsfsize#1#2{0.5#1}
  \centerline{\includegraphics[scale=0.34]{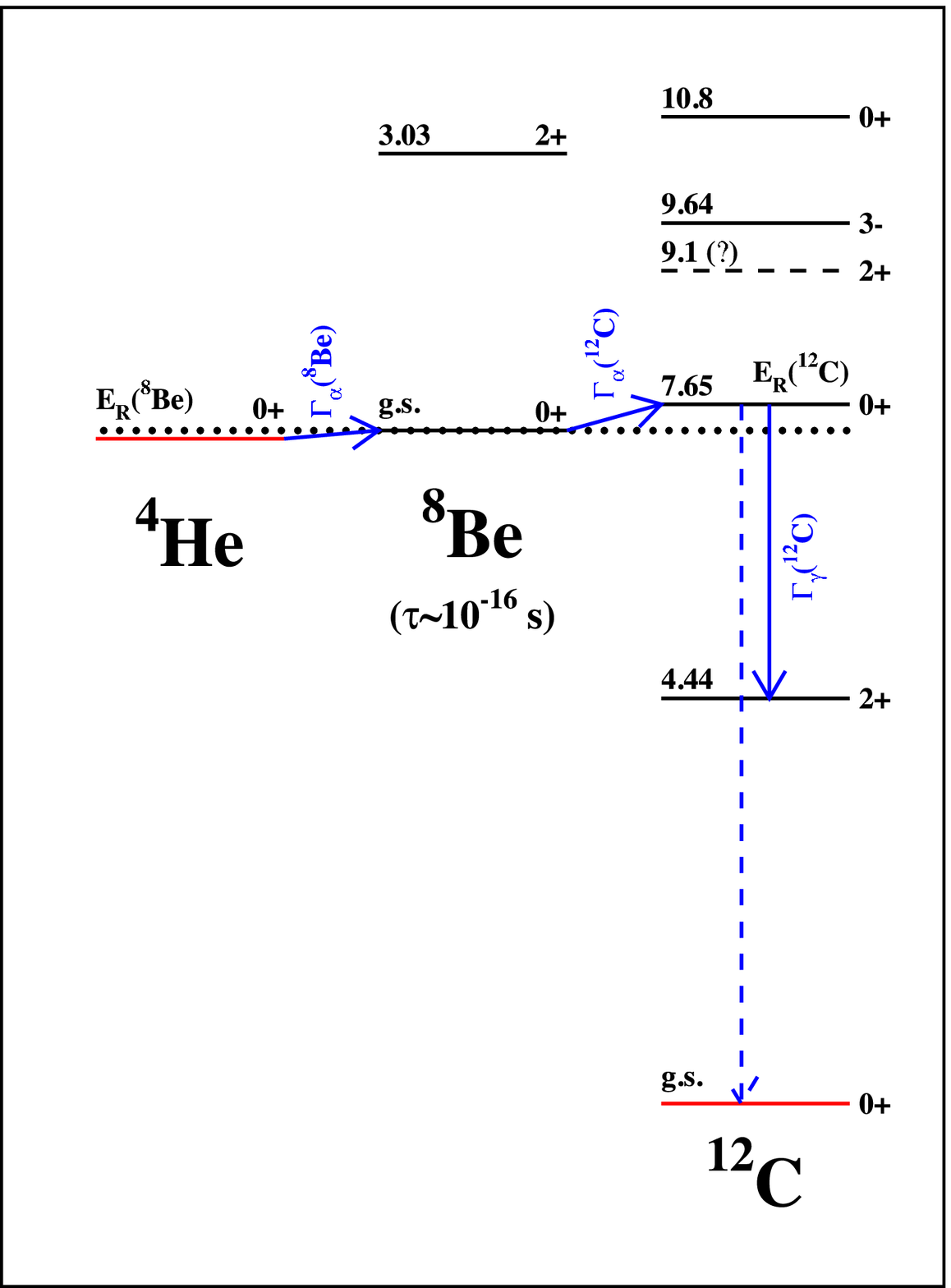}$\qquad$\includegraphics[scale=0.35]{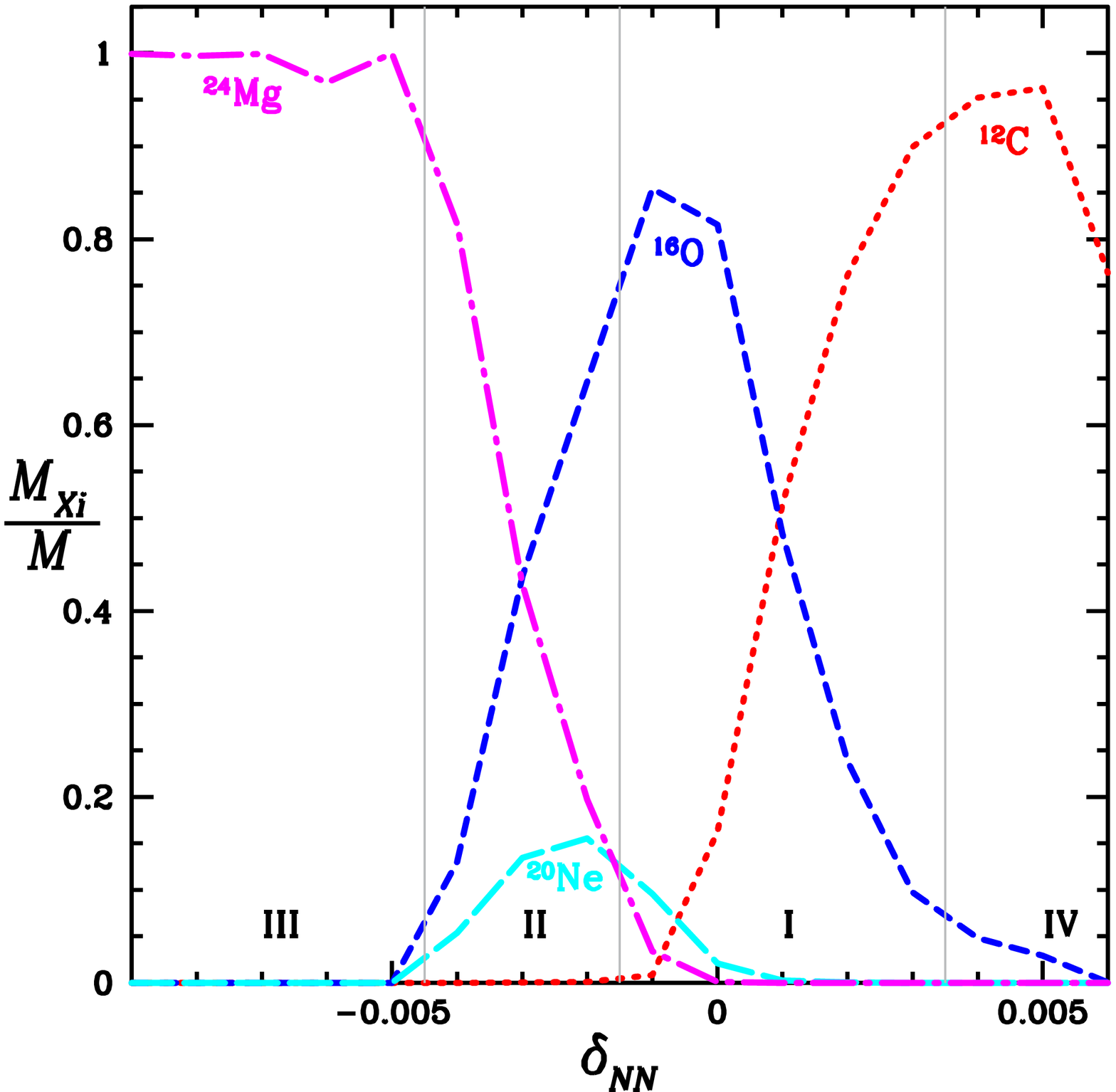}}
  \caption{\it (Left) Level scheme of nuclei participating to the
  $ {}^4{\rm He}(\alpha\alpha,\gamma) {}^{12}{\rm C}$ reaction.
  (Right) Central abundances at the end of the CHe burning as a function of $\delta_{NN}$
  for a $60M_\odot$ star with $Z=0$. From Ref.~\cite{coc3a}.}
  \label{fig-3a1}
\end{figure}

Stars start to accumulate helium produced by the pp-reaction and
the CNO cycle in their core. Furthermore, the products of further
nuclear reactions of helium with either helium or hydrogen lead
to isotopes with $A=5$ or $A=8$, which are highly unstable. In
order to produce elements heavier than $A>7$ by fusion of lighter
isotopes, the stars need to reach high temperatures and densities.
In these conditions, newly produced ${}^{12}{\rm C}$ would almost immediately be
fused further to form heavier elements so that one expects only a
tiny amount of ${}^{12}{\rm C}$ to be produced, in contradiction with the
observed abundances. This led Hoyle~\cite{star-hoyle} to conclude
that a then unknown excited state of the ${}^{12}{\rm C}$ with an
energy close to the $3\alpha$-threshold should exist since such a
resonance would increase the probability that ${}^8{\rm Be}$
captures an $\alpha$-particle. It follows that the production of
${}^{12}{\rm C}$ in stars relies on the three conditions:
\begin{itemize}
 \item the decay lifetime of ${}^8{\rm Be}$, of order
$10^{-16}$~s, is four orders of magnitude longer than the time for
two $\alpha$ particles to scatter, so that a macroscopic amount of
beryllium can be produced, which is sufficient to lead to
considerable production of carbon,
 \item an excited state of ${}^{12}{\rm C}$ lies just above the energy of ${}^8{\rm
 Be}+\alpha$, which allows for
 $$
 {}^4{\rm He} + {}^4{\rm He} \leftrightarrow {}^8{\rm Be},\qquad
 {}^8{\rm Be}+{}^4{\rm He}\leftrightarrow {}^{12}{\rm C}^*
 \rightarrow {}^{12}{\rm C} + 7.367\,{\rm MeV},
 $$
 \item the energy level of ${}^{16}{\rm O}$ at 7.1197~MeV is non
resonant and below the energy of ${}^{12}{\rm C}+\alpha$, of order
7.1616~MeV, which ensures that most of the carbon synthesized is
not destroyed by the capture of an $\alpha$-particle. The
existence of this resonance, the $O_2^+$-state of ${}^{12}{\rm C}$
was actually discovered~\cite{star-cooke} experimentally later,
with an energy of $372\pm4$~keV [today,
$E_{O_2^+}=379.47\pm0.15$~keV], above the ground state of three
$\alpha$-particles (see Fig.~\ref{fig-3a1}).
\end{itemize}

The variation of any constant that would modify the energy of
this resonance would also endanger the stellar nucleosynthesis
of carbon, so that the possibility for carbon production has often
been used in anthropic arguments. Qualitatively, if $E_{O_2^+}$ is increased then
the carbon would be rapidly processed to oxygen since the star
would need to be hotter for the triple-$\alpha$ process to start.
On the other hand, if $E_{O_2^+}$ is decreased, then all
$\alpha$-particles would produce carbon so that no oxygen would be
synthesized. It was estimated~\cite{star-livio} that the carbon
production in intermediate and massive stars is suppressed if the
various of the energy of the resonance is outside the range
$-250$~keV$\la\Delta E_{O_2^+}\la60$~keV, which was further
improved~\cite{star-ober1} to, $-5$~keV$\la\Delta
E_{O_2^+}\la50$~keV in order for the C/O ratio to be larger than
the error in the standard yields by more than 50\%. Indeed, in
such an analysis, the energy of the resonance was changed by hand.
We expect however that if $E_{O_2^+}$ is modified due to the
variation of a constant other quantities, such as the resonance of
the oxygen, the binding energies and the cross-sections will also
be modified in a complex way.

In practice, to draw a constraint on the variation of the
fundamental constants from the stellar production of carbon, one
needs to go through different steps, any of them involving
assumptions,
\begin{enumerate}
 \item to determine the effective parameters, e.g. cross
 sections, which affects the stellar evolution. The simplest choice
 is to modify only the energy of the resonance but it may not be
 realistic since all cross-sections and binding energies should also
 be affected. This requires to use a stellar evolutionary model;
 \item relate these parameters to nuclear parameters. This
 involves the whole nuclear physics machinery;
 \item to relate the nuclear parameters to fundamental constants.
 As for the Oklo phenomenon, it requires to link QCD to nuclear
 physics.
\end{enumerate}

A first analysis~\cite{star-ober3,star-ober4,star-ober1} used a
model that treats the carbon nucleus by solving the 12-nucleon
Schr\"odinger equation using a three-cluster wave-function
representing the three-body dynamics of the ${}^{12}$ state. The
NN interaction was described by the Minnesota
model~\cite{star-min,star-thom} and its strength was modified by multiplying
the effective NN-potential by an arbitrary number $p$. This allows
to relate the energy of the Hoyle level relative to the triple alpha threshold, 
$\varepsilon\equiv Q_{\alpha\alpha\alpha}$, and the gamma width,
$\Gamma_\gamma$, as a function of the parameter $p$, the latter
being almost not affected. The modified $3\alpha$-reaction rate
was then given by
\begin{equation}
 r_\alpha = 3^{3/2}N_\alpha^3\left(\frac{2\pi \hbar^2}{M_\alpha k_{\rm B}T}\right)^3
 \frac{\Gamma}{\hbar} \exp\left[{-\frac{\varepsilon(p)}{k_{\rm
 B}T}}\right],
\end{equation}
where $M_\alpha$ and $N_\alpha$ are the mass and number density of
the $\alpha$-particle, The resonance width
$\Gamma=\Gamma_\alpha\Gamma_\gamma/ (\Gamma_\alpha +
\Gamma_\gamma) \sim \Gamma_\gamma$. This was included in a stellar
code and ran for red giant stars with $1.3$, $5$ and
$20M_\odot$ with solar metallicity up to thermally pulsating asymptotic giant branch~\cite{star-ober3}
and in low, intermediate and high mass ($1.3, 5, 15, 25M_\odot$) with solar metallicity
also up to TP-AGB~\cite{star-ober1} to conclude
that outside a window of respectively 0.5\% and 4\% of the values
of the strong and electromagnetic forces, the stellar production
of carbon or oxygen will be reduced by a factor 30 to 1000.

In order to compute the resonance energy of the beryllium-8 and carbon-12 a microsopic
cluster model was developed~\cite{star-min}.
The Hamiltonian of the system is then  of the form $H=\sum_i^A T({\bf r}_i +\sum_{j<i}^A V({\bf r}_{ij}) $,
where $A$ is the nucleon number, $T$ the kinetic energy and $V$ the NN interaction potential.
In order to implement the variation of the strength of the nuclear interaction with respect to the
electromagnetic interaction, it was taken as
$$
V({\bf r}_{ij}) = V_C({\bf r}_{ij}) + (1+\delta_{NN})V_N({\bf r}_{ij}),
$$
where $\delta_{NN}$ is a dimensionless parameter that describes the
change of the nuclear interaction, $V_N$ being described in Ref.~\cite{star-thom}. 
When $A>4$ no exact solution can be found
and approximate solutions in which the wave function of the beryllium-8 and carbon-12
are described by clusters of respectively 2 and 3 $\alpha$-particle is
well adapted. 

First, $\delta_{NN}$ can be related to the deuterium binding energy as
\begin{equation}
 \Delta B_D/B_D = 5.7701\times \delta_{NN}
\end{equation}
which, given the discussion in \S~\ref{bbn2cste}, allows to relate $\delta_{NN}$
to fundamental constants, as e.g. in Ref.~\cite{cnouv}. Then, the resonance energy of the beryllium-8 and carbon-12
scale as
\begin{equation}
 E_R({}^8{\rm Be}) = (0.09208-12.208\times\delta_{NN})\,{\rm Mev},\quad
 E_R({}^{12}{\rm C}) = (0.2877-20.412\times\delta_{NN})\,{\rm Mev},
\end{equation}
so that the energy of the Hoyle level relative to the triple alpha threshold
is $Q_{\alpha\alpha\alpha}= E_R({}^8{\rm Be})+E_R({}^{12}{\rm C})$.

This was implemenetd~\cite{coc3a,star-new} to population~III stars with typical masses,
15 and 60~$M_\odot$ with zero metallicity, in order to compute the central abundances
at the end of the core He burning. From Fig.~\ref{fig-3a1}, one can distinguish 4 regimes
(I) the star ends the CHe burning phase with a core composed of a mixture of carbon-12 and oxygen-16,
as in the standard case; (II) if the $3\alpha$ rate is weaker, ${}^{12}{\rm C}$ is produced slower,
the raction ${}^{12}{\rm C}(\alpha,\gamma){}^{16}{\rm O}$ becomes efficient earlier so that
the star ends the CHe burning phase with a core composed mostly of oxygen-16; (III) for
weaker rates, the oxygen-16 is further processed to neon-20 and then ${}^{24}{\rm Mg}$
so that the star ends the CHe burning phase with a core composed of ${}^{24}{\rm Mg}$
and (IV) if the $3\alpha$ rate is stronger, the carbon-12 is produced more rapidly and
the star ends the CHe burning phase with a core composed mostly of carbon-12. Typically this imposes
that
\begin{equation}
 -5\times10^{-4}<\delta_{NN}<1.5\times 10^{-3}, \qquad
 -3\times10^{-4}<\Delta B_D/B_D<9\times10^{-3}
\end{equation}
to ensure the ratio C/O to be of order unity.

To finish, a recent study~\cite{star-adam} focus on the existence
of stars themselves, by revisiting the stellar equilibrium when
the values of some constants are modified. In some sense, it can
be seen as a generalization of the work by Gamow~\cite{gamow67a}
to constrain the Dirac model of a varying gravitational
constant by estimating its effect on the lifetime of the Sun. In
this semi-analytical stellar structure model, the effect of the
fundamental constants was reduced phenomenologically to 3
parameters, $G$ which enters mainly on the hydrostatic
equilibrium, $\aem$ which enters in the Coulomb barrier
penetration through the Gamow energy, and a composite parameter
$\mathcal{C}$ which describes globally the modification of the
nuclear reaction rates. The underlying idea is to assume that the
power generated per unit volume, $\varepsilon(r)$, and which
determines the luminosity of the star, is proportional to the
fudge factor $\mathcal{C}$, which would arise from a modification
of the nuclear fusion factor, or equivalently of the cross
section. It thus assumes that all cross-sections are affected is a
similar way. The parameter space for which stars can form and for
which stable nuclear configurations exist was determined, showing
that no fine-tuning seems to be required.\\

This new system is very promising and will provide new informations on the fundamental constants
at redshifts smaller than $z\sim15$ where no constraints exist
at the moment, even though drawing a robust constraint
seems to be difficult at the moment. In particular, an underlying limitation
arises from the fact that the composition of the interstellar
media is a mixture of ejecta from stars with different masses and
it is not clear which type of stars contribute the most the carbon
and oxygen production. Besides, one would need to include rotation
and mass loss~\cite{meynet}. As for the Oklo phenomenon, another
limitation arises from the complexity of nuclear physics.

\subsection{Cosmic Microwave Background}\label{subsec34}

The CMB radiation is composed of photons emitted at the time of
the recombination of hydrogen and helium when the universe was
about 300,000 years old [see e.g.
Ref.~\cite{peteruzanbook} for details on the physics
of the CMB]. This radiation is observed to be a black-body with a
temperature $T_0=2.723$~K with small anisotropies of order of the
$\mu$K. The temperature fluctuation in a direction
$(\vartheta,\varphi)$ is usually decomposed on a basis of
spherical harmonics as
\begin{equation}\label{cmb1}
 \frac{\delta T}{T}(\vartheta,\varphi)=\sum_{\ell}\sum_{m=-\ell}^{m=+\ell}a_{\ell
  m}Y_{\ell m}(\vartheta,\varphi).
\end{equation}
The angular power spectrum multipole $C_\ell=\langle \vert
a_{lm}\vert^2 \rangle$ is the coefficient of the decomposition of
the angular correlation function on Legendre polynomials. Given a
model of structure formation and a set of cosmological parameters,
this angular power spectrum can be computed and compared to
observational data in order to constrain this set of
parameters.\\

The CMB temperature anisotropies mainly depend on three constants:
$G$, $\aem$ and $m_{\rm e}$.\\

The gravitational constant enters in the Friedmann equation and in
the evolution of the cosmological perturbations. It has mainly
three effects~\cite{cmb-G1} that are detailed in \S~\ref{subsecGcmb}.
$\aem$, $m_{\rm e}$ affect the dynamics of the recombination.
Their influence is complex and must be computed numerically. We
can however trace their main effects since they mainly modify the
CMB spectrum through the change in the differential optical depth
of photons due to the Thomson scattering
\begin{equation}\label{cmb2}
 \dot\tau=x_{\rm e}n_{\rm e}c\sigma_{\rm T}
\end{equation}
which enters in the collision term of the Boltzmann equation
describing the evolution of the photon distribution function and
where $x_{\rm e}$ is the ionization fraction (i.e. the number
density of free electrons with respect to their total number
density $n_{\rm e}$).

The first dependence arises from the Thomson scattering
cross-section given by
\begin{equation}\label{cmb3}
 \sigma_{\rm T}=\frac{8\pi}{3}\frac{\hbar^2}{m_{\rm e}^2c^2}\aem^2
\end{equation}
and the scattering by free protons can be neglected since $m_{\rm
e}/m_{\rm p}\sim5\times10^{-4}$.

The second, and more subtle dependence, comes from the ionization
fraction. Recombination proceeds via 2-photon emission from the
$2s$ level or via the Ly-$\alpha$ photons which are redshifted out
of the resonance line~\cite{cmb-peebles68} because recombination
to the ground state can be neglected since it leads to immediate
reionization of another hydrogen atom by the emission of a
Ly-$\alpha$ photons. Following Refs.~\cite{cmb-peebles68,cmb-ma-b}
and taking into account, for the sake of simplicity, only the
recombination of hydrogen, the equation of evolution of the
ionization fraction takes the form
\begin{equation}\label{eecmb1}
 \frac{\dd x_{\rm e}}{\dd t}={\cal C}\left[\beta\left(1-x_{\rm e}\right)
  \hbox{exp} \left(-\frac{B_1-B_2}{k_{_{\rm B}}T_M}\right)
  -{\cal R}n_{\rm p}x_{\rm e}^2\right],
\end{equation}
where $T_M$ is the temperature. At high redshift, $T_M$ is
identical to the one of the photons $T_\gamma=T_0(1+z)$ but
evolves according to
\begin{equation}\label{e_cmbT}
 \frac{\dd T_M}{\dd t} = -\frac{8\sigma_{\rm T} a_R}{3 m_{\rm e}}
 T_R^4\frac{x_{\rm e}}{1+x_{\rm e}}(T_M - T_\gamma) - 2 H T_M
\end{equation}
where the radiation constant $a_R=4\sigma_{\rm SB}/c$ with
$\sigma_{\rm SB} =k^4_{_{\rm B}}\pi^2/(60\pi c^2\hbar^3)$ the
Stefan-Boltzmann constant. In Eq.~(\ref{eecmb1}), $B_n=-E_I/n^2$
is the energy of the $n$th hydrogen atomic level, $\beta$ is the
ionization coefficient, ${\cal R}$ the recombination coefficient,
${\cal C }$ the correction constant due to the redshift of
Ly-$\alpha$ photons and to 2-photon decay and $n_p=n_e$ is the
number density of protons. $\beta$ is related to ${\cal R}$ by the
principle of detailed balance so that
\begin{equation}\label{cmb5}
\beta={\cal R}\left(\frac{2\pi m_{\rm e} k_{_{\rm
B}}T_M}{h^2}\right)^{3/2}\hbox{exp}\left(-\frac{B_2}{k_{_{\rm
B}}T_M}\right).
\end{equation}
The recombination rate to all other excited levels is
$$
{\cal R}=\frac{8\pi}{c^2}\left(\frac{k_{_{\rm B}}T}{2\pi m_{\rm
e}}\right)^{3/2} \sum_{n,l}^*(2l+1)\hbox{e}^{B_n/k_{_{\rm
B}}T}\int_{B_n/k_{_{\rm B}}T}^\infty \sigma_{nl}\frac{y^2\dd
y}{\hbox{e}^y-1}
$$
where $\sigma_{nl}$ is the ionization cross-section for the
$(n,l)$ excited level of hydrogen. The star indicates that the sum
needs to be regularized and the $\aem$-, $m_{\rm e}$-dependence of
the ionization cross-section is complicated to extract. It can
however be shown to behave as $\sigma_{nl}\propto\aem^{-1}m_{\rm
e}^{-2}f(h\nu/B_1)$. Finally, the factor ${\cal C}$ is given by
\begin{equation}\label{cmb7}
{\cal
C}=\frac{1+K\Lambda_{2s}(1-x_e)}{1+K(\beta+\Lambda_{2s})(1-x_e)}
\end{equation}
where $\Lambda_{2s}$ is the rate of decay of the $2s$ excited
level to the ground state via 2 photons; it scales as $m_{\rm
e}\aem^8$. The constant $K$ is given in terms of the Ly-$\alpha$
photon $\lambda_{\alpha}=16\pi\hbar/(3m_{\rm e}\aem^2c)$ by
$K=n_p\lambda_\alpha^3/(8\pi H)$ and scales as $m_{\rm
e}^{-3}\aem^{-6}$.

In summary, both the temperature of the decoupling and the
residual ionization after recombination are modified by a
variation of $\aem$ or $m_{\rm e}$. This was first discussed in
Ref.~\cite{cmb-bat,cmb-kap}. The last scattering surface can
roughly be determined by the maximum of the visibility function
$g=\dot\tau\exp(-\tau)$ which measures the differential
probability for a photon to be scattered at a given redshift.
Increasing $\aem$ shifts $g$ to a higher redshift at which the
expansion rate is faster so that the temperature and $x_e$
decrease more rapidly, resulting in a narrower $g$. This induces a
shift of the $C_\ell$ spectrum to higher multipoles and an
increase of the values of the $C_\ell$. The first effect can be
understood by the fact that pushing the last scattering surface to
a higher redshift leads to a smaller sound horizon at decoupling.
The second effect results from a smaller Silk damping.

Most studies have introduced those modification in the RECFAST
code~\cite{cmb-seager} including similar equations for the
recombination of helium. Our previous analysis shows that the
dependences in the fundamental constants have various origins,
since the binding energies $B_i$ scale has $m_{\rm e}\aem^2$, $\sigma_T$ as $\aem^2m_{\rm e}^{-2}$, 
$K$ as $m_{\rm e}^{-3}\aem^{-6}$, the ionisation coefficients $\beta$ as $\aem^3$, the
transition frequencies as $m_{\rm e}\aem^{2}$, the Einstein coefficients as $m_{\rm e}\aem^{5}$,
the decay rates $\Lambda$ as $m_{\rm e}\aem^{8}$ and $\cal{R}$ has complicated
dependence which roughly reduces to $\aem^{-1}m_{\rm e}^{-2}$.
Note that a change in the fine-structure constant and in the mass of the
electron are degenerate according to $\Delta\aem\approx0.39\Delta m_{\rm e}$
but this degeneracy is broken for multipoles higher than
1500~\cite{cmb-bat}. In
earlier works~\cite{cmb-han,cmb-kap} it was approximated by the
scaling ${\cal R}\propto\aem^{2(1+\xi)}$ with $\xi\sim0.7$.\\

The first studies~\cite{cmb-han,cmb-kap} focused on the
sensibility than can be reached by WMAP\footnote{{\tt
http://map.gsfc.nasa.gov/}} and Planck\footnote{{\tt
http://astro.estec.esa.nl/SA-general/Projects/Planck/}}. They
concluded that they should provide a constraint on $\aem$ at
recombination, i.e. at a redshift of about $z\sim1,000$, with a
typical precision $\vert\Delta\aem/\aem\vert\sim10^{-2}-10^{-3}$.

The first attempt~\cite{cmb-avelino00} to actually set a
constraint was performed on the first release of the data by
BOOMERANG and MAXIMA. It concluded that a value of $\aem$ smaller
by a few percents in the past was favoured but no definite bound
was obtained, mainly due to the degeneracies with other
cosmological parameters. It was later
improved~\cite{cmb-avelino01} by a joint analysis of BBN and CMB
data that assumes that only $\aem$ varies and that included 4 cosmological parameters
($\Omega_\mat,\Omega_\baryon,h,n_s)$ assuming a universe with Euclidean spatial section, leading to
$-0.09<\Delta\aem<0.02$ at 68\% confidence level. A similar analysis~\cite{cmb-landau01},
describing the dependence of a variation of the fine-structure constant
as an effect on recombination the redshift of which was modelled to scale as
$z_* = 1080[1+2\Delta\aem/\aem]$, set the constraint
$-0.14<\Delta\aem<0.02$, at a $2\sigma$ level, assuming a spatially flat cosmological
models with adiabatic primordial fluctuations that. The effect of
reionisation was discussed in Ref.~\cite{cmb-mart}. These works
assume that only $\aem$ is varying but, as can been seen from
Eqs.~(\ref{cmb1}-\ref{cmb7}), assuming the electron mass constant.

With the WMAP first year data, the bound on the
variation of $\aem$ was sharpened~\cite{cmb1-2003} to
$-0.05<\Delta\aem/\aem<0.02$, 
after marginalizing over the remaining cosmological parameters 
($\Omega_\mat h^2,\Omega_\baryon h^2,\Omega h^2,n_s,\alpha_s,\tau)$ assuming
a universe with Euclidean spatial sections. Restricting to a model with a vanishing running
of the spectral index ($\alpha_s\equiv\dd n_s/\dd\ln k=0$), it gives
$-0.06<\Delta\aem/\aem<0.01$,
at a 95\% confidence level. In particular it shows
that a lower value of $\aem$ makes $\alpha_s=0$ more compatible with the data. This bounds were obtained
without using other cosmological data sets. This constraint was
confirmed by the analysis of Ref.~\cite{cmb2-2006}, which got
$-0.097<\Delta\aem\aem<0.034$, with the WMAP-1yr data alone and
$- 0.042<\Delta\aem/\aem<0.026$, 
at a 95\% confidence level, when combined with constraints on the
Hubble parameter from the HST Hubble Key project. 

The analysis of the WMAP-3yr data allows to improve~\cite{cmb3-2007} this bound
to $- 0.039<\Delta\aem/\aem$$<0.010$,
at a 95\% confidence level, assuming ($\Omega_\mat,\Omega_\baryon,h, n_s,z_{\rm re},A_s$) 
for the cosmological parameters ($\Omega_\Lambda$ being derived from
the assumption $\Omega_K=0$, as well as $\tau$ from the reionisation redshift, $z_{\rm re}$) 
and using both temperature and polarisation data ($TT$, $TE$, $EE$). 

The WMAP 5-year data were analyzed,
in combination with the 2dF galaxy redshift survey, assuming that
both $\aem$ and $m_{\rm e}$ can vary and that the universe was spatially Euclidean. Letting
6 cosmological parameters  [($\Omega_\mat h^2,\Omega_\baryon h^2,\Theta,\tau, n_s,A_s$), $\Theta$
being the ratio between the sound horizon and the angular distance at decoupling] and 2 constants vary they, it
was concluded~\cite{scoccola,scoccola2} $- 0.012<\Delta\aem/\aem<0.018$ and $- 0.068<\Delta m_{\rm e}/m_{\rm e}<0.044$,
the bounds fluctuating slightly depending on the choice of the recombination scenario.
A similar analyis~\cite{wmap-alpha} not including $m_{\rm e}$ gave $- 0.050<\Delta\aem/\aem<0.042$,
which can be reduced by taking into account some further prior from the HST data. Including polarisation data
data from ACBAR, QUAD and BICEP,  it was also obtained~\cite{menegoni}
$- 0.043<\Delta\aem/\aem<0.038$ at 95\% C.L. and $- 0.013<\Delta\aem/\aem<0.015$
including HST data, also at 95\% C.L. Let us also emphasize the work by Ref.~\cite{martins37}
trying to include the variation of the Newton constant by assuming that $\Delta\aem/\aem=Q\Delta G/G$,
$Q$ being a constant and the investigation of Ref.~\cite{naka00} taking into account
$\aem$, $m_{\rm e}$ and $\mu$, $G$ being kept fixed. Considering  ($\Omega_\mat,\Omega_\baryon,h, n_s,\tau$) 
for the cosmological parameters they concluded from WMAP-5 data ($TT$, $TE$, $EE$) that
$-8.28\times10^{-3}<\Delta\aem/\aem<1.81\times10^{-3}$ and $-0.52<\Delta\mu/\mu<0.17$

The analysis of Refs.~\cite{scoccola,scoccola2} was updated~\cite{scoccola3} to the WMAP-7yr data,
including polarisation and SDSS data. It leads to
$- 0.025<\Delta\aem/\aem<-0.003$ and $0.009<\Delta m_{\rm e}/m_{\rm e}<0.079$ at a 1$\sigma$ level.

The main limitation of these analysis lies in the fact that the
CMB angular power spectrum depends on the evolution of both the
background spacetime and the cosmological perturbations. It
follows that it depends on the whole set of cosmological
parameters as well as on initial conditions, that is on the shape
of the initial power spectrum, so that the results will always be
conditional to the model of structure formation. The constraints
on $\aem$ or $m_{\rm e}$ can then be seen mostly as constraints on
a delayed recombination. A strong constraint on the variation of
$\aem$ can be obtained from the CMB only if the cosmological
parameters are independently known. Ref.~\cite{cmb1-2003}
forecasts that CMB alone can determine $\aem$ to a maximum
accuracy of 0.1\%.

\begin{table}[htb]
\begin{center}
{\small
\begin{tabular}{p{1.7 cm} llc}
\hline\hline
 Constraint  &  Data  & Comment   & Ref. \\
($\aem\times10^2$)  & &   & \\
 \hline
  $[-9,2]$ &   BOOMERanG-DASI-COBE + BBN & BBN with $\aem$ only& \cite{cmb-avelino01}\\
   & & ($\Omega_\mat,\Omega_\baryon,h, n_s$) & \\
  $[-1.4,2]$ & COBE-BOOMERanG-MAXIMA& ($\Omega_\mat,\Omega_\baryon,h, n_s$)&  \cite{cmb-landau01} \\
  $[-5,2]$ &   WMAP-1 & ($\Omega_\mat h^2,\Omega_\baryon h^2,\Omega_\Lambda h^2,\tau, n_s,\alpha_s$) & \cite{cmb1-2003}\\
  $[-6,1]$&   WMAP-1 & same + $\alpha_s=0$ &  \cite{cmb1-2003}\\
  $[-9.7,3.4]$&   WMAP-1 & ($\Omega_\mat,\Omega_\baryon,h, n_s,\tau,m_{\rm e}$) &  \cite{cmb2-2006}\\
  $[-4.2,2.6]$&   WMAP-1 + HST &  same &  \cite{cmb2-2006}\\
   $[-3.9,1.0]$&   WMAP-3 (TT,TE,EE) + HST& ($\Omega_\mat,\Omega_\baryon,h, n_s,z_{\rm re},A_s$)   &  \cite{cmb3-2007}\\
  $[-1.2,1.8]$&   WMAP-5 + ACBAR + CBI + 2df&  ($\Omega_\mat h^2,\Omega_\baryon h^2,\Theta,\tau, n_s,A_s,m_{\rm e}$)   &  \cite{scoccola}\\ 
  $[-1.9,1.7]$&   WMAP-5 + ACBAR + CBI + 2df&  ($\Omega_\mat h^2,\Omega_\baryon h^2,\Theta,\tau, n_s,A_s,m_{\rm e}$)   &  \cite{scoccola2}\\
   $[-5.0,4.2]$&   WMAP-5 + HST&  ($\Omega_\mat h^2,\Omega_\baryon h^2,h,\tau, n_s,A_s$)   &  \cite{wmap-alpha}\\      
    $[-4.3,3.8]$&   WMAP-5 + ACBAR + QUAD + BICEP&  ($\Omega_\mat h^2,\Omega_\baryon h^2,h,\tau, n_s$)   &  \cite{menegoni}\\ 
   $[-1.3,1.5]$&   WMAP-5 + ACBAR + QUAD + BICEP+HST&  ($\Omega_\mat h^2,\Omega_\baryon h^2,h,\tau, n_s$)  &  \cite{menegoni}\\  
   $[-0.83,0.18]$&   WMAP-5 (TT,TE,EE)&  ($\Omega_\mat h^2,\Omega_\baryon h^2,h,\tau, n_s,A_s,m_{\rm e},\mu$)   &  \cite{naka00}\\
   $[-2.5,-0.3]$&   WMAP-7 + $H_0$ + SDSS&  ($\Omega_\mat h^2,\Omega_\baryon h^2,\Theta,\tau, n_s,A_s,m_{\rm e}$)   &  \cite{scoccola3}\\ 
\hline\hline
\end{tabular}
\caption{\it Summary of the latest constraints on the variation of
fundamental constants obtained from the analysis of cosmological data and more
particularly of CMB data. All assume $\Omega_K=0$.}
\label{tab-cmb}
}
\end{center}
\end{table}

\subsection{21 cm}\label{subsec55}

After recombination, the CMB photons are redshifted and their
temperature drops as $(1+z)$. The baryons however are prevented
from cooling adiabatically since the residual amount of free
electrons, that can couple the gas to the radiation through
Compton scattering, is too small. It follows that the matter
decouples thermally from the radiation at a redshift of order
$z\sim200$.

The intergalactic hydrogen atoms after recombination are in their
ground state which hyperfine-structure splits into a singlet and a
triple states ($1s_{1/2}$ with $F=0$ and $F=1$ respectively, see
\S~III.B.1 of FCV~\cite{jpu-revue}). It was recently
proposed~\cite{21cm-1} that the observation of the 21~cm emission
can provide a test on the fundamental constants. We refer to
Ref.~\cite{21cm-2} for a detailed review on 21~cm.

The fraction of atoms in the excited (triplet) state versus the
ground (singlet) state is conventionally related by the spin
temperature $T_{\rm s}$ defined by the relation
\begin{equation}
 \frac{n_t}{n_s} = 3 \exp\left(-\frac{T_*}{T_{\rm s}}\right)
\end{equation}
where $T_*\equiv hc/(\lambda_{21}k_{\rm B})=68.2$~mK is the
temperature corresponding to the 21 cm transition and the factor 3
accounts for the degeneracy of the triplet state (note that this
is a very simplified description since the assumption of a unique
spin temperature is probably not correct~\cite{21cm-2}. The
population of the two states is determined by two processes, the
radiative interaction with CMB photons with a wavelength of
$\lambda_{21}=21.1$~cm (i.e. $\nu_{21}=1420$~MHz) and
spin-changing atomic collision. The evolution of the spin
temperature is thus dictated by
\begin{equation}\label{e_21Ts}
 \frac{\dd T_{\rm s}}{\dd t} = 4C_{10}\left(\frac{1}{T_{\rm s}}-
                                    \frac{1}{T_{\rm g}}\right)T_{\rm s}^2
                                +(1+z)HA_{10}\left(\frac{1}{T_{\rm s}}-
                                    \frac{1}{T_\gamma}\right)\frac{T_\gamma}{T_*}
\end{equation}
The first term corresponds to the collision desexcitation rate
from triplet to singlet and the coefficient $C_{10}$ is decomposed
as
$$
 C_{10} = \kappa_{10}^{HH}n_p + \kappa_{10}^{eH}x_{\rm e}n_p
$$
with the respective contribution of H-H and $e$-H collisions. The
second term corresponds to spontaneous transition and $A_{10}$ is
the Einstein coefficient. The equation of evolution for the gas
temperature $T_{\rm g}$ is given by Eq.~(\ref{e_cmbT}) with
$T_M=T_{\rm g}$ (we recall that we have neglected the contribution
of helium) and the electronic density satisfies
Eq.~(\ref{eecmb1}).

It follows~\cite{21cm-1} that the change in the brightness
temperature of the CMB at the corresponding wavelength scales as
$T_{\rm b}\propto A_{12}/\nu_{21}^2$. Observationally, we can
deduce the brightness temperature from the brightness $I_\nu$,
that is the energy received in a given direction per unit area,
solid angle and time, defined as the temperature of the black-body
radiation with spectrum $I_\nu$. Thus $k_{\rm B}T_{\rm b} \simeq
I_\nu c^2/2\nu^2$. It has a mean value, $\bar T_{\rm b}(z_{\rm
obs})$ at various redshift where $1+z_{\rm obs} = \nu_{21}^{\rm
today}/\nu_{\rm obs}$. Besides, as for the CMB, there will also be
fluctuation in $T_{\rm b}$ due to imprints of the cosmological
perturbations on $n_p$ and $T_{\rm g}$. It follows that we also
have access to an angular power spectrum $C_\ell(z_{\rm obs})$ at
various redshift (see Ref.~\cite{21cm-3} for details on this
computation).

Both quantities depend on the value of the fundamental constants.
Beside the same dependencies of the CMB that arise from the
Thomson scattering cross-section, we have to consider those
arising from the collision terms. In natural units, the Einstein
coefficient scaling is given by
$A_{12}=\frac23\pi\aem\nu_{21}^3m_{\rm e}^{-2}\sim
2.869\times10^{-15}\,{\rm s}^{-1}$. It follows that it scales as
$A_{10}\propto g_{\rm p}^3\mu^3\aem^{13}m_{\rm e}$. The brightness
temperature depends on the fundamental constant as $T_{\rm
b}\propto g_{\rm p}\mu\aem^{5}/m_{\rm e}$. Note that the signal can also
be affected by a time variation of the gravitational constant
through the expansion history of the universe. Ref.~\cite{21cm-1} (see
also Ref.~\cite{21cm-2} for further discussions),
focusing only on $\aem$, showed that this was the dominant effect
on a variation of the fundamental constant (the effect on $C_{10}$
is much complicated to determine but was argued to be much
smaller). It was estimated that a single station telescope like
LWA\footnote{http://lwa.unm.edu} or
LOFAR\footnote{http://www.lofar.org} can lead to a constraint of
the order of $\Delta\aem/\aem\sim0.85\%$, improving to $0.3\%$ for
the full LWA. The fundamental challenge for such a measurement is
the substraction of the foreground.

The 21 cm absorption signal in a available on a band of redshift
typically ranging from $z\lesssim1000$ to $z\sim20$, which is between the
CMB observation and the formation of the first stars, that is
during the so called ``dark age''. It thus offers an interesting
possibility to trace the constraints on the evolution of the
fundamental constants between the CMB epoch and the quasar
absorption spectra.

As for CMB, cosmological parameters since a change of 1\% in
respectively the baryon density  or the Hubble parameter implies a
2\% (resp. 3\%)on the mean bolometric temperature. The effect on
the angular power spectrum have been estimated but still require an
in depth analysis along the lines of e.g.~\cite{21cm-3}. It is
motivating since $C_{\ell}(z_{\rm obs})$ is expected to depend on
the correlators of the fundamental constants e.g.
$\langle\aem(\bx,z_{\rm obs})\aem(\bx',z_{\rm obs})\rangle$ and
thus in principle allows to study their fluctuation, even though
it will also depend on the initial condition, e.g. power spectrum,
of the cosmological perturbations.

In conclusion, the 21cm observation opens a observational window
on the fundamental at redshifts ranging typically from 30 to 100,
but full in-depth analysis is still required (see Refs.~\cite{21com,21reply}
for a critical discussion of this probe). 

\subsection{Big bang nucleosynthesis}

\subsubsection{Overview}\label{secbbnoverview}

The amount of $^{4}{\rm He}$ produced during the big bang
nucleosynthesis is mainly determined by the neutron to proton
ratio at the freeze-out of the weak interactions that interconvert
neutrons and protons. The result of Big Bang nucleosynthesis (BBN)
thus depends on $G$, $\aw$, $\aem$ and $\as$ respectively through
the expansion rate, the neutron to proton ratio, the
neutron-proton mass difference and the nuclear reaction rates,
besides the standard parameters such as e.g. the number of
neutrino families.

The standard BBN scenario~\cite{bbn-cyburt2,peteruzanbook} proceeds in three main
steps:
\begin{enumerate}
\item for $T>1$~MeV, ($t<1$~s) a first stage during which the
neutrons, protons, electrons, positrons an neutrinos are kept in
statistical equilibrium by the (rapid) weak interaction
\begin{eqnarray}\label{bbn0}
&&n\longleftrightarrow p+e^-+\bar\nu_e,\qquad
n+\nu_e\longleftrightarrow  p+e^-,\qquad n+e^+\longleftrightarrow
p+\bar\nu_e.
\end{eqnarray}
As long as statistical equilibrium holds, the neutron to proton
ratio is
\begin{equation}
(n/p)=\hbox{e}^{-Q_{\rm np}/k_{_{\rm B}}T}
\end{equation}
where $Q_{\rm np}\equiv (m_{\rm n}-m_{\rm p})c^2=1.29$~MeV. The
abundance of the other light elements is given
by~\cite{peteruzanbook}
\begin{eqnarray}
Y_A&=&g_A\left(\frac{\zeta(3)}{\sqrt{\pi}}\right)^{A-1}2^{(3A-5)/2}A^{5/2}
     \left[\frac{k_{_{\rm B}}T}{m_{\rm N}c^2}\right]^{3(A-1)/2}
      \eta^{A-1}Y_{\rm p}^ZY_{\rm n}^{A-Z}\hbox{e}^{B_A/k_{_{\rm B}}T},
\end{eqnarray}
where $g_A$ is the number of degrees of freedom of the nucleus
$_Z^A{\rm X}$, $m_{\rm N}$ is the nucleon mass, $\eta$ the
baryon-photon ratio and $B_A\equiv(Zm_{\rm p}+(A-Z)m_{\rm
n}-m_A)c^2$ the binding energy. \item Around $T\sim0.8$~MeV
($t\sim2$~s), the weak interactions freeze out at a temperature
$T_{\rm f}$ determined by the competition between the weak
interaction rates and the expansion rate of the universe and thus
roughly determined by $\Gamma_{_{\rm w}}(T_{\rm f})\sim H(T_{\rm
f})$ that is
\begin{equation}
\gfermi^2(k_{_{\rm B}}T_{\rm f})^5\sim\sqrt{GN_*}(k_{_{\rm
B}}T_{\rm f})^2
\end{equation}
where $\gfermi$ is the Fermi constant and $N_*$ the number of
relativistic degrees of freedom at $T_{\rm f}$. Below $T_{\rm f}$,
the number of neutrons and protons change only from the neutron
$\beta$-decay between $T_{\rm f}$ to $T_{\rm N}\sim0.1$~MeV when
$p+n$ reactions proceed faster than their inverse dissociation.
 \item For $0.05$~MeV$<T<0.6$~MeV ($3\,{\rm s}<t<6\,{\rm min}$),
the synthesis of light elements occurs only by two-body reactions.
This requires the deuteron to be synthesized ($p+n\rightarrow D$)
and the photon density must be low enough for the
photo-dissociation to be negligible. This happens roughly when
\begin{equation}\label{n0}
\frac{n_{\rm d}}{n_\gamma}\sim\eta^2\exp(-B_D/T_{\rm N})\sim 1
\end{equation}
with $\eta\sim3\times10^{-10}$. The abundance of $^4{\rm He}$ by
mass, $Y_{\rm p}$, is then well estimated by
\begin{equation}\label{n1}
Y_{\rm p}\simeq2\frac{(n/p)_{\rm N}}{1+(n/p)_{\rm N}}
\end{equation}
with
\begin{equation}\label{n2}
(n/p)_{\rm N}=(n/p)_{\rm f}\exp(-t_{\rm N}/\tau_{\rm n})
\end{equation}
with $t_{\rm N}\propto G^{-1/2}T_{\rm N}^{-2}$ and $\tau_{\rm
n}^{-1}=1.636\gfermi^2(1+3g_A^2)m_{\rm e}^5/(2\pi^3)$, with
$g_A\simeq1.26$ being the axial/vector coupling of the nucleon.
Assuming that $B_D\propto\as^2$, this gives a dependence $t_{\rm
N}/\tau_{\rm p}\propto G^{-1/2}\as^2\gfermi^2$.
 \item The abundances of the light element abundances, $Y_i$, are then
obtained by solving a series of nuclear reactions
$$
 \dot Y_i = J - \Gamma Y_i,
$$
where $J$ and $\Gamma$ are time-dependent source and sink terms.
\end{enumerate}

From an observational point of view, the light elements abundances
can be computed as a function of $\eta$ and compared to their
observed abundances. Fig.~\ref{fig-bbn} summarizes the
observational constraints obtained on helium-4, helium-3,
deuterium and lithium-7. On the other hand, $\eta$ can be
determined independently from the analysis of the cosmic microwave
background anisotropies and the WMAP data~\cite{wmap} have led to
to the conclusion that
$$
 \eta = \eta_{_{\rm WMAP}} = (6.14\pm0.25)\times10^{-10}.
$$
This number being fixed, all abundances can be computed. At
present, there exists a discrepancy between the predicted
abundance of lithium-7 based on the WMAP results~\cite{cocbbn,cocnew} for
$\eta$, ${}^7{\rm Li}/{\rm H}=(5.14\pm0.50)\times10^{-10}$
and its values measured in metal-poor halo stars in our
Galaxy~\cite{bbnbonif}, ${}^7{\rm Li}/{\rm
H}=(1.26\pm0.26)\times10^{-10}$ which is factor 3 lower, at
least~\cite{bbn-cyburt} (see also Ref.~\cite{spite2}), than the predicted value. No solution to
this {\em Lithium-7} problem is known. A back of the envelope
estimates shows that we can mimic a lower $\eta$ parameter, just
by modifying the deuterium binding energy, letting $T_N$
unchanged, since from Eq.~(\ref{n0}), one just need $\Delta
B_D/T_{\rm N}\sim -\ln 9$ so that the effective $\eta$ parameter,
assuming no variation of constant, is three times smaller than
$\eta_{_{\rm WMAP}}$. This rough rule of thumb explains that the
solution of the lithium-7 problem may lie in a possible variation
of the fundamental constants (see below for details).

\subsubsection{Constants everywhere...}

In complete generality, the effect of varying constants on the BBN
predictions is difficult to model because of the intricate
structure of QCD and its role in low energy nuclear reactions. A
solution is thus to proceed in {\em two steps}, first by
determining the dependencies of the light element abundances on
the BBN parameters and then by relating those parameters to the
fundamental constants.

The analysis of the previous section, that was restricted to the
helium-4 case, clearly shows that the abundances will depend on:
(1) $\ag$ which will affect the Hubble expansion rate at the time
of nucleosynthesis in the same way as extra-relativistic degrees
of freedom do, so that it modifies the freeze-out time $T_{\rm
f}$. This is the only gravitational sector parameter. (2)
$\tau_{\rm n}$, the neutron lifetime dictates the free neutron
decay and appears in the normalisation of the proton-neutron
reaction rates. It is the only weak interaction parameter and it
is related to the Fermi constant $\gfermi$, or equivalently the
Higgs vev. (3) $\aem$, the fine-structure constant. It
enters in the Coulomb barriers of the reaction rates through the
Gamow factor, in all the binding energies. (4) $Q_{\rm np}$, the
neutron-proton mass difference enters in the neutron-proton ratio
and we also have a dependence in
(5) $m_{\rm N}$ and $m_{\rm e}$ and (6) the binding energies.

Clearly all these parameters are not independent but their
relation is often model-dependent. If we focus on helium-4, its
abundance mainly depends on $Q_{\rm np}$, $T_{\rm f}$ and $T_{\rm N}$
(and hence mainly on the neutron lifetime, $\tau_{\rm n}$). Early
studies (see \S~III.C.2 of FVC~\cite{jpu-revue}) generally focused
on one of these parameters. For instance, Kolb {\em et
al.}~\cite{bbnkolb} calculated the dependence of primordial
${}^4{\rm He}$ on $G$, $\gfermi$ and $Q_{\rm np}$ to deduce that the
helium-4 abundance was mostly sensitive in the change in $Q_{\rm np}$
and that other abundances were less sensitive to the value of
$Q_{\rm np}$, mainly because $^4{\rm He}$ has a larger binding energy;
its abundances is less sensitive to the weak reaction rate and
more to the parameters fixing the value of $(n/p)$. To extract the
constraint on the fine-structure constant, they decomposed
$Q_{\rm np}$ as $Q_{\rm np}=\aem Q_\alpha+\beta Q_\beta$ where the first
term represents the electromagnetic contribution and the second
part corresponds to all non-electromagnetic contributions.
Assuming that $Q_\alpha$ and $Q_\beta$ are constant and that the
electromagnetic contribution is the dominant part of $Q$, they
deduced that $|\Delta\aem/\aem|<10^{-2}$. Campbell and
Olive~\cite{bbnco} kept track of the changes in $T_{\rm f}$ and
$Q_{\rm np}$ separately and deduced that $\frac{\Delta Y_{\rm
p}}{Y_{\rm p}}\simeq\frac{\Delta T_{\rm
f}}{T_{\rm f}}-\frac{\Delta Q_{\rm np}}{Q_{\rm np}}$ while more
recently the analysis~\cite{landaubbn} focused on $\aem$ and $v$. \\

Let us now see how the effect of all these parameters are now
accounted for in BBN codes.

Bergstr\"om {\em et al.}~\cite{bbnberg} started to focus on the
$\aem$-dependence of the thermonuclear rates.  In the
non-relativistic limit, it is obtained as the thermal average of
the product of the cross, the relative velocity and the the number
densities. Charged particles must tunnel through a Coulomb barrier
to react. Changing $\aem$ modifies these barriers and thus the
reaction rates. Separating the Coulomb part, the low-energy cross
section can be written as
\begin{equation}\label{bir}
 \sigma(E)=\frac{S(E)}{E}\hbox{e}^{-2\pi\eta(E)}
\end{equation}
where $\eta(E)$ arises from the Coulomb barrier and is given in
terms of the charges and the reduced mass $M_r$ of the two
interacting particles as
\begin{equation}
 \eta(E)=\aem Z_1Z_2\sqrt{\frac{M_r c^2}{2E}}.
\end{equation}
The form factor $S(E)$ has to be extrapolated from experimental
nuclear data but its $\aem$-dependence as well as the one of the
reduced mass were neglected.  Keeping all other constants
fixed, assuming no exotic effects and taking a lifetime of 886.7~s
for the neutron, it was deduced that $\left|{\Delta\aem}/
{\aem}\right| <2\times10^{-2}$. This analysis was then
extended~\cite{bbnnollet} to take into account the
$\aem$-dependence of the form factor to conclude that
$$
 \sigma(E) = \frac{2\pi\eta(E)}{\exp^{2\pi\eta(E)}-1}
   \simeq 2\pi\aem
   Z_1Z_2\sqrt{\frac{M_rc^2}{c^2}}\exp^{-2\pi\eta(E)}.
$$
Ref.~\cite{bbnnollet} also took into a account (1) the effect that
when two charged particles are produced they must escape the
Coulomb barrier. This effect is generally weak because the
$Q_i$-values (energy release) of the different reactions are
generally larger than the Coulomb barrier at the exception of two
cases, ${}^3{\rm He}(n,p){}^3{\rm H}$ and ${}^7{\rm
Be}(n,p){}^7{\rm Li}$. The rate of these reactions must be
multiplied by a factor $(1+a_i\Delta\aem/\aem)$. (2) The radiative
capture (photon emitting processes) are proportional to $\aem$
since it is the strength of the coupling of the photon and nuclear
currents. All these rates need to be multiplied by
$(1+\Delta\aem/\aem)$. (3) The electromagnetic contribution to all
masses was taken into account, which modify the $Q_i$-values as
$Q_i\rightarrow Q_i+ q_i\Delta\aem/\aem)$. For helium-4 abundance
these effects are negligible since the main $\aem$-dependence
arises from $Q_{\rm np}$. Equiped with these modifications, it was
concluded that ${\Delta\aem}/{\aem}=-0.007^{+0.010}_{-0.017}$
using only deuterium and helium-4 since the lithium-7 problem was
still present.

Then the focus fell on the deuterium binding energy, $B_D$.
Flambaum and Shuryak~\cite{oklo-12,oklo-14,bbn-dimi,bbn-dimi2}
illustrated the sensitivity of the light element abundances on
$B_D$. Its value mainly sets the beginning of the nucleosynthesis,
that is of $T_{\rm N}$ since the temperature must low-enough in
order for the photo-dissociation of the deuterium to be negligible
(this is at the origin of the deuterium bottleneck). The
importance of $B_D$ is easily understood by the fact that the
equilibrium abundance of deuterium and the reaction rate
$p(n,\gamma){\rm D}$ depends exponentially on $B_D$ and on the
fact that the deuterium is in a shallow bound state. Focusing on
the $T_{\rm N}$-dependence, it
was concluded~\cite{oklo-12} that $\Delta B_D/B_D<0.075$.\\

This shows that the situation is more complex and that one cannot
reduce the analysis to a single varying parameter. Many studies
then tried to determinate the sensitivity to the variation of many
independent parameters.

\begin{figure}[hptb]
  \def\epsfsize#1#2{0.5#1}
  \centerline{\includegraphics[scale=0.45]{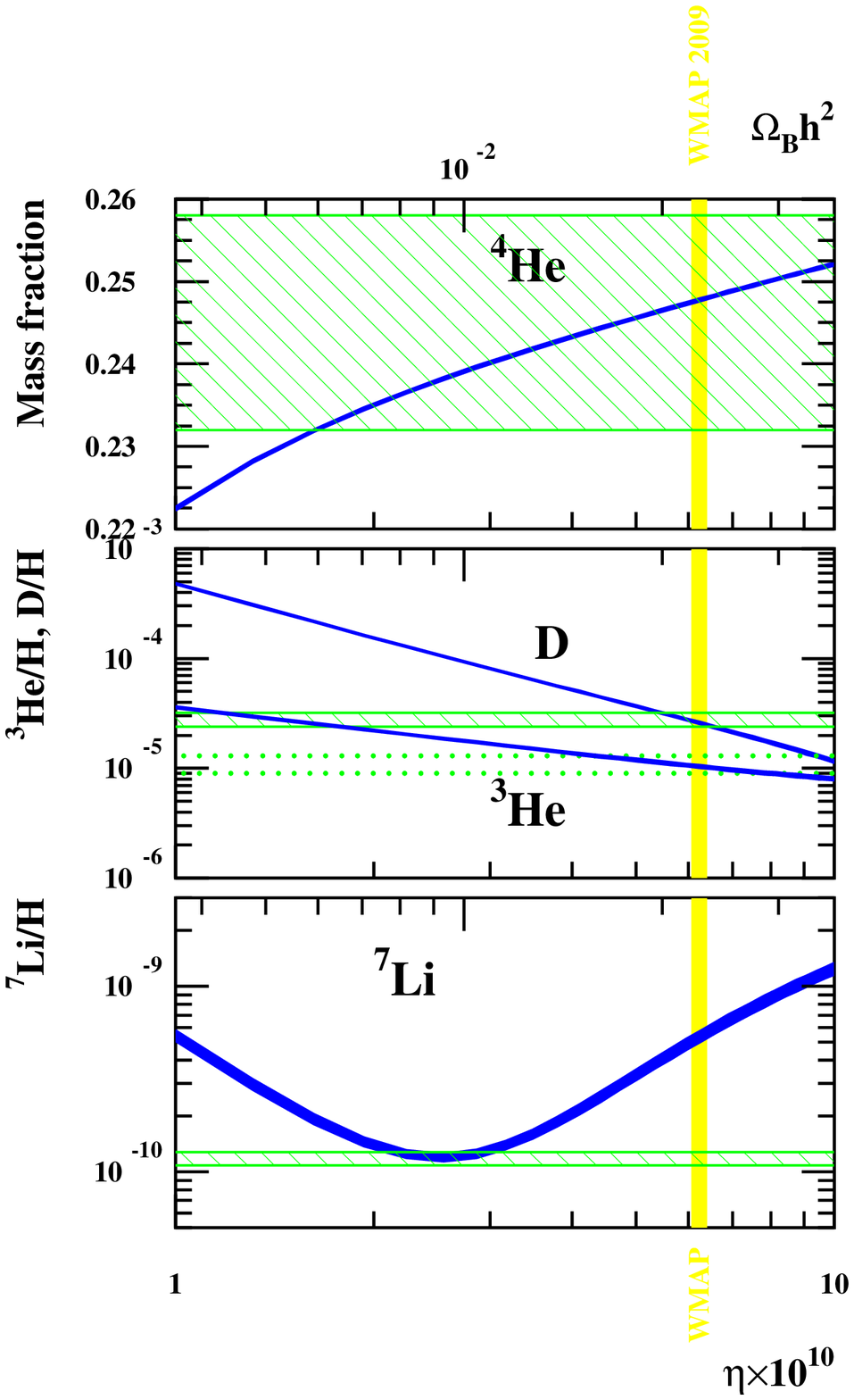}$\qquad$
  \includegraphics[scale=0.45]{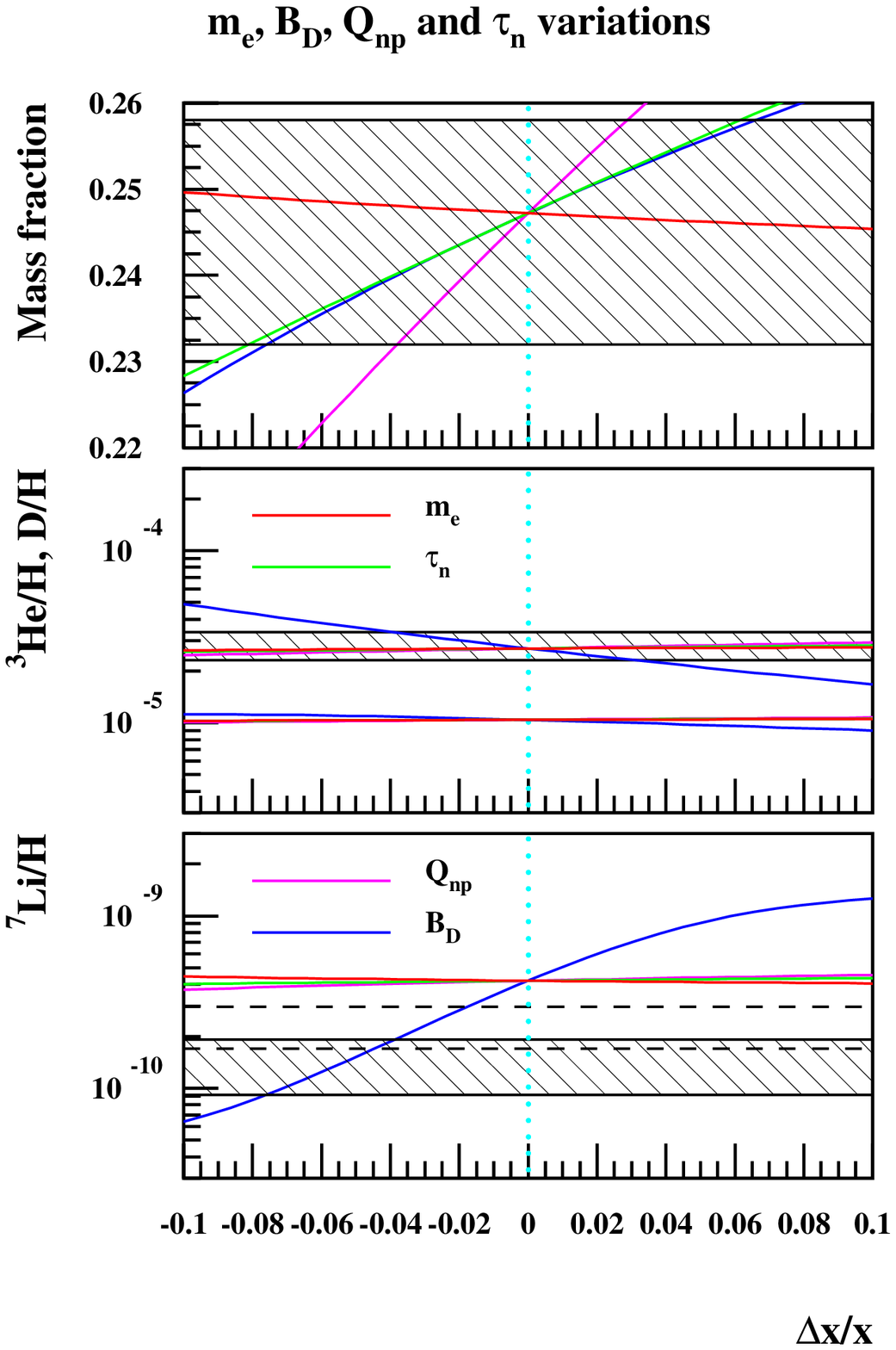}}
  \caption{\it (Left): variation of the light element abundances
 in function of $\eta$ compared to the spectroscopic abundances.
 The vertical line depicts the constraint obtained on $\eta$ from
 the study of the cosmic microwave background data. The lithium-7
 problem lies in the fact that $\eta_{_{\rm spectro}}<\eta_{_{\rm WMAP}}$. From
 Ref.~\cite{cocnew}.
 (right): Dependence of the light element abundance on the independent variation of
 the BBN parameters, assuming $\eta=\eta_{_{\rm WMAP}}$. From Ref.~\cite{couv}}
  \label{fig-bbn}
\end{figure}

The sensitivity of the helium-4 abundance to the variation of 7
parameters was first investigated by M\"uller
\etal~\cite{bbn-muller} considering the dependence on the
parameters $\{X_i\}\equiv\{G,\aem,v,m_{\rm e},\tau_{\rm n},Q_{\rm np},$
$B_D\}$ independently,
$$
\Delta \ln Y_{\rm He} = \sum_i c_i^{(X)}\Delta \ln X_i
$$
and assuming $\Lambda_{\rm QCD}$ fixed (so that the 7 parameters
are in fact dimensionless quantities). The $c_i^{(X)}$ are the
sensitivities to the BBN parameters, assuming the six others
fixed. It was concluded that $Y_{\rm He}\propto
\aem^{-0.043}v^{2.4}m_{\rm e}^{0.024}\tau_{\rm n}^{0.24}Q_{\rm np}^{-1.8}
B_D^{0.53} G^{0.405}$ for independent variations. They further
related $(\tau_{\rm n},Q_{\rm np},B_D)$ to $(\aem,v,m_{\rm e},m_{\rm N},m_{\rm d}-m_{\rm u})$,
as we shall discuss in the next section.

This was generalized by Landau \etal~\cite{bbn-landau} up to
lithium-7 considering the parameters $\{\aem,\gfermi,$
$\Lambda_{\rm QCD},\Omega_b h^2\}$, assuming $G$ constant where
the variation of $\tau_{\rm n}$ and the variation of the masses
where tied to these parameters but the effect on the binding
energies were not considered.

Coc \etal~\cite{cnouv} considered the effect of a variation of
$(Q_{\rm np}, B_D, \tau_{\rm n}, m_{\rm e})$ on the abundances of the light
elements up to lithium-7, neglecting the effect of
$\aem$ on the cross-section. Their dependence on the independent
variation of each of these parameters is depicted on
Fig.~\ref{fig-bbn}. It confirmed the result of Refs.~\cite{oklo-12,olive2-rio}
that the deuterium binding energy is the most sensitive parameter.
From the helium-4 data alone, the bounds
\begin{equation}
 -8.2\times10^{-2}\la\frac{\Delta\tau_{\rm n}}{\tau_{\rm n}}\la6\times10^{-2},
 \quad
 -4\times10^{-2}\la\frac{\Delta Q_{\rm np}}{Q_{\rm np}}\la2.7\times10^{-2},
\end{equation}
and
\begin{equation}
-7.5\times10^{-2}\la\frac{\Delta B_D}{B_D}\la6.5\times10^{-2},
\end{equation}
at a 2$\sigma$ level, were set (assuming $\eta_{_{\rm WMAP}}$).
The deuterium data set the tighter constraint
$-4\times10^{-2}\la\Delta\ln B_D\la3\times10^{-2}$. Note also on
Fig.~\ref{fig-bbn} that the lithium-7 abundance can be brought in
concordance with the spectroscopic observations provided that
$B_D$ was smaller during BBN
$$
-7.5\times10^{-2}\la\frac{\Delta B_D}{B_D}\la-4\times10^{-2},
$$
so that $B_D$ may be the most important parameter to resolve the
lithium-7 problem. The effect of the quark mass on the binding energies
was described in Ref.~\cite{BBNberengut}. They then concluded
that a variation of $\Delta m_{\rm q}/m_{\rm q}=0.013\pm0.002$
allows to reconcile the abundance of lithium-7 and the value
of $\eta$  deduced from WMAP.

This analysis was extended~\cite{bbn-dent} to incorporate the
effect of 13 independent BBN parameters including the parameters
considered before plus the binding energies of deuterium, tritium,
helium-3, helium-4, lithium-6, lithium-7 and beryllium-7. The
sensitivity of the light element abundances to the independent
variation of these parameters is summarized in Table~I of
Ref.~\cite{bbn-dent}. These BBN parameters were then related to
the same 6 ``fundamental'' parameters used in
Ref.~\cite{bbn-muller}.

All these analysis demonstrate that the effects of the BBN
parameters on the light element abundances are now under
control. They have been implemented in BBN codes and most
results agree, as well as with semi-analytical estimates. As long
as these parameters are assume to vary independently, no
constraints sharper than $10^{-2}$ can be set. One should also not
forget to take into account standard parameters of the BBN
computation such as $\eta$ and the effective number of
relativistic particle.

\subsubsection{From BBN parameters to fundamental constants}\label{bbn2cste}

To reduce the number parameters, we need to relate the BBN
parameters to more fundamental ones, keeping in mind that
this can usually be done only in a model-dependent way. We shall
describe some of the relations that have been used in many studies.
They mainly concern $Q_{\rm np}$, $\tau_{\rm n}$ and $B_D$.\\

At lowest order, all dimensional parameters of QCD, e.g. masses,
nuclear energies etc., are to a good approximation simply
proportional to some powers of $\Lambda_{\rm QCD}$. One needs to
go beyond such a description and takes the effects of the masses of
the quarks into account.

$Q_{\rm np}$ can be expressed in terms of the mass on the quarks u and
d and the fine-structure constant as
$$
 Q_{\rm np} = a\aem\Lambda_{\rm QCD} +(m_{\rm d}-m_{\rm u}),
$$
where the electromagnetic contribution today is
$(a\aem\Lambda_{\rm QCD})_0=-0.76$~MeV and therefore the quark
mass contribution today is $(m_{\rm d}-m_{\rm u})=2.05$~\cite{gasser82} so
that
\begin{equation}
 \frac{\Delta Q_{\rm np}}{Q_{\rm np}} = -0.59\frac{\Delta\aem}{\aem} +
 1.59\frac{\Delta(m_{\rm d} - m_{\rm u})}{(m_{\rm d} - m_{\rm u})}.
\end{equation}
All the analysis cited above agree on this dependence.

The neutron lifetime can be well approximated by
$$
 \tau_{\rm n}^{-1} = \frac{1+3g_A^2}{120\pi^3}\gfermi^2
 m_{\rm e}^5\left[\sqrt{q^2-1}(2q^4-9q^2-8) + 15\ln\left(q+\sqrt{q^2-1} \right)
 \right],
$$
with $q\equiv Q_{\rm np}/m_{\rm e}$ and $\gfermi=1/\sqrt{2}v^2$. Using the
former expression for $Q_{\rm np}$ we can express $\tau_{\rm n}$ in
terms of $\aem$, $v$ and the u, d and electron masses. It follows
\begin{equation}
 \frac{\Delta\tau_{\rm n}}{\tau_{\rm n}} = 3.86\frac{\Delta\aem}{\aem} +
   4\frac{\Delta v}{v} + 1.52\frac{\Delta m_{\rm e}}{m_{\rm e}}
   -10.4\frac{\Delta(m_{\rm d} - m_{\rm u})}{(m_{\rm d} - m_{\rm u})}.
\end{equation}
Again, all the analysis cited above agree on this dependence.

Let us now turn to the binding energies, and more particularly to
$B_D$ that, as we have seen, is a crucial parameter. This is one
the better known quantities in the nuclear domain and it is
experimentally measured to a precision better than
$10^{-6}$~\cite{bbn-bdmes}. Two approaches have been followed.
\begin{itemize}
 \item{\it Pion mass}. A first route is to use the dependence of
 the binding energy on the pion mass~\cite{bbnpi1,bbnpi2}, which is related to the u and d
 quark masses by
 $$m_\pi^2 =  m_{\rm q} \langle \bar uu+\bar dd\rangle f_\pi^{-2}\simeq
 \hat m\Lambda_{\rm QCD},$$
 where $ m_{\rm q} \equiv\frac12(m_{\rm u}+m_{\rm d})$ and assuming that the
 leading order of $\langle \bar uu+\bar dd\rangle f_\pi^{-2}$
 depends only on $\Lambda_{\rm QCD}$, $f_\pi$ being the pion decay constant.
 This dependence was parameterized~\cite{nnbyoo} as
 $$
 \frac{\Delta B_D}{B_D} = -r\frac{\Delta m_\pi}{m_\pi},
 $$
 where $r$ is a fitting parameter found to be between 6~\cite{bbnpi1}
 and 10~\cite{bbnpi2}. Prior to this result, the analysis of Ref.~\cite{oklo-12}
 provides two computations of this dependence which respectively
 lead to $r=-3$ and $r=18$ while, following the same lines, Ref.~\cite{bbn-landau2}
 got $r=0.082$.

 Ref.~\cite{bbn-muller}, following the computations of
 Ref.~\cite{bbn-pudliner}, adds an electromagnetic contribution $-0.0081\Delta\aem/\aem$
 so that
 \begin{equation}
 \frac{\Delta B_D}{B_D} = -\frac{r}{2}\frac{\Delta
 m_{\rm q}}{m_{\rm q}}-0.0081\frac{\Delta\aem}{\aem},
 \end{equation}
 but this latter contribution has not been included in other works.
 \item{\it Sigma model}. In the framework of the Walecka model,
 where the potential for the nuclear forces keeps only the
 $\sigma$ and $\omega$ meson exchanges,
 $$
 V=-\frac{g_s^2}{4\pi r}\exp(-m_\sigma r) + \frac{g_v^2}{4\pi r}\exp(-m_\omega
 r),
 $$
 where $g_s$ and $g_v$ are two coupling constants. Describing
 $\sigma$ as a SU(3) singlet state, its mass was related to the
 mass of the strange quark. In this way one can hope to take into
 account the effect of the strange quark, both on the nucleon mass and
 the binding energy. In a second step $B_D$ is related to
 the meson and nucleon mass by
 $$
  \frac{\Delta B_D}{B_D} = -48\frac{\Delta m_\sigma}{m_\sigma}
     +50\frac{\Delta m_\omega}{m_\omega}+ 6\frac{\Delta m_{\rm N}}{m_{\rm N}}
 $$
 so that ${\Delta B_D}/{B_D}\simeq-17{\Delta
 m_{\rm s}}{m_{\rm s}}$~\cite{oklo-14}. Unfortunately, a complete treatment of
 all the nuclear quantities on $m_{\rm s}$ has not been performed yet.

\end{itemize}
The case of the binding energies of the other elements has been
less studied. Ref.~\cite{bbn-dent} follows a route similar than
for $B_D$ and relates them to pion mass and assumes that
$$
 \frac{\partial B_i}{\partial m_\pi}=
 f_i(A_i-1)\frac{B_D}{m_\pi}r\simeq-0.13f_i(A_i-1),
$$
where $f_i$ are unknown coefficients assumed to be of order unity
and $A_i$ is  the number of nucleons. No other estimates has been
performed. Other nuclear potentials (such as Reid 93 potential, Nijmegen potential, Argonne $v18$ potential and
Bonn potential) have been
used in Ref.~\cite{civitarese} to determine the dependence
of $B_D$ on $v$ and agree with previous studies.\\

These analysis allow to reduce all the BBN parameter to the
physical constants $(\aem,v,m_{\rm e},m_{\rm d}-m_{\rm u},m_{\rm q})$ and $G$ that is not
affected by this discussion. This set can be further reduce since
all the masses can be expressed in terms of $v$ as $m_i=h_iv$,
where $h_i$ are Yukawa couplings.

To go further, one needs to make more assumption, such as grand
unification, or by relating the Yukawa coupling of the top to $v$
by assuming that weak scale is determined by dimensional
transmutation~\cite{cnouv}, or that the varition of the constant
is induced by a string dilaton~\cite{bbnco}. At each step, one
gets more stringent constraints, which can reach the
$10^{-4}$~\cite{bbn-dent} to $10^{-5}$~\cite{cnouv} level but
indeed more model-dependent!

\subsubsection{Conclusion}

Primordial nucleosynthesis offers a possibility to test almost all
fundamental constants of physics at a redshift of $z\sim10^8$. It is thus very rich but indeed
the effect of each constant is more difficult to disentangle. The
effect of the BBN parameters has been quantified with precision
and they can be constrained typically at a $10^{-2}$ level, and in
particular it seems that the most sensitive parameter is the
deuterium binding energy.

The link with more fundamental parameters is better understood but
the dependence of the deuterium binding energy still left some
uncertainties and a good description of the effect of the strange
quark mass is missing.

We have not considered the variation of $G$ in this section. Its
effect is disconnected from the other parameters. Let us just
stress that assuming the BBN sensitivity on $G$ by just modifying
its value may be misleading. In particular $G$ can vary a lot
during the electron-positron annihilation so that the BBN
constraints can in general not be described by an effective speed-up
factor~\cite{couv,bbn-Gpichon}.

\section{The gravitational constant}\label{section4}

The gravitational constant was the first constant whose constancy
was questioned~\cite{dirac37}. From a theoretical point of view,
theories with a varying gravitational constant can be designed to
satisfy the equivalence principle in its weak form but not in its
strong form~\cite{will-book}. Most theories of gravity that violate the strong
equivalence principle predict that the locally measured
gravitational constant may vary with time.

The value of the gravitational constant is
$G=6.674 28(67)\times10^{-11}\, {\rm m}^3\cdot {\rm kg}^{-1}\cdot {\rm s}^{-2} $ 
so that its relative standard uncertainty fixed by the CODATA\footnote{The
CODATA is the COmmittee on Data for Science and Technology, see
{\tt http://www.codata.org/}.} in 2006 is 0.01\%. Interestingly,
the disparity between different experiments led, in 1998, to a
temporary increase of this uncertainty to 0.15\%~\cite{G-marko},
which demonstrates the difficulty in measuring the value of this
constant. This explains partly why the constraints on the time
variation are less stringent than for the other constants.

A variation of the gravitational constant, being a pure
gravitational phenomenon, does not affect the local physics, such
as e.g. the atomic transitions or the nuclear physics. In particular,
it is equivalent at stating that the masses of all particles are
varying in the same way to that their ratios remain constant.
Similarly all absorption lines will be shifted in the same way.
It follows that most constraints are obtained from systems in which gravity
is non-negligible, such as the motion of the bodies of the
Solar system, astrophysical and cosmological
systems. They are mostly related in the comparison of a
gravitational time scale, e.g. period of orbits, to a
non-gravitational time scale. It follows that in general the constraints
assume that the values of the other constants are fixed. Taking
their variation into account would add degeneracies and make the
constraints cited below less stringent.

We refer to \S~IV~of~FVC~\cite{jpu-revue} for earlier constraints
based e.g. on the determination of the Earth surface temperature,
which roughly scales as $G^{2.25}M_\odot^{1.75}$ and gives a
constraint of the order of $|\Delta G/G|<0.1$~\cite{gamow67a}, or
on the estimation of the Earth radius at different geological
epochs.

\subsection{Solar systems constraints}

Monitoring the orbits of the various bodies of the Solar system offers a possibility
to constrain deviations from general relativity, and in particular the time variation of $G$. This accounts for
comparing a gravitational time scale (related to the orbital motion) and an
atomic time scale and it is thus assumed that the variation of
atomic constants is negligible on the time of the experiment. 

The first constraint arises from the Earth-Moon system. A time variation of $G$
is then related to a variation of the mean motion ($n=2\pi/P$) of the orbit of the
Moon around the Earth.  A decrease in $G$ would induce both the Lunar mean distance and
period to increase. As long as the gravitational binding energy is negligible, one has
\begin{equation}
 \frac{\dot P}{P} = -2\frac{\dot G}{G}.
\end{equation}
Earlier constraints rely on paleontological data and ancient eclipses obervations
(see \S~IV.B.1~of~FVC~\cite{jpu-revue}) and none of them are very reliable.
A main difficulty arises from tidal dissipation that also causes the mean distance and orbital period to increase (for
tidal changes $2\dot n/n+3\dot a /a=0$), but not as in the same ratio as for $\dot G$.

The Lunar Laser Ranging (LLR) experiment has measured the relative position
of the Moon with respect to the Earth with an accuracy of the order of 1~cm over 3 decades. An early analysis
of this data~\cite{williams76} assuming a Brans-Dicke theory of gravitation gave that
$|\dot G/G| \leq 3\times10^{-11}\,{\rm yr}^{-1}$. It was improved~\cite{muller91} 
by using 20 years of observation to get $|\dot G/G| \leq 1.04\times10^{-11}\,{\rm yr}^{-1}$,
the main uncertainty arising from Lunar tidal acceleration. With, 24 years of data, one
reached~\cite{G-llr0}  $|\dot G/G| \leq 6\times10^{-12}\,{\rm yr}^{-1}$ and
finally, the latest analysis of the Lunar laser ranging
experiment~\cite{uff3} increased the constraint to
\begin{equation}
 \left.\frac{\dot G}{G}\right|_0 = (4\pm9)\times10^{-13}\,{\rm yr}^{-1}.
\end{equation}

Similarly, Shapiro \etal~\cite{shapiro} compared radar-echo time delays between
Earth, Venus and Mercury with a caesium atomic clock between 1964 and
1969. The data were fitted to the theoretical equation of motion for
the bodies in a Schwarzschild spacetime, taking into account the
perturbations from the Moon and other planets. They concluded that
$|\dot G/G|<4\times10^{-10}\,{\rm yr}^{-1}$.
The data concerning Venus cannot be used due to
imprecision in the determination of the portion of the planet
reflecting the radar. This was improved to
$|\dot G/G|<1.5\times10^{-10}\,{\rm yr}^{-1}$
by including Mariner 9 and Mars orbiter data~\cite{reasenberg78}.  
The analysis was further extended~\cite{shapiro90}  to
give
$\dot G/G=(-2\pm10)\times10^{-12}\,{\rm yr}^{-1}$.
The combination of Mariner 10 an Mercury and Venus ranging data
gives~\cite{anderson91}
\begin{equation}
 \left.\frac{\dot G}{G}\right|_0=(0.0\pm2.0)\times10^{-12}\,{\rm yr}^{-1}.
\end{equation}

Reasenberg {\em et al.}~\cite{reasenberg79} considered the 14 months data obtained
from the ranging of the Viking spacecraft and deduced, assuming a
Brans-Dicke theory, $|\dot G/G|<10^{-12}\,{\rm yr}^{-1}$.
Hellings {\em et al.}~\cite{hellings83}
using all available astrometric data and in particular the ranging
data from Viking landers on Mars deduced that
\begin{equation}
 \left.\frac{\dot G}{G}\right|_0 =(2\pm4)\times10^{-12}\,{\rm yr}^{-1}.
\end{equation}
The major contribution to the uncertainty is due to the modeling of
the dynamics of the asteroids on the Earth-Mars range.  Hellings {\em
et al.}~\cite{hellings83} also tried to attribute their result to a time
variation of the atomic constants.  Using the same data but a
different modeling of the asteroids, Reasenberg~\cite{reasenberg83} got
$|\dot G/G|<3\times10^{-11}\,{\rm yr}^{-1}$, which was then improved by Chandler {\em et al.}~\cite{chandler93} to
$|\dot G/G|<10^{-11}\,{\rm yr}^{-1}$.

\subsection{Pulsar timing}

Contrary to the Solar system case, the dependence of the gravitational
binding energy cannot be neglected while computing the time variation
of the period. Here two
approaches can be followed; either one sticks to a model
(e.g. scalar-tensor gravity) and compute all the effects in this model
or one has a more phenomenological approach and tries to put some
model-independent bounds.

Eardley~\cite{G-eardley75} followed the first route and discussed the effects of a
time variation of the gravitational constant on binary pulsar in the
framework of the Brans-Dicke theory. In that case, both a dipole
gravitational radiation and the variation of $G$ induce a periodic
variation in the pulse period.  Nordtvedt~\cite{nordtvedt90} showed that the
orbital period changes as
\begin{equation}
\frac{\dot P}{P}=-\left[2+\frac{2(m_1c_1+m_2c_2)+3(m_1c_2+m_2c_1)}{m_1+m_2}
\right]\frac{\dot G}{G}
\end{equation}
where $c_i\equiv\delta\ln m_i/\delta\ln G$. He concluded that for the pulsar
PSR~1913+16 ($m_1\simeq m_2$ and $c_1\simeq c_2$) one gets
\begin{equation}\label{pn}
\frac{\dot P}{P}=-\left[2+5c\right]\frac{\dot G}{G},
\end{equation}
the coefficient $c$ being model dependent.  As another
application, he estimated that $c_{_{\rm
Earth}}\sim-5\times10^{-10}$, $c_{_{\rm Moon}}\sim-10^{-8}$ and
$c_{_{\rm Sun}}\sim-4\times10^{-6}$ justifying the formula used in the
Solar system.

Damour {\em et al.}~\cite{damour88} used the timing data of the binary pulsar
PSR~1913+16. They implemented the effect of the time variation of $G$
by considering the effect on $\dot P/P$. They defined, in a phenomenological way, that $\dot
G/G=-0.5\delta\dot P/P$, where $\delta\dot P$ is the part of the
orbital period derivative that is not explained otherwise (by
gravitational waves radiation damping). This theory-independent
definition has to be contrasted with the theory-dependent result
(\ref{pn}) by Nordtvedt~\cite{nordtvedt90}. They got
\begin{equation}
\dot G/G=(1.0\pm2.3)\times10^{-11}\,{\rm yr}^{-1}.
\end{equation}
Damour and Taylor~\cite{damour91} then reexamined the data of PSR~1913+16 and established the
upper bound
\begin{equation}
\dot G/G<(1.10\pm1.07)\times10^{-11}\,{\rm yr}^{-1}.
\end{equation}
Kaspi {\em et al.}~\cite{kaspi} used data from PSR~B1913+16 and PSR~B1855+09
respectively to get
\begin{equation}
\dot G/G=(4\pm5)\times10^{-12}\,{\rm yr}^{-1}
\end{equation}
and
\begin{equation}
\dot G/G=(-9\pm18)\times10^{-12}\,{\rm yr}^{-1},
\end{equation}
the latter case being more ``secure'' since the orbiting companion is
not a neutron star.

All the previous results concern binary pulsars but isolated ones can
also be used.  Heintzmann and Hillebrandt~\cite{heitzmann75} related the
spin-down of the pulsar JP1953 to a time variation of $G$. The
spin-down is a combined effect of electromagnetic losses, emission
of gravitational waves, possible spin-up due to matter accretion.
Assuming that the angular momentum is conserved so that
$I/P=$constant, one deduces that
\begin{equation}
 \left.\frac{\dot P}{P}\right\vert_G=\left(\frac{\dd\ln I}{\dd\ln G}\right)
\frac{\dot G}{G}.
\end{equation}
The observational spin-down can be decomposed as
\begin{equation}
 \left.\frac{\dot P}{P}\right\vert_{_{\rm obs}}= \left.\frac{\dot P}{P}\right\vert_{_{\rm mag}}
+ \left.\frac{\dot P}{P}\right\vert_{_{\rm GW}}+ \left.\frac{\dot P}{P}\right\vert_G.
\end{equation}
Since ${\dot P}/{P}_{_{\rm mag}}$ and ${\dot P}/{P}_{_{\rm GW}}$ are
positive definite, it follows that ${\dot P}/{P}_{_{\rm obs}}\geq{\dot
P}/{P}_G$ so that a bound on $\dot G$ can be inferred if the main
pulse period is the period of rotation.
Heintzmann and Hillebrandt~\cite{heitzmann75} then modelled the pulsar by a polytropic
$(P\propto\rho^n$) white dwarf and deduced that ${\dd\ln
I}/{\dd\ln G}=2-3n/2$ so that $\vert\dot G/G\vert<10^{-10}\,{\rm yr}^{-1}$.
Mansfield~\cite{mansfield76} assumed a relativistic degenerate, zero temperature
polytropic star and got that, when $\dot G<0$,
$0\leq-{\dot G}/{G}<6.8\times10^{-11}\,{\rm yr}^{-1}$
at a $2\sigma$ level. He also noted that a positive $\dot G$ induces a
spin-up counteracting the electromagnetic spin-down which can provide
another bound if an independent estimate of the pulsar magnetic field
can be obtained.  Goldman~\cite{goldman90}, following Eardley~\cite{G-eardley75}, used the
scaling relations $N\propto G^{-3/2}$ and $M\propto G^{-5/2}$ to
deduce that $2{\dd\ln I}/{\dd\ln G}=-5+3{\dd\ln I}/{\dd\ln N}$. He
used the data from the pulsar PSR~0655+64 to deduce that the rate of decrease
of $G$ was smaller than
\begin{equation}\label{goldman}
  0\leq-\dot G/G<5.5\times10^{-11}\,{\rm yr}^{-1}.
\end{equation}
The analysis~\cite{G-pulsar2} of 10 years high pecision timing data on the millisecond
pulsar PSR~J0437-4715 has allowed to improve the constraint to
\begin{equation}\label{goldman}
  \vert\dot G/G\vert<2.3\times10^{-11}\,{\rm yr}^{-1}.
\end{equation}

Recently, it was argued~\cite{jofre,G-neutronstar3} that a variation of $G$ would induce a departure
of the neutron star matter from $\beta$-equilibrium, due to the
changing hydrostatic equilibrium. This would force non-equilibrium $\beta$-processes to occur, 
which release energy that is invested partly in neutrino emission and partly in heating the stellar interior. 
Eventually, the star arrives at a stationary state in which the temperature remains nearly constant, 
as the forcing through the change of $G$ is balanced by the ongoing reactions. 
Comparing the surface temperature of the nearest millisecond pulsar, 
PSR J0437-4715, inferred from ultraviolet observations, two upper limits for this variation were obtained, 
$|\dot G/G| < 2 \times 10^{-10}$ yr$^{-1}$, direct Urca reactions operating in the neutron star core are allowed, and 
$|\dot G/G| < 4 \times 10^{-12}$ yr$^{-1}$, considering only modified Urca reactions. This was
extended in Ref.~\cite{G-neutronstar} in order to take into account the correlation between the surface temperatures
and the radii of some old neutron stars to get
$\vert\dot G/G\vert < 2.1 \times 10^{-11}$~yr$^{-1}$.

\subsection{Stellar constraints}

Early works, see \S~IV.C of FVC~\cite{jpu-revue}, studied the
Solar evolution in presence of a time varying gravitational
constant, concluding that under the Dirac hypothesis, the original
nuclear resources of the Sun would have been burned by now. This
results from the fact that an increase of the gravitational
constant is equivalent to an increase of the star density (because
of the Poisson equation).

The idea of using stellar evolution to constrain the
possible value of $G$ was originally proposed by Teller~\cite{teller48}, 
who stressed that the evolution of a star was
strongly dependent on $G$. The luminosity of a main
sequence star can be expressed as a function of NewtonÕs
gravitational constant and its mass by using
homology relations~\cite{gamow67a,teller48}. In the particular case
that the opacity is dominated by free-free transitions,
Gamow~\cite{gamow67a} found that the luminosity of the star
is given approximately by $L\propto  G^{7.8} M^{5.5}$. In the
case of the Sun, this would mean that for higher values
of $G$, the burning of hydrogen will be more efficient
and the star evolves more rapidly, therefore
we need to increase the initial content of hydrogen
to obtain the present observed Sun. In a numerical
test of the previous expression, Delg'Innocenti \etal~\cite{G-globular} 
found that low-mass stars evolving from the
Zero Age Main Sequence to the red giant branch satisfy
$L\propto G^{5.6}M^{4.7}$, which agrees to within 10\% of
the numerical results, following the idea that Thomson
scattering contributes significantly to the opacity
inside such stars. Indeed, in the case of the opacity
being dominated by pure Thomson scattering, the luminosity
of the star is given by $L\propto G^4M^3$. It follows
from the previous analysis that the evolution of the
star on the main sequence is highly sensitive to the
value of $G$.

The driving idea behind the stellar constraints is that a secular
variation of $G$ leads to a variation of the gravitational
interaction. This would affect the hydrostatic equilibrium of the
star and in particular its pressure profile. In the case of
non-degenerate stars, the temperature, being the only control
parameter, will adjust to compensate the modification of the
intensity of the gravity. It will then affect the nuclear reaction
rates, which are very sensitive to the temperature, and thus the
nuclear time scales associated to the various processes. It
follows that the main stage of the stellar evolution, and in
particular the lifetimes of the various stars, will be modified.
As we shall see, basically two types of methods have been
used, the first in which on relate the variation of $G$ to
some physical characteristic of a star (luminosity,
effective temperature, radius), and a second
in which only a statistical measurement of the change of $G$
can be infered. Indeed, the first class of methods are more
reliable and robust but is usually restricted to nearby stars.
Note also that they usually require to have a precise
distance determination of the star, which may depend
on $G$.

\subsubsection{Ages of globular clusters}

The first application of these idea has been performed with
globular clusters. Their ages, determined for instance from the
luminosity of the main-sequence turn-off, have to be compatible
with the estimation of the age of the Galaxy. This gives the
constraint~\cite{G-globular}
\begin{equation}
\dot G/G=(-1.4\pm2.1)\times10^{-11}\,{\rm yr}^{-1}.
\end{equation}

The effect of a possible time dependence
of $G$ on luminosity has been studied in
the case of globular cluster H-R diagrams but has not
yielded any stronger constraints than those relying on
celestial mechanics

\subsubsection{Solar and stellar sysmology}

A side effect of the change of luminosity is a change in the depth of
the convection zone so that the inner
edge of the convecting zone changes its location. This induces a modification of the vibration
modes of the star and particularly to the acoustic waves, i.e
$p$-modes~\cite{demarque94}. \\

{\em  Helioseismology}. This waves are observed for our star, the Sun,
and heliosysmology allows to determine the sound speed
in the core of the Sun and, together with an equation of state,
the central densities and abundances of helium and hydrogen.
Demarque {\em et al.}~\cite{demarque94}
considered an ansatz in which $G\propto t^{-\beta}$ and showed that
$|\beta|<0.1$ over the last $4.5\times 10^9$ years, which corresponds
to $\vert\dot G/G\vert<2\times10^{-11}\,{\rm yr}^{-1}$.
Guenther {\em et al.}~\cite{guenther95} also showed that $g$-modes could provide
even much tighter constraints but these modes are up to now very
difficult to observe. Nevertheless, they concluded, using the claim of
detection by Hill and Gu~\cite{hill90}, that $\vert\dot G/G\vert<4.5\times10^{-12}\,{\rm yr}^{-1}$.
Guenther {\em et al.}~\cite{guenther98} then compared
the $p$-mode spectra predicted by different theories with varying
gravitational constant to the observed spectrum obtained by a network
of six telescopes and deduced that
\begin{equation}\label{gsun3}
  \left\vert\dot G/G\right\vert<1.6\times10^{-12}\,{\rm yr}^{-1}.
\end{equation}
The standard Solar model depends on few parameters and $G$ plays a
important role since stellar evolution is dictated by the balance
between gravitation and other interactions. Astronomical observations
determines $G M_\odot$ with an accuracy better than $10^{-7}$ and a variation of $G$ with
$GM_\odot$ fixed induces a change of the pressue
($P=GM_\odot^2/R_\odot^2$) and density
($\rho=M_\odot/R_\odot^3$). 
The experimental uncertainties in $G$ between different experiments have important
implications for helioseismology.  In particulat the uncertainties for the standard solar
model lead to a range in the value of the sound speed in the nuclear region
that is as much as 0.15\% higher than the inverted helioseismic sound speed~\cite{lopesilk}. While a
lower value of $G$ is preferred for the standard model, any definite prediction is masked
by the uncertainties in the solar models available in the literature.
Ricci and Villante~\cite{rv02} studied the
effect of a variation of $G$ on the density and pressure profile of
the Sun and concluded that present data cannot constrain $G$ better
than $10^{-2}\%$. It was also shown~\cite{lopesilk}  that the information provided by the
neutrino experiments is quite significant because it
constitutes an independent test of $G$ complementary
to the one provided by helioseismology.\\

{\em  White dwarfs}. The observation of the period of non-radial pulsations
of white dwarf allows to set similar constraints.  White dwarfs represent the final
stage of the stellar evolution for stars with a mass smaller to about $10M_\odot$.
Their structure is supported against gravitational collapse by the pressure of degenerate
electrons. It was discovered that some white dwarfs are variable stars and in fact
non-radial pulsator. This opens the way to use seismological techniques to
investigate their internal propoerties. In particular, their non-radial oscillations
is mostly determined by the Brunt-V\"ais\"al\"a frequency
$$
 N^2 = g \frac{\dd\ln P^{1/\gamma_1}/\rho}{\dd r}
$$
where $g$ is the gravitational acceleration, $\Gamma_1$ the first adiabatic exponent
and $P$ and $\rho$ the pressure and density (see e.g. Ref.~\cite{G-white1}
for a white dwarf model taking into account a varying $G$). A variation of $G$ induces a modification
of the degree of degeneracy of the white dwarf, hence on the frequency $N$ as well
as the cooling rate of the star, even though this is thought
to be negligible at the luminosities where white dwarfs are pulsationally unstable\cite{G-white2}. 
Using the observation of G117-B15A that has been
monitored during 20 years, it was concluded~\cite{G-puls1} that
\begin{equation}
 -2.5\times10^{-10}\,{\rm yr}^{-1}<\dot G/G< 4.0\times10^{-11}\,{\rm yr}^{-1},
\end{equation}
at a 2$\sigma$-level. The same observations were reanalyzed in Ref.~\cite{G-white2} to obtain
\begin{equation}
 \vert\dot G/G\vert< 4.1\times10^{-11}\,{\rm yr}^{-1}.
\end{equation}

\subsubsection{Late stages of stellar evolution and supernovae}

A variation of $G$ can influence the white dwarf cooling and the light curves
ot Type~Ia supernovae.

Garcia-Berro {\em et al.}~\cite{gbwd} considered the effect of a variation
of the gravitational constant on the cooling of white dwarfs and on
their luminosity function. As first pointed out by Vila~\cite{vila76}, the
energy of white dwarfs, when they are cool enough, is entirely of gravitational and thermal origin
so that a variation of $G$ will induce a modification of their energy
balance and thus of their luminosity. Restricting to cold white dwarfs with luminosity smaller than
ten Solar luminosity, the luminosity can be related to the star
binding energy $B$ and gravitational energy, $E_{_{\rm grav}}$, as
\begin{equation}
L=-\frac{\dd B}{\dd t}+\frac{\dot G}{G}E_{_{\rm grav}}
\end{equation}
which simply results from the hydrostatic equilibrium. Again, the
variation of the gravitational constant intervenes via the Poisson
equation and the gravitational potential. The cooling process is
accelerated if $\dot G/G<0$ which then induces a shift in the position
of the cut-off in the luminosity function.
Garcia-Berro {\em et al.}~\cite{gbwd} concluded that
\begin{equation}\label{gsun}
  0\leq-\dot G/G<(1\pm1)\times10^{-11}\,{\rm yr}^{-1}.
\end{equation}
The result depends on the details of the cooling theory, on
whether the C/O white dwarf is stratified or not and on hypothesis
on the age of the galactic disk. For instance, with no
stratification of the C/O binary mixture, one would require $\dot
G/G=-(2.5\pm0.5)\times10^{-11}\,{\rm yr}^{-1}$ if the Solar
neighborhood has a value of 8~Gyr (i.e. one would require a
variation of $G$ to explain the data). In the case of the standard
hypothesis of an age of 11~Gyr, one obtains that $0\leq-\dot
G/G<3\times10^{-11}\,{\rm yr}^{-1}$.\\

The late stages of stellar evolution are governed by the Chandrasekhar
mass $(\hbar c/G)^{3/2}m_{\rm n}^{-2}$ mainly determined by the
balance between the Fermi pressure of a degenerate electron gas and
gravity. 

Simple analytical models of the light curves
of Type~Ia supernovae predict that the peak of luminosity is proportional
to the mass of nickel synthetized. In a good approximation, it is
a fixed fraction of the Chandrasekhar mass. In models allowing for
a varying $G$, this would induce a modification of the luminosity distance-redshift
relation~\cite{G-sn1,gaztanaga02,cmb-G1}. It was however shown that this effect
is small. Note that it will be degenerate with the cosmological parameters. In particular,
the Hubble diagram is sensitive to the whole history of $G(t)$ between the highest
redshift observed and today so that one needs to rely on a better defined
model, such as e.g. scalar-tensor theory~\cite{cmb-G1} (the effect
of the Fermi constant was also considered in Ref.~\cite{sn1GF}).

In the case of Type~II supernovae, the Chandrasekhar mass also gouvernes
the late evolutionary stages of massive stars, including the formation of neutron stars.
Assuming that the mean neutron star mass is given by the
Chandrasekhar mass, one expects that $\dot G/G=-2\dot M_{_{\rm
NS}}/3M_{_{\rm NS}}$. Thorsett~\cite{thorsett96} used the observations of five
neutron star binaries for which five Keplerian parameters can be
determined (the binary period $P_b$, the projection of the orbital
semi-major axis $a_1\sin i$, the eccentricity $e$, the time and
longitude of the periastron $T_0$ and $\omega$) as well as the
relativistic advance of the angle of the periastron $\dot \omega$.
Assuming that the neutron star masses vary slowly as $M_{_{\rm
NS}}=M_{_{\rm NS}}^{(0)}-\dot M_{_{\rm NS}} t_{_{\rm NS}}$, that their
age was determined by the rate at which $P_b$ is increasing (so that
$t_{NS}\simeq2P_b/\dot P_b$) and that the mass follows a normal
distribution, Thorsett~\cite{thorsett96} deduced that, at $2\sigma$,
\begin{equation}\label{gsun4}
  \dot G/G=(-0.6\pm4.2)\times10^{-12}\,{\rm yr}^{-1}.
\end{equation}

\subsubsection{New developments}

It has recently been proposed that the variation of $G$ inducing a modification
of binary's binding energy, it should affect the gravitational wave luminosity, hence
leading to corrections in the chirping frequency~\cite{yunes}. For instance,
it was estimated that a LISA observation of an equal-mass inspiral event with total 
redshifted mass of $10^5M_\odot$ for three years should be able to measure $\dot{G}/G$ 
at the time of merger to better than $10^{-11}/$yr. This method paves the way
to constructing constraints in a large band of redshifts as well as in different directions in the sky,
which would be an unvaluable constraint for many models.

More speculative is the idea~\cite{G-grb} that a variation of $G$ can lead a neutron to
enter into the region where strange or hydrid stars are the true ground state. This
would be associated to Gamma-Ray-Burst that are claimed to be able to
reach the level of $10^{-17}/$yr on the time variation of $G$.

\subsection{Cosmological constraints}

Cosmological observations are more difficult to use in order to set
constraints on the time variation of $G$. In particular, they require
to have some ideas about the whole history of $G$ as a function of
time but also, as the variation of $G$ reflects an extension of General relativity,
it requires to modify all equations describing the evolution
(of the universe and of the large scale structure) in a consistent way.
We refer to Refs.~\cite{jpu-model,jpu-revu3,ugrg} for a discussion of the use of cosmological
data to constrain deviations from general relativity.

\subsubsection{Cosmic microwave background}\label{subsecGcmb}

A time-dependent gravitational constant will have mainly three
effects on the CMB angular power spectrum (see 
Ref.~\cite{cmb-G1} for discussions in the framework 
of scalar-tensor gravity in
which $G$ is considered as a field):
\begin{enumerate}
 \item The variation of $G$ modifies the Friedmann equation and
therefore the age of the Universe (and, hence, the sound horizon). For
instance, if $G$ is larger at earlier time, the age of the Universe is
smaller at recombination, so that the peak structure is shifted
towards higher angular scales.
 \item The amplitude of the Silk damping is modified.  At small scales,
viscosity and heat conduction in the photon-baryon fluid produce a
damping of the photon perturbations. The damping scale
is determined by the photon diffusion length at recombination, and
therefore depends on the size of the horizon at this epoch, and hence,
depends on any variation of the Newton constant throughout the
history of the Universe.
\item The thickness of the last scattering surface is modified. In the
same vein, the duration of recombination is modified by a variation of
the Newton constant as the expansion rate is different. It is well
known that CMB anisotropies are affected on small scales because the
last scattering ``surface'' has a finite thickness. The net effect is
to introduce an extra, roughly exponential, damping term, with the
cutoff length being determined by the thickness of the last scattering
surface. When translating redshift into time (or length), one has to
use the Friedmann equations, which are affected by a variation of the
Newton constant.  The relevant quantity to consider is the visibility
function $g$. In the limit of an infinitely thin last scattering
surface, $\tau$ goes from $\infty$ to $0$ at recombination epoch. For
standard cosmology, it drops from a large value to a much smaller one,
and hence, the visibility function still exhibits a peak, but it is much
broader.
\end{enumerate}

In full generality, the variation of $G$ on the CMB temperature anisotropies
depends on many factors: (1) modification of the background equations
and the evolution of the universe, (2) modification of the perturbation
equations, (3) whether the scalar field inducing the time variaiton
of $G$ is negligible or not compared to the other matter components,
(4) on the time profile of $G$ that has to be determine to be consistent
with the other equations of evolution. This explains why it is
very difficult to state a definitive constraint. For instance, in
the case of scalar-tensor theories (see below), one has two arbitrary
functions that dictate the variation of $G$. As can be seen
e.g. from Ref.~\cite{cmb-G1,G-cmb}, the profiles and effects on the CMB
can be very different and difficult to compare. Indeed, the effects described
above are also degenerate with a variation of the cosmological parameters.

In the case of Brans-Dicke theory, one just has a single constant parameter
$\omega_{_{\rm BD}}$ characterizing the deviation from general relativity and the time
variation of $G$. It is thus easier to compare the different constraints.
Chen and Kamionkowski~\cite{chen99} showed that
CMB experiments such as WMAP will be able to constrain these theories
for $\omega_{_{\rm BD}}<100$ if all parameters are to be determined by
the same CMB experiment, $\omega_{_{\rm BD}}<500$ if all parameters
are fixed but the CMB normalization and $\omega_{_{\rm BD}}<800$ if
one uses the polarization. For the Planck mission these numbers are
respectively, 800, 2500 and 3200. Ref.~\cite{acqua} concluded
from the analysis of WMAP, ACBAR, VSA and CBI, and galaxy
power spectrum data from 2dF, that  $\omega_{_{\rm BD}}>120$,
in agreement with the former analysis of Ref.~\cite{G-cmb}.
An analysis~\cite{wmapBD}
indictates that The ÒWMAP-5yr dataÓ and the Òall
CMB dataÓ both favor a slightly non-zero (positive) $\dot G/G$ but with the
addition of the SDSS poser spectrum data, the best-fit value is back to zero,
concluding that $-0.083<\Delta G/G<0.095$ between recombination and today,
which corresponds to $-1.75\times 10^{-12}\,{\rm yr}^{-1} < \dot G/G < 1.05\times 10^{-12}\,{\rm yr}^{-1}$.

From a more phenomenoloical prospect, some works modelled
the variation of $G$ with time in a purely ad-hoc way, for
instance~\cite{cmb-cc} by assuming a linear evolution with time or
a step function.

\subsubsection{BBN}

As explained in details in section~\ref{secbbnoverview}, changing the
value of the gravitational constant affects the freeze-out temperature
$T_{\rm f}$. A larger value of $G$ corresponds to a higher expansion rate.
This rate is determined by the combination $G\rho$ and
in the standard case the Friedmann equations imply that $G\rho t^2$ is
constant.  The density $\rho$ is determined by the number $N_*$ of
relativistic particles at the time of nucleosynthesis so that
nucleosynthesis allows to put a bound on the number of neutrinos
$N_\nu$. Equivalently, assuming the number of neutrinos to be three,
leads to the conclusion that $G$ has not varied from more than 20\%
since nucleosynthesis.  But, allowing for a change both in $G$ and
$N_\nu$ allows for a wider range of variation.  Contrary to the fine
structure constant the role of $G$ is less involved.

The effect of a varying $G$ can be described, in its most
simple but still useful form, by introducing a speed-up factor, $\xi=H/H_{GR}$,
that arises from the modification of the value of the gravitational
constant during BBN. Other approaches
considered the full dynamics of the problem but restricted
themselves to the particular class of Jordan-Fierz-Brans-Dicke 
theory~\cite{accetta90,arai87,barrow78,casas92,clifton05,dg91,rothman82,yang79}
(Casas {\em et al.}~\cite{casas92}
concluded from the study of helium and deuterium that
$\omega_{_{\rm BD}}>380$ when $N_\nu=3$ and $\omega_{_{\rm BD}}>50$ when
$N_\nu=2$.), 
of a massless dilaton with a quadratic
coupling~\cite{couv,coc4a,bbn-Gpichon,santiago97} or to a general massless dilaton~\cite{serna96}. It
should be noted that a combined analysis of BBN and
CMB data was investigated in Refs.~\cite{copi,kneller03}. The former
considered $G$ constant during BBN while the latter focused
on a nonminimally quadratic coupling and a runaway
potential. It was concluded that from the BBN in conjunction with WMAP
determination of $\eta$ set that $\Delta G/G$ has to be smaller than 20\%.
We however stress that the dynamics of the field can
modify CMB results (see previous section) so that one needs to be careful while
inferring $\Omega_\baryon$ from WMAP unless the scalar-tensor theory
has converged close to general realtivity at the time of decoupling.

In early studies, Barrow~\cite{barrow78} assumed that $G\propto t^{-n}$ and obtained from
the helium abundances that $-5.9\times10^{-3}<n<7\times10^{-3}$ which
implies that $|{\dot G}/{G}|<(2\pm9.3)\,h\times 10^{-12}\,{\rm yr}^{-1}$,
assuming a flat universe.  This corresponds in terms of the
Brans-Dicke parameter to $\omega_{_{\rm BD}}>25$.  Yang {\em et al.}~\cite{yang79}
included the deuterium and lithium to improve
the constraint to $n<5\times10^{-3}$ which corresponds to
$\omega_{_{\rm BD}}>50$. It was further improved by
Rothman and Matzner~\cite{rothman82} to $|n|<3\times10^{-3}$ implying
$|{\dot G}/{G}|<1.7\times 10^{-13}\,{\rm yr}^{-1}$.
Accetta {\em et al.}~\cite{accetta90} studied the dependence of the abundances of D, $^3{\rm He}$,
$^4{\rm He}$ and $^7{\rm Li}$ upon the variation of $G$ and concluded
that $-0.3<{\Delta G}/{G}<0.4$
which roughly corresponds to $|\dot G/G|<9\times10^{-13}\,{\rm yr}^{-1}$.
All these investigations assumed that the other constants are kept
fixed and that physics is unchanged. Kolb {\em et al.}~\cite{bbnkolb} assumed
a correlated variation of $G$, $\aem$ and $\gfermi$ and got a bound on
the variation of the radius of the extra-dimensions.

Although the uncertainty in the helium-4 abundance has been argued to
be significantly larger that what was assumed in the past~\cite{osli},
interesting bounds can still be  derived~\cite{bbn-cyburt2}. In particular
translating the bound on extra relativistic degress of freedom ($-0.6<\delta N_\nu<0.82$)
to a constraint on the speed-up factor ($0.949<\xi<1.062$), it was concluded~\cite{bbn-cyburt2},
since $\Delta G/G=\xi^2-1=7\delta N_\nu/43$, that
\begin{equation}
 -0.10<\frac{\Delta G}{G}<0.13.
\end{equation}

The relation between the speed-up factor, or an extra number of
relativistic degrees of freedom, with a variation of $G$ is only approximate
since it assumes that the variation of $G$ affects only the Friedmann
equation by a renormalisation of $G$. This is indeed accurate only when the
scalar field is slow-rolling. For instance~\cite{couv}, the speed-up factor
is given (with the notations of Section~\ref{subsecST}) by
$$
 \xi = \frac{A(\varphi_*)}{A_0}\frac{1+\alpha(\varphi_*)\varphi_*'}{\sqrt{1-\varphi_*^{2\prime}/3}}\frac{1}{\sqrt{1+\alpha_0^2}}
$$
so that
\begin{equation}
 \xi^2 = \frac{G}{G_0}\frac{(1+\alpha(\varphi_*)\varphi_*')^2}{(1+\alpha^2)(1-\varphi_*^{2\prime}/3)},
\end{equation}
so that $\Delta G/G_0=\xi^2-1$ only if $\alpha\ll1$ (small deviation from general relativity)
and $\varphi_*'\ll 1$ (slow rolling dilaton).
The BBN in scalar-tensor theories was investigated~\cite{couv,bbn-Gpichon} in the case of
a two-parameter family involving a non-linear scalar field-matter coupling function. They
concluded that even in the cases where before BBN the scalar-tensor
theory was far from general relativity, BBN enables to set quite tight
constraints on the observable deviations from general relativity
today. In particular, neglecting the
cosmological constant, BBN imposes $\alpha_0^2 < 10^{-6.5} 
\beta^{-1} (\Omega_\mat h^2 / 0.15)^{-3/2}$ when $\beta > 0.5$
(with the definitions introduced below Eq.~(\ref{matter*})).

\section{Theories with varying constants}\label{section6}

As explained in the introduction, Dirac postulated that $G$
varies  as the inverse of the cosmic time. Such an hypothesis
is indeed not a theory since the evolution of $G$ with time
is postulated and not derived from an equation of 
evolution\footnote{Note that Dirac hypothesis can also be achieved by assuming
that $e$ varies as $t^{1/2}$. Indeed this reflects a choice of
units, either atomic or Planck units. There is however a
difference: assuming that only $G$ varies violates the strong
equivalence principle while assuming a varying $e$ results in a
theory violating the weak equivalence principle. It does not
mean we are detecting the variation of a dimensionful constant but
simply that either $e^2/\hbar c$ or $Gm_{\rm e}^2/\hbar c$ is
varying. This shows that many implementation of this
idea are a priori possible.} 
consistent with the other field equations, that have to take
into account that $G$ is no more a constant (in particular
in a Lagrangian formulation one needs to take into account
that $G$ is no more constant when varying.

The first implementation of Dirac's phenomenological idea into a
field-theory framework (i.e. modifying Einstein gravity and
incorporating non-gravitational forces and matter) was proposed by
Jordan~\cite{jordan37}. He realized that the constants have to
become dynamical fields and proposed the action
\begin{eqnarray}
S=\int\sqrt{-g}\dd^4\bx\phi^\eta\left[R-\xi\left(\frac{\nabla\phi}{\phi}
\right)^2-\frac{\phi}{2}F^2 \right],
\end{eqnarray}
$\eta$ and $\xi$ being two parameters. It follows that both $G$
and the fine-structure constant have been promoted to the status
of a dynamical field.

Fierz~\cite{fierz56} realized that
with such a Lagrangian, atomic spectra will be space-time-dependent,
and he proposed to fix $\eta$ to the value -1 to prevent such a
space-time dependence.  This led to the definition of a one-parameter
($\xi$) class of scalar-tensor theories in which only
$G$ is assumed to be a dynamical field. This
was then further
explored by Brans and Dicke~\cite{brans61} (with the change of notation
$\xi\rightarrow \omega$).  
In this Jordan-Fierz-Brans-Dicke theory the gravitational
constant is replaced by a scalar field which can vary both in space
and time.  It follows that, for cosmological solutions, $G\propto
t^{-n}$ with
$n^{-1}=2+3\omega_{_{\rm BD}}/2$. Einstein gravity is thus recovered when
$\omega_{_{\rm BD}}\rightarrow\infty$.  This kind of theory was further
generalized to obtain various functional dependencies for $G$ in the
formalisation of scalar-tensor theories of gravitation (see
e.g. Damour and Esposito-Far\`ese~\cite{damour92} or Will~\cite{will-book}).

\subsection{Introducing new fields: generalities}

\subsubsection{The example of scalar-tensor theories}\label{subsecST}

Let us start to remind how the standard general relativistic
framework can be extended to make $G$ dynamical
on the example of scalar-tensor theories, in
which gravity is mediated not only by a massless spin-2 graviton
but also by a spin-0 scalar field that couples universally to
matter fields (this ensures the universality of free fall). In the
Jordan frame, the action of the theory takes the form
\begin{eqnarray}\label{actionJF}
  S &=&\int \frac{\dd^4 x }{16\pi G_*}\sqrt{-g}
     \left[F(\varphi)R-g^{\mu\nu}Z(\varphi)\varphi_{,\mu}\varphi_{,\nu}
        - 2U(\varphi)\right]+ S_{\rm matter}[\psi;g_{\mu\nu}]
\end{eqnarray}
where $G_*$ is the bare gravitational constant. This action
involves three arbitrary functions ($F$, $Z$ and $U$) but only two
are physical since there is still the possibility to redefine the
scalar field. $F$ needs to be positive to ensure that the graviton
carries positive energy. $S_{\rm matter}$ is the action of the
matter fields that are coupled minimally to the metric
$g_{\mu\nu}$. In the Jordan frame, the matter is universally
coupled to the metric so that the length and time as measured by
laboratory apparatus are defined in this frame.

The variation of this action gives the following field equations
\begin{eqnarray}
F(\varphi) \left(R_{\mu\nu}-{1\over2}g_{\mu\nu}R\right)
&=& 8\pi G_* T_{\mu\nu}
+ Z(\varphi) \left[\partial_\mu\varphi\partial_\nu\varphi
- {1\over 2}g_{\mu\nu}
(\partial_\alpha\varphi)^2\right]
\nonumber\\
&&+\nabla_\mu\partial_\nu F(\varphi) - g_{\mu\nu}\Box F(\varphi)
- g_{\mu\nu} U(\varphi)\ ,
\label{einstein}\\
2Z(\varphi)~\Box\varphi &=&
-{dF\over d\varphi }\,R - {dZ\over d\varphi }\,(\partial_\alpha\varphi )^2
+ 2 {dU\over d\varphi }\ ,
\label{BoxPhi}\\
\nabla_\mu T^\mu_\nu &=& 0\ ,
\label{matter}
\end{eqnarray}
\label{2.2}
where $T \equiv T^\mu_\mu$ is the trace of the
matter energy-momentum tensor $T^{\mu\nu} \equiv (2/\sqrt{-g})\times
\delta S_m/\delta g_{\mu\nu}$.  As expected~\cite{ellisu},
we have an equation which reduces to the standard Einstein equation
when $\varphi$ is constant and a new equation to describe the
dynamics of the new degree of freedom while the conservation equation
of the matter fields is unchanged, as expected from the weak equivalence principle.\\

It is useful to define an Einstein frame action through a
conformal transformation of the metric $g_{\mu\nu}^* =
F(\varphi)g_{\mu\nu}$. In the following all quantities labelled by
a star (*) will refer to Einstein frame. Defining the field
$\varphi_*$ and the two functions $A(\varphi_*)$ and
$V(\varphi_*)$ (see e.g. Ref.~\cite{gefpolar}) by
$$
 \left(\frac{\dd\varphi_*}{\dd\varphi}\right)^2
              = \frac{3}{4}\left(\frac{\dd\ln F(\varphi)}{\dd\varphi}\right)^2
                  +\frac{1}{2F(\varphi)},
                  \quad
 A(\varphi_*) = F^{-1/2}(\varphi),\quad
 2V(\varphi_*)= U(\varphi) F^{-2}(\varphi),
$$
the action (\ref{actionJF}) reads as
\begin{eqnarray}
 S &=& \frac{1}{16\pi G_*}\int \dd^4x\sqrt{-g_*}\left[ R_*
        -2g_*^{\mu\nu} \partial_\mu\varphi_*\partial_\nu\varphi_*
        - 4V\right]+ S_{\rm matter}[A^2g^*_{\mu\nu};\psi].
\end{eqnarray}
The kinetic terms have been diagonalized so that the spin-2 and
spin-0 degrees of freedom of the theory are perturbations of
$g^*_{\mu\nu}$ and $\varphi_*$ respectively. In this frame
the field equations are given by
\begin{eqnarray}
R^*_{\mu\nu} - {1\over 2} R^* g^*_{\mu\nu} &=& 8\pi G_* T^*_{\mu\nu}
+ 2 \partial_\mu\varphi_*\partial_\nu\varphi_* -
g^*_{\mu\nu}(g_*^{\alpha\beta}
\partial_\alpha\varphi_*\partial_\beta\varphi_*)
- 2 V(\varphi) g^*_{\mu\nu}\ ,
\label{einstein*}\\
\Box_*\varphi_* &=& -4\pi G_*\alpha(\varphi_*)~T_*
+dV(\varphi)/d\varphi_*\ ,
\label{Boxvarphi}\\
\nabla^*_\mu T^\mu_{*\nu} &=& \alpha(\varphi_*)~T_*
\partial_\nu\varphi_*\ ,
\label{matter*}
\end{eqnarray}
with $\alpha\equiv \dd\ln A/\dd\varphi_*$ and $\beta\equiv\dd\alpha/\dd\varphi_*$.
In this version, the Einstein equations are not modified, but since the theory
can now be seen as the theory in which all the mass are varying in the
same way, there is a source term to the conservation equation. This shows
that the same theory can be interpreted as a varying $G$ theory or a
universally varying mass theory, but remember that whathever its form
the important parameter is the dimensionless quantity $Gm^2/\hbar c$.\\

The action~(\ref{actionJF}) defines an effective gravitational
constant $G_{\rm eff} = G_*/F = G_*A^2$. This constant does not
correspond to the gravitational constant effectively measured in a
Cavendish experiment. The Newton constant measured in this
experiment is 
\begin{equation}\label{gcav}
 G_{\rm cav} = G_*A_0^2(1+\alpha_0^2)
                    = \frac{G_*}{F}\left(1 + \frac{F_\phi^2}{2F + 3 F_\phi^2} \right)
\end{equation} 
 where the
first term, $G_*A_0^2$ corresponds to the exchange of a graviton
while the second term $G_*A_0^2\alpha_0^2$ is related to the long
range scalar force. The gravitational constant depends on the
scalar field and is thus dynamical.

This illustrates the main features that will appear in any such models:
(i) new dynamical fields appear (here a scalar field), (ii) some
constant will depend the value of this scalar field (here $G$ is
a function of the scalar field). It follows that the Einstein equations
will be modified and that there will exist a new equation dictating the
propagation of the new degree of freedom.

In this particular example, the coupling of the scalar field is universal
so that no violation of the universality of free fall is expected. The deviation 
from general relativity can be quantified in terms of
the post-Newtonian parameters, which can be expressed in terms of the
values of $\alpha$ and $\beta$ today as
\begin{equation}
 \gamma^{\rm PPN} - 1 = -\frac{2\alpha_0^2}{1+\alpha^2_0},\qquad
 \beta^{\rm PPN} - 1 =\frac{1}{2}
 \frac{\beta_0\alpha_0^2}{(1+\alpha_0^2)^2}.
\end{equation}
This expression are valid only if the field is light on the Solar system
scales. If this is not the case then this conclusions may be changed~\cite{cameleon}.
The Solar system constraints imply $\alpha_0$ to be very small,
typically $\alpha_0^2<10^{-5}$ while $\beta_0$ can still be large.
Binary pulsar observations~\cite{gefpul2,gefpul} impose that
$\beta_0>-4.5$. The time variation of $G$ is then related to
$\alpha_0$, $\beta_0$ and the time variation of the scalar
field today
\begin{equation}\label{Gdotst}
 \frac{\dot G_{\rm cav}}{G_{\rm cav}} =2\alpha_0\left(1+\frac{\beta_0}{1+\alpha_0^2} 
 \right)\dot\varphi_{*0}.
\end{equation}
This example shows that the variation of the constant and the deviation from general relativity
quantified in terms of the PPN parameters are of the same magnitude, because they
are all driven by the same new scalar field.\\

The example of scalar-tensor theories is also very illustrative to show how
deviation from general relativity can be fairly large in the early universe while
still being compatible with Solar system constraints.  It relies on the
attraction mechanism toward general relativity~\cite{dn1,dn2}.
 
Consider the simplest model  of a massless dilaton ($V(\varphi_*) = 0$) with
quadratic coupling ($\ln A = a =\frac{1}{2}\beta\varphi_*^2$). Note that
the linear case correspond to a Brans-Dicke theory with a fixed deviation
from general relativity. It follows that $\alpha_0 = \beta\varphi_{0*}$ and
$\beta_0  = \beta$.
As long as $V=0$, the Klein-Gordon equation can be
rewritten in terms of the variable $p=\ln a$ as
\begin{eqnarray}\label{kgqq}
 \frac{2}{3-\varphi_*^{'2}}\varphi_*''
 +(1-w)\varphi_*' =-\alpha(\varphi_*)(1-3w).
\end{eqnarray}
As emphasized in Ref.~\cite{dn1}, this is the equation of motion
of a point particle with a velocity dependent inertial mass,
$m(\varphi_*)=2/(3-\varphi_*^{'2})$ evolving in a potential
$\alpha(\varphi_*)(1-3w)$ and subject to a damping force,
$-(1-w)\varphi_*'$. During the cosmological evolution the field is
driven toward the minimum of the coupling function. If $\beta>0$,
it drives $\varphi_*$ toward 0, that is $\alpha\rightarrow0$, so
that the scalar-tensor theory becomes closer and closer to general
relativity. When $\beta<0$, the theory is driven way from general
relativity and is likely to be incompatible with local tests
unless $\varphi_*$ was initially arbitrarily close from 0.\\

It follows that the deviation from general relativity remains constant
during the radiation era (up to threshold effects in the
early universe~\cite{cocbbn,bbn-Gpichon} and quantum effects~\cite{copu})
and the theory is then attracted toward general relativity during the matter
era. Note that it implies that postulating a linear or inverse variation of
$G$ with cosmic time is actually not realistic in this class of models.
Since the theory is fully defined, one can easily compute various
cosmological observables (late time dynamics~\cite{msu},
CMB anisotropy~\cite{cmb-G1}, weak lensing~\cite{carlowl}, 
BBN~\cite{couv,coc4a,bbn-Gpichon}) in a consistent way and confront them with data.

\subsubsection{Making other constants dynamical}

Given this example, it seems a priori simple to cook up
a theory that will describe a varying fine-structure constant by
coupling a scalar field to the electromagnetic Faraday tensor as
\begin{equation}
 S = \int\left[\frac{R}{16\pi G} - 2(\partial_\mu\phi)^2
 -\frac{1}{4}B(\phi)F_{\mu\nu}^2 \right]\sqrt{-g}\dd^4 x
\end{equation}
so that the fine-structure will evolve according to
$\alpha=B^{-1}$.
Such an simple implementation may however have dramatic
implications. In particular, the contribution of the
electromagnetic binding energy to the mass of any nucleus can be
estimated by the Bethe-Weiz\"acker formula so that
$$
 m_{(A,Z)}(\phi) \supset 98.25\,\alpha(\phi)\,\frac{Z(Z-1)}{A^{1/3}}\,\hbox{MeV}.
$$
This implies that the sensitivity of the mass to a variation of
the scalar field is expected to be of the order of
\begin{equation}
 f_{(A,Z)} = \partial_\phi m_{(A,Z)}(\phi) \sim 10^{-2}
 \frac{Z(Z-1)}{A^{4/3}} \alpha'(\phi).
\end{equation}
It follows that the level of the violation of the universality of
free fall is expected to be of the level of $
\eta_{12}\sim10^{-9}X(A_1,Z_1;A_2,Z_2)(\partial_\phi\ln B)^2_0$.
Since the factor $X(A_1,Z_1;A_2,Z_2)$ typically ranges as
$\mathcal{O}(0.1-10)$, we deduce that $(\partial_\phi\ln B)_0$ has
to be very small for the Solar system constraints to be satisfied.
It follows that today the scalar field has to be very close to the
minimum of the coupling function $\ln B$.
This is indeed very simplistic because we only take into account the
effect of the electromagnetic binding energy (see \S~\ref{subsec22}).\\

Let us also note that such a simple coupling cannot be eliminated
by a conformal rescaling $g_{\mu\nu}=A^2(\phi)g_{\mu\nu}^*$ since
$$
 \int B(\phi)g^{\mu\rho}g^{\mu\nu}F_{\nu\sigma}F_{\rho\sigma}\sqrt{-g}\dd^4 x
 \longrightarrow
 \int B(\phi)A^{D-4}(\phi)g_*^{\mu\rho}g_*^{\mu\nu}F_{\nu\sigma}F_{\rho\sigma}\sqrt{-g_*}\dd^4 x
$$
so that the action is invariant in $D=4$ dimensions.\\

This example shows that we cannot couple a field blindly to e.g. the
Faraday tensor to make the fine-structure constant dynamics and that
some mechanism for reconciling this variation with local constraints,
and in particular the university of free fall, will be needed.

\subsection{High-energy theories and varying constants}

\subsubsection{Kaluza-Klein}

Such coupling terms naturally appear when
compactifying a higher-dimensional theory. As an example, let us
recall the compactification of a 5-dimensional Einstein-Hilbert
action (Ref.~\cite{peteruzanbook}, chapter~13)
$$
 S=\frac{1}{12\pi^2 G_5}\int\bar R\sqrt{-\bar g}\dd^5 x.
$$
Decomposing the 5-dimensional metric $\bar g_{AB}$ as
$$
 \bar g_{AB} = \left(
\begin{array}{cc}
  g_{\mu\nu}+\frac{A_\mu A_\nu}{M^2}\phi^2 & \frac{A_\mu}{M}\phi^2\\
  \frac{A_\nu}{M}\phi^2 & \phi^2 \\
\end{array}
 \right),
$$
where $M$ is a mass scale, we obtain
\begin{equation}\label{KKaction}
 S=\frac{1}{16\pi G_*}\int\left(R - \frac{\phi^2}{4M^2}F^2\right)\phi\sqrt{-g}\dd^4
 x,
\end{equation}
where the 4-dimensional gravitational constant is $G_*=3\pi
G_5/4\int\dd y$. The scalar field couples explicitly to the
kinetic term of the vector field and cannot be eliminated by a
redefinition of the metric: again, this is the well-known conformal
invariance of electromagnetism in four dimensions. Such a term
induces a variation of the fine-structure constant as well as a
violation of the universality of free-fall. Such dependencies of
the masses and couplings are generic for higher-dimensional
theories and in particular string theory. It is actually one of
the definitive predictions for string theory that there exists a
dilaton, that couples directly to matter~\cite{taylor88}
and whose vacuum expectation value determines the string
coupling constants~\cite{witten84}.

In the models by Kaluza~\cite{kaluza21} and Klein~\cite{klein1926} the 5-dimensional spacetime 
was compactified assuming that one spatial
extra-dimension $S^1$, of radius $R_{_{\rm KK}}$.
It follows that any field $\chi(x^\mu,y$) can be Fourier transformed along the compact
dimension (with coordinate $y$), so that, from a 4-dimensional point of view, it gives rise to a tower of of
fields $\chi^{(n)}(x^\mu)$ of mas $m_{\rm n}=n R_{KK}$. At energies small compared
to $R_{KK}^{-1}$ only the $y$-independent part of the field remains and the physics
looks 4-dimensional.

Assuming that the action~(\ref{KKaction}) corresponds to the Jordan frame action,
as the coupling $\phi R$ may suggest, it follows that the gravitational constant and
the Yang-Mills coupling associated with the vector field $A^\mu$ must
scale as 
\begin{equation}
 G\propto \phi^{-1}, \qquad
 g_{YM}^{-2} \propto \phi^2/G \propto \phi^3.
\end{equation}
Note that the scaling of $G$ with $\phi$ (or time)  is not the one of the gravitational
constant that would be measured in a Cavendish experiment since
Eq.~(\ref{gcav}) tells us that $G_{\rm cav}\propto G_*\phi^{-1}\left(1+\frac{1}{2\phi+3}\right)$.

This can be generalized to the case of $D$ extra-dimensions~\cite{cremmer77} to
\begin{equation}\label{kkDdim}
G\propto \phi^{-D},\quad
\alpha_i(m_{_{\rm KK}})=K_i(D)G\phi^{-2}
\end{equation}
where the constants $K_i$ depends only on the dimension and
topology of the compact space~\cite{weinberg83b} so that the only
fundamental constant of the theory is the mass scale $M_{4+D}$
entering the $4+D$-dimensional theory. A theory on
${\cal M}_4\times {\cal M}_D$ where ${\cal M}_D$ is a
$D$-dimensional compact space generates a low-energy quantum field
theory of the Yang-Mills type related to the isometries of ${\cal
M}_D$ [for instance Ref.~\cite{witten81} showed that for $D=7$, it can
accommodate the Yang-Mills group $SU(3)\times SU(2)\times U(1)$].
The two main problems of these theories are that one cannot
construct chiral fermions in four dimensions by compactification
on a smooth manifold with such a procedure and that gauge theories
in five dimensions or more are not renormalisable.

In such a framework the variation of the gauge couplings and of the gravitational
constant arises from the variation of the size of the extra-dimensions so that
one can derives stronger constraints that by assuming independent variation,
but at the expense of being more model-dependent. Let us mention the
works by Marciano~\cite{marciano84}   and Wu and Wang~\cite{wu86} 
in which the structure constants
at lower energy are obtained by the renormalisation group.

Ref.\cite{bbnkolb} used the variation (\ref{kkDdim}) to
constrain the time variation of the radius of the extra-dimensions
during primordial nucleosynthesis to conclude that$|\Delta
R_{_{\rm KK}}/R_{_{\rm KK}}|<1\%$. Ref.~\cite{barrow87} took the effects of
the variation of $\as\propto R_{_{\rm KK}}^{-2}$  
and deduced from the helium-4 abundance that
$|\Delta R_{_{\rm KK}}/R_{_{\rm KK}}|<0.7\%$ and $|\Delta R_{_{\rm
KK}}/R_{_{\rm KK}}|<1.1\%$ respectively for $D=2$ and $D=7$
Kaluza-Klein theory and that $|\Delta R_{_{\rm KK}}/R_{_{\rm
KK}}|<3.4\times10^{-10}$ from the Oklo data. An analysis
of most cosmological data (BBN, CMB, quasar etc..) assuming that
the extra-dimension scales as $R_0(1+\Delta t^{-3/4})$ and
$R_0[1+\Delta](1-\cos\omega(t-t_0)$ concluded that $\Delta$ has
to be smaller tha $10^{-16}$ and $10^{-8}$ respectively~\cite{landauKK},
while Ref.~\cite{lichu}  assumes that gauge fields and matter fields
can propagate in the bulk. Ref.~\cite{aguilar} evaluated the effect
of such a couple variation of $G$ and the structures constants on distant
supernova data, concluding that a variation similar to the one reported
in Ref.~\cite{q-webprl01} would make the distant supernovae brighter.

\subsubsection{String theory}

There exist five anomaly free, supersymmetric perturbative string
theories respectively known as type I, type IIA, type IIB, SO(32)
heterotic and $E_8\times E_8$ heterotic theories (see e.g.
Ref.~\cite{polchinski97}). One of the definitive predictions of these theories
is the existence of a scalar field, the dilaton, that couples directly
to matter~\cite{taylor88} and whose vacuum expectation
value determines the string coupling constant~\cite{witten84}. There
are two other excitations that are common to all perturbative string
theories, a rank two symmetric tensor (the graviton) $g_{\mu\nu}$ and
a rank two antisymmetric tensor $B_{\mu\nu}$. The field content then
differs from one theory to another. It follows that the 4-dimensional
couplings are determined in terms of a string scale and various
dynamical fields (dilaton, volume of compact space, \ldots). When the
dilaton is massless, we expect {\it three} effects: (i) a scalar
admixture of a scalar component inducing deviations from general
relativity in gravitational effects, (ii) a variation of the couplings
and (iii) a violation of the weak equivalence principle. Our purpose is
to show how the 4-dimensional couplings are related to the string mass
scale, to the dilaton and the structure of the extra-dimensions mainly
on the example of heterotic theories.

To be more specific, let us consider an example. The two {\it
heterotic theories} originate from the fact that left- and
right-moving modes of a closed string are independent. This
reduces the number of supersymmetry to $N=1$ and the quantization
of the left-moving modes imposes that the gauge group is either
$SO(32)$ or $E_8\times E_8$ depending on the fermionic boundary
conditions. The effective tree-level action is
\begin{eqnarray}\label{het}
S_{H}&=&\int\dd^{10}{\bf x}\sqrt{-g_{10}}\hbox{e}^{-2\Phi}
         \left[M_{_{H}}^8\left\lbrace R_{10}+4\Box\Phi-4(\nabla\Phi)^2
         \right\rbrace-\frac{M_{_{H}}^6}{4}F_{AB}F^{AB}
         +\ldots\right].
\end{eqnarray}
When compactified on a 6-dimensional Calabi-Yau space, the effective
4-dimensional action takes the form
\begin{eqnarray}\label{het4}
S_{H}&=&\int\dd^{4}{\bf x}\sqrt{-g_{4}}\phi
\left[M_{_{H}}^8\left\lbrace R_{4}+\left(\frac{\nabla\phi}{\phi}\right)^2
-\frac{1}{6}\left(\frac{\nabla V_6}{V_6}\right)^2\right\rbrace-\frac{M_{_{H}}^6}{4}F^2\right]+\ldots
\end{eqnarray}
where $\phi\equiv V_6\hbox{e}^{-2\Phi}$ couples identically to the
Einstein and Yang-Mills terms. It follows that
\begin{equation}
M_4^2=M_{_{H}}^8\phi,\qquad
g^{-2}_{_{\rm YM}}=M_{_{H}}^6\phi
\end{equation}
at tree-level. Note that to reach this conclusion, one has to
assume that the matter fields (in the `dots' of Eq.~(\ref{het4})
are minimally coupled to $g_4$; see e.g. Ref.~\cite{maeda88}).

The strongly coupled SO(32) heterotic string theory is equivalent to
the weakly coupled type I string theory.  {\it Type I superstring}
admits open strings, the boundary conditions of which divide the
number of supersymmetries by two. It follows that the tree-level
effective bosonic action is $N=1$, $D=10$ supergravity which takes the
form, in the string frame,
\begin{eqnarray}
S_{I}&=&\int\dd^{10}{\bf x}\sqrt{-g_{10}}M_{_{I}}^6\hbox{e}^{-\Phi}
        \left[\hbox{e}^{-\Phi}
        M_{_{I}}^2R_{10}-\frac{F^2}{4}+\ldots\right]
\end{eqnarray}
where the dots contains terms describing the dynamics of the dilaton,
fermions and other form fields. At variance with (\ref{het}), the
field $\Phi$ couples differently to the gravitational and Yang-Mills
terms because the graviton and Yang-Mills fields are respectively
excitation of close and open strings. It follows that $M_I$ can be
lowered even to the weak scale by simply having $\exp\Phi$ small
enough. Type I theories require $D9$-branes for consistancy. When
$V_6$ is small, one can use T-duality (to render $V_6$ large, which
allows to use a quantum field theory approach) and turn the $D9$-brane
into a $D3$-brane so that
\begin{eqnarray}
S_{I}&=&\int\dd^{10}{\bf x}\sqrt{-g_{10}}
\hbox{e}^{-2\Phi}M_{_{I}}^8R_{10}-\int\dd^{4}{\bf x}\sqrt{-g_{4}}\hbox{e}^{-\Phi}
\frac{1}{4}F^2+\ldots
\end{eqnarray}
where the second term describes the Yang-Mills fields localized on the
$D3$-brane. It follows that
\begin{equation}
M_4^2=\hbox{e}^{-2\Phi}V_6M_{_{I}}^8,\qquad
g^{-2}_{_{\rm YM}}=\hbox{e}^{-\Phi}
\end{equation}
at tree-level. If one compactifies the $D9$-brane on a 6-dimensional
orbifold instead of a 6-torus, and if the brane is localized at an
orbifold fixed point, then gauge fields couple to fields $M_i$ living
only at these orbifold fixed points with a (calculable) tree-level
coupling $c_i$ so that
\begin{equation}
M_4^2=\hbox{e}^{-2\Phi}V_6M_{_{I}}^8,\qquad
g^{-2}_{_{\rm YM}}=\hbox{e}^{-\Phi}+c_iM_i.
\end{equation}
The coupling to the field $c_i$ is a priori non universal.  At
strong coupling, the 10-dimensional $E_8\times E_8$ heterotic
theory becomes M-theory on $R^{10}\times S^1/Z_2$~\cite{horava96}. 
The gravitational field propagates in the
11-dimensional space while the gauge fields are localized on two
10-dimensional branes.

At one-loop, one can derive the couplings by including Kaluza-Klein
excitations to get~\cite{dudas00}
\begin{equation}
g^{-2}_{_{\rm YM}}=M_{_{H}}^6\phi-\frac{b_a}{2}(RM_{_H})^2+\ldots
\end{equation}
when the volume is large compared to the mass scale and in that case
the coupling is no more universal. Otherwise, one would get a more
complicated function. Obviously, the 4-dimensional effective
gravitational and Yang-Mills couplings depend on the considered
superstring theory, on the compactification scheme but in any case
they depend on the dilaton.

As an example, Ref.~\cite{maeda88} considered the ($N=1, D=10$)-supergravity model
derived from the heterotic superstring theory in the low energy limit
and assumed that the 10-dimensional spacetime is compactified on a
6-torus of radius $R(x^\mu)$ so that the effective 4-dimensional
theory described by (\ref{het4}) is of the Brans-Dicke type with
$\omega=-1$.  Assuming that $\phi$ has a mass $\mu$,  and
couples to the matter fluid in the universe as $S_{_{\rm
matter}}=\int\dd^{10}{\bf x}\sqrt{-g_{10}}\exp(-2\Phi){\cal L}_{_{\rm
matter}}(g_{10})$, the reduced 4-dimensional matter action is
\begin{equation}
S_{_{\rm matter}}=\int\dd^{4}{\bf
x}\sqrt{-g}\phi{\cal L}_{_{\rm matter}}(g).
\end{equation}
The cosmological evolution of $\phi$ and $R$ can then be computed
to deduce that $\dot\aem/\aem\simeq10^{10}$ $(\mu/1\,{\rm
eV})^{-2}\,{\rm yr}^{-1}$. Ref.~ considered the same model but assumed that
supersymmetry is broken by non-perturbative effects such as
gaugino condensation. In this model, and contrary to 
Ref.~\cite{maeda88}, $\phi$ is stabilized and the variation of the
constants arises mainly from the variation of $R$ in a runaway
potential.\\

To conclude, superstring theories offer a natural theoretical framework to
discuss the value of the fundamental constants since they become
expectation values of some fields. This is a first step towards
their understanding but yet, no complete and satisfactory
mechanism for the stabilization of the extra-dimensions and dilaton
is known. 

It has paved the way to various models that we 
detail in \S~\ref{subsec81}. 

\subsection{Relations between constants}

There are different possibilities to relate the variations of different
constants. First, in quantum field theory, we have to take into
account the running of coupling constants with energy and
the possibilities of grand unification to relate them. It will also give
a link between the QCD scale, the coupling constants and
the mass of the fundamental particles (i.e. the Yukawa couplings
and the Higgs vev). Second, one can
compute the binding energies and the masses of the proton,
neutron and different nuclei in terms of the gauge couplings
and the quark masses. This step involves QCD and nuclear
physics. Third, one can relate the gyromagnetic factor
in terms of the quark masses. This is particularly important
to interpret the constraints from the atomic clocks and the
QSO spectra. This allows to set stronger constraints on the
varying parameters at the expense of a model-dependence.

\subsubsection{Implication of gauge coupling unification}\label{subsecGUT}

The first theoretical implication of high-energy physics arises 
from the unification of the non-gravitational interactions. In these
unification schemes,  the three standard model coupling constants
derive from one unified coupling constant.

In quantum field, the calculation of scattering processes include higher order
corrections of the coupling constants related to loop corrections that
introduce some intergrals over internal 4-momenta. Depending on the theory, these
integrals may be either finite or diverging as the logarithm or power law
of a UV cut-off. In a class of theories, called renormalizable, among which the
standard model of particule physics, the physical quantities calculated at any order
do not depend on the choice of the cut-off scale. But the result may depend
on $\ln E/m$ where $E$ is the typical energy scale of the process.
It follows that the values of the coupling constants of the standard model depend on
the energy at which they are measured (or of the process in which
thay are involved). This running arises
from the screening due to the existence of virtual particles which are
polarized by the presence of a charge. The renomalization group
allows to compute the dependence of a coupling constants as
a function of the energy $E$ as
$$
 \frac{\dd g_i(E)}{\dd\ln E}=\beta_i(E),
$$
where the beta functions, $\beta_i$, depend on the gauge group and
on the matter content of the theory and may be expended in powers
of $g_i$.
For the SU(2) and U(1) gauge couplings of the standard model, they are given by
$$
 \beta_2(g_2)=-\frac{g_2^3}{4\pi^2}\left(\frac{11}{6} - \frac{n_g}{3}\right),\qquad
  \beta_1(g_1)=+\frac{g_1^3}{4\pi^2} \frac{5n_g}{9} 
$$
where $n_g$ is the number of generations for the fermions. 
We remind that the fine-structure constant is defined in the
limit of zero momentum transfer so that cosmological variation of $\aem$ are
independent of the issue of the renormalisation group depence.
For the SU(3) sector,
with fundamental Dirac fermion representations, 
$$
 \beta_3(g_3)=-\frac{g_3^3}{4\pi^2}\left(\frac{11}{4} - \frac{n_f}{6}\right),
$$
$n_f$ being the number of quark flavours with mass smaller than $E$. The negative sign
implies that (1) at large momentum transfer the coupling decreases and loop corrections become
less and less significant: QCD is said to be asymptotically free; (2) integrating the renormalisation group
equation for $\alpha_3$ gives
$$
 \alpha_3(E)= \frac{6\pi}{(33-n_f)\ln(E/\Lambda_c)}
$$ 
so that it diverges as the energy scale approaches $\Lambda_c$ from above,
that we decided to call $\Lambda_{\rm QCD}$.
This scale characterises all QCD properties and in particular
the masses of the hadrons are expected to be proportional to
$\Lambda_{\rm QCD}$ up to corrections of order $m_{\rm q}/\Lambda_{\rm QCD}$.

It was noticed quite early that these relations imply that the weaker gauge coupling
becomes stronger at high energy, while the strong coupling becomes weaker so
that one can thought the three non-gravitational interactions may have
a single common coupling strength above a given energy. This is the driving
idea of Grand Unified Theories (GUT) in which one introduces a mechanism
of symmetry-breaking from a higher symmetry group, such e.g. as SO(10) or SU(5),  
at high energies. It has two important consequences for our present considerations.
First there may exist algebraic relations between the Yukawa couplings of the
standard model. Second, the  structure constants of the standard model
unify at an energy scale $M_U$
\begin{equation}
 \alpha_1(M_U)= \alpha_2(M_U)= \alpha_3(M_U)\equiv  \alpha_U(M_U).
\end{equation}
We note that the electroweak mixing angle, i.e. the can also be time dependent
parameter, but only for $E\not=M_U$ since at  $E=M_U$, it is fixed by the symmetry
to have the value $\sin^2\theta=3/8$, from which we deduce that
$$
\aem^{-1}(M_Z)=\frac{5}{3}\alpha_1^{-1}(M_Z) + \alpha_2^{-1}(M_Z).
$$
It follows from the renormalisation group relations that
\begin{equation}
 \alpha_i^{-1}(E)=\alpha_i^{-1}(M_U) - \frac{b_i}{2\pi}\ln\frac{E}{M_U},
\end{equation}
where the beta-function coefficients are given by $b_i=(41/10,-19/6,7)$ for
the standard model (or below the SUSY scale $\Lambda_{\rm SUSY}$) and
by $b_i=(33/5,1,-3)$ for $N=1$ supersymmetric theory.
Given a field decoupling at $m_{\rm th}$, one has 
$$
\alpha_i^{-1}(E_-)=\alpha_i^{-1}(E_+) - \frac{b^{(-)}_i}{2\pi}\ln\frac{E_-}{E_+}
 - \frac{b^{({\rm th})}_i}{2\pi}\ln\frac{m_{\rm th}}{E_+}
$$
where $b^{({\rm th})}_i=b^{(+)}-b^{(-)}$ with $b^{(+/-)}$ the beta-function coefficients
respectively above and below the mass threshold, with tree-level matching
at $m_{\rm th}$. In the case of multiple thresholds, one must sum the different
contributions. The existence of these thresholds implies that the running
of $\alpha_3$ is complicated since it depends on the masses of heavy quarks and
coloured superpartner in the case of supersymmetry. For non-supersymmetric
theories,  the low-energy expression
of the QCD scale is 
\begin{equation}\label{QCDscale}
 \Lambda_{\rm QCD} = E\left(\frac{m_{\rm c}m_{\rm b}m_{\rm t}}{E}\right)^{2/27}
 \exp\left(-\frac{2\pi}{9\alpha_3(E)} \right)
\end{equation}
for $E>m_{\rm t}$. This implies that the variation of Yukawa couplings, gauge couplings, Higgs
vev and $\Lambda_{\rm QCD}/ M_{\rm P}$ are correlated. A second set of relations arises in models 
in which the weak scale is determined by dimensional transmutation~\cite{transmut,transmut2}. 
In such cases, the Higss vev is related to the Yukawa constant of the top quark by
\begin{equation}
 v = M_p\exp\left(-\frac{8\pi^2 c}{h_{\rm t}^2} \right),
\end{equation}
where $c$ is a constant of order unity. This would imply that
$\delta\ln v=S\delta\ln h$ with $S\sim160$~\cite{cnouv}.\\

The first consequences of this unification were investigated in Refs.~\cite{cf1,cf2,langacker} where
the variation of the 3 coupling constants was reduced to the one of $\alpha_U$ and
$M_U/ M_{\rm P}$. It was concluded that, setting 
\begin{equation}
 R\equiv\delta\ln\Lambda_{\rm QCD}/\delta\ln\aem,
\end{equation}
$R\sim 34$ with a stated accuracy of about 20\%~\cite{langacker0,langacker} (assuming only $\alpha_U$ can vary), $R=38\pm6$~\cite{cf1} 
and then $R=46$~\cite{cf2,cf3}, the difference arising from the quark
masses and their associated thresholds. However, these results implicitely assume that the
electroweak symmetry breaking and supersymmetry breaking mechanisms, as well as the
fermion mass generation, are not affected by the variation of the unified coupling.
It was also mentioned in Ref.~\cite{cf2} that $R$ can reach $-235$
in unification based on SU(5) and SO(10). The large value of $R$ arises from the
exponential dependence of $\Lambda_{\rm QCD}$ on $\alpha_3$.
In the limit in which the quark masses are set to zero,
the proton mass, as well as all other hadronic
masses are proportional to $\Lambda_{\rm QCD}$, i.e.
$m_{\rm p}\propto\Lambda_{\rm QCD}(1+{\cal O}(m_{\rm q}/\Lambda_{\rm QCD}))$. Ref.~\cite{langacker}
further relates the Higgs vev to $\aem$ by $\dd\ln v/\dd\ln\aem\equiv\kappa$
and estimated that $\kappa\sim70$ so that, assuming that the variation of the Yukawa
couplings is negligible, it could be concluded that 
$$
\delta \ln \frac{m}{\Lambda_{\rm QCD}}\, \sim 35 \delta\ln\aem,
$$
for the quark and electron masses. This would also implies that the variation of $\mu$ and
$\aem$ are correlated, still in a very model-dependent way, typically one can conclude~\cite{cnouv}
that
$$
 \frac{\delta\mu}{\mu} =-0.8R \frac{\delta\aem}{\aem} +0.6(S+1) \frac{\delta h}{h},
$$
with $S\sim160$. The running of $\alpha_U$ can be extrapolated to the Planck mass, $ M_{\rm P}$. Assumiung $\alpha_U( M_{\rm P})$ fixed
and letting $M_U/ M_{\rm P}$ vary, it was concluded~\cite{dine} that $R=2\pi(b_U+3)/[9\aem(8b_U/3-12)]$ where $b_U$ is
the beta-function coefficient describing the running of $\alpha_U$. This shows that a variation of
$\aem$ and $\mu$ can open a windows on GUT theories. A similar analysis~\cite{dent03} assuming that
electroweak symmetry breaking was triggered by nonperturbative effects in such a way that $v$
and $\alpha_U$ are related, concludes that ${\delta\mu}/{\mu} =(13\pm7){\delta\aem}/{\aem}$
in a theory with soft SUSY breaking and ${\delta\mu}/{\mu} =(-4\pm5){\delta\aem}/{\aem}$ otherwise.\\

From a phenomenological point of view, Ref.~\cite{dent1} 
making an assumption of proportionality with fixed ``unification
coefficients" assumes that the variations of the constants
at a given redshift $z$ depend on a unique evolution factor $\ell(z)$ and that the variation of all
the constants can be derived from those of the unification mass scale (in Planck units), $M_U$, the
unified gauge coupling $\alpha_U$, the Higgs vev, $v$ and in the case of supersymmetric theories
the soft supersymmetry breaking mass, $\tilde m$. Introducing the coefficients $d_i$ by
$$
\Delta\ln\frac{M_U}{ M_{\rm P}} = d_M\ell,\quad
\Delta\ln\alpha_U = d_U\ell,\quad
\Delta\ln\frac{v}{M_U} = d_H\ell,\quad
\Delta\ln\frac{\tilde m}{ M_{\rm P}} = d_S\ell,
$$
($d_S=0$ for non-supersymmetric theories)
and assuming that the masses of the standard model fermions all vary with $v$ so
that the Yukawa couplings are assumed constant, it was shown that the variations
of all constants can be related to $(d_M,d_U,d_H,d_S)$ and $\ell(z)$, using the renormalisation
group equations (neglecting the effects induced by the variation of $\alpha_U$ on the
RG running of fermion masses).
This decomposition is a good approximation provided that the time variation is slow, which is actually backed up by
the existing constraints, and homogeneous in space (so that it may not be applied as such in the
case a chameleon mechanism is at work~\cite{braxm}).

This allowed to define 6 classes of scenarios: (1) varying gravitational constant ($d_H=d_S=d_X=0$)
in which only $M_U/ M_{\rm P}$ or equivalently $G\Lambda^2_{\rm QCD}$ is varying; (2) varying unified
coupling $(d_U=1,d_H=d_S=d_M=0)$; (3) varying Fermi scale defined by $(d_H=1,d_U=d_S=d_M=0)$
in which one has $\dd\ln\mu/\dd\ln\aem=-325$; (4) varying Fermi scale and SUSY-breaking
scale $(d_S=d_H=1,d_U=d_M=0)$ and for which $\dd\ln\mu/\dd\ln\aem=-21.5$; (5) varying
unified coupling and Fermi scale $(d_X=1, d_H=\tilde\gamma d_X, d_S=d_M=0)$
and for which $\dd\ln\mu/\dd\ln\aem=(23.2-0.65\tilde\gamma)/(0.865+0.02\tilde\gamma)$; 
(6) varying unified coupling and Fermi scale with SUSY $(d_X=1, d_S\simeq d_H=\tilde\gamma d_X, d_M=0)$
and for which $\dd\ln\mu/\dd\ln\aem=(14-0.28\tilde\gamma)/(0.52+0.013\tilde\gamma)$. 

Each scenario can be compared to the existing constraints to get sharper bounds on them~\cite{bbn-dent,dent1,dent2,bbn-muller}
and emphasize that the correlated variation between different constants (here $\mu$ and $\aem$)
depends strongly on the theoretical hypothesis that are made.

\subsubsection{Masses and binding energies}\label{subsubmass}

The previous section described the unification of the gauge couplings. When we consider
``composite" systems such as proton, neutron, nuclei or even planets and stars, we need to 
compute their mass, which requires to determine their binding energy.
As we have already seen, the electromagnetic binding energy induces a direct dependence on $\aem$
and can be evaluated using e.g. the Bethe-Weiz\"acker formula~(\ref{bethe}).
The dependence of the masses on the quark masses, via nuclear interactions,
and the determination of the nuclear binding energy are especially difficult to estimate.

In the chiral limit of QCD in which all quark masses are negligible compared to $\Lambda_{\rm QCD}$
all dimensionful quantities scale as some power of $\Lambda_{\rm QCD}$. For instance,
concerning the nucleon mass, $m_{\rm N}=c\Lambda_{\rm QCD}$ with $c\sim3.9$ being computed
from lattice QCD. This predicts a mass of order 860~MeV, smaller than the observed
value of 940~MeV. The nucleon mass can be computed in chiral perturbation theory and
expressed in terms of the pion mass as~\cite{leinweber}
$ m_{\rm N} = a_0 + a_2m_\pi^2 + a_4m_\pi^4 + a_6 m_\pi^6 +\sigma_{N\pi} + \sigma_{\Delta\pi} +\sigma_{\rm tad}$
(where all coefficients of this expansion are defined in Ref.~\cite{leinweber}),
which can be used to show~\cite{clock-muq} that the nucleon mass is scaling as
\begin{equation}
 m_{\rm N} \propto\Lambda_{\rm QCD}X_{\rm q}^{0.037}X_{\rm s}^{0.011}.
\end{equation}
It was further extanded~\cite{oklo-14} by using a sigma model to infer that
$ m_{\rm N} \propto\Lambda_{\rm QCD}X_{\rm q}^{0.045}X_{\rm s}^{0.19}$.
This two examples explicitely show the strong dependence in the nuclear modelling.\\

To go further and determine the sensitivity of the mass of a nucleus to the
various constant,
$$
m(A,Z)=Zm_{\rm p}+(A-Z)m_{\rm n}+Zm_{\rm e}+E_{_{\rm S}}+E_{_{\rm EM}}
$$
one should determine the strong binding energy [see related discussion below
Eq.~(\ref{mass})] in function of the atomic number $Z$ and the mass number $A$.

The case of the deuterium binding energy$B_D$  has been discussed in different
ways (see \S~\ref{bbn2cste}). Many modelisations have been performed. 
A first route relies on the use of the dependence of $B_D$ on
the pion mass~\cite{bbnpi1,bbnpi2,bbn-pudliner,nnbyoo}, which can then be related to
$m_{\rm u}$, $m_{\rm d}$ and $\Lambda_{\rm QCD}$. A second avenue is
to use a sigma model in the framework of the Walecka model~\cite{waleka} in which
the potential for the nuclear forces keeps only the  $\sigma$, $\rho$ and $\omega$ meson exchanges~\cite{oklo-14}. 
We also emphasize that the deuterium is only produced during BBN, as it is too weakly
bound to survive in the regions of stars where nuclear processes take place. The fact that
we do observe deuterium today sets a non-trivial constraint on the constants by imposing that
the deuterium remains stable from BBN time to today. Since it is weakly bound, it
is also more sensitive to a variation of the nuclear force compared to the electromagnetic force.
This was used in Ref.~\cite{dentfair} to constrain the variation of the nuclear strength
in a sigma-model.

For larger nuclei, the situation is more complicated since there is no
simple modelling. For large mass number $A$, the strong binding energy
can be approximated by the liquid drop model
\begin{equation}\label{eq.li-drop}
 \frac{E_{_{\rm S}}}{A} = a_V -\frac{a_S}{A^{1/3}} - a_A\frac{(A-2Z)^2}{A^2} + a_P
 \frac{(-1)^A+(-1)^Z}{A^{3/2}}
\end{equation}
with$(a_V,a_S,a_A,a_P)=(15.7,17.8,23.7,11.2)$~MeV~\cite{drop}. It has also 
been suggested~\cite{lilley} that the nuclear binding energy can be
expressed as
\begin{equation}
 E_{_{\rm S}} \simeq A a_3 + A^{2/3} b_3\qquad \hbox{with}\qquad
 a_3 = a_3^{\rm chiral\, limit} + m^2_\pi\frac{\partial a_3}{\partial m_\pi^2}.
\end{equation}
In the chiral limit, $a_3$ has a non-vanishing
limit to which we need to add a contribution scaling like $m^2_\pi\propto\Lambda_{\rm QCD}m_{\rm q}$. 
Ref.~\cite{lilley} also pointed out that the delicate balance between attractive and repulsive nuclear
interactions~\cite{waleka} implies that the binding energy of nuclei is expected
to depend strongly on the quark masses~\cite{donomass}. Recently, a fitting
formula derived from effective field theory and based of the semi-empirical formula
derived in Ref.~\cite{furnstahl} was proposed~\cite{damourdono} as
\begin{equation}
 \frac{E_{_{\rm S}}}{A} = -\left(120-\frac{97}{A^{1/3}}\right)\eta_S +\left(67-\frac{57}{A^{1/3}}\right)\eta_V + \ldots
\end{equation}
where $\eta_S$ and $\eta_V$ are the strength of respectively the scalar (attractive) and vector (repulsive)
nuclear contact interactions normalized to their actual value. These two parameters need to be
related to the QCD parameters~\cite{donomass}. We also refer to Ref.~\cite{fw2} for
the study of the dependence of the binding of light ($A\leq 8$) nuclei on possible variations of hadronic
masses, including meson, nucleon, and nucleon-resonance masses.

These expressions allow to compute the sensitivity coefficients that enter in
the decomposition of the mass [see Eq.~(\ref{mdephi})]. They also emphasize
one of the most difficult issue concerning the investigation about constant
related to the intricate structure of QCD and its role in low energy nuclear
physics, which is central to determine the masses of nuclei and the binding
energies, quantities that are particularly important for BBN, the universality of free fall
and stellar physics.

\subsubsection{Gyromagnetic factors}

The constraints arising from the comparison of atomic clocks (see \S~\ref{subsec31})
involve the fine-structure constant $\aem$, the proton-to-electron mass ratio
$\mu$ and various gyromagnetic factors. It is important to relate these
factors to fundamental constants.

The proton and neutron gyromagnetic factors are respectively given by
$g_{\rm p}=5.586$ and $g_{\rm n}=-3.826$ and are expected
to depend on $X_{\rm q}=m_{\rm q}/\Lambda_{\rm QCD}$~\cite{flambq}.
In the chiral limit in which $m_{\rm u}=m_{\rm d}=0$, the nucleon
magnetic moments remain finite so that one could have thought that
the finite quark mass effects should be small. However, it is
enhanced by $\pi$-meson loop corrections which are
proportional to $m_\pi\propto\sqrt{m_{\rm q}\Lambda_{\rm QCD}}$.
Following Ref.~\cite{leinweber}, this dependence can be described by the
approximate formula
$$
 g(m_\pi) = \frac{g(0)}{1+ a m_\pi + b m_\pi^2}.
$$
The coefficients are given by $a=(1.37,1.85)/$GeV and $b=(0.452,0.271)/$GeV$^2$
respectively for the proton an neutron. This lead~\cite{flambq} to
$g_{\rm p}\propto m_\pi^{-0.174}\propto X_{\rm q}^{-0.087}$ and
$g_{\rm n}\propto m_\pi^{-0.213}\propto X_{\rm q}^{-0.107}$.
This was further extended in Ref.~\cite{clock-muq} to take into the depence
with the strange quark mass $m_{\rm s}$ to obtain
\begin{equation}
 g_{\rm p}\propto X_{\rm q}^{-0.087}X_{\rm s}^{-0.013},\qquad
 g_{\rm n}\propto X_{\rm q}^{-0.118}X_{\rm s}^{0.0013}.
\end{equation}

This allows to express the results of atomic clocks (see \S~\ref{sec-clock-phy}) in terms of $\aem$,
$X_{\rm q}$, $X_{\rm s}$ and $X_{\rm e}$. Similarly, for the constants constrained
by QSO observation, we have (see Table~\ref{tab-quasar})
\begin{eqnarray}
 x &\propto&\aem^2X_{\rm q}^{-0.087}X_{\rm s}^{-0.013},\nonumber\\
 y&\propto&\aem^2X_{\rm q}^{-0.124}X_{\rm s}^{-0.024}X_{\rm e},\nonumber\\
 \bar\mu&\propto& X_{\rm q}^{-0.037}X_{\rm s}^{-0.011}X_{\rm e},\nonumber\\
 F&\propto&\aem^{3.14}X_{\rm q}^{-0.0289}X_{\rm s}^{0.0043}X_{\rm e}^{-1.57},\nonumber\\
 F'&\propto&\aem^2X_{\rm q}^{0.037}X_{\rm s}^{0.011}X_{\rm e}^{-1},\nonumber\\
 G&\propto&\aem^{1.85}X_{\rm q}^{-0.0186}X_{\rm s}^{0.0073}X_{\rm e}^{-1.85},
\end{eqnarray}
once the scaling of the nucleon mass as $m_{\rm N}\propto\Lambda_{\rm QCD}
X_{\rm q}^{0.037}X_{\rm s}^{0.011}$ (see \S~\ref{subsubmass}). This shows that the 7 observable
quantities that are constrained by current QSO observations can be reduced
to only 4 parameters.

\subsection{Models with varying constants}\label{subsec81}

The models that can be constructed are numerous and cannot be all
reviewed here. We thus focus in the string dilaton model in \S~\ref{subsub1}
and then discuss the chameleon mechanism in  \S~\ref{subsub2} and
the Bekenstein framework in \S~\ref{subsub3}.

\subsubsection{String dilaton and Runaway dilaton models}\label{subsub1}

Damour and Polyakov~\cite{damour94a,damour94b} argued that the effective action
for the massless modes taking into account the full string loop
expansion should be of the form
\begin{eqnarray}
S&=&\int\dd^4{\bf x}\sqrt{-\hat g}\left[M_s^2
\left\lbrace B_g(\Phi)\hat R+4B_\Phi(\Phi)\left[\hat {\Box}\Phi
-(\hat\nabla\Phi)^2\right]
\right\rbrace-B_F(\Phi)\frac{k}{4}\hat F^2\right.\nonumber\\
&&\left.-B_\psi(\Phi)\bar{\hat
\psi}\hat\dslash\hat\psi+\ldots\right]
\end{eqnarray}
in the string frame, $M_s$ being the string mass scale. The functions
$B_i$ are not known but can be expanded 
(from the genus expansion of string theory) in the limit $\Phi\rightarrow-\infty$ as
\begin{equation}\label{ans}
B_i(\Phi)=\hbox{e}^{-2\Phi}+c^{(i)}_0+c^{(i)}_1\hbox{e}^{2\Phi}+
c^{(i)}_2\hbox{e}^{4\Phi}
+\ldots
\end{equation}
where the first term is the tree level term.
It follows that these functions can
exhibit a local maximum. After a conformal transformation
($g_{\mu\nu}=CB_g\hat g_{\mu\nu},
\psi=(CB_g)^{-3/4}B_\psi^{1/2}\hat\psi$), the action in Einstein frame
takes the form
\begin{eqnarray}
S&=&\int\frac{\dd^4{\bf x}}{16\pi G}\sqrt{-g}\left[
R-2(\nabla\phi)^2-\frac{k}{4}B_F(\phi)F^2-\bar
\psi\dslash\psi+\ldots\right]
\end{eqnarray}
where the field $\phi$ is defined as
$$
 \phi\equiv\int \left[\frac{3}{4}\left(\frac{B_g'}{B_g}\right)^2+2\frac{B_\Phi'}{B_\Phi} +
 2\frac{B_\Phi'}{B_g}\right]\dd\Phi.
$$
It follows that the Yang-Mills coupling behaves as
$g^{-2}_{_{\rm YM}}=kB_F(\phi)$. This also implies that the QCD mass scale
is given by
\begin{equation}\label{qcd}
\Lambda_{_{\rm QCD}}\sim M_s(CB_g)^{-1/2}\hbox{e}^{-8\pi^2kB_F/b}
\end{equation}
where $b$ depends on the matter content. It follows that the mass of
any hadron, proportional to $\Lambda_{_{\rm QCD}}$ in first
approximation, depends on the dilaton, $m_A(B_g, B_F,\ldots)$.

If, as allowed by
the anstaz (\ref{ans}), $m_A(\phi)$ has a minimum $\phi_m$
then the scalar field will be driven toward this minimum during
the cosmological evolution. However if the various coupling
functions have different minima then the minima of $m_A(\phi)$
will depend on the particle $A$.  To avoid
violation of the equivalence principle at an unacceptable level,
it is thus necessary to assume that
all the minima coincide in  $\phi=\phi_m$, which can be implemented by setting
$B_i=B$. This can be realized by assuming that $\phi_m$ is a special
point in field space, for instance it could be associated
to the fixed point of a $Z_2$ symmetry of the $T$- or $S$-duality~\cite{lilley}.

Expanding $\ln B$ around its maximum $\phi_m$ as $\ln
B\propto-\kappa(\phi-\phi_m)^2/2$, Damour and Polyakov~\cite{damour94a,damour94b} 
constrained the set of parameters $(\kappa,\phi_0-\phi_m)$ using the
different observational bounds. This toy model allows one to address
the unsolved problem of the dilaton stabilization, to study all the
experimental bounds together and to relate them in a quantitative
manner (e.g. by deriving a link between equivalence-principle
violations and time-variation of $\aem$). This model was compared to 
astrophysical data in Ref.~\cite{landaudp} to conclude that 
$\vert\Delta\phi\vert<3.4\kappa 10^{-6}$.

An important feature of this model lies in the fact that at lowest order
the masses of all nuclei are proportional to $\Lambda_{\rm QCD}$ so that
at this level of approximation, the coupling is universal and the theory reduces
to a scalar-tensor theory and there will be no violation of the universality of free fall.
It follows that the deviation from general relativity 
are characterized by the PPN parameters
$$
 \gamma^\ppn-1\simeq -2f^2_A=-2\beta_s^2\kappa^2\Delta\phi_0^2,\qquad
 \beta^\ppn-1\simeq\frac{1}{2}f^2_A\frac{\dd f_A}{\dd\phi}=\frac{1}{2}\beta_s^3\kappa^3\Delta\phi_0^2
$$
with 
\begin{equation}
 f_A=\frac{\partial\ln \Lambda_{\rm QCD}(\phi)}{\partial\phi}
 =-\left[\ln\frac{M_s}{m_A}+\frac{1}{2} \right]\frac{\dd\ln B}{\dd\phi}
 \equiv -\beta_s\frac{\dd\ln B}{\dd\phi}=\beta_s\kappa\Delta\phi_0
\end{equation} 
with $\Delta\phi_0=\phi_0-\phi_m$ and $\beta_s\sim40$~\cite{damour94a}. 
The variation of the gravitational constant is, from Eq.~(\ref{Gdotst}), simply
$$
\frac{\dot G}{G} =2f_A\dot\phi_0 = -2\left[\ln\frac{M_s}{m_A}
        +\frac{1}{2} \right]\frac{\dd\ln B}{\dd\phi}\dot\phi_0.
$$
The value of $\dot\phi_0=H_0\phi_0'$ is obtained from the Klein-Gordon 
equation~(\ref{kgqq}) and is typically given by $\phi_0'=-Z\beta_s\kappa H_0\Delta\phi_0$
were $Z$ is a number that depends on the equation of state of the fluid
dominating the matter content of the universe in the last $e$-fold  and the
cosmological parameters so that
\begin{equation}
 \left.\frac{\dot G}{G}\right\vert_0 =2f_A\dot\phi_0 = -2ZH_0\beta^2_s\kappa^2\Delta\phi_0^2.
\end{equation}
The factor $Z$ is model-dependent and another way to estimate $\dot\phi_0$ is
to use the Friedmann equations which imply that $\dot\phi_0=H_0
\sqrt{1+q_0-\frac{3}{3}\Omega_{\rm m0}}$ where $q$ is the deceleration parameter.

When one considers the effect of the quark masses and binding energies,
various composition-dependent effects appear. First, the fine-structure
constant scales as $\aem\simeq B^{-1}$ so that
\begin{equation}
 \left.\frac{\dot \alpha}{\alpha}\right\vert_0 =\kappa\Delta\phi_0\dot\phi_0
 = -ZH_0\beta_s\kappa^2\Delta\phi^2_0.
\end{equation}
The second effect is, as pointed out earlier, a violation of the universality of
free fall. In full generality, we expect that
\begin{equation}\label{mdephi}
  m_A(\phi)= N\Lambda_{\rm QCD}(\phi)\left[
  1 + \sum_{\rm q}\epsilon^q_A\frac{m_{\rm q}}{\Lambda_{\rm QCD}} + \epsilon^{\rm EM}_A\aem
  \right].
\end{equation}
Using an expansion of the form~(\ref{mAZ}),
it was concluded that 
\begin{equation}
 \eta_{AB}=\kappa^2(\phi_0-\phi_m)^2\left[
  C_B\Delta\left(\frac{B}{M}\right)+ C_D\Delta\left(\frac{D}{M}\right)+ C_E\Delta\left(\frac{E}{M}\right)
 \right]
\end{equation}
with $B=N+Z$, $D=N-Z$ and $E=Z(Z-1)/(N+Z)^{1/3}$ and where the value 
of the parameters $C_i$ are model-dependent.

It follows from this model that:
\begin{itemize}
 \item The PPN parameters, the time variation of $\alpha$ and $G$ today and the
 violation of the university of free-fall all scale as $\Delta\phi_0^2$.
 \item The field is driven toward $\phi_m$ during the cosmological evolution,
 a point at which the scalar field decouples from the matter field. The
 mechanism is usually called {\it the least coupling principle}.
 \item Once the dynamics for the scalar field is solved, $\Delta\phi_0$ can be
 related to $\Delta\phi_i$ at the end of inflation. Interestingly, this quantity can
 be expressed in terms of amplitude of the density contrast at the end of inflation,
 that is to the energy scale of inflation.
 \item The numerical estimations~\cite{damour94a} indicate that $\eta_{U,H}\sim
 -5.4\times10^{-5}(\gamma^\ppn-1)$ showing that in such a class
 of models, the constraint on $\eta\sim10^{-13}$ implies $1-\gamma^\ppn\sim
 2\times10^9$ which is a better constraint that the one obtained direcly.
\end{itemize}

This model was extended~\cite{damourrunaway} the a case
where the coupling functions have a smooth finite limit for infinite
value of the bare string coupling, so that $B_i=C_i+{\cal O}({\rm
e}^{-\phi}$), folling Ref.~\cite{gasperini02}. The dilaton runs away toward its attractor at infinity
during a stage of inflation. The late time dynamics of the scalar field
is similar as in quintessence models, so that the model can also explain
the late time acceleration of the cosmic expansion.
The amplitude of residual dilaton
interaction is related to the amplitude of the primordial density
fluctuations and it induces a variation of the fundamental
constants, provided it couples to dark matter or dark energy. It is
concluded that, in this framework, the largest allowed variation of
$\aem$ is of order $2\times10^{-6}$, which is reached for a violation
of the universality of free fall of order $10^{-12}$ and it was
established that
\begin{equation}\label{leastuff}
\left.\frac{\dot\aem}{\aem}\right\vert_0\sim\pm10^{-16}\sqrt{1+q_0-\frac{3}{2}\Omega_{\rm m0}}
\sqrt{10^{12}\eta}\,{\rm yr}^{-1},
\end{equation}
where the first square-root arises from the computation of $\dot\phi_0$.
The formalism was also used to discuss the time variation
of $\aem$ and $\mu$~\cite{chibarun}.\\

The coupling of the dilaton to the standard model fields was further
investigated in Refs.~\cite{dado1,dado2}. Assuming that
the heavy quarks and weak gauge bosons
have been integrated out and that
the dilaton theory has been matched to the light fields
below the scale of the heavy quarks, the coupling of the
dilaton has been parameterised by 5 parameters:
$d_e$ and $d_g$ for the couplings to the electromagnetic and gluonic
field-strength terms, and $d_{m_e}$, $d_{m_u}$ and $d_{m_d}$ for the couplings
to the fermionic mass terms so that the interaction Lagrangian
is reduces to a linear coupling (e.g. $\propto d_e\phi F^2$ for the coupling
to electromagnetism etc.) It follows that $\Delta\aem/\aem = d_e\kappa\phi$
for the fine structure constant, $\Delta\Lambda_{\rm QCD}/\Lambda_{\rm QCD} = d_d\kappa\phi$ 
for the strong sector and $\Delta m_i/m_i = d_{m_i}\kappa\phi$
for the masses of the fermions. These parameters can be constrained by the
test of the equivalence principle in the Solar system [see \S~\ref{subsec22}].\\

In these two string-inspired scenarios, the amplitude of the variation of the constants is
related to the one of the density fluctuations during inflation and the
cosmological evolution.

\subsubsection{The Chameleon mechanism}\label{subsub2}

A central property of the least coupling principle, that is at the heart of the
former models, is that all coupling functions have the same minimum so that
the effective potential entering the Klein-Gordon equation for the dilaton
has a well-defined minimum.

It was realized~\cite{cameleon} that if the dilaton has a coupling $A^2(\phi)$ to matter while
evolving in a potential $V(\phi)$ the source of the Klein-Gordon equation~(\ref{kgqq})
has a an effective potential
$$
 V_{\rm eff} = V(\phi) + A^2(\phi)\rho.
$$
In the case where $V$ is a decreasing function of $\phi$, e.g. a runaway potential,
and the coupling is an increasing function, e.g. $A^2 =\exp\beta\phi/M_{\rm P}$, the
effective potential has a minimum the value of which depends on the matter
local density $\rho$ (see also Ref.~\cite{ekow}). The field is thus expected to be massive on Earth where the
density is high and light in space in the Solar system. It follows that the experiment
on the universality of free fall in space may detect violations of the universality
of free fall larger than the bounds derived by laboratory experiments~\cite{cameleon2,cameleon3}.
It follows (1) that the constraints on the time variation of the constants today can be
relaxed if such a mechanism is at work and (2) that is the constants depend on the local
value of the chameleon field, their value will be environment dependent and 
will be different on Earth and in space.

The cosmological variation of $\aem$ in such model was investigated in Ref.~\cite{brax,brax2}.
Models based on the Lagrangian~(\ref{olive}) and exhibiting the chameleon mechanism
were investigated in Ref.~\cite{op2}. 

The possible shift in the value of $\mu$ in the Milky Way (see \S~\ref{subsec23mw}) was related~\cite{mu-lev,levmusp2,levmusp2b} to
the model of Ref.~\cite{op2} to conclude that such a shift was compatible with this model.

\subsubsection{Bekenstein and related models}\label{subsub3}

Bekenstein~\cite{beken1,beken2} introduced a theoretical framework in which only  the electromagnetic
sector was modified by the introduction of a dimensionless scalar field $\epsilon$ so that all electric charges
vary in unison $e_i=e_{0i}\epsilon(x^\alpha$) so that only $\aem$ is assumed to
possibly vary. 

To avoid the arbitrariness in the definition of $\epsilon$, which can be rescaled by 
a constant factor while $e_{0i}$ is inversely rescales, it was postulated that the dynamics
of $\epsilon$ be invariant under global rescaling so that its action should be of the form
\begin{equation}
 S_\epsilon = -\frac{\hbar c}{2l^2}\int \frac{g^{\mu\nu}\partial_\mu\epsilon\partial_\nu\epsilon}{\epsilon^2}
 \sqrt{-g}\dd^4x,
\end{equation}
$l$ being a constant length scale. Then, $\epsilon$ is assumed to enter all electromagnetic
interaction via $e_iA_\mu\rightarrow e_{0i}\epsilon A_\mu$ where $A_\mu$ is the usual
electromagnetic potential and the gauge invariance is then preserved only if 
$\epsilon A_\mu\rightarrow \epsilon A_\mu + \lambda_{,\mu}$ for any scalar
function $\lambda$. It follows that the the action for the electromagnetic sector
is the standard Maxwell action
\begin{equation}
 S_\epsilon = -\frac{1}{16\pi}\int F^{\mu\nu}F_{\mu\nu}
 \sqrt{-g}\dd^4x,
\end{equation}
for the generalized Faraday tensor
\begin{equation}
 F_{\mu\nu} =\frac{1}{\epsilon}\left[
 (\epsilon A_\nu)_{,\mu} -  (\epsilon A_\mu)_{,\nu}. 
 \right]
\end{equation}
To finish the gravitational sector is assumed to be described by the standard
Einstein-Hilbert action. Finally, the matter
action for point particles of mass $m$ takes the form
$S_m=\sum\int[-mc^2+(e/c)u^\mu
A_\mu]\gamma^{-1}\delta^3(x^i-x^i(\tau))\dd^4\bx$ where $\gamma$ is the
Lorentz factor and $\tau$ the proper time. Note that the Maxwell equation
becomes
\begin{equation}\label{beken4}
\nabla_\mu\left(\epsilon^{-1}F^{\mu\nu}\right)=4\pi j^\nu
\end{equation}
which is the same as for electromagnetism in
a material medium with dielectric constant $\epsilon^{-2}$ and
permeability $\epsilon^2$ (this was the original description proposed
by Fierz~\cite{fierz56} and Lichn\'erowicz~\cite{lichne}; see also Dicke~\cite{dicke64}).

It was proposed~\cite{sbm} to rewrite this theory by introducing the two fields
$$
 a_\mu \equiv \epsilon A_\mu, \qquad
 \psi\equiv \ln\epsilon
$$
so that the theory takes the form
\begin{equation}
  S= \frac{c^3}{16\pi g}\int R\sqrt{-g}\dd^4x 
        -\frac{1}{16\pi}\int\hbox{e}^{-2\psi} f^{\mu\nu}f_{\mu\nu}\sqrt{-g}\dd^4x 
       -\frac{1}{8\pi\kappa^2} \int(\partial_\mu\psi)^2\sqrt{-g}\dd^4x 
\end{equation}
with $\kappa=l/(4\pi\hbar c)$ and $f_{\mu\nu}$ the Faraday tensor
associated with $a_\mu$. The model was further extended to include a potential
for $\psi$~\cite{barrowli} and to include the electroweak theory~\cite{shaw}.

As discussed previously, this class of models predict a violation of the 
universality of free fall and, from Eq.~(\ref{eq.inertie2}), it is expected that the
anomalous acceleration is given by  $\delta{\bf a}=-M^{-1}(\partial E_{_{\rm
EM}}/\partial\epsilon)\nabla\epsilon$.

From the confrontation of the local and cosmological constraints
on the variation of $\epsilon$ Bekenstein~\cite{beken1} concluded, given his
assumptions on the couplings, that $\aem$ ``{\it is a parameter, not a
dynamical variable}'' (see however Ref.~\cite{beken2} and then
Ref.~\cite{kraise}). 
This problem was recently bypassed by Olive and
Pospelov~\cite{olive01} who generalized the model to allow additional coupling
of a scalar field $\epsilon^{-2}=B_F(\phi)$ to non-baryonic dark matter
(as first proposed in Ref.~\cite{damour90}) and cosmological
constant, arguing that in supersymmetric dark matter, it is natural to expect that
$\phi$ would couple more strongly to dark matter than to baryon. For
instance, supersymmetrizing Bekenstein model, $\phi$ will get a
coupling to the kinetic term of the gaugino of the form
$M_*^{-1}\phi\bar\chi\partial\chi$. Assuming that the gaugino
is a large fraction of the stable lightest supersymmetric particle,
the coupling to dark matter would then be of order $10^3-10^4$ times
larger. Such a factor could almost reconcile the constraint arising
from the test of the universality of free fall with the order of
magnitude of the cosmological variation. This generalization of
the Bekenstein model relies on an action of the form
\begin{eqnarray}\label{olive}
S&=&\frac{1}{2}M_4^2\int R\sqrt{-g}\dd^4\bx
-\int\left[\frac{1}{2}M_*^2\partial_\mu\phi\partial^\mu\phi+\frac{1}{4}
   B_F(\phi)
    F_{\mu\nu}F^{\mu\nu}\right]\sqrt{-g}\dd^4\bx
    \\
 &-&\int\left\lbrace\sum\bar
   N_i[i\dslash-m_iB_{N_i}(\phi)]N_i+\frac{1}{2}\bar\chi\partial\chi
   +M_4^2B_\Lambda(\phi)\Lambda+\frac{1}{2}M_\chi
    B_\chi(\phi)\chi{}^T\chi
    \right\rbrace\sqrt{-g}\dd^4\bx\nonumber
\end{eqnarray}
where the sum is over proton
[$\dslash=\gamma^\mu(\partial_\mu-ie_0A_\mu)$] and neutron
[$\dslash=\gamma^\mu\partial_\mu$]. The functions $B$ can be expanded
(since one focuses on small variations of the fine-structure constant
and thus of $\phi$) as $B_X=1+\zeta_X\phi+ \xi_X\phi^2/2$. It follows
that $\aem(\phi)={e_0^2}/{4\pi B_F(\phi)}$ so that
$\Delta\aem/\aem=\zeta_F\phi+(\xi_F-2\zeta_F^2)\phi^2/2$.  This
framework extends the analysis of Ref.~\cite{beken1} to a 4-dimensional
parameter space ($M_*,\zeta_F,\zeta_m,\zeta_\Lambda$). It contains the
Bekenstein model ($\zeta_F=-2$, $\zeta_\Lambda=0$,
$\zeta_m\sim10^{-4}\xi_F$), a Jordan-Brans-Dicke model ($\zeta_F=0$,
$\zeta_\Lambda=-2\sqrt{2/2\omega+3}$, $\xi_m=-1/\sqrt{4\omega+6}$), a
string-like model ($\zeta_F=-\sqrt{2}$, $\zeta_\Lambda=\sqrt{2}$,
$\zeta_m=\sqrt{2}/2$) so that $\Delta/\aem/\aem=3$) and
supersymmetrized the Bekenstein model ($\zeta_F=-2$, $\zeta_\chi=-2$,
$\zeta_m=\zeta_\chi$ so that $\Delta\aem/\aem\sim5/\omega$). In all
the models, the universality of free fall sets a strong constraint on
$\zeta_F/\sqrt{\omega}$ (with $\omega\equiv M_*/2M_4^2$) and the
authors showed that a small set of models may be compatible with
a vairation of $\aem$ from quasar data while being compatible the
equivalence principle tests. A similar analysis~\cite{marra} concluded
that such model can reproduce the variation of $\aem$ from quasar
while being compatible with Oklo and meteorite data. Note that
under this form, the effective theory is very similar to the one
detailed in \S~\ref{subsub2}.

This theory was also used~\cite{beken3} to study the spacetime structure around 
charged black-hole, which corresponds to an extension of dilatonic charged black hole.
It was concluded that a cosmological growth of $\aem$ would decrease the
black-hole entropy but with half the rate expected from the earlier analysis~\cite{bha1,bah2}.

\subsubsection{Other ideas}

Let us mention without details other theoretical models which can accomodate varying
constants:
\begin{itemize}
 \item Models involving a late time phase transition in the electromagnetic
sector~\cite{chako,anchot};
\item Braneworld models~\cite{aguilar,bw0,byrne,librane,palma} or extra-dimensions~\cite{steinh};
\item Model with pseudo-scalar couplings~\cite{flamst};
 \item Growing neutrino models~\cite{gnu1,gnu2} in which the neutrino masses
are a function of a scalar field, that is also responsible for the late time
acceleration of the universe. In these models the neutrinos freeze
the evolution of the scalar field when they become non-relativistic
while its evolution is similar as in quintessence when the neutrinos are ultra-relativistic;
 \item Models based on discrete quantum gravity~\cite{gambini} or on loop
 quantum gravity in which the Barbero-Immirzi parameter controls the minimum
eigenvalue of the area operator and could be promoted to a field, leading to a classical coupling of Einstein gravity
with a scalar-field stress-energy tensor~\cite{lqg1,lqg2}
 \item ``varying speed of light'' models for which we refer to the review~\cite{vsl} and our
previous analysis~\cite{ellisu} for a critical view;
\item Quintessence models with a non-minimal coupling of the quintessence field~\cite{avelino06,anchor03,chiba2001,copQ,doran,fujii0,lee,lee2,nunes09,marra,parkinson,xyz3}
[see discussion \S~\ref{subseccosmo}];
\item Holographic dark energy models with non-minimal couplings~\cite{granda}
\end{itemize}

\section{Spatial variations}\label{subsec23}

The constraints on the variation of the fundamental constant that we have
described so far are
mainly related to their cosmological evolution so that, given the
Copernican principle, they reduce to constrains on the time variation
of the fundamental constants. Indeed, spatial variations can also
occur. They may be used to set constraints in two regimes:
\begin{itemize}
 \item On cosmological scales, the fields dictating the variation of the constants
 have fluctuations that can let their imprint in some cosmological observables.
 \item On local scales (e.g. Solar system or Milky Way) the fields at
 the origin of the variation of the constants are sourced by the local matter
 distribution so that one expect that the constants are not homogeneous on these
 scales.
\end{itemize}

\subsection{Local scales}\label{subsuba}

In order to determine the profile of the constant in the Solar system, let us assume that
their value is dictated by the value of a scalar field. As in \S~\ref{subsub1}, we can
assume that at lowest order the profile of the scalar field will be obtained from
the scalar-tensor theory, taking into account that all masses scale as 
$\Lambda_{\rm QCD}(\phi_*)$ where $\phi_*$ is the value of the field in
the Einstein frame.

\subsubsection{Generalities}

We restrict to the weakly self-gravitating ($V_*/c^2\ll 1$) and slow moving ($T^{01}\ll T^{00}$)
localized material systems and follow Ref.~\cite{damour92}. Using 
harmonic coordinates, defined with respect to the metric $g_*$, the Einstein frame metric
can be expanded as
$$
 g_{00}^* = -\exp\left(-2\frac{V_*}{c^2}\right)+{\mathcal O}(c^{-6}),\qquad
 g_{0i}^* = -\frac{4}{c^3}V_i^*+{\mathcal O}(c^{-5}),\qquad
 g_{ij}^* = -\exp\left(2\frac{V_*}{c^2}\right)\delta_{ij}+{\mathcal O}(c^{-6}),
$$
so that Eqs.~(\ref{einstein*}-\ref{Boxvarphi}) take the form
\begin{equation}\label{daleq}
   \Box_{*}V_* = -4\pi G_*\sigma_*+{\mathcal O}(c^{-4}),\quad
    \Box_{*}V^i_* = -4\pi G_*\sigma^i_*+{\mathcal O}(c^{-2}),\quad
     \Box_{*}\phi_* = -4\pi G_*\frac{S}{c^2}+{\mathcal O}(c^{-6})
\end{equation}
where $\Box_*$ is the flat d'Alembertian and where the scalar
field has been assumed to be light so that one can neglect its
potential. The source terms can be expressed
in terms of the matter stress-energy tensor in the Einstein frame as
$$
 \sigma_*c^2=T^{00}_* + T^{ii}_*,\qquad
 \sigma^i_* = T^{0i}_*,\qquad
 Sc^2=-\alpha(\phi_*)(T^{00}_* - T^{ii}_*).
$$
Restricting to the static case with a single massive point source, the only
non-vanishing source terms are $\sigma_*(\br)=M_*\delta^{3}(\br_*)$ and
$S(\br)=-\alpha(\phi_*)M_*\delta^{3}(\br_*)$ so that the set of equations reduces
to two Poisson equations
\begin{equation}\label{daleq}
   \Delta_{*}V_* = -4\pi G_*M_*\delta^{3}(\br_*)+{\mathcal O}(c^{-4}),\qquad
      \Delta_{*}\phi_* = 4\pi\frac{G_*M_*}{c^2}\delta^{3}(\br_*)+{\mathcal O}(c^{-6}).
\end{equation}
This set of equations  can be solved in the of the retarded Green function.
It follows that the Einstein frame gravitational potential is
$ V_*(r_*) = {G_*M_*}/{r_*}$.
The equation for $\phi_*$ can be solved iteratively, since at lowest order in $G_*/c^2$
it has solution
$$
 \phi_* = \phi_1(r_*) \equiv \phi_0 - \frac{\alpha_0}{c^2}V_*(r_*).
$$
This can be used to determine the Jordan frame metric and the variation
of the scalar field in function of the Jordan frame coordinates. It follows that
at lowest order  the Newton potential and the scalar field are given by
\begin{equation}
 \Phi_N =  \frac{GM}{r},\qquad
  \phi_* = \phi_1(r) \equiv \phi_0 - \alpha_0\frac{\Phi_N(r)}{c^2},
\end{equation}
where we have neglected the corrections $-\alpha(\phi)(\phi-\phi_0)$ for the
gravitational potential which, given the Solar system
constraints on $\alpha_0$, is a good approximation.

Now let us consider any constant $\alpha_i$ function of $\phi$. Its profile is
thus given by
$\alpha_i(r) = \alpha_i(\phi_0) - \alpha_0\alpha'_i(\phi_0){\Phi_N(r)}/{c^2}$
so that
\begin{equation}
 \frac{\Delta\alpha_i}{\alpha_i}(r)=  - s_i(\phi_0)\alpha_0\frac{\Phi_N(r)}{c^2}
\end{equation}
where $s_i(\phi_0)$ is the sensitivity of the constant $\alpha_i$ to a variation
of the scalar field, $s_i\equiv\dd\ln\alpha_i/\dd\phi$. For laboratory in orbit
on an elliptic trajectory,
$$
 r = \frac{a(1-e^2)}{1+e\cos\psi}, \qquad
 \cos\psi = \frac{\cos E - e}{1- e\cos E},\qquad
 t = \sqrt{\frac{a^3}{GM}}(E - e\sin E)
$$
where $a$ is the semi-major axis, $e$ the excentricity and $\psi$ the true anomaly.
It follows that
$$
  \frac{\Delta\alpha_i}{\alpha_i}(a,\psi) = -s_0\alpha_0\frac{GM}{ac^2}
  -s_0\alpha_0\frac{GM}{ac^2}e\cos\psi +{\mathcal O}(e^2).
$$
The first term represents the variation of the mean value of the constant
on the orbit compared with its cosmological value. This shows that 
local terrestrial and Solar system experiments do measure the
effects of the cosmological variation of the constants~\cite{damour92,shaw,shaw2}.
The second term is a seasonal modulation and it is usually 
parameterized~\cite{local1} as
\begin{equation}\label{defki}
  \left.\frac{\Delta\alpha_i}{\alpha_i}\right\vert_{\rm seasonal} = k_i\frac{\Delta\Phi_N}{c^2},
\end{equation}
defining the parameters $k_i$.

\subsubsection{Solar system scales}

The parameters $k_i$ can be constrained from laboratory measurements on Earth.
Since $e\simeq0.0167$ for the Earth orbit, the signal should have a peak-to-peak
amplitude of $2{GMe}/{ac^2}\sim3.3\times10^{-10}$ on a period of 1~year. This shows that
the order of magnitude of the constraints will be roughly of $10^{-16}/10^{-10}\sim10^{-6}$
since atomic clocks reach an accuracy of the order of $10^{-16}$.
The data of Refs.~\cite{clock-fortier07} and~\cite{bauch} lead respectively
to the two constraints~\cite{local1}
\begin{equation}
    k_{\aem}+0.17k_{e}=(-3.5\pm6)\times10^{-7},\qquad
   \vert k_{\aem}+0.13k_{e}\vert<2.5\times10^{-5},
\end{equation}
for $\aem$ and $m_{\rm e}/\Lambda_{\rm QCD}$ respectively.
The atomic dyprosium experiment~\cite{clock-cingoz} allowed to set the constraint~\cite{ferrel}
\begin{equation}
 k_{\aem}=(-8.7\pm6.6)\times10{-6},
\end{equation}
which, combined with the previous constraints, allows to conclude that
\begin{equation}
   k_{e}=(4.9\pm3.9)\times10^{-5}, \qquad
   k_{q}=(6.6\pm5.2)\times10^{-5},
\end{equation}
for $m_{\rm e}/\Lambda_{\rm QCD}$ and $m_{\rm q}/\Lambda_{\rm QCD}$ respectively.
Ref.~\cite{clock-blatt}, using the comparison of caesium and a strontium clocks
derived that
\begin{equation}\label{eq:ka}
   k_{\aem}+0.36k_{e}=(1.8\pm3.2)\times10^{-5},
\end{equation}
which, combined with
measurement of H-maser~\cite{ashby}, allow to set the 3 constraints
\begin{equation}\label{eq:blatt}
 k_{\aem}=(2.5\pm3.1)\times10^{-6}, \qquad
  k_{\mu}=(-1.3\pm1.7)\times10^{-5}, \qquad
   k_{q}=(-1.9\pm2.7)\times10^{-5}.
\end{equation}
Ref.~\cite{barrowkk,sbkk} reanalyzed the data by Ref.~\cite{clock-peik04}
to conclude that $ k_{\aem}+0.51k_{\mu}=(7.1\pm3.4)\times10^{-6}$.
Combined with the constraint~(\ref{eq:ka}), it led to
\begin{equation}\label{eq:barrow}
   k_{\mu}=(3.9\pm3.1)\times10^{-6}, \qquad
   k_{q}=(0.1\pm1.4)\times10^{-5}.
\end{equation}
Ref.~\cite{barrowkk} also  used the data of Ref.~\cite{clock-rosen08} to conclude
\begin{equation}
 k_{\aem}=(-5.4\pm5.1)\times10^{-8}.
\end{equation}
All these constraints use the sensitivity coefficients computed in Refs.~\cite{q-calc2,kappa-tedesco}.
We refer to Ref.~\cite{jenkins} as an unexplained seasonal variation that demonstrated
the difficulty to interpret phenomena.

Such bounds can be improved by comparing clocks on Earth and onboard
of satellites~\cite{local1,clock-ACES,maleki} while the observation of atomic spectra near the Sun
can lead to an accuracy of order unity~\cite{local1}. A space mission with atomic
clocks onboard and sent to the Sun could reach an
accuracy of $10^{-8}$~\cite{maleki,wolf09}. 

\subsubsection{Milky Way}\label{subsec23mw}

An attempt~\cite{mu-lev,molaro3} to constrain $k_\mu$ from emission lines due to 
ammonia in interstellar clouds of the Milky Way led to the conclusion that $k_\mu\sim1$,
by considering different transitions in different environements.
This is in contradiction with the local constraint~(\ref{eq:blatt}).
This may result from rest frequency uncertainties or it would require
that a mechanism such as chameleon is at work
(see \S~\ref{subsub2}) in order to be compatible with local constraints. The analysis
was based on an ammonia spectra atlas of 193 dense protostellar and prestellar
cores of low masses in the Perseus molecular cloud, comparison of
N$_2$H$^+$ and N$_2$D$^+$ in the dark cloud L183.

A second analysis~\cite{levmusp2} using high resolution spectral observations
of molecular core in lines of NH$_3$, HC$_3$N and $N_2$H$^+$ with 3 radio-telescopes
showed that $|\Delta\mu/\mu|<3\times10^{-8}$ between the cloud environement and the local
laboratory environment. An offset was however measured that could be interpreted as a variation of $\mu$
of amplitude $\Delta\bar\mu/\bar\mu=(2.2\pm0.4_{\rm stat}\pm0.3_{\rm sys})\times10^{-8}$. A
second analysis~\cite{levmusp2b} map four molecular cores L1498, L1512, L1517, and L1400K selected from the previous sample in order to estimate systematic effects due to possible velocity gradients. The measured velocity offset, once 
expressed in terms of $\Delta\bar\mu$, gives $\Delta\bar\mu= (26 \pm 1_{\rm stat} \pm 3_{\rm sys})\times10^{-9}$. 

A similar analysis~\cite{levFsp} based on the fine-structure transitions in atomic carbon
C{\sc i} and low-laying rotational transitions in $^{13}$CO probed the spatial
variation of $F=\aem^2\mu$ over the Galaxy. It concluded that
\begin{equation}
 |\Delta F'/F'| < 3.7\times10^{-7}
\end{equation}
between high (terrestrial) and low (interstellar) densities of baryonic matter. Combined with
the previous constraint on $\mu$ it would imply that $ |\Delta\aem/\aem| < 2\times10^{-7}$.
This was updated~\cite{levrio} to $ |\Delta F'/F'| < 2.3\times10^{-7}$ so that $ |\Delta\aem/\aem| < 1.1\times10^{-7}$.

Since extragalactic gas clouds have densities similar to those in the interstellar medium, these
bounds give an upper bound on a hypothetic chameleon effect which are much below
the constraints obtained on time variations from QSO absorption spectra.

\subsection{Cosmological scales}\label{subsubb}

During inflation, any light scalar field develop super-Hubble fluctuations
of quantum origin, with an almost scale invaraint power spectrum (see
chapter~8 of Ref.~\cite{peteruzanbook}). It follows that if
the fundamental constants depend on such a field, their value must
fluctuate on cosmological scales and have a non-vanishing correlation function.
More important these fluctuations can be correlated with the metric perturbations.

In such a case, the fine-structure constant will behave as $\aem=\aem(t)
+\delta\aem(\bx,t)$, the fluctuations being a stochastic variable. As we have
seen earlier, $\aem$ enters the dynamics of recombination, which would
then become patchy. This has several consequences for the CMB anisotropies.
In particular, similarly to weak gravitational lensing, it will modify the mean power spectra
(this is a negligible effect)  and induce a curl component (B mode) to the polarization~\cite{sigur}.
Such spatial fluctuations also induce non-Gaussian temperature and polarization correlations 
in the CMB~\cite{sigur,pub}. Such correlations have not allowed to set observational
constraints yet but they need to be included foe consistency, see e.g. the
example of CMB computation in scalar-tensor theories~\cite{cmb-G1}.
The effect on large the scale structure was also studied in Refs.~\cite{balss,motablss}
and the Keck/HIRES QSO absorption spectra showed~\cite{q-murphyAD} that the correlation
function of the fine-structure constant is consistent on scales ranging between 0.2 and 
13~Gpc.

Recently, it has been claimed~\cite{berenspa,webspace} that the fine structure constant
may have a dipolar variation that would explain consistently the data from the Southern
and Northern hemispheres (see \S~\ref{MMQSO}). Let assume a constant, $X$ say, depend
on the local value of a dynamical scalar field $\phi$. The value of $X$ at the observation point
is compared to its value here and today,
$$
\Delta X/X_0 \equiv X(\phi)/X(\phi_0) -1.
$$
Decomposing the scalar field as $\phi=\phi_0+\Delta\phi$, one gets
that $\Delta X/X_0=s_X(\phi)\Delta\phi$, with $s_X$ defined in Eq.~(\ref{def_sphi}).
Now the scalar field can be decomposed into a background and perturbations
as $\phi = \bar\phi(t)+\delta\phi(\bx,t)$ where the background value depends
only on $t$ because of the Copernican hypothesis. It follows that
\begin{equation}
 \frac{\Delta X(\bx,t)}{X_0}= s_X(\bar\phi)[\bar\phi(t)-\phi_0] +
 \lbrace s_X(\bar\phi)+s'_X(\bar\phi)[\bar\phi(t)-\phi_0] \rbrace\delta\phi(\bx,t)
 \equiv s_X(\bar\phi)\Delta\bar\phi + {\cal{S}}_X(\bar\phi)\delta\phi(\bx,t).
\end{equation}
The first term of the r.h.s. depends only on time while the second is space-time
dependent. It is also expected that the second term in the curly brackets is
negligible with respect to the first, i.e. ${\cal{S}}_X(\bar\phi)\sim s_X(\bar\phi)$.
It follows that one needs $\delta\phi(\bx,t)$ not to be small compared
to the background evolution term $\Delta\bar\phi$ for the spatial
dependence to dominate over the large scale time dependence. This can be
achieved for instance if $\phi$ is a seed field whose mean value is frozen.
Because of statistical isotropy, and in the same way as for CMB anisotropies
(see e.g. Ref.~\cite{peteruzanbook}), one can express the equal-time angular 
power spectrum of $\Delta X/X_0$ for two events on our past lightcone as
\begin{equation}
 \left\langle \frac{\Delta X(\bn_1,r,t)}{X_0}\frac{\Delta X(\bn_2,r,t)}{X_0}\right\rangle =
 \sum_\ell\frac{2\ell+1}{4\pi}C_\ell^{(XX)}(z) P_\ell(\bn_1\cdot \bn_2).
\end{equation}
If $\delta\phi$ is a stochastic field characterized by its power spectrum,
$\langle \delta\phi(\bk_1,t)\delta\phi(\bk_2,t)\rangle=P_\phi(k,t)\delta(\bk_1+\bk_2)$ in Fourier space,
then
\begin{equation}
 C_\ell^{(XX)}(z) = \frac{2}{\pi}{\cal{S}}^2_X[\bar\phi(z)] \int P_\phi(k,z)j_\ell[k(\eta_0-\eta)] k^2\dd k,
\end{equation}
$j_\ell$ being a spherical Bessel function. For instance, if $P_\phi\propto k^{n_s-1}$
where $n_s$ is a spectral index, $n_s=1$ corresponding to scale invariance, one gets
that $\ell(\ell+1)C_\ell^{(XX)}\propto\ell^{n_s-1}$ on large angular scales. The comparison of the
amplitude of the angular correlation and the isotropic (time) variation is
model-dependent and has not yet been investigated.

Another possibility would be that the Copernican principle is not fully statisfied,
such as in various void models. Then the background value of $\phi$ would depend
e.g. on $r$ and $t$ for a spherically symmetric spacetime (such as a Lema\^{\i}tre-Tolman-Bondi
spacetime). This could give rise to a dipolar modulation of the constant if the observer (us)
is not located at the center of the universe. Note however that such a cosmological
dipole would also reflect itself e.g. on CMB anisotropies. Similar possibilities are also
offered within the chameleon mechanism where the value of the scalar field depends
on the local matter density (see \S~\ref{subsub2}).\\

More speculative, is the effect that such fluctuations can have during preheating after
inflation since the decay rate of the inflaton in particles may fluctuate on large
scales~\cite{modfluc1,modfluc2}.

\subsection{Implication for the universality of free fall}\label{subsec22}

As we have seen in the previous sections, the tests of the universality of free fall
is central in contraining the model involving variations of the fundamental constants.

From Eqs.~(\ref{eq.inertie2}), the amplitude of the violation of the universality of free fall is 
given by $\eta_{AB}$ which takes the form
$$
 \eta_{AB}=\frac{1}{g_N}\sum_i\left\vert f_{Ai} - f_{Bi} \right\vert \vert\nabla\alpha_i\vert.
$$
In the case in which the variation of the constants arises from the same scalar
field, the analysis of \S~\ref{subsuba} implies that $\nabla\alpha_i$ can be
related to the gravitational potential by $\vert\nabla\alpha_i\vert
=\alpha_i s_i(\phi)\alpha_{\rm ext} g_N$ so that
\begin{equation}
 \eta_{AB}=\sum_i\left\vert f_{Ai} - f_{Bi} \right\vert s_i(\phi)\alpha_i\alpha_{\rm ext}
  =\sum_i\left\vert \lambda_{Ai} - \lambda_{Bi} \right\vert s_i(\phi)\alpha_{\rm ext}.
\end{equation}
This can be expressed in  terms of the sensitivity coefficient $k_i$ defined in Eq.~(\ref{defki})
as
\begin{equation}
 \eta_{AB}  = \sum_i\left\vert \lambda_{Ai} - \lambda_{Bi} 
                      \right\vert k_i,
\end{equation}
since $\vert\nabla\alpha_i\vert
=\alpha_i k_i g_N$. This shows that each experiment will yield a constraint
on a linear combination of the coefficients $k_i$ so that one requires at
least as many independent pairs of test bodies as the number of constants
to be constrained.

While the couplings to mass number, lepton number and the electroamgnetic
binding energy have been considered~\cite{damourmat} [see the example of \S~\ref{subsub1}]
the coupling to quark masses remains a difficult issue. In particular,
the whole difficulty lies in the determination of the coefficients $\lambda_{ai}$ [see
\S~\ref{subsubmass}]. In the formalism developed in Refs.~\cite{dado1,dado2},
see \S~\ref{subsub1}, one can relate the expected deviation
from the universality of free fall to the 5 parameters $d$ and get constraints
on $D_{\hat m}\equiv d_g^*(d_{\hat m}-d_g)$ and $D_e\equiv d_g^*d_e$ where
$d_g^*\equiv d_g+0.093(d_{\hat m}-d_g)+0.00027d_e$. For instance,
Be-Ti E\"otWash experiment and LRR experiment respectively imply
$$
 \left\vert  D_{\hat m}+ 0.22D_e\right\vert < 5.1\times 10^{-11},\qquad
  \left\vert  D_{\hat m}+ 0.28D_e\right\vert < 24.6\times 10^{-11}.
$$
This shows that while the Lunar experiment has a
slightly better differential-acceleration sensitivity, the
laboratory-based test is more sensitive to the dilaton coefficients
because of a greater difference in the dilaton
charges of the materials used, and of the fact that only
one-third of the Earth mass is made of a different material.\\ 

The link between the time variation of fundamental constants
and the violation of the universality of free fall have been 
discussed by Bekenstein~\cite{beken1} in the framework described
in \S~\ref{subsub2} and by Damour-Polyakov~\cite{damour94a,damour94b}
in the general framework described in \S~\ref{subsub1}. In all
these models, the two effects are triggered by a scalar field. It
evolves according to a Klein-Gordon equation
($\ddot\phi+3H\dot\phi+m^2\phi+\ldots=0$), which implies that $\phi$ is
damped as $\dot\phi\propto a^{-3}$ if its mass is much smaller than
the Hubble scale. Thus, in order to be varying during the last Hubble
time, $\phi$ has to be very light with typical mass $m\sim
H_0\sim10^{-33}$~eV. As a consequence,
$\phi$ has to be very weakly coupled to the standard model fields to avoid
a violation of the universality of free fall. 

This link was revisited in Ref.~\cite{chiba2001,dvaliZ,wetterich02}
in which the dependence of $\aem$ on the scalar field responsible
for its variation is expanded as
\begin{equation}
 \aem=\aem(0)+\lambda\frac{\phi}{M_4}+{\cal O}\left(\frac{\phi^2}{M_4^2} \right).
\end{equation}
The cosmological observation from QSO spectra implies that
$\lambda\Delta\phi/M_4\sim 10^{-7}$ at best during the last Hubble time.  
Concentrating only on the electromagnetic binding energy contribution to
the proton and of the neutron masses, it was concluded that
a test body composed of $n_{\rm n}$ neutrons and $n_{\rm p}$ protons will
be characterized by a sensitivity
${\lambda}(\nu_{\rm p}B_{\rm p}+\nu_{\rm n}B_{\rm n})/{m_{\rm N}}$
where $\nu_{\rm n}$ (resp. $\nu_{\rm p}$) is the ratio of neutrons
(resp. protons) and where it has been assumed that $m_{\rm n}\sim
m_{\rm p}\sim m_{\rm N}$. Assuming\footnote{For copper $\nu_{\rm
p}=0.456$, for uranium $\nu_{\rm p}=0.385$ and for lead $\nu_{\rm
p}=0.397$.}  that $\nu_{{\rm n,p}}^{\rm Earth}\sim1/2$ and using that
the compactness of the Moon-Earth system $\partial\ln(m_{\rm
Earth}/m_{\rm Moon})/\partial\ln\aem\sim10^{-3}$, one gets
$\eta_{12}\sim10^{-3}\lambda^2$.  Dvali and Zaldarriaga~\cite{dvaliZ}
obtained the same result by considering that $\Delta\nu_{{\rm
n,p}}\sim6\times10^{-2}-10^{-1}$. This implies that $\lambda<10^{-5}$
which is compatible with the variation of $\aem$ if
$\Delta\phi/M_4>10^{-2}$ during the last Hubble period.
From the cosmology one can deduce that
$(\Delta\phi/M_4)^2\sim (\rho_\phi+P_\phi)/\rho_{\rm total}$. If
$\phi$ dominates the matter content of the universe, $\rho_{\rm
total}$, then $\Delta\phi\sim M_4$ so that $\lambda\sim 10^{-7}$
whereas if it is sub-dominant $\Delta\phi\ll M_4$ and $\lambda\gg
10^{-7}$. In conclusion $10^{-7}<\lambda<10^{-5}$.
This explicits the tuning on the parameter $\lambda$. Indeed, 
an important underlying approximation is that the $\phi$-dependence arises only
from the electromagnetic self-energy. This analysis was extended in
Ref.~\cite{dentuff} who included explicitely the electron and related
the violation of the universality of free fall to the
variation of $\mu$.

In a similar analysis~\cite{wetterich02}, the scalar field is responsible for both
a variation of $\aem$ and for the acceleration of the universe. Assuming its
equation of state is $w_h\not=1$, one can express its time variation (as
long as it has a standard kinetic term) as
$$
 \dot\phi= H\sqrt{3\Omega_\phi(1+w_h)}.
$$
It follows that the expected violation of the universality of
free fall is related to the time variation of $\aem$ today by
$$
 \eta=-1.75\times10^{-2}\left(\frac{\partial\ln\aem}{\partial z}\right)^2_{z=0}
 \frac{(1+\tilde Q)\Delta\frac{Z}{Z+N}}{\Omega_\phi^{(0)}(1+w_h^{(0)})},
$$
where $\tilde Q$ is a parameter taking into account the influence of the mass
ratios. Again, this shows that in the worse case in which the
Oklo bound is saturated (so that $\partial\ln\aem/\partial z\sim10^{-6}$), one
requires $1+w_h^{(0)}\gtrsim10^{-2}$ for $\eta\lesssim10^{-13}$, hence
providing a string bond between the dark energy equation of state and
the violation of the universality of free fall. This was extended in Ref.~\cite{dent2}
in terms of the phenomenological model of unification presented
in \S~\ref{subsecGUT}. In the
case of the string dilaton and runaway dilaton models, one reaches a similar
conclusion [see Eq.~(\ref{leastuff}) in \S~\ref{subsub1}].
A similar result~\cite{msu}
was obtained in the case of pure scalar-tensor theory, relating
the equation of state to the post-Newtonian parameters. In all these
models, the link between the local constraints and the cosmological
constraints arise from the fact that local experiments constrain the upper value of
$\dot\phi_0$, which quantify both the deviation of its equation of state from $-1$
and the variation of the constants. It was conjectured that most realistic quintessence
models suffer from such a problem~\cite{braxm2}.\\

One question concerns the most sensitive probes of the equivalence principle.
This was investigated in Ref.~\cite{dentuff} in which the coefficients
$\lambda_{Ai}$ are estimated using the model~(\ref{eq.li-drop}). It was concluded
that they are 2-3 order of magnitude over cosmic clock bounds. However,
Ref.~\cite{dentuff2} concluded that the most sensitive probe depends on the unification relation that exist between
the different couplings of the standard model. Ref.~\cite{sbkk} concluded similarly that
the univerality of free fall is more constraining that the seasonal variations.
The comparison with QSO spectra is more difficult since it involves the dynamics of the field between
$z\sim1$ and today. To finish, let us stress that these results may be changed significantly
if a chameleon mechanism is at work.

\section{Why are the constants just so?}\label{section5}

The numerical values of the fundamental constants are not determined by the laws
of nature in which they appear. One can wonder why they have the values
we observe. In particular, as pointed by many authors (see below), the constants
of nature seem to be fine tuned~\cite{leslie}. Many physicists take this fine-tuning to be
an explanandum that cries for an explanans, hence following Hoyle~\cite{hoyle1}
who wrote that ``one must at least have a modicum of curiosity about the strange dimensionless 
numbers that appears in physics.''

\subsection{Universe and multiverse approaches}

Two possible lines of explanation are usually envisioned: a {\it design or consistency hypothesis} and 
an {\it ensemble hypothesis},
that are indeed not incompatible together. The first hypothesis includes the possibility that
all the dimensionless parameters in the ``final'' physical
theory will be fixed by a condition of consistency or an external cause. In the ensemble hypothesis,
the universe we observe is only a small part of the totality of physical existence, usually called
the multiverse. This structure needs not be fine-tuned and shall be sufficiently large and
variegated so that it can contain as a proper part a universe like the one we observe the fine-tuning
of which is then explained by an {\it observation selection effect}~\cite{borstrom}.

These two possibilities send us back to the 
large number hypothesis by Dirac~\cite{dirac37} that has been used as an early motivation to investigate
theories with varying constants. The main concern was the existence of some large
ratios between some combinations of constants. As we have seen in \S~\ref{subsecGUT},
the running of coupling constants with energy, dimensional transmutation or
relations such as Eq.~(\ref{QCDscale}) have opened a way to a rational explanation of
very small (or very large) dimensional numbers. 
This follows the ideas developped by Eddington~\cite{eddi1,eddi2} aiming at deriving the
values of the constants from consistency relations, e.g. he proposed to link the fine-structure constant
to some algebraic structure of spacetime.
Dicke~\cite{dicke61} pointed out another possible explanation to the
origin of Dirac  large numbers: the density of the universe is
determined by its age, this age being related to the time needed to
form galaxies, stars, heavy nuclei... This led Carter~\cite{carter74}
to argue that these numerical coincidence should not be a surprise and that
conventional physics and cosmology could have been used to predict them, 
at the expense of using the anthropic principle.

The idea of such a structure called the {\it multiverse} has attracted a lot of
attention in the past years and we refer to Ref.~\cite{carrbook} for
a more exhaustive account of this debate. While many versions
of what such a multiverse could be, one of them finds its route
in string theory. In 2000, it was realized~\cite{boussoP} that vast numbers of discrete
choices, called flux vacua, can be obtained in compactifying superstring
theory. The number of possibilities is estimated to range between
$10^{100}$ and $10^{500}$, or maybe more. No principle is
yet known to fix which of these vaua is chosen.
Eternal inflation offers a possibility to populate these vacua and to generate
an infinite number of regions in which the parameters, initial conditions but
also the laws of nature or the number of spacetime dimensions can vary from
one universe to another, hence being completely contingent.
It was later suggested by Susskind~\cite{sus03} that the anthropic principle may
actually constrain our possible locations in this vast string landscape.
This is a shift from the early hopes~\cite{kane} that M-theory may conceivably predict all the
fundamental constants uniquely.

Indeed such a possibility radically changes the way we approach the question of the
relation of these parameters to the underlying fundamental theory since we now expect
them to be distributed randomly in some range. Among this range of parameters
lies a subset, that we shall call the {\it anthropic range}, which allow for universe
to support the existence of observers. This range can be determined by asking
ourselves how the world would change if the values of the constants were changed,
hence doing {\it counterfactual cosmology}. This is however very restrictive since
the mathematical form of the law of physics may be changed as well and we are restricting
to a local analysis in the neighborhood of our observed universe. The determination
of the anthropic region is not a prediction but just a characterisation of the
sensitivity of ``our'' universe to a change of the fundamental constants {\em ceteris paribus}.
Once this range is determined, one can ask the general question of quantifying the probability
that we observe a universe as ours, hence providing a probabilistic prediction. This involves the use of the
anthropic principle, which expresses the fact what we observe are not just
observations but observations made by us, and requires to state what an
observer actually is~\cite{neal}.

\subsection{Fine-tunings and determination of the anthropic range}

As we have discussed in the previous sections, the outcome of many physical
processes are strongly dependent  on the value of the fundamental constants. 
One can always ask the scientific question of what would change in the
world around us if the values of some constants were changed, hence doing some
counterfactual cosmology in order to determine the range within which the universe
would have developed complex physics and chemistry, what
is usually thought to be a prerequisit for the emergence of complexity and life (we emphasize
the difficulty of this exercice when it goes beyond small and local deviations from our
observed universe and physics, see e.g. Ref.~\cite{harnik} for a possibly life supporting
universe without weak interaction).
In doing so, one should consider the fundamental parameters entering our physical
theory but also the cosmological parameters. 

First there are several constraints that the fundamental parameters listed in Table~\ref{tab-list}
have to satisfy in order for the universe to allow for complex physics and chemistry.
Let us stress, in a non-limitative way, some examples.
\begin{itemize}
\item It has been noted that the stability of the proton requires $m_{\rm d}-m_{\rm u}\gtrsim\aem^{3/2}m_{\rm p}$. 
The anthropic bounds on $m_{\rm d}$, $m_{\rm u}$ and $m_{\rm e}$ (or on the
Higgs vev) arising from the
existence of nuclei, the di-neutron and the di-proton cannot form a bound state, the deuterium is
stable have been investigated in many works~\cite{ag1,ag2,damourdono,dentfair,dono2,dono3,hogan00,hogan1},
even allowing for nuclei made of more than 2 baryon species~\cite{jaffeq}.
Typically, the existence of nuclei imposes that $m_{\rm d}+m_{\rm u}$ and $v$ cannot
vary by more that 60\% from their observed value in our universe.
\item If the difference of the neutron and proton masses where less that about 1~MeV, the neutron 
would become stable and hydrogen would be unstable~\cite{rozental88,hogan2} so that helium would have
been the most abundant at the end of BBN so that the whole history of the formation and
burning of stars would have been different. It can be deduced that~\cite{hogan00} one needs
$m_{\rm d}-m_{\rm u}-m_{\rm e}\gtrsim1.2$~MeV so that the universe does not become all neutrons;
$m_{\rm d}-m_{\rm u}+m_{\rm e}\lesssim3.4$~MeV for the $pp$ reaction to be exothermic and
$m_{\rm e}>0$ leading to a finite domain.
\item A coincidence emerges from the existence of stars with convective and radiative envelopes, since
it requires~\cite{carr79} that $\ag\sim\aem^{20}$.
It arises from the fact that the typical mass that seprates these two behaviour is roughly $\ag^{-2}\aem^{10}m_{\rm p}$
while the masses of star span a few decades aroung $\ag^{-3}m_{\rm p}$. Both stars seem to be needed
since only radiative stars can lead to supernovae, required to disseminate heavy elements, while only
convective stars may generate winds in their early phase, which may be associated with formation of rocky planets. This relation
while being satisfied numerically in our universe cannot be explained from fundamental principles.
\item Similarly, it seems that for neutrinos to eject the envelope of a star in a supernovae explosion, one requires~\cite{carr79}
$ \ag\sim\aw^{4}$.
\item As we discussed in \S~\ref{secstellar}, the production of carbon seems to imply that the relative
strength of the nuclear to electromagnetic interaction must be tuned typically at the 0.1\% level.
\end{itemize}

Other coincidences involve also the physical properties, not only of the physical theories, but also of
our universe, i.e. the cosmological parameters summarized in Table~\ref{tab-cosmo}. Let us
remind some examples
\begin{itemize}
 \item The total density parameter $\Omega$ must lie within an order of magnitude of unity. If it were
 much larger the universe will have recollapsed rapidly, on a time scale muxh shorter that the main-sequence star
 lifetime. If it were to small, density fluctuations would have frozen before galaxies could form. Typically
 one expects $0.1<\Omega_0<10$. Indeed, most inflationary scenarios lead to $\Omega_0\sim1$ so that
 this may not be anthropically determined but in that case inflation should last sufficiently long
 so that this could lead to a fine tuning on the parameters of the inflationary potential.
 \item The cosmological constant was probably the first one to be questioned in an anthropical way~\cite{weinberg87}.
 Weinberg noted that if $\Lambda$ is too large, the universe will start accelerating before structures had time
 to form. Assuming that it does not dominate the matter content of the universe before the redshift $z_*$
 at which earliest galaxy are formed, one concludes that $\rho_V=\Lambda/8\pi G<(+z_*)\rho_{\mat0}$.
 Weinberg~\cite{weinberg87} estimated $z_*\sim4.5$ and concluded that ``if it is the anthropic
 principle that accounts for the smallness of the cosmological constant, then we would expect the vacuum energy
 density $\rho_v\sim(10-100)\rho_{\mat0}$ because there is no anthropic reason for it
 to be smaller''. Indeed, the observations indicate  $\rho_v\sim2\rho_{\mat0}$
 \item Tegmark and Rees~\cite{tegrees} have pointed out that the amplitude of the initial density
 perturbation, $Q$ enters into the calculation and determined the anthropic region in the
 plane $(\Lambda,Q)$. This demonstrates the importance of determinating the parameters
 to include in the analysis.
 \item Different time scales of different origin seem to be comparable: the radiative cooling,
 galactic halo virialization, time of cosmological constant dominance, the age of the universe
 today. These coincidence were interpreted as an anthropic sign~\cite{bousso}.
\end{itemize}
These are just a series of examples. For a multi-parameter study of the anthropic bound, 
we refer e.g. to Ref.~\cite{tegmark} and to Ref.~\cite{hall} for a general anthropic
investigation of the standard model parameters.

\subsection{Anthropic predictions}

The determination of the anthropic region for a set of parameters is in no way a prediction
but simply a characterisation of our understanding of a physical phenomenon $P$ that
we think is important for the emergence of observers. It reflects that, the condition
$C$ stating that the constants are in some interval, $C\Longrightarrow P$, is equivalent
to $!P\Longrightarrow !C$.

The anthropic principle~\cite{carter74} states that ``what we can expect to observe must be restricted
by the conditions necessary for our presence as observers''. It has received many interpretations
among which the {\it weak anthropic principle} stating that ``we must be prepared to take account of the fact that our
location in the universe in necessarily priviledged to the extent of being compatible with our existence
as observers'', which is a restriction of the Copernican principle oftenly used in cosmology,
and the {\it strong anthropic principle} according to which ``the universe (and hence the fundamental parameters
on which it depends) must be such as to admit the creation of observers within it at some stage.''
(see Ref.~\cite{barrowtip} for further discussions and a large bibliography on the subject).

One can then try to determine the probability that an observer measure the value $x$ of the constant $X$
(that is a random variable fluctuating in the multiverse and the density of observers depend
on the local value of $X$). According to Bayes theorem, 
$$
P(X=x|{\rm obs})\propto P({\rm obs}|X=x) P(X=x).
$$
$P(X=x)$ is the prior distribution which is related to the volume of those parts of the universe in which
$X=x$ at $\dd x$. $P({\rm obs}|X=x)$ is proportional to the density of observers that are going
to evolve when $X=x$. $P(X=x|{\rm obs})$ then gives the probability that a randomly selected observer is located in a region
where $X=x\pm \dd x$. It is usually rewritten as~\cite{vilenkin}
$$
 P(x)\dd x = n_{\rm obs}(x)P_{\rm prior}\dd x.
$$
This higlights the difficulty to make a prediction. First, one has no idea of how
to compute $n_{\rm obs}(x)$. When restricting to the cosmological constant,
one can argue~\cite{vilenkin} that $\Lambda$ does not affect microphysics
and chemistry and then estimate $n_{\rm obs}(x)$ by the fraction of
matter clustered in giant galaxies and that can be computed from a model
of structure formation. This may not be a good approximation when other
constants are allowed to vary and it needs to be defined properly. Second, 
$P_{\rm prior}$ requires an explicit model of multiverse that would generate
sub-universes with different values $x_i$ (continuous or discrete) for $x$.
A general argument~\cite{weinberg89} states that if  the range over which $X$ 
varies in the multiverse is large compared to the anthropic region $X\in[X_{\rm min},X_{\rm max}]$ 
one can postulate  that $P_{\rm prior}$ is flat on $[X_{\rm min},X_{\rm max}]$. Indeed, such
a statement requires a measure in the space of the constants (or of the theories) that are allowed to vary.
This is a strong hypothesis which is difficult to control. In particular if  $P_{\rm prior}$
peaks outside of the anthropic domain, it would predict that the constants should lie on the
boundary of the antropic domain~\cite{ruba}. It also requires that there are sufficently enough
values of $x_i$ in the antrhopic domain, i.e. $\delta x_i\ll X_{\rm max}-X_{\rm min}$.
Garriga and Vilenkin~\cite{garvi} stressed that the hypothesis of a flat $P_{\rm prior}$ for the cosmological
constant may not hold in various Higgs models, and that the weight can lower the mean viable
value. To finish, one want to consider $P(x)$ as the probability that a random observer measures
the value $x$. This relies on the fact that we are a typical observer and we are implicitely making
a self sampling hypothesis. It requires to state in which class of observers we are
supposed to be typical (and the final result may depend on this choice~\cite{neal}) and
this hypothesis leads to conlusions such as the doomsday argument that have be
debated actively~\cite{borstrom,neal}.\\

This approach to the understanding of the observed values of the fundamental constants
(but also of the initial conditions of our universe)
by resorting to the actual existence of a multiverse populated by different ``low-energy''
theory of some ``mother'' microscopic theory allows to explain the observed fine-tuning
by an observational selection effect. It also sets a limit to the Copernican principle stating
that we do not leave in a particular position in space since we have to leave in a region
of the multiverse where the constants are inside the anthropic bound. Such an approach
is indeed not widely accepted and has been criticized in many 
ways~\cite{aguirre,ellismulti,ellismulti2,dpage,stoeger,vaas,stark}.

Among the issues to be answered before such an approach becomes more rigorous,
let us note: (1) what is the shape of the string landscape;
(2) what constants should we scan.
It is indeed important
to distinguish the parameters that are actually fine-tuned in order to determine those that we should
hope to explain in this way~\cite{wilc,wilczek07}. Here theoretical physics is indeed important since it should
determine which of the numerical coincidences are coincidences and which are
expected for some unification or symmetry reasons; (3) How is the landscape populated;
(4) what is the measure to be used in order and what is the correct way to compute anthropically
conditioned probabilities.

While considered as not following the standard scientific approach, this is the only existing
window on some understanding the value of the fundamental constants.

\section{Conclusions}\label{sectionconcl}

The study of fundamental constants has witnessed tremendous progresses in the past years.
In a decade, the constraints on their possible space and time varitions have flourished. They
have reached higher precision and new systems, involving different
combinations of constants and located at different redshifts, have been considered. This has improved
our knowledge on the equivalence principle and allowed to test it on astrophysical and
cosmological scales. We have reviewed them in \S~\ref{section3} and \S~\ref{section4}.
We have emphasized the experimental observational progresses expected in the coming years such
as the E-ELT, radio observations, atomic clocks in space, or the use of gravitational waves.

From a theoretical point of view, we have described in \S~\ref{section6} the high-energy models that predict
such variation, as well as the link with the origin of the acceleration of the universe. In all these
cases, a spacetime varying fundamental constant reflects the existence of an almost massless
field that couples to matter. This will be at the origin of a violation of the universality of free
fall and thus of utmost importance for our understanding of gravity and of the
domain of validity of general relativity. Huge progresses have been made in the understanding
of the coupled variation of different constants. While more model-dependent, this allows to
set stronger constraints and eventually to open a observational window on unification
mechanisms.

To finish, we have discussed in \S~\ref{section5} the ideas that try to understand the value of the fundamental constant.
While considered as borderline with respect to the standard physical approach, it reveals the necessity
of considering a universe larger than our own, and called the multiverse. It will also give us a hint
on our location in this structure in the sense that the anthropic principle limits the Copernican principle
at the basis of most cosmological models. We have stressed the limitations of this approach and
the ongoing debate on the possibility to make it predictive.

To conclude, the puzzle about the large numbers pointed ou by Dirac has led to a better
understanding of the fundamental constants and of their roles in the laws of physics.
They are now part of the general tests of general relativity, as well as 
a Breadcrumbs to understand the origin of the acceleration of
the universe and to more speculative structures, such as a multiverse structure, and 
possibly a window on string theory.

\section*{Acknowledgements}

I would like to thank all my collaborators on this topic, Alain
Coc, Pierre Descouvemont, Sylvia Ekstr\"om, George Ellis, Georges
Meynet, Nelson Nunes, Keith Olive and Elisabeth Vangioni as well as
B\'en\'edicte Leclercq and Roland Lehoucq.

I also thank many colleagues for sharing their thoughts on the
subject with me, first at the Institut d'Astrophysique de Paris,
Luc Blanchet, Michel Cass\'e, Gilles Esposito-Far\`ese, Bernard
Fort, Guillaume Faye, Jean-Pierre Lasota, Yannick Mellier, Patrick Petitjean; in
France, Francis Bernardeau, S\'ebastien Bize, Fran\c{c}oise
Combes, Thibault Damour, Nathalie Deruelle, Christophe Salomon,
Carlo Schimd, Peter Wolfe; and to finish worldwide, John Barrow, Thomas Dent, Victor Flambaum,
Bala Iyer, Lev Kofman, Paolo Molaro, David Mota, Michael Murphy, Jeff Murugan, Cyril Pitrou,
Anan Srianand, Gabriele Veneziano, John Webb, Amanda Weltman, Christof Wetterich.
To finish, I thank Clifford Will for motivating me to write this review.

This work was supported by a PEPS-PTI grant from CNRS (2009-2011)
and the PNCG (2010) but,
despite all our efforts, has not been supported by the French-ANR.

\appendix
\section{Notations}

\subsection{Constants}

The notations and numerical values of the constants used in this review are
summarized in Table~\ref{tab-list} and Table~\ref{tab-list2}.

\subsection{Sensibility coefficients}

The text introduces several sensitivity coefficients. We recall their definition here.
\begin{itemize}
 \item Given an observable $O$ the value of which depends on a set of
 primary parameters $G_k$, the sensitivity of the measured value of $O$
 to these parameters is
 \begin{equation}
  \frac{\dd \ln O}{\dd \ln G_k} = c_k.
 \end{equation}
The value of the quantitoes $c_k$ requires a physical description of the
system.
\item the parameters $G_k$ can be related to a set of fundamental constant
$\alpha_i$ and we define
  \begin{equation}
  \frac{\dd \ln G_k}{\dd \ln\alpha_i} = d_{ki}.
 \end{equation}
 The computation of the coefficients $d_{ki}$ requires to specify the
 theoretical framework and depends heavily on our knowledge of
 nuclear physics and the assumptiuons on unification.
\item A particular sets of parameters $d_{ki}$
 has been sibgled out for the sensitivity of the mass of a body $A$
 to a variation of the fundamental constants
  \begin{equation}
  \frac{\dd \ln m_A}{\dd \alpha_i} = f_{Ai}.
 \end{equation}
 One also introduces
 \begin{equation}
  \frac{\dd \ln m_A}{\dd \ln\alpha_i} = \lambda_{Ai}
 \end{equation}
 so that 
 $$
  \lambda_{Ai} = \alpha_if_{Ai}.
 $$
\item In models where the variation of the fundamental constants are
induced by the variation of a scalar field with define
  \begin{equation}\label{def_sphi}
  \frac{\dd \ln\alpha_i}{\dd \phi} = s_{i}(\phi).
 \end{equation} 
\item In this class of models the variation of the constants can be related
to the gravitational potential by
  \begin{equation}
  \frac{\dd \ln\alpha_i}{\dd \Phi_N} = k_{i}.
 \end{equation}  
 \end{itemize}

\subsection{Background cosmological spacetime}

We consider that the spacetime is describe by a manifold ${\mathcal M}$
with metric $g_{\mu\nu}$ with signature $(-,+,+,+)$. In the
case of a Minkowsky spacetime $g_{\mu\nu}= \eta_{\mu\nu}$.

In the cosmological context, we will describe the universe by a Friedmann-Lema\^{\i}tre
spacetime with metric
\begin{equation}
 \dd s^2 = - \dd t^2 + a^2(t)\gamma_{ij}\dd x^i\dd x^j
\end{equation}
where $t$ is the cosmique time, $a$ the scale factor and $\gamma_{ij}$
the metric on the constant time hypersurfaces. The Hubble function is defined
as $H\equiv \dot a/a$. We also define the redshift by the relation
$1+z=a_0/a$, with $a_0$ the scale factor evaluated today.

The evolution of the scale factor is dictated by the Friedmann equation
\begin{equation}
 H^2= \frac{8\pi G}{3}\rho - \frac{K}{a^2} + \frac{\Lambda}{3},
\end{equation}
where $\rho= \_i\rho_i$ is the total energy density of the matter components
in the universe. Assuming the species $i$ has a constant equation of
state $w_i=P_i/\rho_i$, each component evolves as $\rho_i= \rho_{i0}(1+z)^{2(1+w_i)}$.
The Friedmann equation can then be rewritten as
\begin{equation}
 \frac{H^2}{H_0^2}= \sum \Omega_i (1+z)^{3(1+w_i)} +  \Omega_K(1+z)^2 + \Omega_\Lambda,
\end{equation}
with the densitiy parameters defined by
\begin{equation}
 \Omega_i\equiv \frac{8\pi G\rho_{i0}}{3H_0^2},\qquad
  \Omega_i\equiv -\frac{K}{3H_0^2},\qquad
 \Omega_\Lambda\equiv \frac{\Lambda}{3H_0^2}. 
\end{equation}
They clearly satisfy $\sum \Omega_i +  \Omega_K+ \Omega_\Lambda=1$.

Concerning the properties of the cosmological spacetime, I follow the
notations and results of Ref.~\cite{peteruzanbook}.

\newpage
\bibliography{refs}

\end{document}